\documentclass{article}

\usepackage{PRIMEarxiv}

\usepackage[utf8]{inputenc} %
\usepackage[T1]{fontenc}    %
\usepackage{hyperref}       %
\usepackage{url}            %
\usepackage{booktabs}       %
\usepackage{amsfonts}       %
\usepackage{nicefrac}       %
\usepackage{microtype}      %
\usepackage{fancyhdr}       %
\usepackage{graphicx}       %

\usepackage{amsmath}
\usepackage{amssymb}
\usepackage{bm}
\usepackage{mathtools}
\usepackage{enumerate}
\usepackage[authoryear, round]{natbib}
\usepackage{multirow}
\usepackage{xcolor}

\newcommand{\e}{\mathbb{E}}

\title{Smoothing splines with some discontinuities}

\author{
  Matteo Pelagatti \\
  University of Milano-Bicocca, Italy \\
  \texttt{matteo.pelagatti@unimib.it} \\
  \And
  Paolo Maranzano \\
  University of Milano-Bicocca, Italy \\
  \texttt{paolo.maranzano@unimib.it} \\
  \& \\
  Fondazione Eni Enrico Mattei (FEEM) \\
  Milano, Italy \\
  \AND
  Giulio Cesare Gatto Ronchieri \\
  University of Milano-Bicocca, Italy \\
}

\begin{document}
\maketitle

\begin{abstract}
We generalize cubic smoothing splines to signals with a finite number of discontinuities at unknown positions. Rather than changing the roughness penalty, we deform the domain over which it is computed: every interval between consecutive observations can be stretched by a non-negative warp, subject to a total budget, and the optimizer spends the budget only where a jump makes stretching cheaper than bending. The estimator inherits the state-space form of smoothing splines, so it can be fitted by maximum likelihood with the Kalman filter, and it also admits a fast alternating algorithm that solves a banded linear system and takes a projected-gradient step on the warps. The budget is selected by an extended Bayesian information criterion whose constant is calibrated by simulation. In a Monte Carlo comparison the new estimators reconstruct a piecewise-smooth signal more accurately than the cubic smoothing spline with discontinuities of Storath and Weinmann, at the cost of more false alarms. Two energy applications illustrate the method: the temperature response of electricity demand in Northern Italy, and retail fuel prices, whose discontinuities, changes in excise duties, are dated by public record. The three supplements follow the references.
\end{abstract}


\section{Introduction}\label{sec:intro}
Data recorded over an ordered index often behave smoothly for long stretches punctuated by a handful of sharp, isolated changes: a policy shock in a price series, a phase transition in a physical process, a saturation threshold in a demand curve. Cubic smoothing splines \citep{Wah1978, WecAns1983}, and their equivalent state-space signal-extraction formulation \citep{HarKoo2000}, recover the smooth part of such signals well, but they are twice continuously differentiable by construction: the roughness penalty that makes them smooth also prevents them from reproducing a discontinuity, which is either washed out over several observations or, if the penalty is relaxed everywhere, accommodated at the price of overfitting the noise elsewhere. What is needed is an estimator that is smooth almost everywhere and sharp exactly where the data demand it, without the analyst having to specify in advance where.

This paper proposes such an estimator. The idea is to leave the classical roughness penalty unchanged and to deform the domain over which it is computed. Every pair of consecutive observations is allowed a non-negative \emph{domain warp} $\gamma_i\ge0$, an amount of fictitious domain length inserted between them, subject to a total budget $M$, and the roughness of the spline is penalized on the warped domain. Where the signal is smooth, warping buys nothing, the optimal $\gamma_i$ is zero and the fit coincides with an ordinary smoothing spline. Where the signal jumps, the same vertical change can be spread over a longer warped interval at a much lower curvature cost, so the optimizer concentrates the budget on the few intervals where a jump is present. Strictly speaking, the fitted function remains continuous: it approximates a discontinuity much as a steep sigmoid approximates a step function. Because the mechanism is a reparameterization of the domain, the estimator inherits the state-space representation of smoothing splines, which we use for maximum likelihood estimation via the Kalman filter, and it also admits a fast alternating algorithm that avoids numerical likelihood maximization altogether.

The idea of penalizing discontinuities directly, rather than smoothing through them, goes back to the weak rod and weak plate models of \citet{BlaZis1987} and to the Mumford--Shah functional \citep{MumSha1989}, which penalize the roughness of a piecewise-smooth function together with the number of its discontinuities and lead to a combinatorial search over candidate discontinuity sets. The estimator closest to ours, and our benchmark throughout, is the cubic smoothing spline with discontinuities (CSSD) of \citet{StoWei2024}, which specializes the weak-rod functional to cubic splines and non-equidistant data and solves it exactly by dynamic programming. Wavelet shrinkage \citep{DonJoh1994} denoises jumpy signals by thresholding multiresolution coefficients, but it targets equidistant designs and does not produce a piecewise-smooth curve. The fused lasso \citep{TibEtal2005} and trend filtering \citep{KimEtal2009, Tib2014} penalize the $\ell_1$ norm of successive differences of the fitted values or of their derivatives, a convex problem with automatic knot placement, but the knots are a by-product of a coefficient-level penalty rather than of an explicit, interpretable deformation. Free-knot splines \citep{Koo1997} and two-stage procedures, which first locate breaks with a changepoint detector such as PELT \citep{KilFeaEck2012} and then smooth between them, decouple detection from estimation. Our formulation keeps the roughness penalty unchanged and reparameterizes the metric in which it is computed, so that detection and estimation are solved jointly, through a single continuous, non-negative vector $\bm\gamma$.

Section~\ref{sec:methodology} introduces the domain-warping formulation, its state-space form, the alternating algorithm and the tuning of the method. Section~\ref{sec:simulations} compares the estimators with CSSD and with a classical smoothing spline by Monte Carlo simulation, and Section~\ref{sec:applications} applies them to two energy-related data sets, chosen to cover the two kinds of data the method is meant for. The first application investigates the temperature response of electricity demand, a cross-sectional relation estimated on hourly observations ordered by temperature rather than by time, irregularly spaced and with many ties, whose discontinuities are thresholds of the physical system and are not known in advance. The second uses weekly time series of retail fuel prices, serially dependent and close to a stochastic trend, in which changes in the excise duty shift the level of the prices at dates that are a matter of public record, so that the estimated discontinuities can be checked against a ground truth that real data rarely offer.
Section~\ref{sec:conclusion} concludes. Three supplements accompany the paper: the Computational Supplement collects derivations and implementation details, the Calibration Supplement the tuning of the method, and the Application Supplement additional results for the two applications.

\section{Methodology}\label{sec:methodology}

\subsection{The main idea}\label{sec:idea}
Let $(x_i,y_i)$, $i=1,\dots,n$, be the observations, with $x_1<x_2<\dots<x_n$, and let $\delta_i=x_{i+1}-x_i$ be the length of the $i$-th interval of the physical domain (i.e., the physical interval between consecutive knots). We introduce a vector of non-negative domain warps $\bm\gamma=(\gamma_1,\dots,\gamma_{n-1})^\top$ and define the length of the $i$-th interval in a latent domain as
\[
  \tau_i=\delta_i+\gamma_i .
\]
If $\gamma_i=0$ the physical and latent intervals coincide; if $\gamma_i>0$ the interval $[x_i,x_{i+1})$ is stretched. On each interval the function is a cubic polynomial $f_i(s)$ in the latent coordinate $s\in[0,\tau_i]$, which is linked to the physical coordinate by the linear dilation $s_i(x)=(x-x_i)(1+\gamma_i/\delta_i)$, so that the function observed in the physical domain is $f(x)=f_i(s_i(x))$ for $x_i\le x<x_{i+1}$, $i=1,\dots,n-1$. A function that is smooth in the latent domain can thus be steep in the physical one: the dilation compresses a smooth latent transition into a short physical interval, where it appears as a jump if $\gamma_i$ is large.

The estimator minimizes
\begin{equation}\label{eq:objective}
	J(f, \bm{\gamma}) = \underbrace{\sum_{i=1}^n (y_i - f(x_i))^2}_{\text{Distance from observations}} + \lambda \underbrace{\sum_{i=1}^{n-1} \int_{0}^{\tau_i} \left( \frac{d^2 f_i(s)}{ds^2} \right)^2 ds}_{\text{Total } \text{roughness } \text{energy } \\ \text{in the dilated domain}}\qquad\text{s.t.}\quad \gamma_i\ge0,\quad \sum_{i=1}^{n-1}\gamma_i\le M,
\end{equation}
which balances the distance from the observations against the roughness of the function measured in the latent domain; $\lambda$ is the smoothing constant and $M$ the warping budget.

The mechanism becomes transparent when one looks at the roughness penalty of a single interval, which we call its energy by analogy with the bending energy of an elastic beam. For a cubic segment covering a height difference $\Delta_i=f_{i+1}-f_i$, the energy is dominated, for short intervals, by a proportional term such that
\[
  \int_{0}^{\tau_i} \bigl(f_i''(s)\bigr)^2 ds \;\propto\; \frac{(\Delta y_i)^2}{\tau_i^3} \;=\; \frac{(\Delta y_i)^2}{(\delta_i + \gamma_i)^3}
\]
(see Section~\ref{sec:atw} and the exact statement is in Section~\ref{A-sec:energy} of the Computational Supplement).

\begin{figure}[!ht]
\centering
\includegraphics[width=\textwidth]{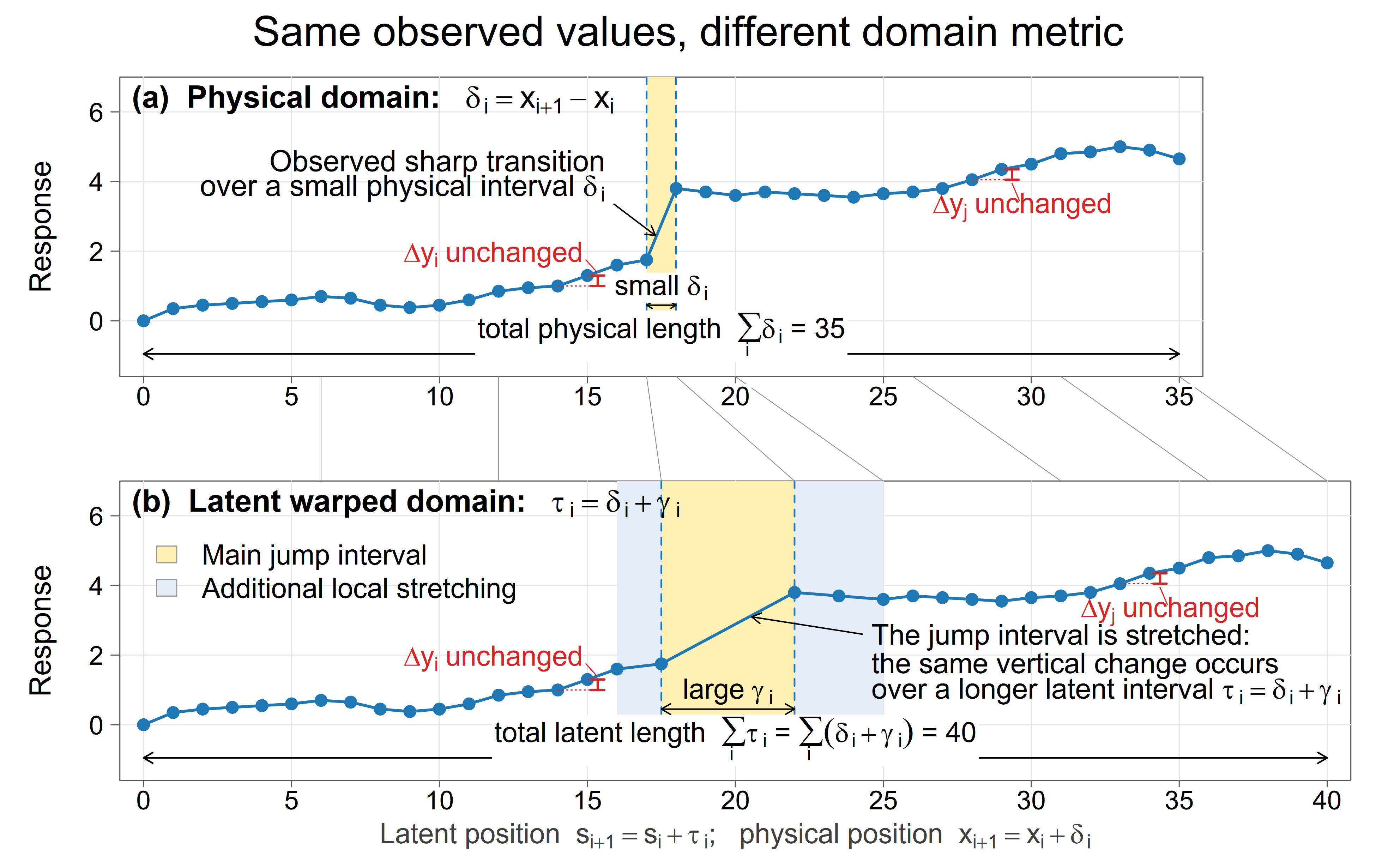}
\caption{Domain warping. The same observations on the physical domain (a) and on the latent warped domain (b), at the same horizontal scale. Away from the jump the intervals are unchanged and the grey connectors are vertical; at the jump, positive warps stretch the interval ($\tau_i=\delta_i+\gamma_i$, in yellow, with a lighter stretching of the neighboring intervals in blue), so that the same vertical change is completed over a longer latent interval at a lower curvature cost. The latent domain is longer than the physical one by the total warping $\sum_i\gamma_i$.}
\label{fig:example_timewarp}
\end{figure}

The key intuition behind the proposed approach is that a discontinuity is difficult for a classical cubic smoothing spline because the spline is forced to explain a large change in the response over a very short interval of the physical domain. On the one hand, in regions of discontinuity, if $|\Delta y_i|$ is large while $\delta_i$ is small, the standard energy term would explode ($\to \infty$); a large change over a short physical interval is therefore very expensive for an ordinary smoothing spline, which is forced to over-smooth it. Instead, our algorithm allocates a large positive $\gamma_i$ that dilates the denominator $(\delta_i + \gamma_i)^3$ and lets the curve complete the same transition as a smooth, low-energy curve in the latent domain (the penalty cost is then drastically reduced), so that a discontinuity in the physical domain is represented as a smooth transition in a locally stretched domain (Figure~\ref{fig:example_timewarp}). On the other hand, in regions of continuity where the signal evolves gradually, extra domain length brings little reduction of energy and the optimal $\gamma_i$ stays at or near zero, the latent interval equals the physical interval ($\tau_i \approx \delta_i$), so that the fit behaves as a standard cubic smoothing spline. The budget constraint $\sum_{i=1}^{n-1}\gamma_i\leq M$ is what makes the solution sparse: it prevents the whole domain from being stretched uniformly and forces the warping onto the few intervals where it is most valuable, so that large values of $\hat\gamma_i$ are data-driven indicators of jumps or sharp regime changes.

The design points are taken to be distinct. A tie would make $\delta_i$ vanish, and with it the latent length of an interval that carries no warp, so that neither the roughness penalty of that interval nor the transition of the state-space form of Section~\ref{sec:statespace} would be defined; more fundamentally, a warp on an interval of zero length would be a discontinuity between responses recorded at the same value of the covariate, which no data can identify. Tied observations, common when the covariate is recorded at a coarse resolution, are therefore collapsed into their mean at each distinct value, weighted by the number of replicates. The reduction leaves the fitted curve unchanged, but the within-group scatter that it discards still informs the observation variance, and the information criterion of Section~\ref{sec:tuning} counts the candidate locations on the distinct values and the observations on the raw data (Section~\ref{A-sec:ties} of the Computational Supplement).

\subsection{A signal extraction formulation}\label{sec:statespace}
Smoothing splines can also be obtained by signal extraction \citep{Wah1978, WecAns1983}: the cubic smoothing spline is $\hat f_t=\e[f(x_t)\mid Y_1=y_1,\dots,Y_n=y_n]$ when the data are generated by $Y_t=f(x_t)+\varepsilon_t$, $\varepsilon_t\sim\mathrm{NID}(0,\lambda\sigma^2)$, $d^2f/dx^2=\sigma w(x)$, with $w(x)$ a continuous-time standard white noise and diffuse initial conditions for $f(x_1)$ and its derivative. This model has the state-space form \citep[see, e.g.,][]{HarKoo2000}
\begin{equation}\label{eq:stateSpace}
    \begin{aligned}
	Y_t &= f(x_t) + \varepsilon_t, \quad  \varepsilon_t \sim \mathrm{NID}(0, \lambda\sigma^2) \\
	\begin{bmatrix}
		f(x_{t+1}) \\
		f'(x_{t+1})
	\end{bmatrix}
	&=
	\begin{bmatrix}
		1 & \delta_t \\
		0 & 1
	\end{bmatrix}
	\begin{bmatrix}
		f(x_{t}) \\
		f'(x_{t})
	\end{bmatrix}
	+
	\begin{bmatrix}
		\eta_t \\
		\zeta_t
	\end{bmatrix},
	\quad
		\begin{bmatrix}
		\eta_t \\
		\zeta_t
	\end{bmatrix}
	\sim
	\mathrm{NID}\left(
	\begin{bmatrix}
		0 \\
		0
	\end{bmatrix},
	\sigma^2
	\begin{bmatrix}
		\frac{\delta_t^3}{3} & \frac{\delta_t^2}{2} \\
		\frac{\delta_t^2}{2} & \delta_t
	\end{bmatrix}
	\right)
\end{aligned}
\end{equation}
with $\bigl(f(x_1),f'(x_1)\bigr)^\top\sim\mathrm N(\bm0,\kappa\bm I_2)$, $\kappa\to\infty$. The Kalman filter delivers the likelihood of $(\sigma,\lambda)$, and the state smoother the smoothing spline itself.

Adapting this representation to the latent domain only requires replacing the physical increments $\delta_t$ with their warped counterparts $\tau_t=\delta_t+\gamma_t$. The likelihood is then maximized with respect to $(\sigma,\lambda,\gamma_1,\dots,\gamma_{n-1})$ subject to $\gamma_t\ge0$ and $\sum_t\gamma_t\le M$, a problem for which the algorithm of \citet{Sva2002}, implemented in the NLopt library \citep{NLopt}, is effective. The analytical scores are available in closed form from one pass of the Kalman filter and smoother, so that a gradient of dimension $n+1$ never has to be approximated by finite differences; for $\gamma_t$ they contain an additional term, because $\gamma_t$ enters the transition matrix as well as the disturbance covariance (Section~\ref{A-sec:kalman_scores} of the Computational Supplement). Since the likelihood is not convex, good starting values for $\bm\gamma$ are important, and values close to zero are generally a good choice. We call this estimator SSJ--MLE (smoothing splines with jumps, maximum likelihood); when $\lambda$ is fixed at a supplied value and the likelihood is maximized with respect to the remaining parameters, we call it SSJ--KFS.

\subsection{An alternating domain-warping algorithm}\label{sec:atw}
As an alternative to the numerical maximization of a non-convex likelihood, we minimize~\eqref{eq:objective} for given $\lambda$ and $M$ by block coordinate descent, alternating a spline step, in which $\bm\gamma$ is fixed and $f$ solves a convex quadratic problem, with a warping step, in which $f$ is fixed and $\bm\gamma$ takes a projected-gradient step. We call the resulting estimator SSJ--ATW (alternating domain warping). Each iteration costs $O(n\log n)$ operations.

\paragraph{Spline step ($f$-step).}
In this step, $\bm{\gamma}$ is fixed. Let $\bm z=(f_1,f'_1,f_2,f'_2,\dots,f_n,f'_n)^\top$ collect the values and first derivatives of the function at the observations. By a well-known argument based on Hermite polynomials \citep[see Section~2 of][]{DouEtal1989}, the energy of interval $i$ (i.e., the contribution of the interval $i$ to the total roughness energy) is the quadratic form
\begin{equation}\label{eq:ki}
  k_i:=\int_0^{\tau_i}\biggl(\frac{d^2f_i(s)}{ds^2}\biggr)^2ds=\bm v_i^\top\bm K_i\bm v_i,\qquad \bm v_i=(f_i,f'_i,f_{i+1},f'_{i+1})^\top,
\end{equation}
with the local stiffness matrix
\[ \bm{K}_i = \frac{1}{\tau_i^3}
	\begin{bmatrix}
		12 & 6\tau_i & -12 & 6\tau_i \\
		6\tau_i & 4\tau_i^2 & -6\tau_i & 2\tau_i^2 \\
		-12 & -6\tau_i & 12 & -6\tau_i \\
		6\tau_i & 2\tau_i^2 & -6\tau_i & 4\tau_i^2
	\end{bmatrix}
\]
Adding each $\bm K_i$ to the $4\times4$ diagonal block of rows and columns $2i-1,\dots,2i+2$ of a $2n\times2n$ matrix of zeros gives the roughness penalty matrix $\bm R$, which is symmetric and block tridiagonal. Following the classical approach that goes back to \citet{StrFix1973}, the objective function becomes $(\bm y^*-\bm S\bm z)^\top(\bm y^*-\bm S\bm z)+\lambda\bm z^\top\bm R\bm z$, where $\bm y^*$ has $y_i$ in position $2i-1$ and zeros elsewhere and the selection matrix $\bm S=\mathrm{diag}(1,0,1,0,\dots,1,0)$ retains the function values, and it is minimized by the solution of the banded linear system
\[
  (\bm S+\lambda\bm R)\bm z=\bm y^* .
\]
Because $\bm S+\lambda\bm R$ is block tridiagonal, a block Cholesky decomposition solves the system in $O(n)$ operations (Section~\ref{A-sec:splinestep} of the Computational Supplement), an idea that goes back to \citet{deB1978} and that the recent work of \citet{StoWei2024} also exploits. The output of this step is the updated $f$ (odd elements of $\bm{z}$) and $f'$ (even elements of $\bm{z}$).

\noindent Recall that $\bm{A} = (\bm{S} + \lambda \bm{R})$. To evaluate the complexity of the model (useful for generalized cross-validation or information criteria of Section~\ref{sec:tuning}), calculate the trace of the smoothing matrix $\bm{A}^{-1}\bm{S}=(\bm S+\lambda\bm R)^{-1}\bm S$ (or equivalently the sum of the odd diagonal elements of $\bm{A}^{-1}$), which provides the effective degrees of freedom (EDF):
$$
\text{EDF} = \sum_{i = 1, 3, \ldots, 2n-1} \bar a_{ii},
$$
where $\bar a_{ij}$ denotes the generic element of the matrix $\bm{A}^{-1}$ \citep[see Section 5.4 of][]{ESL}.

\paragraph{Warping step ($\gamma$-step).}
In this step, $f$ and $f'$ are fixed, and we want to update the domain warp vector $\bm{\gamma}$ to reduce the current objective function value. First, we update $\bm{\gamma}$ to reduce the total roughness energy. With $\bm z$ fixed, the energy of interval $i$ has the reduced form
\[
  k_i=\frac{12\Delta_i^2}{\tau_i^3}-\frac{12\Delta_iS_i}{\tau_i^2}+\frac{4P_i}{\tau_i},
  \qquad
  \Delta_i=f_{i+1}-f_i,\quad S_i=f'_i+f'_{i+1},\quad P_i=(f'_i)^2+f'_if'_{i+1}+(f'_{i+1})^2,
\]
that is, it is a Laurent polynomial in $\tau_i$ whose coefficients do not depend on $\tau_i$ (see Section~\ref{A-sec:energy} of the Computational Supplement for derivation and analytical details). Its derivative with respect to the warp,
\begin{equation}\label{eq:grad}
  \nabla_i=\frac{\partial k_i}{\partial\gamma_i}=-36\frac{\Delta_i^2}{\tau_i^4}+24\frac{\Delta_iS_i}{\tau_i^3}-4\frac{P_i}{\tau_i^2},
\end{equation}
gives the gradient step $\bm\gamma-\eta\bm\nabla$, where the learning rate $\eta$ absorbs the factor $\lambda$. Second, to ensure that $\bm{\gamma}$ respects the constraints, the result of the $f$-step is projected onto the feasible set $\{\bm\gamma:\gamma_i\ge0,\ \sum_i\gamma_i\le M\}$, exactly and in $O(n\log n)$ operations, by the sort-based algorithm of \citet{DouEtal2008} (Section~\ref{A-sec:projection} of the Computational Supplement for details on the projection onto the Simplex). The projection has a statistical interpretation: once the gradient has indicated where additional domain length would reduce the energy, the projection redistributes the available budget and sets many components exactly to zero, so that only a few intervals are stretched, and these are expected to be those where a smoothing spline would otherwise pay an excessive curvature penalty.

\noindent The two steps are repeated until the change in the variables falls below a tolerance level.

\subsection{Tuning and implementation}\label{sec:tuning}
Three constants have to be chosen: the smoothing constant $\lambda$, the warping budget $M$ and the constant $\xi$ of the information criterion that selects $M$. We summarize the choices here; the Calibration Supplement gives the details.

\paragraph{Smoothing constant $\lambda$.} SSJ--MLE estimates $\lambda$ by maximum likelihood together with the other parameters. SSJ--KFS and SSJ--ATW take it as given, either from a preliminary SSJ--MLE fit or by $K$-fold cross-validation: the held-out squared error $\mathrm{CV}(\lambda,M)$ is computed on a grid of pairs, $\hat\lambda$ minimizes $\min_M\mathrm{CV}(\lambda,M)$, and the budget is then selected afresh, on the full sample, by the criterion below. When the observations are serially dependent the folds are contiguous blocks rather than random draws \citep{OttMarFas2024}: a held-out observation whose neighbors stay in the training sample is nearly predicted by them, and the criterion is then won by a fit that all but interpolates. The minimization over $M$ is a profiling step, not a selection. Only the few observations near a transition carry information on the budget, and prediction error is nearly indifferent to a few spurious warps, so selecting $M$ by cross-validation declares many more discontinuities than the criterion below at the same accuracy (Section~\ref{C-sec:tuning} of the Calibration Supplement).

\paragraph{Warping budget.} $M$ is selected on a grid by the extended BIC of \citet{CheChe2008},
\begin{equation}\label{eq:ebic}
  \mathrm{EBIC}(\xi)=-2\hat\ell+\mathrm{df}\,\log n+2\,\xi\,k\,\log(n-1),\qquad \xi\in[0,1],
\end{equation}
where $\hat\ell$ is the maximized log-likelihood, $\mathrm{df}$ the effective degrees of freedom, $k$ the number of active warps, and $\xi=0$ returns the BIC. A budget with $k$ active warps also decides which of the $n-1$ intervals carry them, so the model space at size $k$ has $\binom{n-1}{k}$ members and grows with the sample; a criterion that charges only for dimension, such as the AIC or the BIC, reads the gain from searching over this space as evidence and keeps adding discontinuities. The extended term charges for \emph{where} the discontinuities are placed, as distinct from how many there are.

\paragraph{The constant $\xi$.} The consistency result of \citet{CheChe2008} supplies a floor rather than a value, $\xi\ge1-\log n/\{2\log(n-1)\}\approx0.5$, and it is stated for a criterion in which the extended term alone pays for the search, whereas here the effective degrees of freedom grow with every activated warp and the $\mathrm{df}\log n$ term already pays part of it. We therefore calibrate $\xi$ by simulation, on a design separate from the one used for evaluation in Section~\ref{sec:simulations}, with $200$ replications per cell. The design has three scenarios: a fixed smooth signal observed with independent errors on an irregular design (the reference case), serially dependent errors, and a signal that is partly or entirely a stochastic trend, namely an integrated Wiener process, the prior that the smoothing spline assumes. Each scenario is crossed with the sample size ($n$ from $50$ to $400$), the noise level and the number of true discontinuities ($0$ to $5$). The constant is read off two minimax regret rules over the design cells,\footnote{In a design cell, the regret of a value of $\xi$ is the loss it attains there minus the smallest loss attainable in that cell by any $\xi$; the minimax rule selects the $\xi$ whose largest regret across the cells is smallest.} one on the root mean squared error (RMSE) of the fitted curve and one on the absolute error in the number of discontinuities declared, and Table~\ref{tab:xi_fixed} reports their midpoint.

\begin{table}[!htb]
\centering\footnotesize
\caption{Calibrated values of $\xi$ by scenario, estimator and source of $\lambda$ (from SSJ--MLE, or by cross-validation with $K=5$ or $K=20$ folds drawn at random or as contiguous blocks). Each value is the midpoint, rounded up to the grid, of the minimax regret rules in RMSE and in the count error, each pooled over the levels of the scenario ($\psi\ge0.6$ under dependent errors, $\omega=1$ under a stochastic trend; Section~\ref{C-sec:concl} of the Calibration Supplement).}
\label{tab:xi_fixed}
\setlength{\tabcolsep}{6pt}%
\begin{tabular}{ll ccc}
\toprule
Estimator & $\lambda$ from & reference case & dependent errors & stochastic trend \\
\midrule
SSJ--KFS & blocks $K{=}5$ & $0.15$ & $0.30$ & $0.25$ \\
SSJ--KFS & blocks $K{=}20$ & $0.15$ & $0.35$ & $0.25$ \\
SSJ--KFS & random $K{=}5$ & $0.20$ & $0.25$ & $0.25$ \\
SSJ--KFS & random $K{=}20$ & $0.20$ & $0.20$ & $0.20$ \\
SSJ--KFS & MLE & $0.15$ & $0.20$ & $0.30$ \\
\addlinespace
SSJ--ATW & blocks $K{=}5$ & $0.25$ & $0.40$ & $0.35$ \\
SSJ--ATW & blocks $K{=}20$ & $0.30$ & $0.45$ & $0.30$ \\
SSJ--ATW & random $K{=}5$ & $0.30$ & $0.35$ & $0.30$ \\
SSJ--ATW & random $K{=}20$ & $0.30$ & $0.35$ & $0.35$ \\
SSJ--ATW & MLE & $0.30$ & $0.45$ & $0.35$ \\
\addlinespace
SSJ--MLE & MLE & $0.25$ & $0.25$ & $0.45$ \\
\bottomrule
\end{tabular}
\end{table}

Three regularities matter in practice (Section~\ref{C-sec:concl} of the Calibration Supplement). First, the calibrated values lie below the nominal floor without contradicting the theory, because part of the charge per discontinuity is levied through $\mathrm{df}$; for the same reason SSJ--KFS with $\lambda$ from SSJ--MLE, whose smoothing constant is fixed before the budget is chosen and whose search over the budget can therefore buy less, needs a milder penalty than SSJ--MLE itself in every scenario. Second, serial dependence and stochastic trends both call for a heavier penalty than the reference case, through different channels: persistence acts through $\lambda$, and the penalty responds to it when $\lambda$ is cross-validated on contiguous blocks but not with random folds, whereas a stochastic trend leaves the reconstruction unaffected and inflates the number of discontinuities declared. The value depends on the estimator as much as on the data. Third, a single constant is a compromise across sample sizes: Table~\ref{C-tab:byn_scen} of the Calibration Supplement reports the calibration at each $n$, and Table~\ref{C-tab:rule} a closed-form rule $\xi(n)$ that extends it beyond the sample sizes of the design for SSJ--KFS with $\lambda$ from SSJ--MLE or from contiguous blocks. The values adopted are largely robust to the rule that synthesizes the two criteria (Table~\ref{C-tab:fixed_rob} of the Calibration Supplement).

\paragraph{Declaring discontinuities.} The estimated warps are continuous, and an interval is declared a discontinuity when its warp is large both relative to the total warping of the fit and in absolute terms. Unless otherwise stated, a warp is declared when it exceeds both $5\%$ of the total warping and $30\%$ of the mean sampling interval, a rule that filters out numerically small warps while retaining genuine jumps of very different sizes.

\paragraph{Software.} Both algorithms are implemented in the \texttt{R} package \texttt{jumps}, version 2.0.0 or later, available from CRAN \citep{Rjumps}; the computationally intensive parts are written in \texttt{C++} and interfaced through \texttt{Rcpp} \citep{EddFra2011}.

\section{Simulation experiment}\label{sec:simulations}
To assess how well the estimators recover a piecewise-smooth signal from noisy, irregularly sampled data, we replicate and extend the illustrative simulation of \citet{StoWei2024}. The test signal is
\[
  g(x)=J_1(20x)+x\cdot\mathbf 1_{[0.3,0.4]}(x)-x\cdot\mathbf 1_{[0.6,1]}(x),
\]
where $J_1$ is the Bessel function of the first kind of order one; it has three discontinuities, at $x=0.3$ (jump $+0.3$), $x=0.4$ (jump $-0.4$) and $x=0.6$ (jump $-0.6$), on an oscillating smooth background. For each of three noise levels, $\sigma\in\{0.05,0.10,0.15\}$, we generate $1{,}000$ replications, each with $N=100$ abscissas drawn uniformly on $[0,1]$ and $y_i=g(x_i)+\varepsilon_i$, $\varepsilon_i\sim\mathrm N(0,\sigma^2)$. Every estimator is evaluated on a grid of $500$ equispaced points on $[0,1]$, clamped to the observed range of each replication, so that extrapolation is not scored.

Four estimators are compared. CSSD has the two tuning parameters of its weak-rod penalty chosen by 5-fold cross-validation, as recommended by \citet[Section~3.4]{StoWei2024}, and returns its discontinuity set directly, as part of the solution of the underlying combinatorial problem.\footnote{We ported the original Matlab code of \citet{StoWei2024} to \texttt{R} through an efficient \texttt{C++} implementation based on \texttt{Rcpp} \citep{EddFra2011}.} SSJ--MLE estimates $(\sigma,\lambda)$ by maximum likelihood and SSJ--ATW takes $\lambda$ from 5-fold cross-validation (Section~\ref{sec:tuning}); both select $M$ by the extended BIC at $\xi=0.30$ and declare discontinuities by the rule of Section~\ref{sec:tuning}. This is the reference-case value of Table~\ref{tab:xi_fixed} for SSJ--ATW with $\lambda$ from random 5-fold cross-validation; for SSJ--MLE it is one grid step above the value of the table and equal to the one that the RMSE rule selects on its own (Table~\ref{C-tab:base_all} of the Calibration Supplement), and it keeps the two estimators at an equal penalty. The naive benchmark, which ignores discontinuities altogether, is a cubic smoothing spline with smoothing parameter chosen by generalized cross-validation.

\begin{figure}[htbp]
    \centering
    \includegraphics[width=\linewidth]{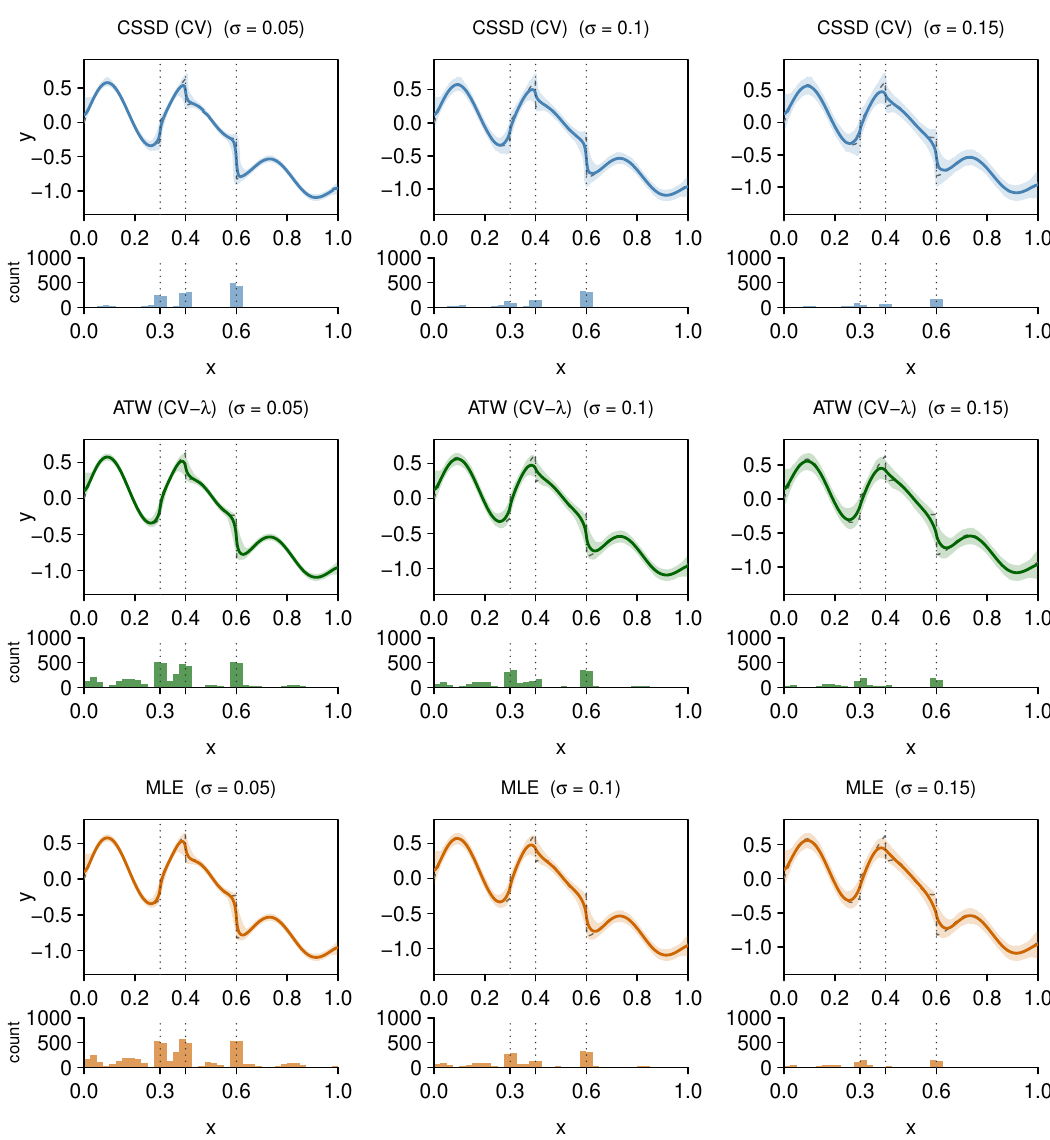}
    \caption{For each of the three discontinuity-aware estimators, the upper panel shows the mean fitted curve (solid) and the pointwise $95\%$ Monte Carlo interval (shaded band) against the true signal (dashed), and the lower panel the histogram of the locations declared to be discontinuities.}
    \label{fig:sim_combined}
\end{figure}

At every noise level SSJ--ATW and SSJ--MLE reconstruct $g$ more accurately than either benchmark (Table~\ref{tab:rmse}), and they stay close to each other: their RMSEs differ by at most 4\%, with SSJ--ATW ahead at $\sigma=0.05$ and SSJ--MLE at $\sigma=0.15$. The two-step algorithm of Section~\ref{sec:atw} thus attains the accuracy of the maximum likelihood fit it approximates. Both improve on the classical smoothing spline, by 7\% to 11\% at $\sigma=0.05$ and by 3\% to 7\% at the other two levels.

\begin{table}[htbp]
\centering
\caption{RMSE of the fitted curve against the true signal $g$, pooled over $1{,}000$ replications and the $500$-point evaluation grid, at three noise levels. All values are multiplied by $10^3$. CSSD is reported twice: over all replications, and excluding those on which its fit diverges between the design points; $n_{\mathrm{ex}}$ counts the latter.}
\label{tab:rmse}
\begin{tabular}{lccc ccc}
\toprule
 & \multicolumn{3}{c}{CSSD} & & & \\
\cmidrule(lr){2-4}
$\sigma$ & all & excl.\ & $n_{\mathrm{ex}}$ & SSJ--ATW & SSJ--MLE & Smooth.\ spline \\
\midrule
0.05 &  \phantom{1}83.9 &  63.4 & 2 &  56.1 &  58.5 &  63.2 \\
0.10 &  \phantom{1}79.1 &  79.1 & 0 &  73.1 &  73.5 &  78.2 \\
0.15 & 102.1 &  95.8 & 2 &  90.9 &  88.0 &  93.3 \\
\bottomrule
\end{tabular}
\end{table}

\begin{figure}[htbp]
    \centering
    \includegraphics[width=\linewidth]{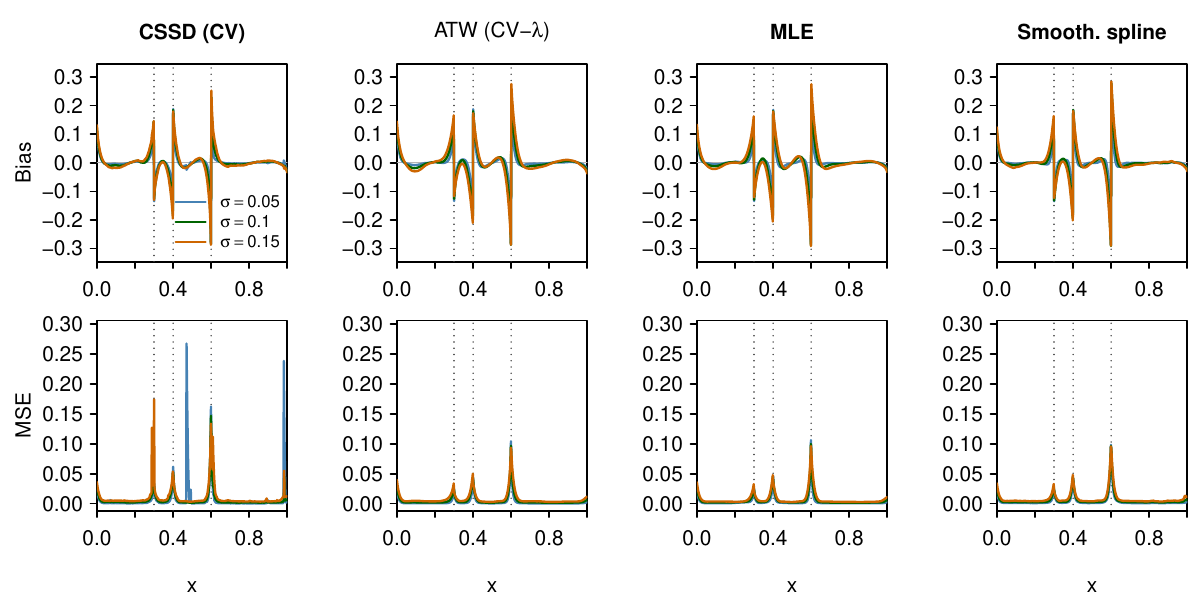}
    \caption{Pointwise bias (top row) and mean squared error (bottom row) of the four estimators, over $1{,}000$ replications, for $\sigma=0.05$ (blue), $0.10$ (green) and $0.15$ (orange). Vertical dotted lines mark the true discontinuities. The two spikes in CSSD's mean squared error near $x=0.47$ and $x=0.98$ are the divergent replications discussed in the text.}
    \label{fig:sim_biasmse}
\end{figure}

The comparison with CSSD requires a qualification, which is why Table~\ref{tab:rmse} reports it twice. Pooled over all replications CSSD is the least accurate of the four estimators, trailing even the smoothing spline, and SSJ--ATW is $33\%$ more accurate at $\sigma=0.05$; but nearly all of the gap is contributed by four replications out of $3{,}000$, two at $\sigma=0.05$ and two at $\sigma=0.15$. On those draws a declared discontinuity isolates two nearly coincident design points into a segment of their own; the spline on that segment is the straight line through them, and the linear extension used to evaluate the fit carries its slope, of order $10^3$, across the segment. The fitted values at the design points remain sane, so the failure appears only between them; we verified it against the reference implementation of \citet{StoWei2024}, and it is a property of the estimator in this sampling regime. These replications, flagged because their RMSE exceeds five times the median of the same estimator at the same noise level (they sit at $8$ to $23$ times the median, the largest unflagged one at $3.7$), produce the two spikes in CSSD's mean squared error near $x=0.47$ and $x=0.98$ in Figure~\ref{fig:sim_biasmse}, far from any jump; away from the three discontinuities, the two divergent replications at $\sigma=0.05$ account for $97\%$ of CSSD's excess mean squared error over SSJ--ATW. With them set aside, SSJ--ATW and SSJ--MLE are between 5\% and 12\% more accurate than CSSD, which is then on a par with the smoothing spline; the medians of the per-replication errors, which no tail can move, agree ($59.5$ for CSSD and $60.0$ for the smoothing spline at $\sigma=0.05$, against $53.5$ and $52.9$ for SSJ--ATW and SSJ--MLE). At the discontinuities themselves the estimators behave alike: at $x=0.6$, the largest jump, the bias of CSSD is $-0.285$ against $-0.286$ for SSJ--ATW, the unavoidable cost of smoothing across a jump, and its mean squared error is larger by a factor of about $1.6$.

The picture changes when the question is where the discontinuities are. The lower panels of Figure~\ref{fig:sim_combined} show the locations declared over the replications, and Table~\ref{tab:detection} the detection rates within $\pm0.02$ of each true break and the false positives. CSSD raises far fewer false alarms, $0.55$ per replication at $\sigma=0.05$ against $2.23$ for SSJ--ATW and $2.92$ for SSJ--MLE, but it also finds fewer of the breaks: at $\sigma=0.05$ it recovers the jump at $x=0.3$ in $45\%$ of the replications, against $83\%$ and $84\%$ for the SSJ estimators, and the one at $x=0.4$ in $58\%$, against $73\%$ and $79\%$; only at the largest jump, $x=0.6$, are the three comparable. As the noise rises, detection power declines for every estimator and the ordering ceases to be clear-cut: by $\sigma=0.15$ no estimator recovers the two smaller jumps reliably. CSSD is conservative and the SSJ estimators are liberal, consistently with a warping vector that is optimized to reduce roughness globally rather than to solve a detection problem, and with a calibration of $\xi$ that weighs the accuracy of the curve as well as the count.

\begin{table}[htbp]
    \centering
    \caption{Discontinuity detection by estimator and noise level. Detection rate: fraction of the $1{,}000$ replications in which a discontinuity is declared within $\pm0.02$ of the true break. False positives: mean number, per replication, of declared locations outside all three windows, and the same figure as a percentage of the $87.0$ candidate locations that lie outside them on average (discontinuities can be declared only between consecutive design points, $N-1=99$ candidates).}
    \label{tab:detection}
    \begin{tabular}{llccccc}
        \toprule
        & & \multicolumn{3}{c}{Detection rate} & \multicolumn{2}{c}{False positives} \\
        \cmidrule(lr){3-5} \cmidrule(lr){6-7}
        Estimator & $\sigma$ & $x=0.3$ & $x=0.4$ & $x=0.6$ & per rep. & rate \\
        \midrule
        \multirow{3}{*}{CSSD}
          & $0.05$ & $45\%$ & $58\%$ & $89\%$ & $0.55$ & $0.63\%$ \\
          & $0.10$ & $20\%$ & $28\%$ & $63\%$ & $0.53$ & $0.61\%$ \\
          & $0.15$ & $14\%$ & $14\%$ & $35\%$ & $0.40$ & $0.46\%$ \\
        \addlinespace
        \multirow{3}{*}{SSJ--ATW}
          & $0.05$ & $83\%$ & $73\%$ & $92\%$ & $2.23$ & $2.56\%$ \\
          & $0.10$ & $49\%$ & $25\%$ & $61\%$ & $1.22$ & $1.40\%$ \\
          & $0.15$ & $23\%$ & $\phantom{0}7\%$ & $30\%$ & $0.61$ & $0.71\%$ \\
        \addlinespace
        \multirow{3}{*}{SSJ--MLE}
          & $0.05$ & $84\%$ & $79\%$ & $93\%$ & $2.92$ & $3.35\%$ \\
          & $0.10$ & $44\%$ & $22\%$ & $55\%$ & $1.07$ & $1.23\%$ \\
          & $0.15$ & $19\%$ & $\phantom{0}4\%$ & $24\%$ & $0.46$ & $0.53\%$ \\
        \bottomrule
    \end{tabular}
\end{table}

In short, CSSD is the more conservative estimator of \emph{where} the discontinuities are, declaring few and being right about nearly all of them, while SSJ--ATW and SSJ--MLE are the more accurate estimators of \emph{what} the signal is doing, and the more sensitive detectors when the signal-to-noise ratio permits. When the discontinuities are few, isolated and large relative to the noise, the three estimators agree, as they do on the fuel prices of Section~\ref{sec:carburanti}.

\section{Applications}\label{sec:applications}

\subsection{Temperature and electricity demand in Northern Italy}\label{sec:elettro}
\begin{figure}[htb]
    \centering
    \includegraphics[width=0.49\linewidth]{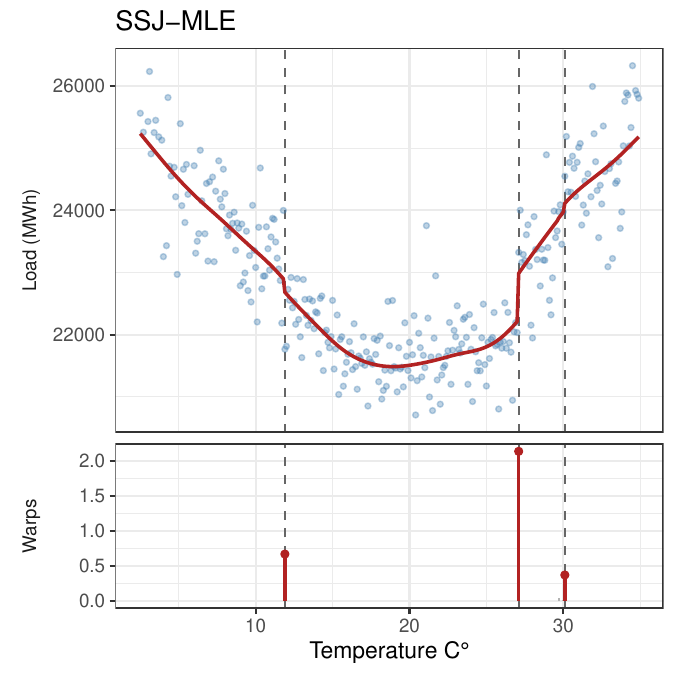}
    \includegraphics[width=0.49\linewidth]{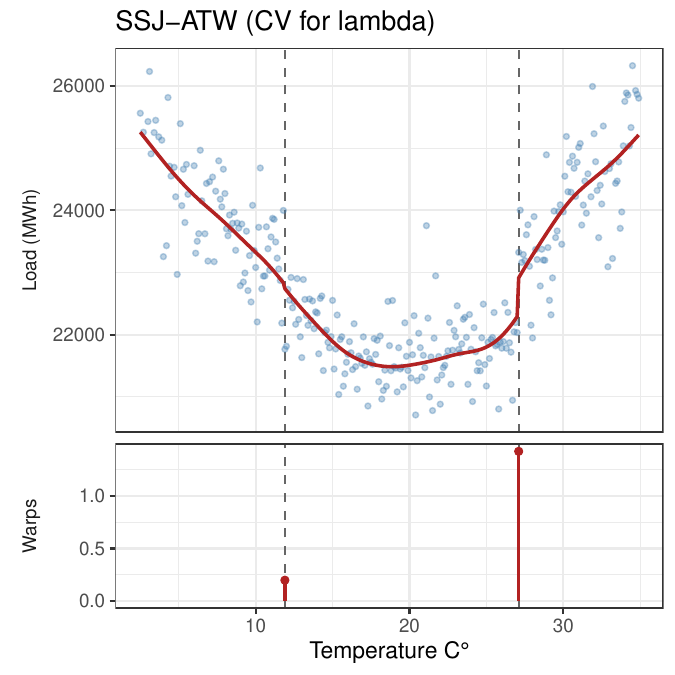}
    \caption{Temperature and electricity demand, with the signal extracted at $\xi=0.30$ by (a) SSJ--MLE, (b) SSJ--ATW with $\lambda$ from 5-fold cross-validation. The band in (a) is $\pm2$ standard errors of the extracted signal, which the state-space form returns and the alternating algorithm does not. The bottom plots show $\gamma_t$ for every active warp.}
    \label{fig:elettro}
\end{figure}

We estimate the relation between the hourly electricity load of Northern Italy, from the day-ahead market data published by the \emph{Gestore Mercati Energetici} (GME, \url{https://www.mercatoelettrico.org/}), and the temperature in Milan, from Open Meteo (\url{https://open-meteo.com/}), between 1 January 2023 and 30 September 2025. To obtain homogeneous consumption behavior we keep the hours from 10~a.m.\ to 3~p.m.\ of Tuesdays to Thursdays and exclude the months with typical Italian vacations, keeping February, March, May, June, July, September, October and November; after dropping the few hours with a demand below $15{,}000$ MWh or a temperature above $35^\circ$C, the sample has $1{,}691$ hourly observations between $2.5$ and $34.8^\circ$C. The temperatures are recorded to the nearest $0.1^\circ$C, so that many hours share the same value; these ties are collapsed into weighted means at the distinct temperatures, as described in Section~\ref{sec:idea} and, in detail, in Section~\ref{A-sec:ties} of the Computational Supplement. The abscissa is the temperature, irregularly spaced, and the relation is a fixed curve observed with errors, the reference case of the calibration; both estimators are therefore run at $\xi=0.30$, as in Section~\ref{sec:simulations}. We discuss the warps using the rule of Section~\ref{sec:tuning}, even though in Figure \ref{fig:elettro} all warps are depicted (in gray those not matching the rule).

SSJ--MLE selects a budget with three active warps, at $11.9^\circ$C, $27.1^\circ$C, and $30.1^\circ$C, while SSJ--ATW (with cross-validated $\lambda$) detects only  $11.9^\circ$C and $27.1^\circ$C. Figure~\ref{fig:elettro} shows that the two fits agree on the shape of the curve and on where it breaks, and differ mainly in how much warping they spend: 3.24 for SSJ--MLE and 1.62 for SSJ--ATW. The state-space form of SSJ-MLE also delivers standard errors.

The discontinuities mark temperatures at which the physical system behind the load changes state. The large jump detected by both methods at $27.1^\circ$C is associated with the temperature at which air conditioning is generally turned on. The break at $11.9^{\circ}$C is an order of magnitude smaller than the one at
$27^{\circ}$C and it has the opposite sign, demand falling as the temperature rises through it. A natural reading is the mirror image of the air-conditioning threshold: space heating in Lombardy is overwhelmingly gas-fired and so largely invisible in electricity demand, but the supplementary electric heating that is used in the shoulder season --- reversible split units run as heat pumps, and electric
convectors in commercial premises --- is switched on below roughly this
temperature. The same behavioral mechanism as at $27^{\circ}$C would then act
on a much smaller share of the load, which is what the relative size of the two
steps suggests. The discontinuity at $30.1^{\circ}$C, detected only by SSJ--MLE, appears to be a change in slope and could be due to many air conditioners already operating at their maximum power, thus, unable to respond to temperatures above $30.1^{\circ}$C.

None of these features can be expressed by the degree-day construct that energy systems conventionally use to summarize the temperature response, two straight lines joined at a base temperature fixed by convention ($20^\circ$C in Italy). The fitted curve attains its minimum at $18.9^\circ$C but stays within $1\%$ of it over the whole interval $[16.0,23.4]^\circ$C, so there is no sharp vertex for the data to locate; its two arms are asymmetric, with about $270$ MWh per degree of cold below $14^\circ$C against $320$ per degree of heat above $24^\circ$C, which a single base temperature cannot serve; and, being continuous, a degree-day representation can only absorb the steps into a steeper slope. The estimator does not have to choose between a kink and a step.

We also applied the CSSD algorithm of \citet{StoWei2024} to the same data, but no discontinuities were detected, so the extracted signal was the same as that of a classical smoothing spline.

\subsection{Retail fuel prices and excise-duty changes in Italy}\label{sec:carburanti}
The second application is a time series in which the location of the discontinuities is known in advance. This is unusual and valuable: change-point procedures are almost always assessed on simulated data, because in real series the truth is not observed. Retail fuel prices are an exception. The excise duty on motor fuels is a per-liter amount fixed by decree, it enters the pump price mechanically, and the date on which a new amount takes effect is a matter of public record, so a change in the duty is a discontinuity whose timing, although not its size, is known exactly.

The last years have provided two such interventions of unusual size. After the invasion of Ukraine the average price of gasoline rose by more than $0.23$ euros per liter in the second week of March 2022, to an all-time high above $2.18$ euros, and the government cut the excise duty by $0.25$ euros per liter from 22 March 2022; the cut was reduced by $0.10$ euros on 1 December 2022 and withdrawn on 1 January 2023. A second cut, of $0.20$ euros per liter, entered the prices on 23 March 2026 and was likewise followed by a staggered restoration. In between, the duties on gasoline and diesel were realigned in two small steps until both reached $0.6729$ euros per liter in January 2026. Table~\ref{tab:excise_dates} lists the ten weeks of the sample in which the duty changed, dated at the first weekly survey in which the new amount appears.\footnote{A decree takes effect on an arbitrary day and the weekly survey is run on Mondays, so a lag of up to one week between the legal date and the price response is allowed when an estimated discontinuity is matched to an event.} Seven dates affect both fuels and three only one, for seventeen events; the two realignment steps are an order of magnitude smaller than the emergency measures and are included in the ground truth without any expectation that they be detectable.

\begin{table}[htbp]
\centering
\small
\caption{Weeks of the estimation sample in which a change in the excise duty on motor fuels entered the surveyed prices, and the size of the change in euros per liter, positive for an increase. A dash means that the duty on that fuel did not change in that week.}
\label{tab:excise_dates}
\setlength{\tabcolsep}{5pt}
\begin{tabular}{lrr@{\qquad}lrr}
\toprule
Week & Gasoline & Diesel & Week & Gasoline & Diesel \\
\midrule
28 Mar 2022 & $-0.250$ & $-0.250$ & 23 Mar 2026 & $-0.200$ & $-0.200$ \\
 5 Dec 2022 & $+0.100$ & $+0.100$ &  4 May 2026 & $+0.150$ & --- \\
 9 Jan 2023 & $+0.150$ & $+0.150$ & 25 May 2026 & --- & $+0.100$ \\
19 May 2025 & $-0.015$ & $+0.015$ &  8 Jun 2026 & --- & $+0.050$ \\
 5 Jan 2026 & $-0.040$ & $+0.040$ &  6 Jul 2026 & $+0.050$ & $+0.050$ \\
\bottomrule
\end{tabular}
\end{table}

We use the weekly survey of the national average consumer prices of petroleum products published as open data by the Italian Ministry of Environment and Energy Security (MASE, \url{https://sisen.mase.gov.it/dgsaie/open-data}), from 5 January 2015 to 27 July 2026, that is, $n=591$ weeks, for gasoline and automotive diesel. For each fuel we smooth two series: the pump price, which a change in the duty moves by construction, and the industrial price, which excludes the excise duty and the value-added tax and therefore reacts to a fiscal measure only to the extent that the measure is not passed on.\footnote{The components satisfy the accounting identity $\text{pump price}=(\text{industrial price}+\text{excise duty})(1+\varphi)$, where $\varphi=0.22$ is the value-added tax rate.} We treat the observations as equally spaced, $x_t=t$, so that $\gamma_t$ is measured in weeks of fictitious time and so is the grid of budgets; a location is declared a discontinuity when $\hat\gamma_t>1$, that is, when the latent interval is more than twice the physical one, and a break is dated at the first observation of the new regime. Weekly prices leave serially dependent residuals, and their level behaves like a stochastic trend rather than a fixed curve, since crude oil, the exchange rate and refining margins accumulate shocks; $\lambda$ is therefore estimated by maximum likelihood and, for SSJ--ATW, also by cross-validation on $K=20$ contiguous blocks of about thirty weeks (Section~\ref{S-sec:supp_app_folds} of the Application Supplement).

These are the two departures of the calibration, for which Table~\ref{tab:xi_fixed} gives values of $\xi$ above the reference case. Because the dates of the interventions are known, the choice can also be checked on the data (see Section~\ref{S-sec:supp_app_sweep} of the Application Supplement). The declared sets are nested in $\xi$, and at the reference-case value the estimators declare a long tail of dates that are not policy events: SSJ--ATW at the maximum likelihood $\lambda$ reports $16$ such weeks on the two pump-price series alongside $11$ of the $17$ events. Raising $\xi$ to $0.45$ removes much of the tail at little cost, leaving $10$ spurious weeks and $9$ events, and beyond $0.45$ nothing more is gained. We therefore set $\xi=0.45$, the upper end of the values that Table~\ref{tab:xi_fixed} gives under the two departures, which a weekly price series combines; Section~\ref{S-sec:supp_app_xi} of the Application Supplement repeats the analysis at $\xi=0.25$, $0.35$ and $0.50$. Four of the seventeen events are recovered by no estimator at any $\xi$: the two realignment steps on either fuel, which move the duty by $1.5$ and $4$ cents against a weekly standard deviation of the pump price several times larger.

\begin{figure}[htbp]
    \centering
    \includegraphics[width=.95\linewidth]{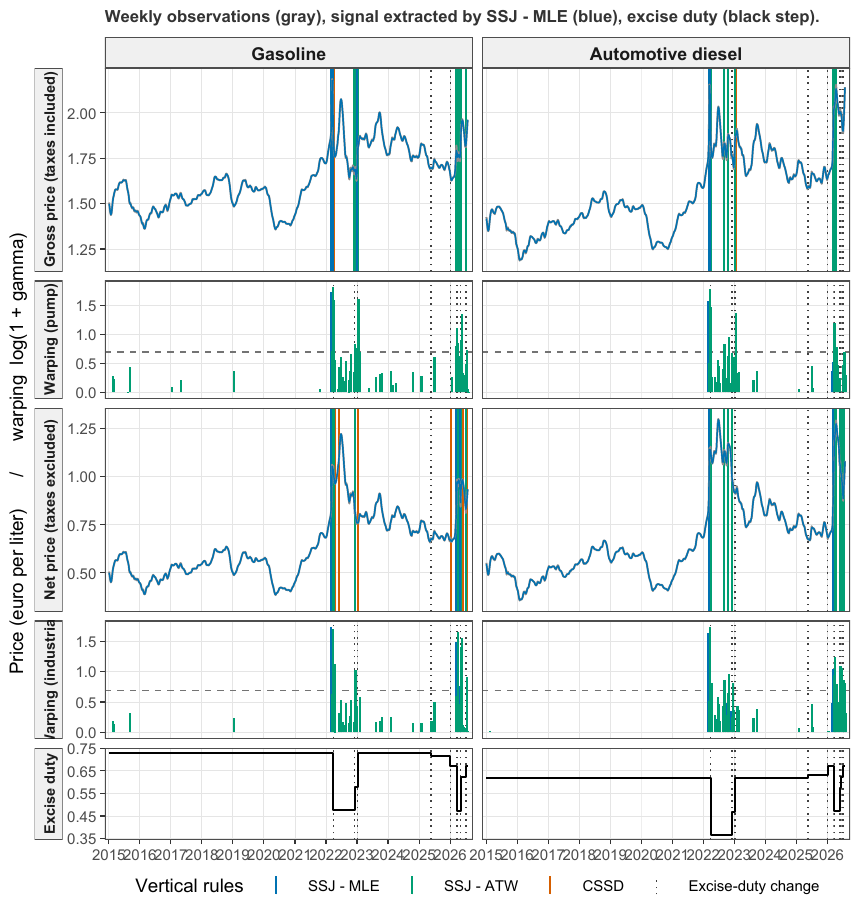}
    \caption{Consumer prices of gasoline (left) and automotive diesel (right) in Italy. Rows one and three: the price paid at the pump (excise duty and VAT included) and the industrial price (taxes excluded); observations (gray lines) and the signal extracted by SSJ--MLE (blue line). Rows two and four: the warping sequences on a $\log(1+\hat\gamma_t)$ scale of SSJ--MLE (left) and SSJ--ATW (right); the dashed rule is the threshold $\hat\gamma_t>1$ above which a location is declared a discontinuity. Bottom row: the excise duty. Colored vertical rules mark the discontinuities found by each estimator, gray dotted rules the weeks in which a change in the excise duty entered the surveyed prices.}
    \label{fig:carburanti}
\end{figure}

Figure~\ref{fig:carburanti} reports the four series and the signals extracted by SSJ--MLE, together with the warping sequences of SSJ--MLE and SSJ--ATW and the discontinuities declared by the two and by CSSD. The estimators are mutually consistent: on gasoline the three return the same dates on both price concepts, and on diesel they differ only by one or two locations at which SSJ--ATW is slightly more liberal. The discontinuities are few, and they sit on the policy dates. On the pump price the largest warp of the whole sample falls on 28 March 2022, the first survey after the emergency cut, with $\hat\gamma_t=3.6$ for gasoline and $5.1$ for diesel, and the second largest on 9 January 2023, the first survey after the duty was restored in full. The one further discontinuity that the estimators isolate on the pump price is not a fiscal event but the market shock of 14 March 2022, two weeks earlier, and it is separated cleanly from the tax change that followed it (Figure~\ref{S-fig:carburanti_zoom} of the Application Supplement magnifies the two episodes).

The estimated jumps are economically interpretable. A cut of $|\Delta A|$ in the duty lowers the pump price by $|\Delta A|(1+\varphi)$ if it is passed on in full, so the $0.25$ euro cut of March 2022 corresponds to a mechanical reduction of $0.305$ euro per liter. The estimated discontinuity is $-0.278$ (SSJ--MLE), $-0.267$ (SSJ--ATW) and $-0.267$ (CSSD) for gasoline, and $-0.284$, $-0.263$ and $-0.267$ for diesel, close to nine tenths of the mechanical effect. The complement is visible in the other series: SSJ--MLE declares no discontinuity on the industrial price at that date, when the industrial price rose by only $0.031$ euro per liter. The restoration of January 2023 gives estimates of $+0.137$ to $+0.168$ against a mechanical $+0.183$, so the transmission is of comparable size in the two directions. The 2026 episode inverts the pattern. The pump price barely reacts, falling by $0.040$ euro per liter for gasoline and $0.010$ for diesel against a mechanical $0.244$, and no discontinuity is declared on it, while a large one appears on the industrial price, estimated at $+0.157$ to $+0.164$ for gasoline and $+0.179$ to $+0.191$ for diesel. Because the cut came on top of a violent market surge, the size of the pass-through is uncertain, but the conclusion that the second intervention reached the pump far less completely than the first is robust (Section~\ref{S-sec:supp_app_2026} of the Application Supplement).

Finally, we remark that a change point in the middle of a long sample is an easy target: near the boundary the spline has almost no data on one side, and the roughness penalty alone must arbitrate between a discontinuity and a steep continuous transition. Because the date of the March 2022 cut is known, this can be put to the test by re-estimating the three procedures on $41$ nested windows in which the edge of the sample is moved toward the cut, from the left and from the right for the cut of March 2022 and from the right for that of March 2026 (Section~\ref{S-sec:supp_app_figs} of the Application Supplement). On the pump price both SSJ estimators recover the March 2022 cut in every window, from either side, and date it exactly; its estimated size stays at $-0.267$ euro per liter from the full sample down to the shortest window, so the boundary neither displaces nor attenuates the jump. CSSD loses the jump in one short window on each side, where it detects a move of about $0.03$ to $0.05$ euro per liter instead. On the industrial price, where the 2022 cut leaves no trace, the estimators find nothing at the policy date in any window, which doubles as a placebo check.

\section{Conclusion}\label{sec:conclusion}
We have generalized the cubic smoothing spline to signals with an unknown, finite number of discontinuities by deforming the domain over which smoothness is measured, rather than by searching over discontinuity sets or penalizing coefficients. Because the mechanism only reparameterizes the metric of the classical roughness penalty, the model inherits the state-space form of smoothing splines, which gives maximum likelihood estimation via the Kalman filter (SSJ--MLE), and it admits a fast alternating algorithm (SSJ--ATW) that alternates a banded linear solve with a projected-gradient update of the warps, without maximizing a non-convex likelihood numerically.

The simulation study and the two applications point to a consistent division of labor with the CSSD benchmark of \citet{StoWei2024}. SSJ--ATW and SSJ--MLE reconstruct the signal more accurately than CSSD and than a classical smoothing spline, and they detect more of the genuine breaks when the signal-to-noise ratio permits; CSSD declares fewer discontinuities and is nearly always right about those it declares. On the fuel prices, where the timing of the discontinuities is a matter of public record, the three estimators agree on the policy dates, and on the pump price the SSJ estimators recover the March 2022 cut, with a stable size, in every boundary window, while CSSD occasionally loses it near the edge of the sample.

Several limitations suggest directions for future work. The optimization over $\bm\gamma$ is non-convex for both SSJ--MLE and SSJ--ATW, so neither is guaranteed to find a global optimum; starting values close to $\bm\gamma=\bm0$ work well in our experiments but are not backed by a formal convergence guarantee. The budget is selected by an extended BIC whose constant is calibrated by simulation and depends on the estimator and on the character of the data; a joint, likelihood-based selection of $(\sigma,\lambda,M)$ remains open. The declaration of a discontinuity is a threshold rule on $\hat\gamma_t$; a sampling distribution for $\hat{\bm\gamma}$, and hence a formal test for a jump at a given location, would put this step on firmer ground and is a natural extension of the analytical scores of the Computational Supplement. Finally, the domain-warping idea is not tied to a one-dimensional domain or to a Gaussian signal: extensions to surfaces, in the spirit of the weak plate model of \citet{BlaZis1987}, and to non-Gaussian observation models within the same state-space machinery are directions we leave for future research.

\section*{Supplementary material}

\begin{description}

\item[Computational Supplement:] Reduced form of the roughness energy of an interval and its gradient, assembly and block-tridiagonal solution of the spline step, projection onto the feasible set of the warps, analytical scores of the Kalman-filter log-likelihood, and the treatment of repeated values of the covariate.

\item[Calibration Supplement:] Tuning of the smoothing constant and of the warping budget, and design, criteria and complete results of the calibration of the extended-BIC constant $\xi$.

\item[Application Supplement:] Additional results for the two applications: the temperature--demand relation over the grid of $\xi$ and the comparison with degree days; for the fuel prices, the check of the calibration on the policy dates, the 2026 episode, the boundary experiment, and the sensitivity to $\xi$ and to the number of cross-validation folds.

\item[R package \texttt{jumps}:] Implementation of SSJ--MLE and SSJ--ATW, version 2.0.0 or later, available from CRAN \citep{Rjumps}.

\item[Replication package:] Code that reproduces every table and figure of the paper and of the supplements, available from the authors upon request. The data are publicly available from the sources given in Section~\ref{sec:applications}.

\end{description}

\bibliographystyle{agsm}
\bibliography{ssj}

\end{document}


\maketitle

\noindent
This document collects the derivations and the implementation details of the two estimators of Section~\ref{P-sec:methodology} of the paper. Section~\ref{sec:energy} derives the reduced form of the roughness energy of one interval and its derivative with respect to the domain warp, on which the warping step of the alternating algorithm (SSJ--ATW) rests. Section~\ref{sec:splinestep} shows how the spline step is assembled and solved in $O(n)$ operations, and Section~\ref{sec:projection} how the warps are projected onto their feasible set. Section~\ref{sec:kalman_scores} derives the analytical scores of the log-likelihood of the state-space form, which make the maximum likelihood estimator (SSJ--MLE) efficient, and Section~\ref{sec:ties} explains how repeated values of the covariate are treated. Equation numbers without the prefix A refer to the paper.

\section{The roughness energy of one interval}\label{sec:energy}
With $\bm v_i=(f_i,f'_i,f_{i+1},f'_{i+1})^\top$ and the local stiffness matrix
\[ \bm{K}_i = \frac{1}{\tau_i^3}
	\begin{bmatrix}
		12 & 6\tau_i & -12 & 6\tau_i \\
		6\tau_i & 4\tau_i^2 & -6\tau_i & 2\tau_i^2 \\
		-12 & -6\tau_i & 12 & -6\tau_i \\
		6\tau_i & 2\tau_i^2 & -6\tau_i & 4\tau_i^2
	\end{bmatrix}
\]
the roughness energy of the $i$-th interval is $k_i=\bm v_i^\top\bm K_i\bm v_i$ (Equation~(\ref{P-eq:ki}) of the paper). Expanding the quadratic form gives
\[
    k_i = \frac{1}{\tau_i^3}\Bigl[\,
      12\bigl(f_i^2 - 2f_if_{i+1} + f_{i+1}^2\bigr)
      + 12\tau_i\bigl(f_if'_i + f_if'_{i+1} - f_{i+1}f'_i - f_{i+1}f'_{i+1}\bigr)
      + 4\tau_i^2\bigl((f'_i)^2 + f'_if'_{i+1} + (f'_{i+1})^2\bigr)
    \,\Bigr].
\]
The three brackets are respectively $(f_{i+1}-f_i)^2=\Delta_i^2$, $(f_i-f_{i+1})(f'_i+f'_{i+1})=-\Delta_iS_i$ and $P_i$, where
\begin{equation}\label{eq:DSP}
  \Delta_i=f_{i+1}-f_i,\qquad S_i=f'_i+f'_{i+1},\qquad P_i=(f'_i)^2+f'_if'_{i+1}+(f'_{i+1})^2,
\end{equation}
so that the energy of the interval admits the reduced form
\begin{equation}\label{eq:kreduced}
  k_i=\frac{12\Delta_i^2}{\tau_i^3}-\frac{12\Delta_iS_i}{\tau_i^2}+\frac{4P_i}{\tau_i}.
\end{equation}
Since $\Delta_i$, $S_i$ and $P_i$ depend on the nodal values only, $k_i$ is a Laurent polynomial in $\tau_i$ whose coefficients do not depend on $\tau_i$. Differentiating it term by term and using $\partial\tau_i/\partial\gamma_i = 1$ gives $\nabla_i$ at once. Notice that the derivative is taken at $\bm z$ fixed, which is what the $\gamma$-step of the block coordinate descent requires, and it is the derivative of the roughness energy alone: the corresponding derivative of the objective function is $\lambda\nabla_i$, so that $\lambda$ is absorbed into the learning rate $\eta$.

\paragraph{Gradient with respect to the warps.} Since $\Delta_i$, $S_i$ and $P_i$ depend on the nodal values only, $k_i$ is a Laurent polynomial in $\tau_i$ whose coefficients do not depend on $\tau_i$. Differentiating it term by term and using $\partial\tau_i/\partial\gamma_i=1$ gives
\begin{equation}\label{eq:gradient}
  \nabla_i=\frac{\partial k_i}{\partial\gamma_i}=-36\frac{\Delta_i^2}{\tau_i^4}+24\frac{\Delta_iS_i}{\tau_i^3}-4\frac{P_i}{\tau_i^2},
\end{equation}
which is Equation~(\ref{P-eq:grad}) of the paper. The derivative is taken at $\bm z$ fixed, which is what the warping step of the block coordinate descent requires, and it is the derivative of the roughness energy alone: the corresponding derivative of the objective function~(\ref{P-eq:objective}) of the paper is $\lambda\nabla_i$, so that $\lambda$ is absorbed into the learning rate $\eta$.

\paragraph{The energy of a short interval.} Section~\ref{P-sec:idea} of the paper states that the energy of an interval is dominated, for short intervals, by a term proportional to $\Delta_i^2/\tau_i^3$, by analogy with the bending energy of an elastic beam. Two precise statements lie behind it, and both follow from Equation~\eqref{eq:kreduced}, with $\Delta_i$ the height difference that the fitted curve actually covers on the interval. If the segment is required to arrive and to leave with zero slope, then $k_i=12\Delta_i^2/\tau_i^3$ exactly, which is the classical stiffness of a beam clamped at both ends. For arbitrary but fixed endpoint derivatives and $\Delta_i\neq0$, the remaining two terms of $k_i$ are of order $\tau_i^{-2}$ and $\tau_i^{-1}$, so that $k_i=12\Delta_i^2\tau_i^{-3}\{1+O(\tau_i)\}$ as $\tau_i\to0$. The proportionality describes how the energy grows as the interval shrinks and is not a lower bound over all cubics through the two points: the minimum of $k_i$ over the endpoint derivatives is zero, attained by the straight line.

\paragraph{Units of the smoothing constant.} The reduced form also shows that the smoothing constant is not scale-free. Each term of Equation~\eqref{eq:kreduced} has the units of $y^2/x^3$ and the fidelity term of the objective function has the units of $y^2$, so $\lambda$ carries the units of $x^3$. This is why the grids of $\lambda$ are specified relative to the cubed range of the abscissa (Calibration Supplement), and it is the same fact that lets $\lambda$ be absorbed into the learning rate of the warping step.

\section{The spline step}\label{sec:splinestep}
In the spline step $\bm\gamma$ is fixed and the objective function is minimized with respect to the state vector $\bm z=(f_1,f'_1,f_2,f'_2,\dots,f_n,f'_n)^\top$ of size $2n$, which contains the values and the first derivatives of the function at the observations. Following the classical approach that goes back to \citet{StrFix1973}, let $\bm y^*$ be the vector of size $2n$ with the $y$ values in the odd positions and zeros in the even positions, and let $\bm S$ be the $2n\times2n$ diagonal selection matrix with diagonal $(1,0,1,0,\dots,1,0)$: applied to $\bm z$, it retains the function values and zeros out the derivatives, as the data inform $f$ but not $f'$.

\paragraph{Assembly of the penalty matrix.} The roughness penalty matrix $\bm R$ is a sparse matrix assembled from the local stiffness matrices $\bm K_i$ of Section~\ref{sec:energy}. Since the elements $(f_i,f'_i)$ of $\bm z$, $i=2,\dots,n-1$, play the role of both the ending point of interval $i-1$ and the starting point of interval $i$, $\bm R$ is built as follows.
\begin{enumerate}
    \item Set all the elements of the $2n\times 2n$ matrix $\bm{R}$ equal to zero.
    \item For $i=1, 2, \ldots, n-1$, set
    \begin{align*}
        &\bm{R}{[(2i-1)\!:\!(2i+2), (2i-1)\!:\!(2i+2)]} = \\
        &\qquad \bm{R}{[(2i-1)\!:\!(2i+2), (2i-1)\!:\!(2i+2)]} + \bm{K}_i,
    \end{align*}
    where $\bm{R}[a\!:\!b, c\!:\!d]$ refers to the submatrix of $\bm{R}$ with (integer) row indices ranging from $a$ to $b$ and column indices from $c$ to $d$.
\end{enumerate}
The matrix $\bm R$ is symmetric and block tridiagonal. The objective function can then be written as
\[
  J(\bm z,\bm\gamma)=(\bm S\bm y^*-\bm S\bm z)^\top(\bm S\bm y^*-\bm S\bm z)+\lambda\bm z^\top\bm R\bm z,
\]
which is minimized by the solution of the linear system
\begin{equation}\label{eq:system}
  (\bm S+\lambda\bm R)\bm z=\bm y^* .
\end{equation}
The output of the step is the updated $f$ (odd elements of $\bm z$) and $f'$ (even elements of $\bm z$).

\subsection{Block-tridiagonal Cholesky decomposition}\label{sec:cholesky}
The matrix $\bm A=\bm S+\lambda\bm R$ has a block-tridiagonal structure with $2\times2$ blocks,
\[
  \bm A=\begin{bmatrix}
    \bm D_1 & \bm U_1 & \bm 0 & \cdots\\
    \bm U_1^\top & \bm D_2 & \bm U_2 & \cdots\\
    \bm 0 & \bm U_2^\top & \bm D_3 & \cdots\\
    \vdots & \vdots & \vdots & \ddots
  \end{bmatrix},
\]
so that, partitioning $\bm z$ and $\bm y^*$ conformably into the $2\times1$ blocks $\bm z_i=(f_i,f'_i)^\top$ and $\bm y_i=(y_i,0)^\top$, system~\eqref{eq:system} is a chain of small systems, each linking a block to its two neighbors. Due to this chain structure, it can be solved in two linear passes, one forward and one backward \citep[see Section~4.5.1 of][]{GolVan2013}; this is essentially Gaussian elimination organized by blocks, the idea exploited by \citet{StoWei2024} and that goes back to \citet{deB1978}.

\paragraph{Forward step.} Set $\tilde{\bm D}_1=\bm D_1$ and $\tilde{\bm y}_1=\bm y_1$ and, for $i=2,\dots,n$, compute the modified diagonal blocks, which are Schur complements, and the modified right-hand sides
\[
  \tilde{\bm D}_i=\bm D_i-\bm U_{i-1}^\top\tilde{\bm D}_{i-1}^{-1}\bm U_{i-1},
  \qquad
  \tilde{\bm y}_i=\bm y_i-\bm U_{i-1}^\top\tilde{\bm D}_{i-1}^{-1}\tilde{\bm y}_{i-1}.
\]
This operation transforms the system into a block upper-triangular one, which can be solved block by block.

\paragraph{Backward step.} Set $\bm z_n=\tilde{\bm D}_n^{-1}\tilde{\bm y}_n$ and, for $i=n-1,\dots,1$, compute
\[
  \bm z_i=\tilde{\bm D}_i^{-1}\bigl(\tilde{\bm y}_i-\bm U_i\bm z_{i+1}\bigr).
\]
Each operation involves $2\times2$ matrices only, so the whole solution costs $O(n)$ operations.

\section{Projection onto the simplex}\label{sec:projection}
In the warping step $f$ and $f'$ are fixed, and the warps are first moved by a gradient step, $\bm\gamma_{temp}=\bm\gamma_{current}-\eta\bm\nabla$, with $\bm\nabla=(\nabla_1,\dots,\nabla_{n-1})^\top$ given by Equation~\eqref{eq:gradient} and learning rate $\eta$. To ensure that the constraints $\bm\gamma_{new}\ge\bm0$ and $\sum_i\gamma_{new,i}\le M$ hold, the result is then projected onto the simplex by the following algorithm, due to \citet{DouEtal2008}.
\begin{enumerate}
	\item Let $\bm{u} = \bm{\gamma}_{temp}$ and set any negative components to 0.
	\item If $\sum u_i \le M$, return $\bm{u}$.
	\item Sort $\bm{u}$ in descending order: $u_1 \ge u_2 \ge \dots$.
	\item Compute the cumulative sum of the sorted $\bm{u}$: $w_k = \sum_{j=1}^k u_j$.
	\item Find the index $\rho$ that is the largest $k$ such that: $$u_k - \frac{w_k - M}{k} > 0$$
	\item Calculate the threshold $\theta$:$$\theta = \frac{w_\rho - M}{\rho}$$
	\item Final update: $\gamma_i = \max(u_i - \theta, 0)$ for $i=1, \ldots, n-1$.
\end{enumerate}
The cost is dominated by the sort, $O(n\log n)$ operations. The spline step and the warping step are repeated until the change in the variables falls below a tolerance level.

\section{Analytical scores of the Kalman-filter log-likelihood}\label{sec:kalman_scores}
Since model~(\ref{P-eq:stateSpace}) of the paper is estimated by exploiting the Kalman filter, in this section we provide the analytical scores of the (penalized) log-likelihood $\ell$ required to make the computations both efficient and precise. We also show that, unlike $\sigma$, the domain-warp parameters $\gamma_t$ require an additional score term beyond the standard formula, because they affect the transition matrix of the state-space form and not only its disturbance covariances.

\paragraph{Scores with respect to the log-likelihood} One can, in fact, adopt the solution presented in~\citet[result~3.2]{KoopShep:scores}. By adapting it to our scenario, given a generic component $\theta$ of the parameter vector $\bm\theta$ we can write
\begin{equation}\label{eq:koopShep}
	\frac{\partial\ell(\bm\theta)}{\partial\theta} = \frac{1}{2} \sum_{t=1}^n\biggl[\,\tr{\biggl(\!(\bm r_t\bm r_t^\top-\bm N_t)\frac{\partial\bm Q_t}{ \partial\theta}\biggr) + (e_t^2-d_t) \frac{\partial\bm H_t}{\partial\theta}}\,\biggr]
\end{equation}
where $\bm H_t$ is the variance matrix of the observation error, and $\bm Q_t$ is the variance matrix of the state error. Formula~(\ref{eq:koopShep}) is valid for any $\theta$ that enters $\bm H_t$ and/or $\bm Q_t$ \emph{only}, leaving the system matrices that drive the state recursion unaffected; we will return to this restriction below.

In our case, the parameter vector is defined as $$ \bm\theta :=
    \begin{bmatrix}
	\sigma & \gamma_1 & \cdots & \gamma_{n-1}
    \end{bmatrix}^\top,
$$ and we write the state and state-disturbance vectors as $\bm\alpha_t := [f(x_t)\ f'(x_t)]^\top$ and $\bm\eta_t := [\eta_t\ \zeta_t]^\top$, so that the state equation in~(\ref{P-eq:stateSpace}) of the paper reads $\bm\alpha_{t+1} = \bm T_t\bm\alpha_t + \bm\eta_t$ with observation matrix $\bm Z := [1\ 0]$ and transition matrix $$ \bm T_t :=
	\begin{bmatrix}
		1 & \tau_t \\
		0 & 1
	\end{bmatrix},
	\qquad \tau_t := \delta_t+\gamma_t. $$ The derivatives of the disturbance variance matrices $$ \bm H_t = \lambda\sigma^2,\qquad \bm Q_t = \sigma^2\bm\Delta_t,\quad \bm\Delta_t :=
    \begin{bmatrix}
        \frac13\tau_t^3 & \frac12\tau_t^2 \\
        \frac12\tau_t^2 & \tau_t \\
    \end{bmatrix}
$$ with respect to $\bm\theta$'s components can be computed promptly: for all $t$ and $s\ne t$ $$ \frac{\partial\bm H_t}{\partial\sigma} = 2\lambda\sigma,\qquad \frac{\partial\bm H_t}{\partial\gamma_t} = 0,\quad \frac{\partial\bm H_s}{\partial\gamma_t} = 0 $$ and $$ \frac{\partial\bm Q_t}{\partial\sigma} = 2\sigma\bm\Delta_t, \qquad \frac{\partial\bm Q_t}{\partial\gamma_t} = \sigma^2
    \begin{bmatrix}
        \tau_t^2 & \tau_t \\
        \tau_t & 1
    \end{bmatrix},\quad
    \frac{\partial\bm Q_s}{\partial\gamma_t} = \bm0. $$ Since $\sigma$ enters only $\bm H_t$ and $\bm Q_t$, formula~(\ref{eq:koopShep}) applies to it unchanged. However $\gamma_t$, through $\tau_t$, \emph{also} enters the transition matrix $\bm T_t$: $$ \frac{\partial\bm T_t}{\partial\gamma_t} = \dot{\bm T} := \begin{bmatrix}0&1\\0&0\end{bmatrix},\qquad \frac{\partial\bm T_s}{\partial\gamma_t} = \bm0 \quad (s\ne t), $$ so formula~(\ref{eq:koopShep}) by itself is \emph{incomplete} for $\gamma_t$: it must be supplemented with an extra term accounting for $\partial\bm T_t/\partial\gamma_t$.

\paragraph{The additional score term induced by the transition matrix} The Kalman filter likelihood is the marginal likelihood of $\bm y$ obtained after integrating out the latent states $\bm\alpha:=(\bm\alpha_1,\ldots,\bm\alpha_n)$ from the complete-data log-likelihood
\begin{equation*}
	\log p(\bm\alpha,\bm y;\bm\theta) = -\frac12\sum_{t=1}^n\Bigl[\log|\bm H_t| + \varepsilon_t^2/H_t\Bigr] -\frac12\sum_{t=1}^{n-1}\Bigl[\log|\bm Q_t| + \bm\eta_t^\top\bm Q_t^{-1}\bm\eta_t\Bigr] + \mathrm{const},
\end{equation*}
with $\varepsilon_t := y_t-\bm Z\bm\alpha_t$ and $\bm\eta_t = \bm\alpha_{t+1}-\bm T_t\bm\alpha_t$. By the Fisher (or Louis) identity that underlies all EM-type score computations, $$ \frac{\partial\ell(\bm\theta)}{\partial\theta} = \e\Bigl[\frac{\partial\log p(\bm\alpha,\bm y;\bm\theta)}{\partial\theta}\Bigm|\bm y\Bigr], $$ where the expectation is taken at the current value of $\bm\theta$. Formula~(\ref{eq:koopShep}) is precisely the result of applying this identity when $\theta$ affects only $\bm H_t,\bm Q_t$. For $\theta=\gamma_t$, only the summand at index $t$ depends on $\theta$, and expanding $\bm\eta_t^\top\bm Q_t^{-1}\bm\eta_t = \bm\alpha_{t+1}^\top\bm Q_t^{-1}\bm\alpha_{t+1} -2\bm\alpha_{t+1}^\top\bm Q_t^{-1}\bm T_t\bm\alpha_t+\bm\alpha_t^\top\bm T_t^\top\bm Q_t^{-1}\bm T_t\bm\alpha_t$ and differentiating with respect to $\gamma_t$ before taking the conditional expectation yields, after simplification,
\begin{equation}\label{eq:koopShepExt}
	\frac{\partial\ell(\bm\theta)}{\partial\gamma_t} = \underbrace{\frac12\tr\Bigl((\bm r_t\bm r_t^\top-\bm N_t)\frac{\partial\bm Q_t}{\partial\gamma_t}\Bigr)}_{\text{as given by~(\ref{eq:koopShep})}} \;+\; \underbrace{\tr\bigl(\bm Q_t^{-1}\dot{\bm T}\,\bm M_t\bigr)}_{\text{omitted by~(\ref{eq:koopShep})}},
\end{equation}
where $\bm M_t := \e[\bm\alpha_t\bm\eta_t^\top\mid\bm y] = \widetilde{\bm P}_{t+1,t}^\top - \widetilde{\bm P}_t\bm T_t^\top,$ $\widetilde{\bm P}_t := \e[\bm\alpha_t\bm\alpha_t^\top\mid\bm y],$ $\widetilde{\bm P}_{t+1,t} := \e[\bm\alpha_{t+1}\bm\alpha_t^\top\mid\bm y].$

The second term in~(\ref{eq:koopShepExt}) is exactly the contribution missed by treating $\gamma_t$ as if it only entered $\bm Q_t$.

To evaluate it in closed form we need the smoothed cross-moment $\bm M_t$, equivalently the smoothed cross-covariance $\Cov(\bm\alpha_t,\bm\eta_t\mid\bm y) = \bm M_t-\widehat{\bm\alpha}_t\widehat{\bm\eta}_t^\top$, where $\widehat{\bm\alpha}_t := \bm a_t+\bm P_t\bm r_{t-1}$ and $\widehat{\bm\eta}_t := \bm Q_t\bm r_t$ are the usual smoothed state and smoothed disturbance ($\bm a_t,\bm P_t$ being the one-step-ahead predicted mean and MSE matrix produced by the Kalman filter). Writing $\bm K_t$ for the Kalman gain and $\bm L_t := \bm T_t-\bm K_t\bm Z$ for the disturbance-smoothing matrix associated with the backward recursions for $\bm r_t,\bm N_t$, the smoothed cross-covariance admits the closed form
\begin{equation}\label{eq:crosscov}
	\Cov(\bm\alpha_t,\bm\eta_t\mid\bm y) = -\bm P_t\bm L_t^\top\bm N_t\bm Q_t,
\end{equation}
which can be obtained from the disturbance- and state-smoothing recursions (cf.\ the lag-one smoothed state covariance of \citealp{ShuSto1982}, here re-expressed without recourse to the filtered-variance sequence) and which we have additionally verified by direct computation against the exact, non-recursive Gaussian conditional covariance of the finite-sample model. Substituting $\bm M_t = \bigl(\widehat{\bm\alpha}_t\bm r_t^\top - \bm P_t\bm L_t^\top\bm N_t\bigr)\bm Q_t$ -- obtained from~(\ref{eq:crosscov}) together with $\widehat{\bm\eta}_t=\bm Q_t\bm r_t$ -- into the second term of~(\ref{eq:koopShepExt}), the factors $\bm Q_t$ and $\bm Q_t^{-1}$ cancel by the cyclic property of the trace: $$ \tr\bigl(\bm Q_t^{-1}\dot{\bm T}\,\bm M_t\bigr) = \tr\Bigl(\dot{\bm T}\bigl(\widehat{\bm\alpha}_t\bm r_t^\top - \bm P_t\bm L_t^\top\bm N_t\bigr)\Bigr) = \widehat\alpha_t^{(2)}r_t^{(1)} - \bigl(\bm P_t\bm L_t^\top\bm N_t\bigr)_{21}, $$ the last equality following from $\dot{\bm T}$ having a single nonzero entry in position $(1,2)$, which selects the $(2,1)$ entry of the matrix it multiplies. No new filter or smoother recursion is required: every quantity above ($\bm P_t,\bm L_t,\bm N_t,\bm r_t,\widehat{\bm\alpha}_t$) is already produced by the Kalman filter and smoother used for $\partial\ell/\partial\sigma$.

\paragraph{The complete score for $\gamma_t$} Granted the notation $$ \bm{r}_t =:
    \begin{bmatrix}
        r_t^{(1)} \\
        r_t^{(2)}
    \end{bmatrix},
    \;\; \bm{N}_t =:
    \begin{bmatrix}
        n_t^{(11)} & n_t^{(12)} \\
        n_t^{(12)} & n_t^{(22)}
    \end{bmatrix},
    \;\; \bm a_t =:
    \begin{bmatrix}
        a_t^{(1)} \\
        a_t^{(2)}
    \end{bmatrix},
    \;\; \bm P_t =:
    \begin{bmatrix}
        p_t^{(11)} & p_t^{(12)} \\
        p_t^{(12)} & p_t^{(22)}
    \end{bmatrix},
$$ and writing $k_t^{(1)},k_t^{(2)}$ for the two components of the Kalman gain $\bm K_t$ (so that $\bm L_t = \begin{bsmallmatrix}1-k_t^{(1)} & \tau_t\\ -k_t^{(2)} & 1\end{bsmallmatrix}$), the score with respect to $\sigma$ is unaffected by the above and remains as in~(\ref{eq:koopShep}), while the complete score with respect to $\gamma_t$ combines the term already given by~(\ref{eq:koopShep}) with the new term derived above:
\begin{equation*}
	\begin{aligned}
		\frac{\partial\ell(\bm\theta)}{\partial\sigma}
        &=
        \sigma\sum_{t=1}^n \Bigl[\bigl(r_t^{(1)}r_t^{(1)}-n_t^{(11)}\bigr)\frac13\tau_t^3 +\bigl(r_t^{(1)}r_t^{(2)}-n_t^{(12)}\bigr)\tau_t^2 \;+ \\
        &
        \qquad\qquad\quad + \bigl(r_t^{(2)}r_t^{(2)}-n_t^{(22)}\bigr) \tau_t\Bigr] + \lambda\sigma\sum_{t=1}^n (e_t^2-d_t) \\[4pt]
        \frac{\partial\ell(\bm\theta)}{\partial\gamma_t}
        &=
        \underbrace{\frac{\sigma^2}{2} \Bigl[\bigl(r_t^{(1)}r_t^{(1)}-n_t^{(11)}\bigr) \tau_t^2 + 2\bigl(r_t^{(1)}r_t^{(2)}-n_t^{(12)}\bigr) \tau_t + \bigl(r_t^{(2)}r_t^{(2)} -n_t^{(22)}\bigr)\Bigr]}_{\text{term already in~(\ref{eq:koopShep})}} \\
        &\qquad + \underbrace{\widehat\alpha_t^{(2)}r_t^{(1)} - p_t^{(12)}\Bigl[\bigl(1-k_t^{(1)}\bigr)n_t^{(11)} - k_t^{(2)}n_t^{(12)}\Bigr] - p_t^{(22)}\Bigl[\tau_t\, n_t^{(11)} + n_t^{(12)}\Bigr]}_{\text{new term, from }\partial\bm T_t/\partial\gamma_t}\\
        &\forall \, t=1, \dots, n-1,
	\end{aligned}
\end{equation*}
where $\widehat\alpha_t^{(2)} = a_t^{(2)} + p_t^{(12)} r_{t-1}^{(1)} + p_t^{(22)} r_{t-1}^{(2)}$ is the smoothed slope at $x_t$.

\paragraph{Scores with respect to the constraint function} Since the constraint function is defined as $$ g(\bm\theta) := \sum_{t=1}^{n-1} \gamma_t $$ its derivatives with respect to $\bm\theta$'s components are trivial: $$ \frac{\partial g(\bm\theta)}{\partial\sigma} = 0, \qquad \frac{\partial g(\bm\theta)}{\partial\gamma_t} = 1 \quad \forall\, t=1,\dots,n-1 $$

\paragraph{With estimation of lambda} By model parameterization we have $\e\,[\varepsilon_t\varepsilon_t^\top] := \lambda\sigma^2$. We can also directly define the variance of the error term $\varepsilon_t$: $$ \sigma_\varepsilon^2 := \e\,[\varepsilon_t\varepsilon_t^\top] $$ This allows us to define the vector of parameters to be estimated as $$ \bm\theta :=
    \begin{bmatrix}
	\sigma & \sigma_\varepsilon & \gamma_1 & \cdots & \gamma_{n-1}
    \end{bmatrix}^\top,
$$ This means that one can now carry out the estimation without a fixed value for $\lambda$. Such estimation returns the estimate for $\sigma_\varepsilon$ and then (if necessary) one can estimate $\lambda$ by $$ \hat\lambda = \frac{\hat\sigma_\varepsilon^2}{\hat\sigma^2} $$

In order to perform the estimation in question, the derivatives of the disturbance variance matrices must be computed. The only change affecting these matrices with respect to the previous case regards $\bm H_t$, which is now defined as $$ \bm H_t = \sigma_\varepsilon^2 $$ Consequently, the derivatives are almost unaffected: for all $t$ and $s\ne t$ $$ \frac{\partial\bm H_t}{\partial\sigma} = 0,\qquad \frac{\partial\bm H_t}{\partial\sigma_\varepsilon} = 2\sigma_\varepsilon,\qquad \frac{\partial\bm H_t}{\partial\gamma_t} = 0,\quad \frac{\partial\bm H_s}{\partial\gamma_t} = 0 $$ and $$ \frac{\partial\bm Q_t}{\partial\sigma} = 2\sigma\bm\Delta_t,\qquad \frac{\partial\bm Q_t}{\partial\sigma_\varepsilon} = \bm0, \qquad \frac{\partial\bm Q_t}{\partial\gamma_t} = \sigma^2
    \begin{bmatrix}
        \tau_t^2 & \tau_t \\
        \tau_t & 1
    \end{bmatrix},\quad
    \frac{\partial\bm Q_s}{\partial\gamma_t} = \bm0 $$ As above, $\sigma_\varepsilon$ enters only $\bm H_t$, so its score follows directly from~(\ref{eq:koopShep}); $\gamma_t$ still requires the additional term derived above, which is unaffected by the reparameterization of $\bm H_t$. The gradient turns out to be
\begin{equation*}
	\begin{aligned}
		\frac{\partial\ell(\bm\theta)}{\partial\sigma}
        &=
        \sigma\sum_{t=1}^n \Bigl[\bigl(r_t^{(1)}r_t^{(1)}-n_t^{(11)}\bigr)\frac13\tau_t^3 +\bigl(r_t^{(1)}r_t^{(2)}-n_t^{(12)}\bigr)\tau_t^2 \;+ \\
        &
        \qquad\qquad\quad + \bigl(r_t^{(2)}r_t^{(2)}-n_t^{(22)}\bigr) \tau_t\Bigr] \\[4pt]
        \frac{\partial\ell(\bm\theta)}{\partial\sigma_\varepsilon}
        &= \sigma_\varepsilon\sum_{t=1}^n (e_t^2-d_t) \\[4pt]
        \frac{\partial\ell(\bm\theta)}{\partial\gamma_t}
        &=
        \underbrace{\frac{\sigma^2}{2} \Bigl[\bigl(r_t^{(1)}r_t^{(1)}-n_t^{(11)}\bigr) \tau_t^2 + 2\bigl(r_t^{(1)}r_t^{(2)}-n_t^{(12)}\bigr) \tau_t + \bigl(r_t^{(2)}r_t^{(2)} -n_t^{(22)}\bigr)\Bigr]}_{\text{term already in~(\ref{eq:koopShep})}} \\
        &\qquad + \underbrace{\widehat\alpha_t^{(2)}r_t^{(1)} - p_t^{(12)}\Bigl[\bigl(1-k_t^{(1)}\bigr)n_t^{(11)} - k_t^{(2)}n_t^{(12)}\Bigr] - p_t^{(22)}\Bigl[\tau_t\, n_t^{(11)} + n_t^{(12)}\Bigr]}_{\text{new term, from }\partial\bm T_t/\partial\gamma_t} \\
        & \forall \, t=1, \dots, n-1
	\end{aligned}
\end{equation*}
and $$ \frac{\partial g(\bm\theta)}{\partial\sigma} = 0, \qquad \frac{\partial g(\bm\theta)}{\partial\sigma_\varepsilon} = 0, \qquad \frac{\partial g(\bm\theta)}{\partial\gamma_t} = 1 \quad \forall\, t=1,\dots,n-1 $$

\section{Repeated values of the covariate}\label{sec:ties}
The construction of Section 2 of the paper is stated for distinct design points,
and a great deal of it stops working if they are not; yet ties are common in
practice, for instance when the covariate is recorded at a coarse resolution, as
the temperatures of Section 4.1 of the paper are. Write the sample as
$(x_i, y_{ij})$, $j = 1, \dots, m_i$, $i = 1, \dots, n$, where
$x_1 < x_2 < \cdots < x_n$ are the \emph{distinct} values of the covariate,
$m_i \ge 1$ is the number of replicates at $x_i$, and $N = \sum_i m_i$ is the
total number of data points. Each may carry a positive weight $w_{ij}$, a point
of weight $w_{ij}$ being treated as the mean of $w_{ij}$ independent
observations; unit weights are the default. In this section $n$ is the number of
distinct values, so that every $\delta_i = x_{i+1} - x_i$ is positive.

\paragraph{Why ties must be removed.}
The warping is indexed by the intervals $\delta_i = x_{i+1} - x_i$ between
consecutive design points, and a tie makes one of them vanish. Two distinct
things then break. Mechanically, the latent length $\tau_i = \delta_i + \gamma_i$
of such an interval is zero whenever it carries no warp, which at the optimum is
the case for all but a few intervals. The state-space form (2) of the paper then
has $\operatorname{Var}(\eta_i) = \sigma^2\tau_i^3/3$,
$\operatorname{Cov}(\eta_i,\zeta_i) = \sigma^2\tau_i^2/2$ and
$\operatorname{Var}(\zeta_i) = \sigma^2\tau_i$, all of which are zero, so the
transition is degenerate; and the stiffness matrix $\boldsymbol{K}_i$ of Section
A1, proportional to $\tau_i^{-3}$, and the gradient (A3) of the warping step are
not defined. More fundamentally, the model is not identified there: a warp placed
on an interval of zero length is a discontinuity between two responses recorded
at the \emph{same} value of the covariate, and no feature of the data can say
which replicate lies on which side of it. Collapsing ties is therefore not a
computational convenience but the removal of coordinates that carry no
information.

\paragraph{The reduction is exact.}
Let $W_i = \sum_j w_{ij}$ be the total weight at $x_i$,
$\bar y_i = W_i^{-1} \sum_j w_{ij} y_{ij}$ the weighted group mean, and
\begin{equation}
  \mathcal{S}_i = \sum_{j=1}^{m_i} w_{ij}\,(y_{ij} - \bar y_i)^{2},
  \qquad
  \mathcal{S} = \sum_{i=1}^{n} \mathcal{S}_i,
  \label{eq:A8}
\end{equation}
the within-group scatter, written $\mathcal{S}$ to distinguish it from the sums
$S_i$ of Section A1 and from the selection matrix $\boldsymbol{S}$ of Section A2.
For every $i$ the elementary identity
$\sum_j w_{ij}(y_{ij} - f_i)^2 = W_i(\bar y_i - f_i)^2 + \mathcal{S}_i$ holds, so
that the weighted version of the objective function (1) of the paper splits as
\begin{equation}
  \sum_{i=1}^{n}\sum_{j=1}^{m_i} w_{ij}\bigl(y_{ij} - f(x_i)\bigr)^{2}
  + \lambda\sum_{i=1}^{n-1} k_i
  \;=\;
  \underbrace{\sum_{i=1}^{n} W_i\bigl(\bar y_i - f(x_i)\bigr)^{2}
  + \lambda\sum_{i=1}^{n-1} k_i}_{\text{$n$ weighted observations}}
  \;+\; \mathcal{S},
  \label{eq:A9}
\end{equation}
with $k_i$ the energy of interval $i$ (Equation (3) of the paper). The penalty is
untouched, because it depends on $f$, on the warps and on the distinct $x_i$
alone, and $\mathcal{S}$ involves neither $f$ nor $\boldsymbol{\gamma}$. The pair
$(\bar y_i, W_i)$ is therefore sufficient for the signal: at a fixed smoothing
constant, the weighted fit on the $n$ collapsed points is \emph{identical} to the
fit on the $N$ raw ones, not an approximation to it.

In the state-space representation the weight enters through the observation
variance, $H_i = \lambda\sigma^2/W_i$, which is to say that a point of weight
$W_i$ is treated as the mean of $W_i$ independent observations. In the spline
step of Section A2 it enters through the data term
$\sum_i W_i(\bar y_i - f(x_i))^2$: the selection matrix $\boldsymbol{S}$ is
replaced by $\boldsymbol{S}_W = \operatorname{diag}(W_1,0,W_2,0,\dots,W_n,0)$,
and system (A4) becomes
\begin{equation*}
  (\boldsymbol{S}_W + \lambda\boldsymbol{R})\,\boldsymbol{z}
  = \boldsymbol{S}_W\,\bar{\boldsymbol{y}}^{*},
\end{equation*}
with $\bar{\boldsymbol{y}}^{*}$ the vector of size $2n$ carrying $\bar y_i$ in
position $2i-1$ and zeros elsewhere. The system is still block tridiagonal, and
of order $2n$ with $n$ the number of distinct values, so the solution of Section
A2.1 applies unchanged. So does the warping step, since the energies $k_i$ and
their gradient depend on the distinct $x_i$ only.

\paragraph{What the reduction discards, and when it matters.}
Equation~\eqref{eq:A9} says that $\mathcal{S}$ is irrelevant to $f$. It does
\emph{not} say that $\mathcal{S}$ is irrelevant. The within-group scatter is not
ancillary for the observation variance $\sigma_\varepsilon^2 = \lambda\sigma^2$:
it is an estimate of $\sigma_\varepsilon^2$ in its own right, carrying
\begin{equation}
  \nu = \sum_{i=1}^{n} (m_i - 1) = N - n
  \label{eq:A10}
\end{equation}
degrees of freedom, one for every data point beyond the first at each site.
Should the collapsed data be supplied directly, with $\mathcal{S}$ computed from
the underlying raw observations rather than from the points passed to the
fitter, $\nu$ counts those observations instead and becomes $\sum_i W_i - n$.

The consequence is a distinction between the two kinds of fitter. Where the fit
does not involve the observation variance, as in the alternating algorithm at a
supplied $\lambda$, there is nothing to restore and the collapsed problem is the
whole problem. Where $\sigma_\varepsilon^2$ is estimated by maximum likelihood,
as in SSJ--MLE, the log-likelihood of the collapsed data is augmented by
\begin{equation}
  -\tfrac{1}{2}\Bigl(\nu \log \sigma_\varepsilon^{2}
  + \mathcal{S}/\sigma_\varepsilon^{2}\Bigr),
  \label{eq:A11}
\end{equation}
together with the corresponding constant, so that the reported likelihood is
that of all the data points and $\hat\sigma_\varepsilon^2$ uses the within-group
scatter as well as the between-group residuals. In the scores of Section A4,
$\boldsymbol{H}_t$ becomes $\lambda\sigma^2/W_t$, or $\sigma_\varepsilon^2/W_t$
when $\lambda$ is estimated, so that the terms in $e_t^2 - d_t$ are divided by
$W_t$; and the correction~\eqref{eq:A11} adds
$-\nu/\sigma_\varepsilon + \mathcal{S}/\sigma_\varepsilon^{3}$ to the score with
respect to $\sigma_\varepsilon$, or $-\nu/\sigma + \mathcal{S}/(\lambda\sigma^3)$
to the score with respect to $\sigma$ when $\lambda$ is fixed.

SSJ--KFS and the fixed-variance fitter do not apply~\eqref{eq:A11}: they report
the likelihood of the collapsed means, and their $\hat\sigma_\varepsilon^2$ is
formed from the between-group residuals alone. Such an estimate remains
consistent as $n$ grows, since the collapsed means have variance
$\sigma_\varepsilon^2/W_i$ and the weighted filter identifies
$\sigma_\varepsilon^2$ from them; but it throws away $\nu$ data points' worth of
information, and under heavy replication, with many replicates at each of
moderately many sites, that is most of the information about
$\sigma_\varepsilon^2$ in the sample. A log-likelihood reported by SSJ--MLE and
one reported by SSJ--KFS are therefore on different footings whenever ties are
present, and are not comparable as they stand.

\paragraph{Consequences for the selection criterion.}
Every count entering the criteria is read off the collapsed problem. The
candidate locations for a discontinuity are the $n-1$ intervals between distinct
values of the covariate, so the dimension of the model space entering the
extended BIC (5) of the paper is $p = n-1$, with $n$ the number of distinct
values and not $N$, and the extended term is $2\xi k \log(n-1)$. The sample size
whose logarithm multiplies the degrees of freedom is $n$ likewise. The
consistency floor of Chen and Chen (2008) is therefore
$1 - \log n / \{2 \log(n-1)\}$, the value of about $0.5$ recalled in Section 2.4
of the paper, unchanged by the presence of replicates.

The effective degrees of freedom are the trace of the $n \times n$ weighted
smoothing matrix, which equals that of the $N \times N$ smoothing matrix of the
raw observations; in the alternating algorithm it is
$\sum_i W_i \{(\boldsymbol{S}_W + \lambda\boldsymbol{R})^{-1}\}_{2i-1,2i-1}$.
The residual sum of squares $\sum_i W_i(\bar y_i - \hat f_i)^2$ that the
criteria are built on holds only between-point variation, $\mathcal{S}$ having
been removed by the collapse, so it carries $n - \mathrm{edf}$ degrees of
freedom and the residual variance is estimated by
$\sum_i W_i(\bar y_i - \hat f_i)^2 / (n - \mathrm{edf})$ --- one degree of
freedom per fitted point, not one per data point. Fitted values are returned on
the $n$ distinct, sorted values of the covariate, so a fit and the data it came
from have different lengths whenever ties are present, and any comparison of the
two has to be made against $\bar y_i$.

\paragraph{Cross-validation.}
Held-out error is weighted in the same way as the fit, so that a fold is scored
by the observations it actually represents: the cross-validation criterion of the
Calibration Supplement becomes
\begin{equation*}
  \mathrm{CV}(\lambda, M) =
  \frac{\sum_{k=1}^{K} \sum_{i \in \mathcal{F}_k}
        W_i\bigl(\bar y_i - \hat f_{\lambda,M}^{(-k)}(x_i)\bigr)^{2}}
       {\sum_{k=1}^{K} \sum_{i \in \mathcal{F}_k} W_i}.
\end{equation*}
Replication requires a choice about the partition that is easy to overlook. If
folds are drawn over the $N$ data points, replicates of the same site land on
both sides of the split, and a held-out point is then evaluated at a covariate
value that the training sample also contains: the criterion estimates the error
of predicting a fresh observation at an \emph{already observed} $x$, and it is
optimistic as an estimate of the error at a new one. If folds are drawn over the
$n$ distinct sites, all the replicates at a site are held out together and the
criterion estimates the latter; it then differs from the held-out error of the
raw observations only by the constant $\mathcal{S}/\sum_i W_i$, and has the same
minimiser. The second is the relevant target when $\lambda$ is being chosen to
recover the signal, and it is what we use; the distinction is immaterial when the
design has no ties.

\bibliographystyle{agsm}
\bibliography{ssj}


\maketitle

\noindent \textit{This document reports the tuning of the smoothing constant $\lambda$ and of the warping budget $M$ and, in full, the calibration of the extended-BIC constant $\xi$, all summarized in Section~\ref{P-sec:tuning} of the paper: the cross-validation of $\lambda$ and the reasons for selecting $M$ by an information criterion, the data generating process and its three scenarios in closed form, the design and the criteria, every table and figure, and the numbers behind every reading. The experiment rests on $200$ replications per design cell; the code that computes it and the code that produces every table and figure of this document are in the replication package.}
\bigskip

\tableofcontents
\newpage

\section{Tuning the smoothing constant and the warping budget}\label{sec:tuning}
This section gives the details of the tuning of $\lambda$ and $M$ summarized in Section~\ref{P-sec:tuning} of the paper; the rest of the document is devoted to the constant $\xi$ of the extended BIC.

\paragraph{An information criterion for the budget.} In the literature and in the applications of classical cubic smoothing splines the smoothing constant is generally determined by cross-validation, and with the state-space form it can also be estimated by Gaussian maximum likelihood. Cross-validation is not a valid way to select the warping budget $M$, however: given the small number of discontinuities expected in the data, most held-out subsamples reveal none of them. $M$ must therefore be chosen by an information criterion, and with a number of candidate parameters of the same order as the number of observations the classical criteria are, as \citet{CheChe2008} point out, ``usually too liberal''.

Selecting the budget is not a choice of model dimension. A budget admits some number $k$ of positive warps, but it also decides \emph{which} of the $n-1$ inter-observation intervals carry them, and the estimator arrives at that subset by searching over all of them. The relevant model space at size $k$ therefore has $\binom{n-1}{k}$ members and grows with the sample: the number of candidate locations is $p=n-1$, so $p\asymp n$. The Akaike and Bayesian criteria charge only for dimension. Both have the form $-2\hat\ell+c(n)\,\mathrm{df}$, with $c(n)=2$ and $c(n)=\log n$ respectively, and neither term depends on how many configurations were examined before the reported one was chosen. That is harmless when the model space is fixed, because the omitted cost is then a constant, but it is not harmless here. The best-fitting arrangement of $k$ warps among $\binom{n-1}{k}$ possibilities fits appreciably better than a $k$-warp arrangement nominated in advance, and the gap widens with $p$; a criterion blind to it reads the selection gain as evidence and keeps adding discontinuities. Formally, the consistency of the Bayesian criterion rests on $p$ being fixed, and it is lost once $p$ grows with $n$. \citet{CheChe2008} restore it by replacing the prior that is uniform over models with one that is uniform over models \emph{of the same size}, which adds the log-cardinality of the class searched. With $\log\binom{n-1}{k}\approx k\log(n-1)$ for small $k$, this yields the extended BIC of Equation~\eqref{eq:ebic_app} in Section~\ref{sec:xi_what}, whose additional term is a charge for \emph{where} the discontinuities were placed, as distinct from how many there are.

Two remarks on our use of these results, that $\xi$ is calibrated on a design separate from the one used for evaluation and that calibrated values below the nominal consistency floor do not contradict the theory, because the $\mathrm{df}\log n$ term already pays part of the search cost, are developed in Section~\ref{sec:xi_what} and Section~\ref{sec:charge}. A third concerns the way $\lambda$ is supplied, which also decides how much the extended term has left to charge. When SSJ--KFS or SSJ--ATW take $\lambda$ from a preliminary fit, one parameter of the model has been fixed before the budget is chosen: the fit is stiffer than one in which $\lambda$ is free and re-estimated at every budget, part of its degrees of freedom have already been given up, and the search over the budget can buy less, so the additional penalty governed by $\xi$ has to be milder. The calibration confirms this in every scenario (Sections~\ref{sec:res_base} and \ref{sec:concl}).

\paragraph{Cross-validation of the smoothing constant.} In the implementation based on the state-space form, SSJ--MLE, $\lambda$ is estimated together with the other parameters by maximum likelihood and the budget is selected by minimizing the extended BIC on a grid of values of $M$. SSJ--KFS and SSJ--ATW instead take the smoothing constant as given. A simple way of supplying a good value is a preliminary SSJ--MLE fit, but this makes the faster estimator depend on the slower one it is meant to approximate; the following $K$-fold cross-validation removes that dependence at a comparable cost. Write $\hat f_{\lambda,M}^{(-k)}$ for the fit obtained from the observations outside fold $k$, at smoothing constant $\lambda$ and warping budget $M$. With the $n$ observations split into $K$ folds $\mathcal F_1,\dots,\mathcal F_K$ of nearly equal size, which partition the sample, the cross-validation criterion on a grid $\Lambda\times\mathcal M$ of candidate pairs is
\begin{equation}\label{eq:cv}
  \mathrm{CV}(\lambda,M)\;=\;\frac1n\sum_{k=1}^K\sum_{i\in\mathcal F_k}\bigl(y_i-\hat f_{\lambda,M}^{(-k)}(x_i)\bigr)^2 .
\end{equation}
Held-out points falling outside the range of their training sample are evaluated at the nearest training abscissa rather than by extrapolation: beyond the data the fit is only extended linearly, and one or two boundary points would otherwise dominate a fold. The estimator is then run on the full sample at $\hat\lambda=\arg\min_\lambda\min_M\mathrm{CV}(\lambda,M)$, and the budget is chosen afresh by the extended BIC.

The minimization over $M$ in Equation~\eqref{eq:cv} is a profiling step, not a selection: its argument is discarded. This asymmetry is deliberate and reflects what the two constants do. The smoothing constant $\lambda$ governs a global bias--variance trade-off to which every held-out observation contributes, and there cross-validation is competitive with maximum likelihood: on the design of Section~\ref{P-sec:simulations} of the paper the two choices of $\lambda$ agree closely and the resulting root mean squared errors (RMSEs) differ by $1\%$ to $4\%$. The budget $M$ governs a sparse, local feature. Only the observations lying in the transition regions carry information about whether it is right, so the effective sample size for that decision grows with the number of discontinuities and not with $n$; and prediction error is nearly indifferent to a few spurious warps, since each displaces the fitted curve over a short stretch of the domain. Selecting $M$ by Equation~\eqref{eq:cv} therefore declares markedly more discontinuities than the information criterion at the same accuracy, which is why it is not used for that purpose.

The folds can be drawn at random or as contiguous blocks. To respect the temporal sequence of the observations, the blocked scheme holds out contiguous stretches of the series and so preserves its temporal structure \citep{OttMarFas2024}. With serially dependent errors this matters: a held-out observation whose neighbors remain in the training sample is nearly predicted by them, the random-fold criterion is then minimized by a fit that all but interpolates, and the selected $\lambda$ collapses (Section~\ref{sec:res_errors}). The simulations of Section~\ref{P-sec:simulations} of the paper use $K=5$ random folds and the applications of Section~\ref{P-sec:applications} use $K=20$; with contiguous blocks $K=20$ is preferable, since holding out a fifth of the series in one stretch over-smooths (Section~\ref{sec:res_errors}).

\paragraph{Grids and computational cost.} The smoothing constant is not scale-free: it carries the units of $x^3$, because the roughness energy of an interval has the units of $y^2/x^3$ while the fidelity term of the objective function has the units of $y^2$ (the reduced form of the energy, which shows it, is derived in the Computational Supplement). The default grid is therefore specified for the scale-free ratio $\lambda/(x_{(n)}-x_{(1)})^3$, nine logarithmically spaced values from $10^{-6}$ to $10^{-2}$, which are multiplied by $(x_{(n)}-x_{(1)})^3$ before use. The warps $\gamma_i$, and with them the budget $M$, are lengths of the domain, so the grid of budgets is expressed in the units of $x$ and is the same one used for the final selection. The procedure costs $K|\Lambda||\mathcal M|$ fits, which is affordable only because the spline step of the alternating algorithm solves a banded system in $O(n)$ operations; within each fold and each $\lambda$ the budgets are traversed in increasing order, each fit warm-started from the previous one.

\section{The constant}\label{sec:xi_what}

All three estimators that carry an information criterion, namely SSJ--MLE, SSJ--KFS and SSJ--ATW, select the warping budget $M$ by the extended Bayesian information criterion of \citet{CheChe2008},
\begin{equation}\label{eq:ebic_app}
  \mathrm{EBIC}(\xi) \;=\; \mathrm{df}\,\log(n) \;-\; 2\hat\ell
  \;+\; 2\,\xi\,k\,\log(n-1),
\end{equation}
with $\hat\ell$ the maximized log-likelihood, $\mathrm{df}$ the effective degrees of freedom, $k$ the number of active warps and $n-1$ the number of candidate discontinuity locations. The strength $\xi\in[0,1]$ is a free constant.

Three considerations rule out the obvious procedure of sweeping a grid of $\xi$, reporting the minimizer and adopting it:
\begin{enumerate}
\item If $\xi$ were chosen on the design of Section~\ref{P-sec:simulations} of the paper and that same design were then used to report performance, the comparison against CSSD would be contaminated: CSSD's constants are cross-validated \emph{within} each replication, ours would have been chosen by inspecting the truth across all of them. The calibration design must be separate from the evaluation design.
\item The root mean squared error (RMSE) as a function of $\xi$ is nearly flat near its minimum, so a point minimizer is largely Monte Carlo noise. A rule is needed that reports what the design resolves rather than the location of a shallow argmin.
\item The consistency condition of \citet{CheChe2008} for $p_n\asymp n^{\kappa}$ candidates is $\xi\ge 1-1/(2\kappa)$. Here the candidates are the $n-1$ inter-observation intervals, so $\kappa=1$ and the nominal floor is $\xi\ge 1-\log(n)/\{2\log(n-1)\}\approx 0.50$. Every value calibrated below lies under that floor, and the discrepancy has to be explained rather than left for the reader to notice.
\end{enumerate}

The three issues are addressed as follows. The first is met by the design: $\xi$ is calibrated on a data generating process of its own, described in Section~\ref{sec:xi_design} and never used to evaluate the estimators, so that the comparison with CSSD in Section~\ref{P-sec:simulations} of the paper is made with a constant fixed in advance. The second is met by the criteria of Section~\ref{sec:xi_rules}: the point minimizer is replaced by two minimax regret rules, whose output is the value of $\xi$ whose worst regret over the design cells is smallest, and every calibrated value is reported with the spread of the cell-wise optima it summarizes, so that a flat valley is visible as such. The third is met in Section~\ref{sec:charge}. $\mathrm{df}$ is the trace of the smoothing matrix, and that trace itself grows with every activated warp, so the $\mathrm{df}\log n$ term of Equation~\eqref{eq:ebic_app} already charges part of the cost of each declared discontinuity. What the theory constrains is the total charge per discontinuity; the extended term supplies only what is left of it, and a calibrated $\xi$ below the nominal floor is consistent with the theory once the part levied through $\mathrm{df}$ is counted.

\section{Data generating process and scenarios}\label{sec:xi_design}

The constant is calibrated by simulation because 1) the data do not identify $\xi$, since it enters only the selection among fits and never a fit; 2) the theory of \citet{CheChe2008} supplies a floor rather than a value; and 3) only on simulated series are the two quantities a choice of $\xi$ has to be scored against, the true signal and the true number of discontinuities, available. The design has to cover the kinds of data the estimators will meet, and it does so with one data generating process and two departure parameters, run as three scenarios.

Scenario 1 is the common boundary of the other two and is reached by the same code. Scenario 2 describes series whose \emph{errors} are persistent and Scenario 3 series whose \emph{level} accumulates shocks; the fuel prices of Section~\ref{P-sec:carburanti} of the paper have both, so the two bracket the answer rather than either supplying it.

The three experiments differ only in two parameters of one data generating process. With $b$ a fixed smooth background, $r$ a draw from an integrated Wiener process and $e$ the errors,

\dgpbox{%
\begin{equation}\label{eq:dgp_all}
  \begin{aligned}
    x_{(1)}\le\cdots\le x_{(n)} \quad&\text{order statistics of $n$ draws from } \mathrm U(0,1),\\[2pt]
    c_j\sim\mathrm U(0.12,0.88),\quad
    s_j &= \varsigma_j\upsilon_j,\quad
    \varsigma_j\sim\mathrm U\{-1,1\},\ \upsilon_j\sim\mathrm U(0.30,0.70),\\[2pt]
    b_i \;=\; \sin(2\pi x_i)&+\tfrac12\sin(5\pi x_i),
    \qquad
    \begin{bmatrix} r_{i+1}\\ r'_{i+1}\end{bmatrix}
    =\begin{bmatrix}1&\delta_i\\0&1\end{bmatrix}
     \begin{bmatrix} r_i\\ r'_i\end{bmatrix}+\bm\nu_i,\\[2pt]
    \bm\nu_i&\sim\mathrm N\!\left(\bm 0,
      \begin{bmatrix}\delta_i^3/3&\delta_i^2/2\\ \delta_i^2/2&\delta_i\end{bmatrix}\right),
      \qquad (r_1,r'_1)=(0,0),\\[2pt]
    g_i \;=\; \sqrt{1-\omega}\;\frac{b_i-\bar b}{\sd(b)}
    \;&+\;\sqrt{\omega}\;\frac{r_i-\bar r}{\sd(r)},
    \qquad
    f_i \;=\; \frac{g_i-\bar g}{\sd(g)},\\[2pt]
    e_1=\sigma u_1,\qquad e_{i+1}&=\phi_i e_i+\sqrt{1-\phi_i^{2}}\,\sigma u_{i+1},
    \qquad \phi_i=\psi^{\,\delta_i/\bar\delta},\\[2pt]
    y_i &= f_i \;+\; \sum_{j=1}^{q}s_j\mathbf 1_{[c_j,1]}(x_i)\;+\;e_i,
  \end{aligned}
\end{equation}
}
with $u_i$ independent standard normal, $\delta_i=x_{i+1}-x_i$ and $\bar\delta=(x_n-x_1)/(n-1)$.

\noindent The abscissas $x$ are uniform draws, sorted, so the spacing is irregular (thus consistent with both cross-sectional and continuous-time objects). At the same increment $\delta_i$ enters the transition of the simulated error and the transition matrix of the state-space form of Section~\ref{P-sec:statespace} of the paper.

\noindent Discontinuity positions and signed sizes are redrawn every replication, kept away from the edges so that a break is a detection problem rather than a boundary problem, and bounded away from zero so that a miss is a failure of the method rather than of the signal-to-noise ratio. Signal, discontinuities and errors are additive in the last line of Equation~\eqref{eq:dgp_all}, and the discontinuities enter as steps that persist to the end of the sample, so the size $s_j$ is the level shift at $c_j$ and not a transient.

\noindent The covariance $[\delta^3/3,\ \delta^2/2;\ \delta^2/2,\ \delta]$ is that of an integrated Wiener process over a gap $\delta$, which by the correspondence of \citet{Wah1978} is the prior the smoothing spline itself places over the signal.

\noindent The signal $g$ is a weighted linear combination of the deterministic component $b$ with weight $\sqrt{1-\omega}$ and the integrated Wiener stochastic process $r$ with weight $\sqrt{\omega}$. Since the weights are the square roots of two shares that sum to one, $\omega\in[0,1]$ is the share of the signal's variance carried by the stochastic trend: at $\omega=0$ the signal is the fixed curve and only the jumps are random, at $\omega=1$ the curve is replaced entirely by a path of the process, and in between the signal has both a recurring shape and a level that wanders. The two components are independent by construction, $b$ being a fixed function of $x$ and $r$ being driven by innovations that do not depend on it.

\noindent The signal is standardized twice. The first standardization, of $b$ and $r$ separately, makes $\omega$ a share of variance: the two components have no common scale (i.e., $b$ is a fixed sinusoid, $r$ a random walk in the slope whose variance grows along the sample and with $n$) so raw weights would confound the mixing proportion with the accident of their relative sizes. The second standardization acts on the mixture $g$ and holds the amplitude constant across $\omega$: $\sd(g)$ is not $1$ unless the two parts happen to be uncorrelated in the realized sample. It matters because the jump sizes $s_j$ and the noise level $\sigma$ are not rescaled, so without it the jump-to-signal and signal-to-noise ratios would drift with $\omega$ and a change in the calibrated $\xi$ could not be attributed to the departure rather than to the jumps having become relatively smaller.

\noindent In the error recursion the two coefficients do separate jobs. The factor $\phi_i=\psi^{\delta_i/\bar\delta}$ is the correlation carried across a gap of length $\delta_i$, equal to $\psi$ exactly at the mean spacing, larger for closer pairs and smaller for farther ones; correlation therefore decays in distance. The factor $\sqrt{1-\phi_i^2}$ holds the marginal variance fixed: by induction $\mathrm{Var}(e_i)=\sigma^2$ for every $i$ and every $\psi$, while $\mathrm{Corr}(e_i,e_{i+1})=\phi_i$. Persistence here is therefore not extra noise but the same noise arranged differently, which is the case in which it is easiest to overlook. In symbols, if $\mathrm{Var}(e_i)=\sigma^2$ then
\[
  \mathrm{Var}(e_{i+1})
  = \phi_i^2\sigma^2 + (1-\phi_i^2)\sigma^2 = \sigma^2,
  \qquad
  \mathrm{Cov}(e_i,e_{i+1}) = \phi_i\sigma^2 ,
\]
so that $\sigma$ is the standard deviation of the errors at every $i$ and every $\psi$, and $\phi_i$ is exactly their correlation.

\noindent Because $\psi$ is defined at the mean spacing while the design is irregular, the autocorrelation on a realized series is not $\psi$ itself. The gaps of $n$ sorted uniform draws are, to a close approximation, exponentially distributed with mean $\bar\delta$, and since $\phi_i$ is convex in the gap the expected lag-one autocorrelation lies above $\psi$, in closed form
\begin{equation}\label{eq:acf1}
  \e[\mathrm{ACF}_1] \;=\; \e\!\left[\psi^{\,\Delta/\bar\delta}\right] \;=\; \frac{1}{1-\log\psi},\qquad \Delta/\bar\delta\sim\mathrm{Exp}(1),
\end{equation}
so that the ten levels of $\psi$ span realized autocorrelations from $0$ to about $0.9$.
Over $300$ draws at $n=200$ the realized mean autocorrelation was $0.44$, $0.64$ and $0.88$ for $\psi=0.3,0.6,0.9$, against $0.454$, $0.662$ and $0.905$ from Equation~\eqref{eq:acf1}.

\paragraph{The three scenarios.}
\begin{enumerate}
    \item \textbf{Scenario 1} sets $\psi=\omega=0$: every $\phi_i$ is zero so $e_i=\sigma u_i$, the stochastic part drops out of $g$, and what remains is the standardized deterministic component $b$ observed with independent errors on an irregular design. This is the reference case, and it describes a cross-sectional experimental design;
    \item \textbf{Scenario 2} sets $\omega=0$ and sweeps $\psi\in\{0,0.1,\ldots,0.9\}$, keeping the fixed signal and turning on the error dependence. It describes anything recorded sequentially, and, because the dependence is in distance rather than in index, also designs that are not temporal, such as a spatial transect;
    \item \textbf{Scenario 3} sets $\psi=0$ and sweeps $\omega\in\{0,0.25,0.5,0.75,1\}$, keeping independent errors and making the signal itself a draw from the prior the spline assumes. By the equivalence of \citet{Wah1978} the smoothing spline is the posterior mean under an integrated Wiener prior. Here it is used in the other direction, to generate signals from exactly what the estimator believes a signal looks like before it has seen any data, so that a moving signal is tested without confounding it with a signal the estimator was never designed for. At $\omega=0$ the signal is a fixed curve and only the jumps are random. For $0<\omega<1$ part of the variance is a recurring shape and part a trend that never repeats, which is what a stable seasonal or diurnal pattern on a slow-moving level looks like. At $\omega=1$ the signal is an integrated random walk \citep[i.e., the local linear trend of unobserved-components models and the model implicit in the Hodrick--Prescott filter]{MarPel2025} so it is non stationary and follows no fixed function, thus a second realization would produce a different curve, and only its scale and the location of the jumps are left to estimate (Section~\ref{sec:sc3}). Also, at $\omega=1$ the estimator's view of the signal is therefore \emph{correct}, which makes it the well-specified case and gives the results of Section~\ref{sec:res_signal} their force.
\end{enumerate}
Closed forms of each scenario are given in Sections~\ref{sec:sc1} to \ref{sec:sc3}.

\subsection{Scenario 1: a fixed smooth signal with independent errors}\label{sec:sc1}
Setting $\psi=0$ the error recursion of Equation~\eqref{eq:dgp_all} collapse (i.e., $\phi_i=\psi^{\delta_i/\bar\delta}=0$) for every $i$, so the error recursion reduces to $e_i=\sigma u_i$; and with $\omega=0$ the stochastic part and the mixture drop out of $g$, leaving the standardized background. Equation~\eqref{eq:dgp_all} becomes

\dgpbox{%
\begin{equation}\label{eq:dgp_s1}
  \begin{aligned}
    y_i \;&=\; \frac{b_i-\bar b}{\sd(b)}
    \;+\; \sum_{j=1}^{q}s_j\mathbf 1_{[c_j,1]}(x_i)
    \;+\; \varepsilon_i,
    \qquad \varepsilon_i\overset{\text{iid}}{\sim}\mathrm N(0,\sigma^2),\\[4pt]
    b_i \;&=\; \sin(2\pi x_i)+\tfrac12\sin(5\pi x_i).
  \end{aligned}
\end{equation}
}

\noindent This is the reference case (i.e., common boundary of the other two scenarios): a fixed deterministic curve, observed with independent errors, on an irregular design. It describes a designed experiment, or a cross-section in which the abscissa orders the observations without inducing dependence among them.

\begin{figure}[p]
\centering
\includegraphics[width=\linewidth]{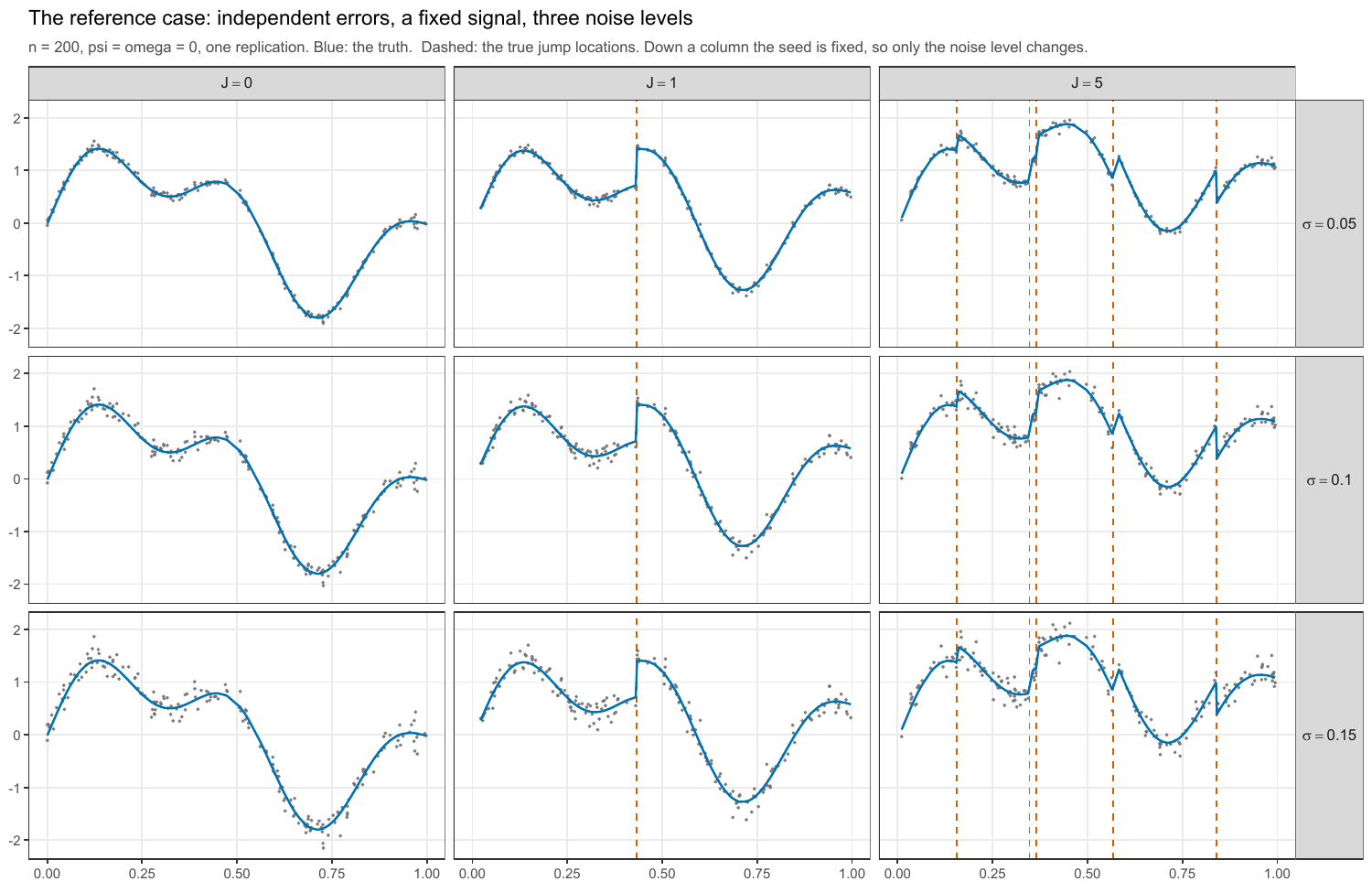}
\caption{\small Scenario 1, Equation~\eqref{eq:dgp_s1}. Rows: the noise level $\sigma$; columns: the number of true discontinuities. Points are the observations, the blue line is the signal, and the dashed rules are the true jump locations. The seed does not depend on $\sigma$, so within a column the abscissas, the signal and the jump configuration are identical going down and only the scatter about the curve changes.}
\label{fig:dgp_base}
\end{figure}

Figure~\ref{fig:dgp_base} shows what it looks like. The scenario has no departure parameter to raise, so the rows carry the one dimension of the factorial that decides how hard the reference problem is: the noise level. At $\sigma=0.05$ the jumps are unmistakable and the curve is traced almost exactly; at $\sigma=0.15$ the smaller of the five is comparable to the scatter, which is what makes the count criterion of Section~\ref{sec:xi_rules} bite in the cells that matter.

\subsection{Scenario 2: serially dependent errors}\label{sec:sc2}
Setting $\omega=0$ and letting $\psi$ vary keeps the fixed signal and turns on the error dependence:

\dgpbox{%
\begin{equation}\label{eq:dgp_s2}
  \begin{aligned}
    y_i \;&=\; \frac{b_i-\bar b}{\sd(b)}
    \;+\; \sum_{j=1}^{q}s_j\mathbf 1_{[c_j,1]}(x_i) \;+\; e_i,\\[4pt]
    e_1 \;&=\; \sigma u_1,
    \qquad
    e_{i+1} \;=\; \phi_i e_i + \sqrt{1-\phi_i^{2}}\,\sigma u_{i+1},
    \qquad \phi_i=\psi^{\,\delta_i/\bar\delta},\\[4pt]
    &\mathrm{Var}(e_i)=\sigma^2 \ \text{ for all } i,
    \qquad \mathrm{Corr}(e_i,e_{i+1})=\phi_i .
  \end{aligned}
\end{equation}}

\noindent $\psi$ is swept over $\{0,0.1,\ldots,0.9\}$, ten levels, for a realized lag-one autocorrelation running from $0$ to about $0.87$. The scenario describes the normal condition of anything recorded sequentially (e.g., weekly prices, daily pollutant concentrations, hourly load, instrument readings) where what is left over once a signal has been removed is itself persistent. Because the dependence is specified in distance rather than in index, it covers designs that are not temporal at all, such as a spatial transect where adjacency means physical proximity.

\paragraph{Locating a real series on the $\psi$ axis.} It is tempting to read the persistence of a series off the autocorrelation of the residuals of a fitted model, and Equation~\eqref{eq:acf1} would then map that number onto $\psi$. On these data this reading does not work, and the reason is worth recording because it recurs whenever smoothing and dependence are estimated together. The residual autocorrelation is not a property of the errors alone: it moves with the amount of smoothing. A fit that over-smooths leaves part of the signal in the residuals and inherits its persistence; a fit that follows the data closely absorbs the persistence into the fitted curve and leaves residuals that are negatively correlated. On the pump price of Section~\ref{P-sec:carburanti} of the paper the same series gives a lag-one residual autocorrelation of about $+0.87$ at a fit with $24$ effective degrees of freedom and of about $-0.20$ at the fit actually selected, which carries $309$. The selected fit is the flexible one, so what its residuals report is not the absence of dependence in the errors but the presence of enough flexibility to have taken it up. Two consequences follow. Persistence in the errors and flexibility in the signal are not separately identified from a single series without further assumptions, so no single number read off one fit can place a series on the $\psi$ axis. The experiment therefore records, at each $\psi$ and at each estimator's own calibrated $\xi$, the residual autocorrelation of the \emph{selected} fit so that a value measured on real data is matched against simulated values rather than against $\psi$ itself.

\begin{figure}[p]
\centering
\includegraphics[width=\linewidth]{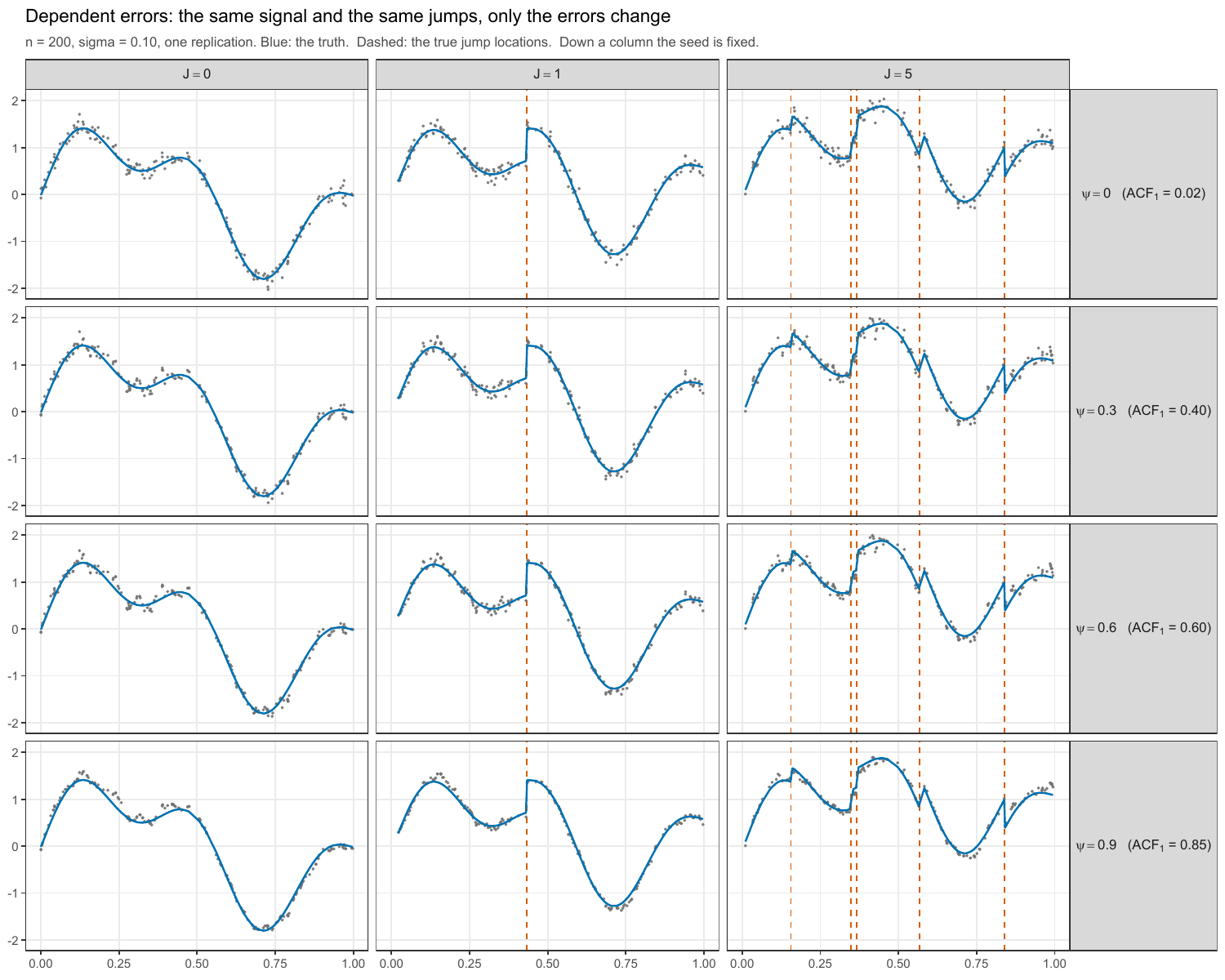}
\caption{\small Scenario 2, Equation~\eqref{eq:dgp_s2} with $\sigma=0.10$. Rows: the persistence $\psi$, with the lag-one autocorrelation it realizes on average. Columns: the number of true discontinuities. Points are the observations, the blue line is the signal, and the dashed rules are the true jump locations. Within a column the seed is fixed, so the abscissas, the standardized shocks and the jump configuration are identical going down and only the correlation of the errors changes.}
\label{fig:dgp_errors}
\end{figure}

Figure~\ref{fig:dgp_errors} depicts some examples drawn from this DGP. Because the marginal variance is held at $\sigma^2$ whatever $\psi$, the scatter does not grow down the rows. Persistence here is not extra noise but the same noise arranged differently. What changes is that the cloud collapses from a symmetric swarm into a slow wobble that hugs the curve, and that is the mechanism behind the failure of random-fold cross-validation under dependence: a held-out point is then nearly predicted by the neighbors left in the training set.

\subsection{Scenario 3: a mixture of deterministic and stochastic trend signal}\label{sec:sc3}
Setting $\psi=0$ and letting $\omega$ vary keeps the errors independent and makes the signal itself random:

\dgpbox{%
\begin{equation}\label{eq:dgp_s3}
  \begin{aligned}
    y_i \;&=\; f_i \;+\; \sum_{j=1}^{q}s_j\mathbf 1_{[c_j,1]}(x_i)
    \;+\; \varepsilon_i,
    \qquad \varepsilon_i\overset{\text{iid}}{\sim}\mathrm N(0,\sigma^2),\\[4pt]
    f_i \;&=\; \frac{g_i-\bar g}{\sd(g)},
    \qquad
    g_i \;=\; \sqrt{1-\omega}\;\frac{b_i-\bar b}{\sd(b)}
          \;+\;\sqrt{\omega}\;\frac{r_i-\bar r}{\sd(r)},\\[4pt]
    &\begin{bmatrix} r_{i+1}\\ r'_{i+1}\end{bmatrix}
     =\begin{bmatrix}1&\delta_i\\0&1\end{bmatrix}
      \begin{bmatrix} r_i\\ r'_i\end{bmatrix}+\bm\nu_i,
      \qquad
      \bm\nu_i\sim\mathrm N\!\left(\bm 0,
      \begin{bmatrix}\delta_i^3/3&\delta_i^2/2\\ \delta_i^2/2&\delta_i\end{bmatrix}\right).
  \end{aligned}
\end{equation}}

\noindent $\omega$ is swept over $\{0,0.25,0.5,0.75,1\}$.

\paragraph{The three regimes.}
\begin{itemize}
    \item At $\omega=0$ the signal is the deterministic curve and the only randomness in the signal is where the jumps fall (e.g., the case of a dose--response relation: a stable link between a covariate and a response, the same whenever and wherever it is measured, which a large enough sample would pin down exactly); there the estimator treats as random something that is fixed, so that the smoothing constant it selects measures how much data are needed to resolve a curve that is genuinely there;
    \item For $0<\omega<1$ part of the signal's variance is periodic and recurring and part is a trend that never repeats (e.g., series with a stable seasonal or diurnal shape riding on a slow-moving level);
    \item At $\omega=1$ the signal is entirely a draw from the stochastic process: in continuous time an integrated Wiener process, in discrete time an integrated random walk \citep[i.e., the local linear trend of unobserved-components models and the model implicit in the Hodrick--Prescott filter ][, so that every series that has ever been HP-filtered is a series someone chose to describe as $\omega=1$]{MarPel2025}. It is not stationary and there is no fixed function that it follows (i.e., a second realization of the same experiment would produce a different curve).
    When $\omega=1$ the signal is a draw from the process the spline assumes, so the estimator's view of it is \emph{correct}: this is the well-specified case rather than a misspecified one (i.e., whatever goes wrong there cannot be blamed on the model being wrong about the signal); the only things left to estimate being the scale $\lambda$, that is, how much of the observed variation is signal, and the location of the jumps.
\end{itemize}

\noindent Two cautions on $\omega$. First, it is not a roughness axis: an integrated Wiener path is continuously differentiable and, standardized, carries no more curvature than $b$ --- the mean absolute second difference runs $0.52$, $0.69$, $0.52$ at $\omega=0,0.5,1$. Second, it induces no dependence: the residual autocorrelation at the calibrated value stays negative and moves toward zero as $\omega$ rises, from $-0.23$ to $-0.10$ on average over the seven arms of Table~\ref{tab:fixed}, so Scenario 2 is not reachable from Scenario 3.

\paragraph{Why a stochastic trend is the right alternative.}
The roughness penalty $\int (f'')^2$ is not a neutral measure of wiggliness: by the equivalence of \citet{Wah1978}, the cubic smoothing spline is the posterior mean of a signal-extraction problem in which $f$ is an integrated Wiener stochastic process (i.e., a process whose second derivative is white noise) observed with independent Gaussian error. Penalizing squared second derivatives and placing that prior on $f$ are the same act seen from two sides; the smoothing constant is the ratio of the variance of the observation error to the variance of the innovations driving the process, and the fitted curve is the conditional expectation of the signal given the data. Section~\ref{P-sec:statespace} of the paper uses this equivalence to obtain the estimator from the Kalman filter. Here it is used to ask whether the estimator believes a signal looks like before it has seen any data, and then to generate signals from exactly that belief. Any other rough alternative (e.g., an arbitrary jagged function, a rougher deterministic curve) would confound two things: whether the calibration survives a moving signal and whether it survives a signal the estimator was never designed for. Drawing from the prior separates them, because on that class the estimator is right. Noticeable, a draw from an integrated Wiener process is continuously differentiable (i.e., its slope is a Brownian motion, so the path has no kinks) but it is nowhere twice differentiable, and the roughness penalty evaluated at such a draw is infinite. This is the ordinary situation in nonparametric smoothing: the penalty is a shrinkage device, not a constraint, and the posterior mean is far smoother than the prior draws from which it is computed. What $\omega$ moves is the class the signal belongs to, from a function with a bounded second derivative to one with none.

\paragraph{The transition equation.}
The third line of Equation~\eqref{eq:dgp_s3} is the exact transition of the state-space model of Section~\ref{P-sec:statespace} of the paper at unit process scale, carrying level and slope; using the exact transition, rather than shocking the slope at unit spacing, is what makes $r$ a legitimate path of the process on an irregular design rather than only on a grid. Indexing the departure by $\lambda$ directly would seem more natural, since a small value indicates a flexible fit in the estimator and would indicate a wandering signal here. However, it does not work: the process has no stationary distribution and its variance grows with the range of $x$, so there is no fixed relationship between $\lambda$ and the amplitude of $f$, and amplitude and persistence would move together along the grid. The mixture of Equation~\eqref{eq:dgp_all} separates them, and a $\lambda$ consistent with a given draw can be read off afterward from a fit rather than imposed as an input.

\begin{figure}[p]
\centering
\includegraphics[width=\linewidth]{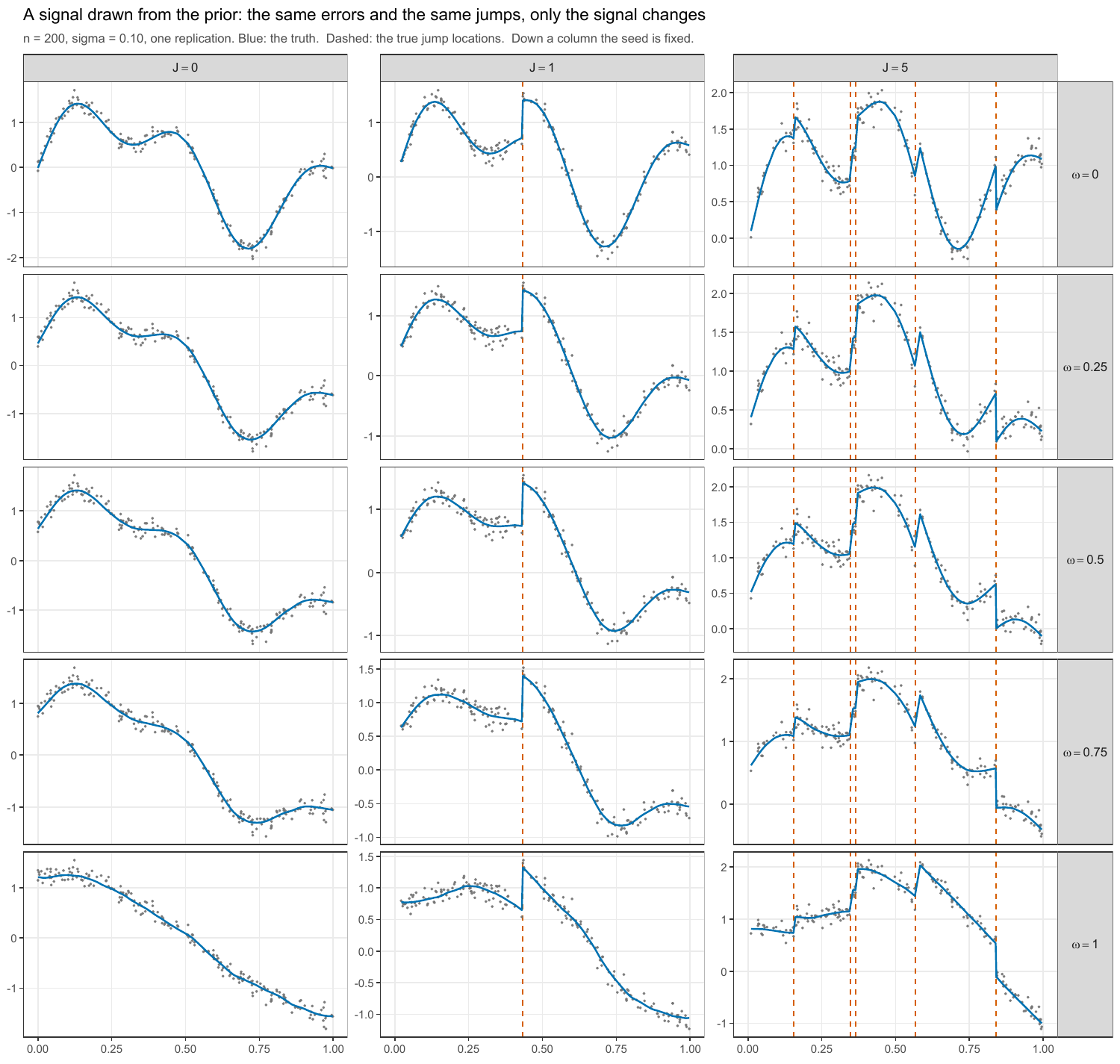}
\caption{\small Scenario 3, Equation~\eqref{eq:dgp_s3} with $\sigma=0.10$. Rows: the share $\omega$ of the signal's variance carried by the integrated Wiener path; columns: the number of true discontinuities. Within a column the seed is fixed, so the errors and the jump configuration are identical going down and only the signal changes.}
\label{fig:dgp_signal}
\end{figure}

Figure~\ref{fig:dgp_signal} shows selected examples of Scenario 3.

\subsection{A joint representation of the three DGPs}\label{sec:whichdata}
\begin{figure}[p]
\centering
\includegraphics[width=\linewidth]{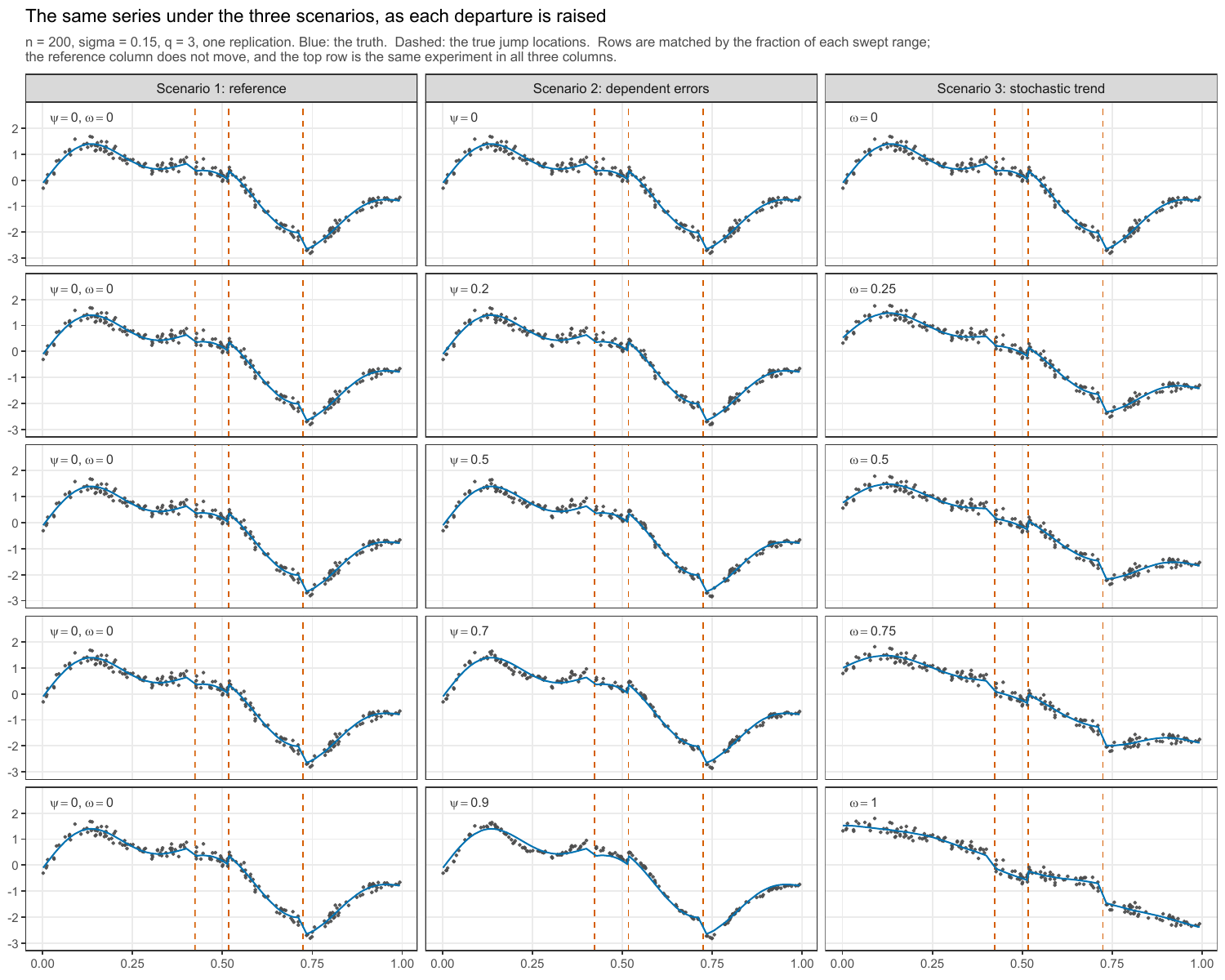}
\caption{\small The three scenarios in one frame, at $q=3$ simulated discontinuities, $n=200$ and $\sigma=0.15$, the largest noise level of the factorial, so that the change in the texture of the errors down the second column is visible at all. Columns are the scenario, rows the departure, matched by the fraction of each swept range, since $\psi$ stops at $0.9$ and $\omega$ at $1$; each panel is labeled with the value it uses. Points are the observations, the blue line is the signal, and the dashed rules are the true jump locations. The seed is the same in every panel, so the abscissas, the standardized shocks and the jump configuration are common to all fifteen; the vertical scale is common as well, since the signal is standardized to unit variance in every scenario.}
\label{fig:dgp_all}
\end{figure}

Figure~\ref{fig:dgp_all} depicts the three scenarios side by side. First column: at $\psi=0$ and $\omega=0$ the data are byte-identical (i.e., the first column repeats unchanged), so the top row is one experiment drawn three times, and then move apart in different directions. In the second column the signal never changes and the observations stop scattering symmetrically about it, so that by $\psi=0.9$ they lie in runs on one side of the curve and then the other. In the third column the errors never change and the signal itself loses its shape, so that by $\omega=1$ what is left is a wandering level on which the three discontinuities are the only sharp features.

\section{Design of the experiment}\label{sec:xi_arms}

The factorial is $n\in\{50,100,200,400\}$, $\sigma\in\{0.05,0.10,0.15\}$ and $q\in\{0,1,3,5\}$, for $48$ cells, crossed with ten levels of $\psi$ or five of $\omega$. Each cell is replicated $200$ times. The allocation is flat across $n$ rather than tapered, and flatly for a reason specific to the choice rule: minimax regret takes the maximum of the regret over cells, so the noisiest cell is the one most likely to decide the answer, and tapering the allocation with $n$ would hand that decision to the large-$n$ cells.

Here $n$ is the length of the series: $50$ is short enough that a single discontinuity is a substantial fraction of the information in the sample, $400$ long enough that the asymptotic arguments begin to hold.
$\sigma$ is the standard deviation of the errors, and since the signal is standardized to unit variance it is also the reciprocal of the signal-to-noise ratio: at $\sigma=0.05$ the discontinuities, whose sizes lie between $0.30$ and $0.70$, stand six to fourteen error standard deviations clear of the noise, and at $\sigma=0.15$ only two to five, so the three levels run from an easy detection problem to a hard one. Finally, $q$ is the number of true discontinuities in the series. $q=0$ is used to measure false discoveries where there is nothing to find, while $q=\{1,3,5\}$ range from a single break to a number large enough that the estimator must spread a fixed budget over several of them.

$\xi$ enters only the selection among grid points and never a fit, so the regularization path over a $25$-point grid of $M$ running to twice the range of $x$ is computed once per replication and every $\xi\in\{0,0.05,\ldots,1\}$ re-selects from it. All values of $\xi$ are therefore compared on identical fits, and the standard error of a difference between two of them is an order of magnitude smaller than that of either level.

SSJ--KFS and SSJ--ATW take $\lambda$ as given, supplied in several ways:
\begin{itemize}
    \item from an SSJ--MLE fit taken at a fixed reference $\xi=0.25$, held constant while $\xi$ is calibrated so that the exercise is not circular,
    \item by cross-validation under two fold layouts (Section~\ref{sec:tuning}), that is, folds are drawn at random and contiguous blocks at $K\in\{5,10,20\}$ each. Random folds leave a held-out point surrounded by its own neighbors, which once the errors are persistent nearly determine it, so the criterion is minimized by a fit that all but interpolates and the selected $\lambda$ collapses. Contiguous blocks remove that adjacency.
\end{itemize}

\section{Criteria}\label{sec:xi_rules}

Two rules are used to identify calibration values, both of the minimax-regret form. Within a cell, the regret of a value of $\xi$ is the loss it attains there minus the smallest loss attainable in that cell by any $\xi$; the rule selects the $\xi$ whose largest regret over the cells is smallest, so that no cell of the design is served badly, which is the guarantee a single constant meant for data of unknown difficulty needs. The two rules differ in the loss: the first is built on the reconstruction of the signal, the second on the number of discontinuities declared. Both are reported because they answer different questions, and because under the departures of Scenarios 2 and 3 they part company; a third criterion, the zero of the signed count error, is used as a check on the second and is defined at the end of this section.

\paragraph{Minimax regret in RMSE.}
Writing $r(\xi,c)$ for the RMSE of $\xi$ in cell $c$, measured against the true signal on a common $500$-point grid, the regret is $\rho(\xi,c)=r(\xi,c)-\min_{\xi'}r(\xi',c)$ and the reported value is $\arg\min_\xi\max_c\rho(\xi,c)$.

\noindent The cell is summarized by the mean over replications. Reporting the per-cell minimizer would overstate what the design resolves and averaging across cells would let the easiest dominate. The replication distribution within a cell is mildly right skewed, so the median was computed as well: at the baseline it moves the calibrated value by at most one grid step on fourteen of the fifteen arms and by three on the fifteenth (SSJ--ATW with contiguous blocks at $K=5$), never downward, which is why it is recorded here and not tabulated.

\paragraph{Minimax regret in the count error.}
The rule above is built on reconstruction error and does not validate whether adding or removing one discontinuity in a long series moves the error very little, so the regret valley is shallow and becomes shallower as the departure grows, until the minimizer is a point drawn out of a plateau. We therefore also apply the minimax regret rule to a loss built on the count. Writing $\ell(\xi,c)=\e\,|\hat k-q|$ for the mean absolute error in the number declared in cell $c$, over the $36$ cells with $q\in\{1,3,5\}$ (the cells with $q=0$ are excluded because there the count error is simply the false-discovery count, which is reported on its own), the calibrated value is $\arg\min_\xi\max_c\{\ell(\xi,c)-\min_{\xi'}\ell(\xi',c)\}$.

\noindent We stress that averaging a signed count error over $q$ would let over-declaration at $q=1$ cancel under-declaration at $q=5$, and under strong dependence the average can then be brought to zero by cancellation, or never reach it at all. The obvious alternative, used below as a check, is the \emph{zero of the signed count error}: writing $\bar e(\xi)$ for the mean, over the $36$ cells with $q>0$, of the number declared minus $q$, it is the value of $\xi$ at which $\bar e$ crosses zero from above, interpolated linearly between grid points, that is, the $\xi$ at which the estimator declares on average the right number; it is undefined when $\bar e(\xi)<0$ at every $\xi$, which happens when the estimator cannot reach the true number at any penalty. The absolute error is defined in every case, and where the signed error does have a zero the two agree to within a step of the grid (over the $209$ combinations of arm and departure level in which the zero exists, out of $225$, the correlation between the two is $0.92$, the median difference $0.03$, and $171$ agree to within one step of the $\xi$ grid).

\paragraph{Why not locations.}
Neither rule asks whether the declared discontinuities are in the right places. While a reader may expect the calibration to rest on that criterion, it cannot: any summary of the confusion matrix between declared and true locations has a denominator that involves the number of true positives, and with $q\in\{0,1,3,5\}$ that number takes so few values that the summary's optimum wanders over the whole grid from one arm to the next. A location score also tolerates a single discontinuity being declared twice at adjacent intervals, which might occur at these values of $\xi$; the count criteria charge for it. Location accuracy is therefore reported in Section~\ref{P-sec:simulations} of the paper as a consequence of the calibration, not used as an input to it.

\section{Results}\label{sec:xi_results}

\subsection{Scenario 1: the baseline}\label{sec:res_base}

\begin{table}[htbp]
\centering\footnotesize
\caption{Scenario 1. The calibrated $\xi$ for each of the fifteen arms under the minimax regret in RMSE and minimax regret in the count error rules, each with the worst-case regret it attains (i.e., on the RMSE scale for the first, in discontinuities for the second). Beneath each calibrated value, in small type, the central half of the same criterion computed cell by cell: it is the spread of the best $\xi$ over the design cells that one pooled number has to cover, not a confidence interval.}
\label{tab:base_all}
\setlength{\tabcolsep}{3pt}
\begin{tabular}{ll cc @{\qquad} cc}
\toprule
Estimator & $\lambda$ from & \multicolumn{2}{c}{minimax regret in RMSE} & \multicolumn{2}{c}{minimax regret in the count error} \\
\cmidrule(lr){3-4}\cmidrule(lr){5-6}
 & & $\xi$ & worst regret & $\xi$ & worst regret \\
\midrule
SSJ--MLE & MLE & \begin{tabular}[t]{@{}c@{}}0.30\\[-1.5pt]{\tiny[.25,.35]}\end{tabular} & 0.0014 & \begin{tabular}[t]{@{}c@{}}0.20\\[-1.5pt]{\tiny[.00,.42]}\end{tabular} & 1.85 \\
\addlinespace
SSJ--KFS & MLE & \begin{tabular}[t]{@{}c@{}}0.15\\[-1.5pt]{\tiny[.10,.20]}\end{tabular} & 0.0018 & \begin{tabular}[t]{@{}c@{}}0.15\\[-1.5pt]{\tiny[.00,.50]}\end{tabular} & 1.82 \\
 & random $K{=}5$ & \begin{tabular}[t]{@{}c@{}}0.20\\[-1.5pt]{\tiny[.15,.25]}\end{tabular} & 0.0028 & \begin{tabular}[t]{@{}c@{}}0.20\\[-1.5pt]{\tiny[.00,.60]}\end{tabular} & 1.73 \\
 & random $K{=}10$ & \begin{tabular}[t]{@{}c@{}}0.20\\[-1.5pt]{\tiny[.14,.25]}\end{tabular} & 0.0024 & \begin{tabular}[t]{@{}c@{}}0.20\\[-1.5pt]{\tiny[.00,.60]}\end{tabular} & 1.68 \\
 & random $K{=}20$ & \begin{tabular}[t]{@{}c@{}}0.20\\[-1.5pt]{\tiny[.15,.25]}\end{tabular} & 0.0025 & \begin{tabular}[t]{@{}c@{}}0.20\\[-1.5pt]{\tiny[.00,.60]}\end{tabular} & 1.64 \\
 & blocks $K{=}5$ & \begin{tabular}[t]{@{}c@{}}0.10\\[-1.5pt]{\tiny[.10,.20]}\end{tabular} & 0.0012 & \begin{tabular}[t]{@{}c@{}}0.15\\[-1.5pt]{\tiny[.00,.76]}\end{tabular} & 1.90 \\
 & blocks $K{=}10$ & \begin{tabular}[t]{@{}c@{}}0.15\\[-1.5pt]{\tiny[.10,.21]}\end{tabular} & 0.0022 & \begin{tabular}[t]{@{}c@{}}0.15\\[-1.5pt]{\tiny[.00,.68]}\end{tabular} & 1.84 \\
 & blocks $K{=}20$ & \begin{tabular}[t]{@{}c@{}}0.15\\[-1.5pt]{\tiny[.10,.25]}\end{tabular} & 0.0025 & \begin{tabular}[t]{@{}c@{}}0.15\\[-1.5pt]{\tiny[.00,.51]}\end{tabular} & 1.63 \\
\addlinespace
SSJ--ATW & MLE & \begin{tabular}[t]{@{}c@{}}0.30\\[-1.5pt]{\tiny[.15,.31]}\end{tabular} & 0.0064 & \begin{tabular}[t]{@{}c@{}}0.25\\[-1.5pt]{\tiny[.00,.56]}\end{tabular} & 2.01 \\
 & random $K{=}5$ & \begin{tabular}[t]{@{}c@{}}0.30\\[-1.5pt]{\tiny[.15,.30]}\end{tabular} & 0.0069 & \begin{tabular}[t]{@{}c@{}}0.25\\[-1.5pt]{\tiny[.00,.44]}\end{tabular} & 1.79 \\
 & random $K{=}10$ & \begin{tabular}[t]{@{}c@{}}0.30\\[-1.5pt]{\tiny[.15,.30]}\end{tabular} & 0.0065 & \begin{tabular}[t]{@{}c@{}}0.30\\[-1.5pt]{\tiny[.00,.42]}\end{tabular} & 1.82 \\
 & random $K{=}20$ & \begin{tabular}[t]{@{}c@{}}0.30\\[-1.5pt]{\tiny[.15,.30]}\end{tabular} & 0.0072 & \begin{tabular}[t]{@{}c@{}}0.30\\[-1.5pt]{\tiny[.00,.41]}\end{tabular} & 1.82 \\
 & blocks $K{=}5$ & \begin{tabular}[t]{@{}c@{}}0.25\\[-1.5pt]{\tiny[.14,.35]}\end{tabular} & 0.0051 & \begin{tabular}[t]{@{}c@{}}0.25\\[-1.5pt]{\tiny[.00,.78]}\end{tabular} & 1.96 \\
 & blocks $K{=}10$ & \begin{tabular}[t]{@{}c@{}}0.25\\[-1.5pt]{\tiny[.15,.30]}\end{tabular} & 0.0087 & \begin{tabular}[t]{@{}c@{}}0.25\\[-1.5pt]{\tiny[.00,.60]}\end{tabular} & 1.82 \\
 & blocks $K{=}20$ & \begin{tabular}[t]{@{}c@{}}0.30\\[-1.5pt]{\tiny[.15,.31]}\end{tabular} & 0.0071 & \begin{tabular}[t]{@{}c@{}}0.25\\[-1.5pt]{\tiny[.00,.46]}\end{tabular} & 1.99 \\
\bottomrule
\end{tabular}
\end{table}

The two rules agree to within one grid step on every arm but SSJ--MLE, for which the count rule sits two steps below the reconstruction rule, at $0.20$ against $0.30$, and the alternative ways of supplying $\lambda$ are nearly indistinguishable. For reconstruction the cells put the best $\xi$ inside a band of median width $0.15$, three steps of the grid, so the pooled value is close to what a typical cell would have chosen on its own.

\noindent The count loss is a different matter: its band has median width $0.56$, more than half the grid, because the cells with one discontinuity want a large penalty and those with five a small one, and the pooled value is the compromise between them; under the departures the bands widen further.

\noindent Notice that SSJ--KFS settles below SSJ--MLE, at $0.15$ against $0.20$ and $0.30$, for a reason that is general rather than procedural. At a given $\lambda$ and budget the two fit the same model, but SSJ--KFS takes $\lambda$ as given, from a preliminary SSJE--MLE fit, while SSJ--MLE re-estimates it at every point of the budget grid. Supplying $\lambda$ fixes one parameter before the budget is chosen: the fit is stiffer, part of its degrees of freedom have already been given up, and the search over the budget can buy less, so the additional penalty that $\xi$ governs has to be milder. The same ordering holds between the sources of $\lambda$ within an estimator: under the count rule, with $\lambda$ from SSJ--MLE the calibrated value is never above the one obtained with a cross-validated $\lambda$, $0.15$ against $0.15$--$0.20$ for SSJ--KFS and $0.25$ against $0.25$--$0.30$ for SSJ--ATW, since a cross-validated constant is chosen on the data at hand and leaves the budget more to buy.

\noindent Every value lies between $0.10$ and $0.30$, so $\xi=1$, the top of the admissible range, is rejected throughout, and so is the nominal floor of $0.50$ derived from \cite{CheChe2008} and discussed in Section~\ref{sec:charge}.

\subsection{Scenario 2: serially dependent errors}\label{sec:res_errors}

\begin{table}[htbp]
\centering\footnotesize
\caption{Scenario 2. The calibrated $\xi$ against the persistence $\psi$ of the errors under the two rules, minimax regret in RMSE and minimax regret in the count error, both pooled over sample size. Beneath each, in small type, the central half of the same criterion computed cell by cell.}
\label{tab:xi_errors_uni}
\setlength{\tabcolsep}{1.5pt}
\begin{tabular}{ll*{6}{c}@{\enspace}*{6}{c}}
\toprule
Estimator & $\lambda$ from & \multicolumn{6}{c}{minimax regret in RMSE} & \multicolumn{6}{c}{minimax regret in the count error} \\
\cmidrule(lr){3-8}\cmidrule(lr){9-14}
 &  & \multicolumn{6}{c}{$\psi$} & \multicolumn{6}{c}{$\psi$} \\
 &  & 0.00 & 0.20 & 0.40 & 0.60 & 0.80 & 0.90 & 0.00 & 0.20 & 0.40 & 0.60 & 0.80 & 0.90 \\
\midrule
SSJ--MLE & MLE & \begin{tabular}[t]{@{}c@{}}0.30\\[-1.5pt]{\tiny[.25,.35]}\end{tabular} & \begin{tabular}[t]{@{}c@{}}0.35\\[-1.5pt]{\tiny[.30,.40]}\end{tabular} & \begin{tabular}[t]{@{}c@{}}0.30\\[-1.5pt]{\tiny[.30,.40]}\end{tabular} & \begin{tabular}[t]{@{}c@{}}0.25\\[-1.5pt]{\tiny[.30,.45]}\end{tabular} & \begin{tabular}[t]{@{}c@{}}0.30\\[-1.5pt]{\tiny[.20,.35]}\end{tabular} & \begin{tabular}[t]{@{}c@{}}0.25\\[-1.5pt]{\tiny[.14,.35]}\end{tabular} & \begin{tabular}[t]{@{}c@{}}0.20\\[-1.5pt]{\tiny[.00,.42]}\end{tabular} & \begin{tabular}[t]{@{}c@{}}0.25\\[-1.5pt]{\tiny[.00,.44]}\end{tabular} & \begin{tabular}[t]{@{}c@{}}0.25\\[-1.5pt]{\tiny[.00,.51]}\end{tabular} & \begin{tabular}[t]{@{}c@{}}0.15\\[-1.5pt]{\tiny[.00,.22]}\end{tabular} & \begin{tabular}[t]{@{}c@{}}0.15\\[-1.5pt]{\tiny[.00,.02]}\end{tabular} & \begin{tabular}[t]{@{}c@{}}0.20\\[-1.5pt]{\tiny[.00,.02]}\end{tabular} \\
\addlinespace
SSJ--KFS & MLE & \begin{tabular}[t]{@{}c@{}}0.15\\[-1.5pt]{\tiny[.10,.20]}\end{tabular} & \begin{tabular}[t]{@{}c@{}}0.30\\[-1.5pt]{\tiny[.24,.31]}\end{tabular} & \begin{tabular}[t]{@{}c@{}}0.35\\[-1.5pt]{\tiny[.25,.35]}\end{tabular} & \begin{tabular}[t]{@{}c@{}}0.30\\[-1.5pt]{\tiny[.25,.40]}\end{tabular} & \begin{tabular}[t]{@{}c@{}}0.25\\[-1.5pt]{\tiny[.25,.35]}\end{tabular} & \begin{tabular}[t]{@{}c@{}}0.30\\[-1.5pt]{\tiny[.25,.40]}\end{tabular} & \begin{tabular}[t]{@{}c@{}}0.15\\[-1.5pt]{\tiny[.00,.50]}\end{tabular} & \begin{tabular}[t]{@{}c@{}}0.25\\[-1.5pt]{\tiny[.00,.66]}\end{tabular} & \begin{tabular}[t]{@{}c@{}}0.20\\[-1.5pt]{\tiny[.00,.70]}\end{tabular} & \begin{tabular}[t]{@{}c@{}}0.15\\[-1.5pt]{\tiny[.00,.48]}\end{tabular} & \begin{tabular}[t]{@{}c@{}}0.15\\[-1.5pt]{\tiny[.00,.55]}\end{tabular} & \begin{tabular}[t]{@{}c@{}}0.20\\[-1.5pt]{\tiny[.00,.26]}\end{tabular} \\
 & random $K{=}5$ & \begin{tabular}[t]{@{}c@{}}0.20\\[-1.5pt]{\tiny[.15,.25]}\end{tabular} & \begin{tabular}[t]{@{}c@{}}0.30\\[-1.5pt]{\tiny[.20,.35]}\end{tabular} & \begin{tabular}[t]{@{}c@{}}0.30\\[-1.5pt]{\tiny[.25,.40]}\end{tabular} & \begin{tabular}[t]{@{}c@{}}0.30\\[-1.5pt]{\tiny[.25,.40]}\end{tabular} & \begin{tabular}[t]{@{}c@{}}0.30\\[-1.5pt]{\tiny[.25,.35]}\end{tabular} & \begin{tabular}[t]{@{}c@{}}0.30\\[-1.5pt]{\tiny[.29,.35]}\end{tabular} & \begin{tabular}[t]{@{}c@{}}0.20\\[-1.5pt]{\tiny[.00,.60]}\end{tabular} & \begin{tabular}[t]{@{}c@{}}0.20\\[-1.5pt]{\tiny[.00,.70]}\end{tabular} & \begin{tabular}[t]{@{}c@{}}0.20\\[-1.5pt]{\tiny[.00,.65]}\end{tabular} & \begin{tabular}[t]{@{}c@{}}0.20\\[-1.5pt]{\tiny[.00,.44]}\end{tabular} & \begin{tabular}[t]{@{}c@{}}0.20\\[-1.5pt]{\tiny[.04,.45]}\end{tabular} & \begin{tabular}[t]{@{}c@{}}0.25\\[-1.5pt]{\tiny[.05,.35]}\end{tabular} \\
 & random $K{=}10$ & \begin{tabular}[t]{@{}c@{}}0.20\\[-1.5pt]{\tiny[.14,.25]}\end{tabular} & \begin{tabular}[t]{@{}c@{}}0.25\\[-1.5pt]{\tiny[.20,.35]}\end{tabular} & \begin{tabular}[t]{@{}c@{}}0.25\\[-1.5pt]{\tiny[.24,.35]}\end{tabular} & \begin{tabular}[t]{@{}c@{}}0.25\\[-1.5pt]{\tiny[.25,.36]}\end{tabular} & \begin{tabular}[t]{@{}c@{}}0.30\\[-1.5pt]{\tiny[.25,.36]}\end{tabular} & \begin{tabular}[t]{@{}c@{}}0.35\\[-1.5pt]{\tiny[.30,.40]}\end{tabular} & \begin{tabular}[t]{@{}c@{}}0.20\\[-1.5pt]{\tiny[.00,.60]}\end{tabular} & \begin{tabular}[t]{@{}c@{}}0.20\\[-1.5pt]{\tiny[.00,.75]}\end{tabular} & \begin{tabular}[t]{@{}c@{}}0.15\\[-1.5pt]{\tiny[.00,.52]}\end{tabular} & \begin{tabular}[t]{@{}c@{}}0.20\\[-1.5pt]{\tiny[.00,.39]}\end{tabular} & \begin{tabular}[t]{@{}c@{}}0.20\\[-1.5pt]{\tiny[.05,.38]}\end{tabular} & \begin{tabular}[t]{@{}c@{}}0.25\\[-1.5pt]{\tiny[.05,.35]}\end{tabular} \\
 & random $K{=}20$ & \begin{tabular}[t]{@{}c@{}}0.20\\[-1.5pt]{\tiny[.15,.25]}\end{tabular} & \begin{tabular}[t]{@{}c@{}}0.25\\[-1.5pt]{\tiny[.20,.35]}\end{tabular} & \begin{tabular}[t]{@{}c@{}}0.25\\[-1.5pt]{\tiny[.20,.35]}\end{tabular} & \begin{tabular}[t]{@{}c@{}}0.25\\[-1.5pt]{\tiny[.25,.35]}\end{tabular} & \begin{tabular}[t]{@{}c@{}}0.35\\[-1.5pt]{\tiny[.25,.36]}\end{tabular} & \begin{tabular}[t]{@{}c@{}}0.35\\[-1.5pt]{\tiny[.29,.40]}\end{tabular} & \begin{tabular}[t]{@{}c@{}}0.20\\[-1.5pt]{\tiny[.00,.60]}\end{tabular} & \begin{tabular}[t]{@{}c@{}}0.20\\[-1.5pt]{\tiny[.00,.75]}\end{tabular} & \begin{tabular}[t]{@{}c@{}}0.15\\[-1.5pt]{\tiny[.00,.41]}\end{tabular} & \begin{tabular}[t]{@{}c@{}}0.15\\[-1.5pt]{\tiny[.00,.41]}\end{tabular} & \begin{tabular}[t]{@{}c@{}}0.20\\[-1.5pt]{\tiny[.00,.40]}\end{tabular} & \begin{tabular}[t]{@{}c@{}}0.25\\[-1.5pt]{\tiny[.05,.35]}\end{tabular} \\
 & blocks $K{=}5$ & \begin{tabular}[t]{@{}c@{}}0.10\\[-1.5pt]{\tiny[.10,.20]}\end{tabular} & \begin{tabular}[t]{@{}c@{}}0.15\\[-1.5pt]{\tiny[.20,.30]}\end{tabular} & \begin{tabular}[t]{@{}c@{}}0.20\\[-1.5pt]{\tiny[.20,.30]}\end{tabular} & \begin{tabular}[t]{@{}c@{}}0.30\\[-1.5pt]{\tiny[.25,.35]}\end{tabular} & \begin{tabular}[t]{@{}c@{}}0.35\\[-1.5pt]{\tiny[.25,.40]}\end{tabular} & \begin{tabular}[t]{@{}c@{}}0.35\\[-1.5pt]{\tiny[.25,.40]}\end{tabular} & \begin{tabular}[t]{@{}c@{}}0.15\\[-1.5pt]{\tiny[.00,.76]}\end{tabular} & \begin{tabular}[t]{@{}c@{}}0.25\\[-1.5pt]{\tiny[.00,.76]}\end{tabular} & \begin{tabular}[t]{@{}c@{}}0.25\\[-1.5pt]{\tiny[.00,.85]}\end{tabular} & \begin{tabular}[t]{@{}c@{}}0.30\\[-1.5pt]{\tiny[.00,.39]}\end{tabular} & \begin{tabular}[t]{@{}c@{}}0.40\\[-1.5pt]{\tiny[.04,.36]}\end{tabular} & \begin{tabular}[t]{@{}c@{}}0.25\\[-1.5pt]{\tiny[.04,.35]}\end{tabular} \\
 & blocks $K{=}10$ & \begin{tabular}[t]{@{}c@{}}0.15\\[-1.5pt]{\tiny[.10,.21]}\end{tabular} & \begin{tabular}[t]{@{}c@{}}0.25\\[-1.5pt]{\tiny[.20,.31]}\end{tabular} & \begin{tabular}[t]{@{}c@{}}0.30\\[-1.5pt]{\tiny[.25,.35]}\end{tabular} & \begin{tabular}[t]{@{}c@{}}0.30\\[-1.5pt]{\tiny[.29,.40]}\end{tabular} & \begin{tabular}[t]{@{}c@{}}0.40\\[-1.5pt]{\tiny[.30,.45]}\end{tabular} & \begin{tabular}[t]{@{}c@{}}0.45\\[-1.5pt]{\tiny[.25,.40]}\end{tabular} & \begin{tabular}[t]{@{}c@{}}0.15\\[-1.5pt]{\tiny[.00,.68]}\end{tabular} & \begin{tabular}[t]{@{}c@{}}0.25\\[-1.5pt]{\tiny[.00,.76]}\end{tabular} & \begin{tabular}[t]{@{}c@{}}0.30\\[-1.5pt]{\tiny[.00,.78]}\end{tabular} & \begin{tabular}[t]{@{}c@{}}0.30\\[-1.5pt]{\tiny[.00,.50]}\end{tabular} & \begin{tabular}[t]{@{}c@{}}0.25\\[-1.5pt]{\tiny[.00,.36]}\end{tabular} & \begin{tabular}[t]{@{}c@{}}0.30\\[-1.5pt]{\tiny[.05,.31]}\end{tabular} \\
 & blocks $K{=}20$ & \begin{tabular}[t]{@{}c@{}}0.15\\[-1.5pt]{\tiny[.10,.25]}\end{tabular} & \begin{tabular}[t]{@{}c@{}}0.25\\[-1.5pt]{\tiny[.20,.31]}\end{tabular} & \begin{tabular}[t]{@{}c@{}}0.30\\[-1.5pt]{\tiny[.25,.36]}\end{tabular} & \begin{tabular}[t]{@{}c@{}}0.35\\[-1.5pt]{\tiny[.25,.40]}\end{tabular} & \begin{tabular}[t]{@{}c@{}}0.40\\[-1.5pt]{\tiny[.30,.45]}\end{tabular} & \begin{tabular}[t]{@{}c@{}}0.45\\[-1.5pt]{\tiny[.30,.41]}\end{tabular} & \begin{tabular}[t]{@{}c@{}}0.15\\[-1.5pt]{\tiny[.00,.51]}\end{tabular} & \begin{tabular}[t]{@{}c@{}}0.25\\[-1.5pt]{\tiny[.00,.68]}\end{tabular} & \begin{tabular}[t]{@{}c@{}}0.30\\[-1.5pt]{\tiny[.00,.63]}\end{tabular} & \begin{tabular}[t]{@{}c@{}}0.30\\[-1.5pt]{\tiny[.00,.45]}\end{tabular} & \begin{tabular}[t]{@{}c@{}}0.25\\[-1.5pt]{\tiny[.00,.31]}\end{tabular} & \begin{tabular}[t]{@{}c@{}}0.25\\[-1.5pt]{\tiny[.05,.35]}\end{tabular} \\
\addlinespace
SSJ--ATW & MLE & \begin{tabular}[t]{@{}c@{}}0.30\\[-1.5pt]{\tiny[.15,.31]}\end{tabular} & \begin{tabular}[t]{@{}c@{}}0.40\\[-1.5pt]{\tiny[.25,.45]}\end{tabular} & \begin{tabular}[t]{@{}c@{}}0.45\\[-1.5pt]{\tiny[.30,.50]}\end{tabular} & \begin{tabular}[t]{@{}c@{}}0.45\\[-1.5pt]{\tiny[.34,.50]}\end{tabular} & \begin{tabular}[t]{@{}c@{}}0.40\\[-1.5pt]{\tiny[.15,.45]}\end{tabular} & \begin{tabular}[t]{@{}c@{}}0.35\\[-1.5pt]{\tiny[.05,.41]}\end{tabular} & \begin{tabular}[t]{@{}c@{}}0.25\\[-1.5pt]{\tiny[.00,.56]}\end{tabular} & \begin{tabular}[t]{@{}c@{}}0.40\\[-1.5pt]{\tiny[.00,.21]}\end{tabular} & \begin{tabular}[t]{@{}c@{}}0.40\\[-1.5pt]{\tiny[.00,.25]}\end{tabular} & \begin{tabular}[t]{@{}c@{}}0.40\\[-1.5pt]{\tiny[.05,.30]}\end{tabular} & \begin{tabular}[t]{@{}c@{}}0.40\\[-1.5pt]{\tiny[.04,.31]}\end{tabular} & \begin{tabular}[t]{@{}c@{}}0.45\\[-1.5pt]{\tiny[.10,.36]}\end{tabular} \\
 & random $K{=}5$ & \begin{tabular}[t]{@{}c@{}}0.30\\[-1.5pt]{\tiny[.15,.30]}\end{tabular} & \begin{tabular}[t]{@{}c@{}}0.40\\[-1.5pt]{\tiny[.25,.36]}\end{tabular} & \begin{tabular}[t]{@{}c@{}}0.45\\[-1.5pt]{\tiny[.30,.45]}\end{tabular} & \begin{tabular}[t]{@{}c@{}}0.40\\[-1.5pt]{\tiny[.30,.45]}\end{tabular} & \begin{tabular}[t]{@{}c@{}}0.35\\[-1.5pt]{\tiny[.29,.45]}\end{tabular} & \begin{tabular}[t]{@{}c@{}}0.30\\[-1.5pt]{\tiny[.05,.40]}\end{tabular} & \begin{tabular}[t]{@{}c@{}}0.25\\[-1.5pt]{\tiny[.00,.44]}\end{tabular} & \begin{tabular}[t]{@{}c@{}}0.35\\[-1.5pt]{\tiny[.00,.30]}\end{tabular} & \begin{tabular}[t]{@{}c@{}}0.35\\[-1.5pt]{\tiny[.04,.35]}\end{tabular} & \begin{tabular}[t]{@{}c@{}}0.35\\[-1.5pt]{\tiny[.05,.30]}\end{tabular} & \begin{tabular}[t]{@{}c@{}}0.20\\[-1.5pt]{\tiny[.09,.35]}\end{tabular} & \begin{tabular}[t]{@{}c@{}}0.40\\[-1.5pt]{\tiny[.05,.36]}\end{tabular} \\
 & random $K{=}10$ & \begin{tabular}[t]{@{}c@{}}0.30\\[-1.5pt]{\tiny[.15,.30]}\end{tabular} & \begin{tabular}[t]{@{}c@{}}0.40\\[-1.5pt]{\tiny[.25,.40]}\end{tabular} & \begin{tabular}[t]{@{}c@{}}0.40\\[-1.5pt]{\tiny[.30,.45]}\end{tabular} & \begin{tabular}[t]{@{}c@{}}0.35\\[-1.5pt]{\tiny[.30,.45]}\end{tabular} & \begin{tabular}[t]{@{}c@{}}0.35\\[-1.5pt]{\tiny[.29,.45]}\end{tabular} & \begin{tabular}[t]{@{}c@{}}0.35\\[-1.5pt]{\tiny[.05,.45]}\end{tabular} & \begin{tabular}[t]{@{}c@{}}0.30\\[-1.5pt]{\tiny[.00,.42]}\end{tabular} & \begin{tabular}[t]{@{}c@{}}0.35\\[-1.5pt]{\tiny[.00,.28]}\end{tabular} & \begin{tabular}[t]{@{}c@{}}0.35\\[-1.5pt]{\tiny[.00,.30]}\end{tabular} & \begin{tabular}[t]{@{}c@{}}0.35\\[-1.5pt]{\tiny[.05,.28]}\end{tabular} & \begin{tabular}[t]{@{}c@{}}0.35\\[-1.5pt]{\tiny[.05,.35]}\end{tabular} & \begin{tabular}[t]{@{}c@{}}0.40\\[-1.5pt]{\tiny[.10,.40]}\end{tabular} \\
 & random $K{=}20$ & \begin{tabular}[t]{@{}c@{}}0.30\\[-1.5pt]{\tiny[.15,.30]}\end{tabular} & \begin{tabular}[t]{@{}c@{}}0.40\\[-1.5pt]{\tiny[.25,.41]}\end{tabular} & \begin{tabular}[t]{@{}c@{}}0.40\\[-1.5pt]{\tiny[.35,.45]}\end{tabular} & \begin{tabular}[t]{@{}c@{}}0.35\\[-1.5pt]{\tiny[.30,.45]}\end{tabular} & \begin{tabular}[t]{@{}c@{}}0.35\\[-1.5pt]{\tiny[.30,.45]}\end{tabular} & \begin{tabular}[t]{@{}c@{}}0.35\\[-1.5pt]{\tiny[.09,.40]}\end{tabular} & \begin{tabular}[t]{@{}c@{}}0.30\\[-1.5pt]{\tiny[.00,.41]}\end{tabular} & \begin{tabular}[t]{@{}c@{}}0.35\\[-1.5pt]{\tiny[.04,.35]}\end{tabular} & \begin{tabular}[t]{@{}c@{}}0.35\\[-1.5pt]{\tiny[.00,.30]}\end{tabular} & \begin{tabular}[t]{@{}c@{}}0.30\\[-1.5pt]{\tiny[.00,.35]}\end{tabular} & \begin{tabular}[t]{@{}c@{}}0.30\\[-1.5pt]{\tiny[.10,.36]}\end{tabular} & \begin{tabular}[t]{@{}c@{}}0.40\\[-1.5pt]{\tiny[.10,.40]}\end{tabular} \\
 & blocks $K{=}5$ & \begin{tabular}[t]{@{}c@{}}0.25\\[-1.5pt]{\tiny[.14,.35]}\end{tabular} & \begin{tabular}[t]{@{}c@{}}0.35\\[-1.5pt]{\tiny[.20,.40]}\end{tabular} & \begin{tabular}[t]{@{}c@{}}0.30\\[-1.5pt]{\tiny[.25,.40]}\end{tabular} & \begin{tabular}[t]{@{}c@{}}0.30\\[-1.5pt]{\tiny[.12,.40]}\end{tabular} & \begin{tabular}[t]{@{}c@{}}0.35\\[-1.5pt]{\tiny[.04,.40]}\end{tabular} & \begin{tabular}[t]{@{}c@{}}0.40\\[-1.5pt]{\tiny[.05,.40]}\end{tabular} & \begin{tabular}[t]{@{}c@{}}0.25\\[-1.5pt]{\tiny[.00,.78]}\end{tabular} & \begin{tabular}[t]{@{}c@{}}0.35\\[-1.5pt]{\tiny[.04,.26]}\end{tabular} & \begin{tabular}[t]{@{}c@{}}0.40\\[-1.5pt]{\tiny[.05,.28]}\end{tabular} & \begin{tabular}[t]{@{}c@{}}0.45\\[-1.5pt]{\tiny[.05,.30]}\end{tabular} & \begin{tabular}[t]{@{}c@{}}0.50\\[-1.5pt]{\tiny[.09,.35]}\end{tabular} & \begin{tabular}[t]{@{}c@{}}0.55\\[-1.5pt]{\tiny[.15,.40]}\end{tabular} \\
 & blocks $K{=}10$ & \begin{tabular}[t]{@{}c@{}}0.25\\[-1.5pt]{\tiny[.15,.30]}\end{tabular} & \begin{tabular}[t]{@{}c@{}}0.35\\[-1.5pt]{\tiny[.24,.40]}\end{tabular} & \begin{tabular}[t]{@{}c@{}}0.35\\[-1.5pt]{\tiny[.30,.45]}\end{tabular} & \begin{tabular}[t]{@{}c@{}}0.40\\[-1.5pt]{\tiny[.30,.45]}\end{tabular} & \begin{tabular}[t]{@{}c@{}}0.45\\[-1.5pt]{\tiny[.28,.45]}\end{tabular} & \begin{tabular}[t]{@{}c@{}}0.45\\[-1.5pt]{\tiny[.05,.45]}\end{tabular} & \begin{tabular}[t]{@{}c@{}}0.25\\[-1.5pt]{\tiny[.00,.60]}\end{tabular} & \begin{tabular}[t]{@{}c@{}}0.35\\[-1.5pt]{\tiny[.00,.22]}\end{tabular} & \begin{tabular}[t]{@{}c@{}}0.40\\[-1.5pt]{\tiny[.00,.31]}\end{tabular} & \begin{tabular}[t]{@{}c@{}}0.50\\[-1.5pt]{\tiny[.00,.33]}\end{tabular} & \begin{tabular}[t]{@{}c@{}}0.55\\[-1.5pt]{\tiny[.05,.35]}\end{tabular} & \begin{tabular}[t]{@{}c@{}}0.60\\[-1.5pt]{\tiny[.14,.40]}\end{tabular} \\
 & blocks $K{=}20$ & \begin{tabular}[t]{@{}c@{}}0.30\\[-1.5pt]{\tiny[.15,.31]}\end{tabular} & \begin{tabular}[t]{@{}c@{}}0.40\\[-1.5pt]{\tiny[.25,.40]}\end{tabular} & \begin{tabular}[t]{@{}c@{}}0.40\\[-1.5pt]{\tiny[.30,.45]}\end{tabular} & \begin{tabular}[t]{@{}c@{}}0.35\\[-1.5pt]{\tiny[.35,.45]}\end{tabular} & \begin{tabular}[t]{@{}c@{}}0.40\\[-1.5pt]{\tiny[.29,.45]}\end{tabular} & \begin{tabular}[t]{@{}c@{}}0.45\\[-1.5pt]{\tiny[.09,.46]}\end{tabular} & \begin{tabular}[t]{@{}c@{}}0.25\\[-1.5pt]{\tiny[.00,.46]}\end{tabular} & \begin{tabular}[t]{@{}c@{}}0.35\\[-1.5pt]{\tiny[.00,.22]}\end{tabular} & \begin{tabular}[t]{@{}c@{}}0.40\\[-1.5pt]{\tiny[.04,.31]}\end{tabular} & \begin{tabular}[t]{@{}c@{}}0.45\\[-1.5pt]{\tiny[.05,.36]}\end{tabular} & \begin{tabular}[t]{@{}c@{}}0.50\\[-1.5pt]{\tiny[.09,.35]}\end{tabular} & \begin{tabular}[t]{@{}c@{}}0.55\\[-1.5pt]{\tiny[.10,.41]}\end{tabular} \\
\bottomrule
\end{tabular}
\end{table}

Persistence substantially raises the value that the count rule asks for: SSJ--ATW with contiguous blocks at $K=20$ moves from $0.25$ at $\psi=0$ to $0.55$ at $\psi=0.9$ and SSJ--KFS from $0.15$ to $0.30$ by $\psi=0.3$, where it settles ($0.25$ at $\psi=0.9$), and the direction is the same for every arm. The reconstruction rule moves in the same direction, from $0.30$ to $0.45$ for SSJ--ATW and from $0.15$ to $0.45$ for SSJ--KFS.

Only contiguous blocks preserve that response. With random folds the movement of the count rule is smaller (SSJ--ATW goes from $0.30$ to $0.40$ and SSJ--KFS from $0.20$ to $0.25$), and it is with random folds that SSJ--ATW stops reaching the true number of discontinuities at all over the middle of the range, whatever the penalty. The mechanism is visible in the selected smoothing constant, which collapses under random folds as $\psi$ grows while the blocked layouts hold it: a held-out point is then nearly predicted by its neighbors in the training set, the criterion rewards a curve that follows the errors, and once $\lambda$ has collapsed the fitted curve absorbs the discontinuities itself, so that no penalty on the count can recover what the smoothing constant has given away. The same collapse is visible in the reconstruction: from moderate persistence onwards the blocked layouts attain the lowest error of any way of choosing $\lambda$, for both estimators and at every $\xi$, after costing a little under independence.

\noindent Regardless of the fold layout, SSJ--KFS suppresses false discoveries at a markedly lower $\xi$ than SSJ--ATW, whose cross-validated arms do not bring the false-discovery rate down anywhere on the grid. The two estimators solve the same problem but their warping sequences are not equally sparse, and this is the simulated counterpart of what Section~\ref{P-sec:carburanti} of the paper finds on the industrial prices, where SSJ--MLE never raises the false alarm at the 2022 policy date and some SSJ--ATW variants do.

\begin{figure}[htbp]\centering
\includegraphics[width=0.90\linewidth]{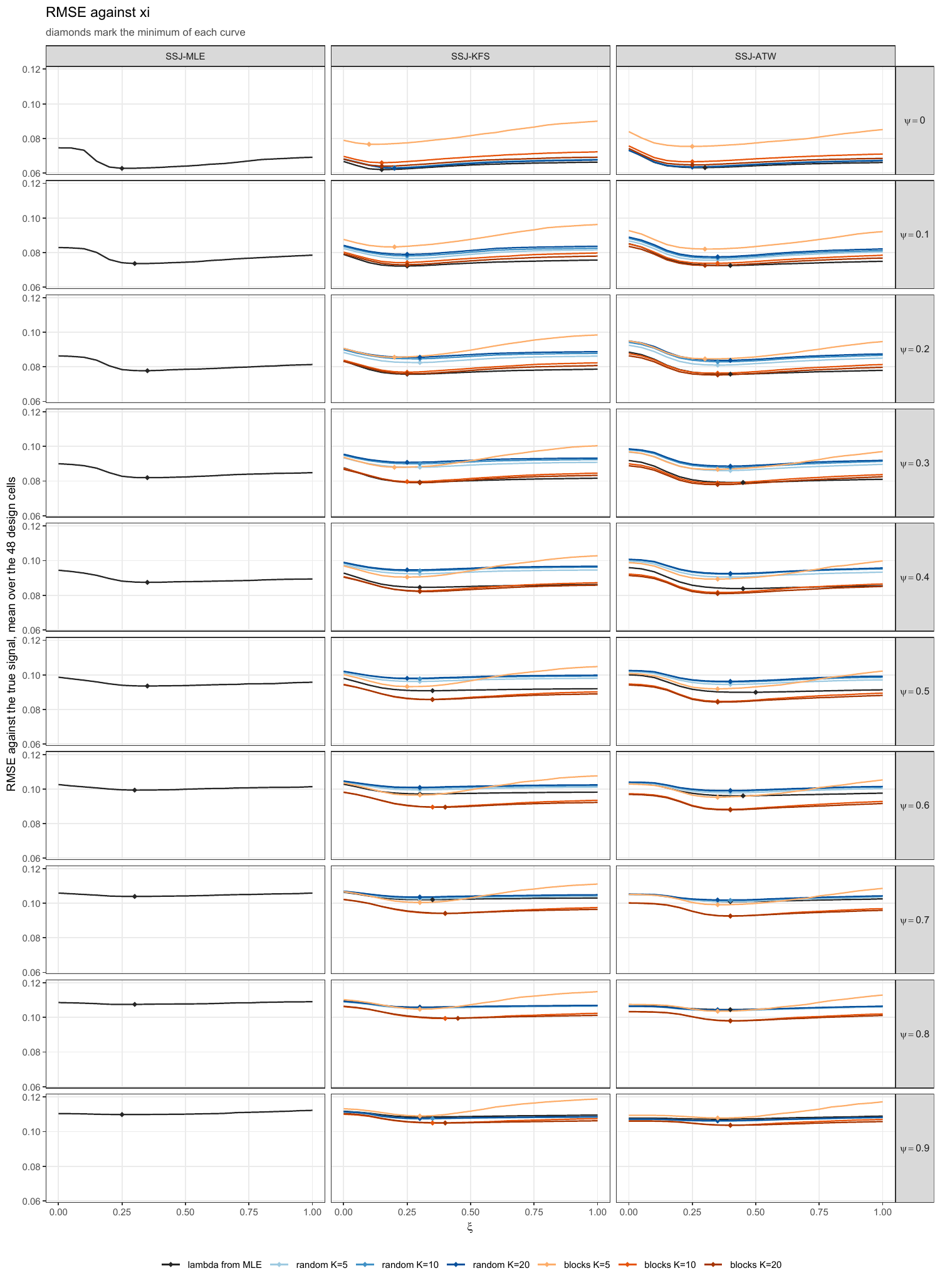}
\caption{Scenario 2, RMSE against the true signal, one level of $\psi$ per row, on a common vertical scale. From $\psi=0.2$ onwards the two block-fold curves at $K=10$ and $K=20$ lie below every other layout, for both estimators and at every $\xi$. This is the clearest single statement of what the fold layout is worth once the errors are dependent.}
\label{fig:rmse_err}
\end{figure}

\begin{figure}[htbp]\centering
\includegraphics[width=0.90\linewidth]{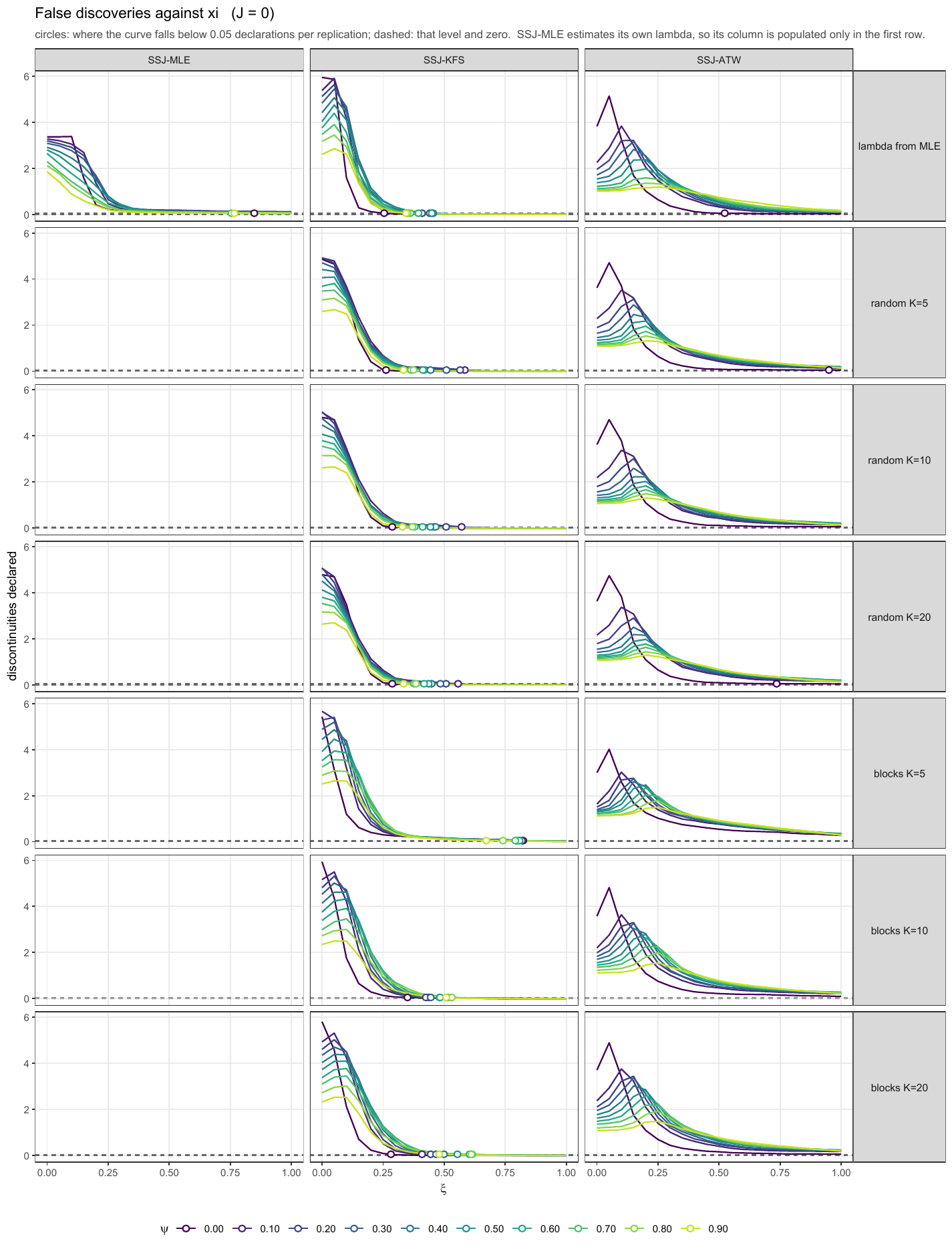}
\caption{Scenario 2, discontinuities declared where there are none; color is $\psi$. Circles mark where a curve falls below $0.05$ declarations per replication. SSJ--KFS reaches that level between $\xi=0.25$ and $0.29$ at $\psi=0$ (later with contiguous blocks at $K=10$ and $K=5$, whose constant over-smooths) and between $0.33$ and $0.51$ at $\psi=0.9$; SSJ--MLE needs $0.85$; and SSJ--ATW reaches it only at $\psi=0$, at $0.52$ with $\lambda$ from maximum likelihood and at $0.74$ and $0.95$ with random folds at $K=20$ and $K=5$, while none of its arms reaches it anywhere on $\xi\le1$ once $\psi>0$. Regardless of the fold layout, SSJ--KFS therefore needs a markedly lower $\xi$ than SSJ--ATW to drive the false-discovery rate to zero, and no admissible penalty makes SSJ--ATW as conservative as SSJ--KFS.}
\label{fig:false_err}
\end{figure}

\begin{figure}[htbp]\centering
\includegraphics[width=0.90\linewidth]{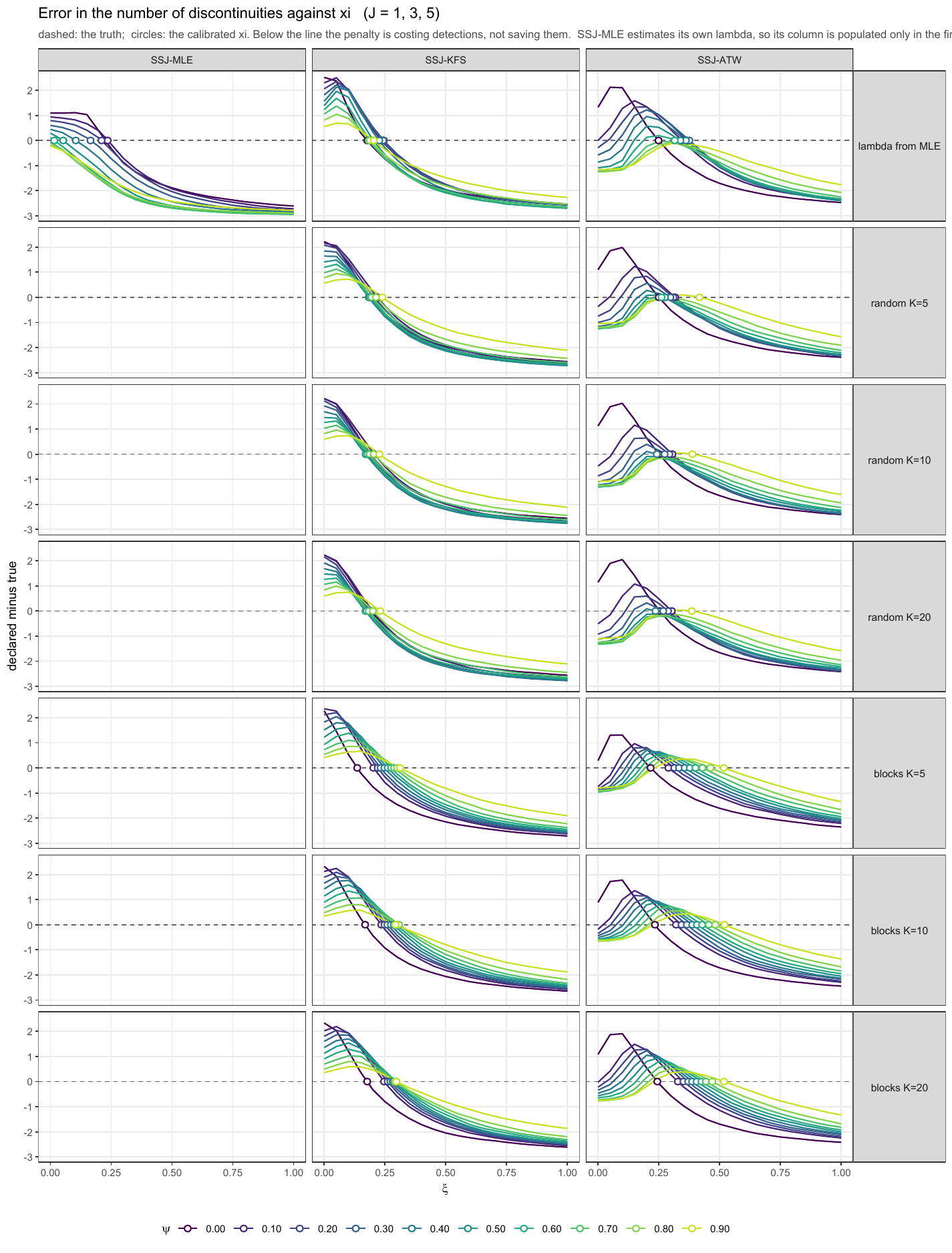}
\caption{Scenario 2, declared minus true over the cells with $q\in\{1,3,5\}$; color is $\psi$, and the circles are the zeros of the curves, not the count rule of Table~\ref{tab:xi_errors_uni}, which sits within a grid step of them where both exist. At $\psi=0$ the zero sits almost exactly at $0.25$ for SSJ--ATW and rises monotonically toward $0.50$ as $\psi$ grows. The SSJ--MLE panel shows the power collapse directly: at $\psi=0.9$ the curve starts below zero at $\xi=0$, so the estimator cannot reach the true number at any penalty.}
\label{fig:bias_err}
\end{figure}

The numbers behind these readings are the following:
\begin{itemize}
\item \textbf{Only contiguous blocks respond to the dependence.} With random folds the movement of the count rule is smaller (i.e., SSJ--ATW goes from $0.30$ to $0.40$ and SSJ--KFS from $0.20$ to $0.25$, against $0.25$ to $0.55$ and $0.15$ to $0.30$ for the corresponding block layouts) and it is with random folds that SSJ--ATW stops reaching the true number of discontinuities at all between $\psi=0.5$ and $\psi=0.8$, whatever the penalty (Section~\ref{sec:xi_rules}). The mechanism is visible in the selected smoothing constant, whose median falls from $10^{-5}$ at $\psi=0$ to $10^{-7}$ by $\psi=0.4$ under random folds while the blocked layouts at $K=10$ and $K=20$ hold at $10^{-5}$ up to $\psi=0.6$ and reach only $10^{-6}$ from $\psi=0.7$. Once $\lambda$ has collapsed the fitted curve absorbs the discontinuities itself, nothing is left for the budget to buy, and no penalty on the count can recover what the smoothing constant has given away.

\item \textbf{The same collapse is visible in the reconstruction.} From $\psi=0.3$ onwards the blocked layouts at $K=10$ and $K=20$ attain a lower RMSE than any other way of choosing $\lambda$, for both estimators and at every $\xi$: at $\psi=0.5$, $0.086$ against $0.098$ with random folds, a gap of more than 10\%. The ordering reverses at $\psi=0$, where blocks cost about 2\%, so contiguous folds are best understood as insurance. Blocks at $K=5$ are the exception at both ends: a fifth of the sample is too long a stretch to hold out, and the criterion is then minimized by a smoothing constant that over-smooths (Figure~\ref{fig:rmse_err}).

\item \textbf{SSJ--MLE loses the ability to detect rather than needing a different penalty.} Its calibrated value in the count error stays between $0.15$ and $0.25$ at every $\psi$, and its minimax-regret value in RMSE between $0.25$ and $0.35$, but its signed count error stops reaching zero from $\psi=0.7$ onwards. At high persistence the estimator cannot reach the true number at any $\xi$, so there is no trade for $\xi$ to make and the count target becomes unattainable. Its row is a diagnosis, not a recommendation.

\item \textbf{SSJ--KFS suppresses false discoveries at a markedly lower $\xi$ than SSJ--ATW, whatever the fold layout.} Taking one spurious discontinuity per twenty replications as the threshold, SSJ--KFS reaches it between $0.25$ and $0.29$ at $\psi=0$ with $\lambda$ from maximum likelihood, from random folds and from blocks at $K=20$ (at $0.35$ with blocks at $K=10$, and not before $\xi=0.8$ with blocks at $K=5$, whose constant over-smooths), and between $0.33$ and $0.51$ at $\psi=0.9$; SSJ--MLE needs $0.85$; and SSJ--ATW reaches it only at $\psi=0$, at $0.52$ with $\lambda$ from maximum likelihood and at $0.74$ and $0.95$ with random folds at $K=20$ and $K=5$, while none of its arms reaches it anywhere on $\xi\le1$ once $\psi>0$ (Figure~\ref{fig:false_err}). The two estimators solve the same problem but their warping sequences are not equally sparse, and no admissible penalty makes SSJ--ATW as conservative as SSJ--KFS. Section~\ref{sec:byq} shows that at the highest levels of persistence it declares fewer discontinuities than there are at every $\xi$.

\item \textbf{The zero of the signed count error moves below the cell-wise zeros as the persistence grows, which identifies cancellation.} At $\psi=0$ the zero of the signed count error defined in Section~\ref{sec:xi_rules} lies within one grid step of the count rule on every one of the fifteen arms, with a median gap of $0.02$. As $\psi$ grows the pooled zero falls below the zeros computed within the individual cells: for SSJ--MLE it is $0.10$ at $\psi=0.4$ and $0.02$ at $\psi=0.6$, whereas the central half of the cells in which a zero exists lies at $0.16$--$0.31$ and $0.10$--$0.28$. A pooled zero below the cell-wise zeros cannot be a low optimum; it is produced by cancellation, the mean count error being brought to zero by over-declaration in the cells with $q=1$ offsetting under-declaration in the cells with $q=5$. The same mechanism removes the zero altogether two levels later, at $\psi\ge0.7$, and it is the reason the rule is built on the absolute count error, which cannot cancel (Figure~\ref{fig:bias_err}).

\item \textbf{The dependence reverses the sample-size pattern.} Under independence the calibrated value falls with $n$; under strong persistence it rises with $n$, because a longer series supplies more observations but not proportionally more independent information, so the penalty required to hold the false-discovery rate rises with the length of the series instead of falling. Section~\ref{sec:byn} reports the values at each sample size.
\end{itemize}

\begin{figure}[htbp]\centering
\includegraphics[width=0.97\linewidth]{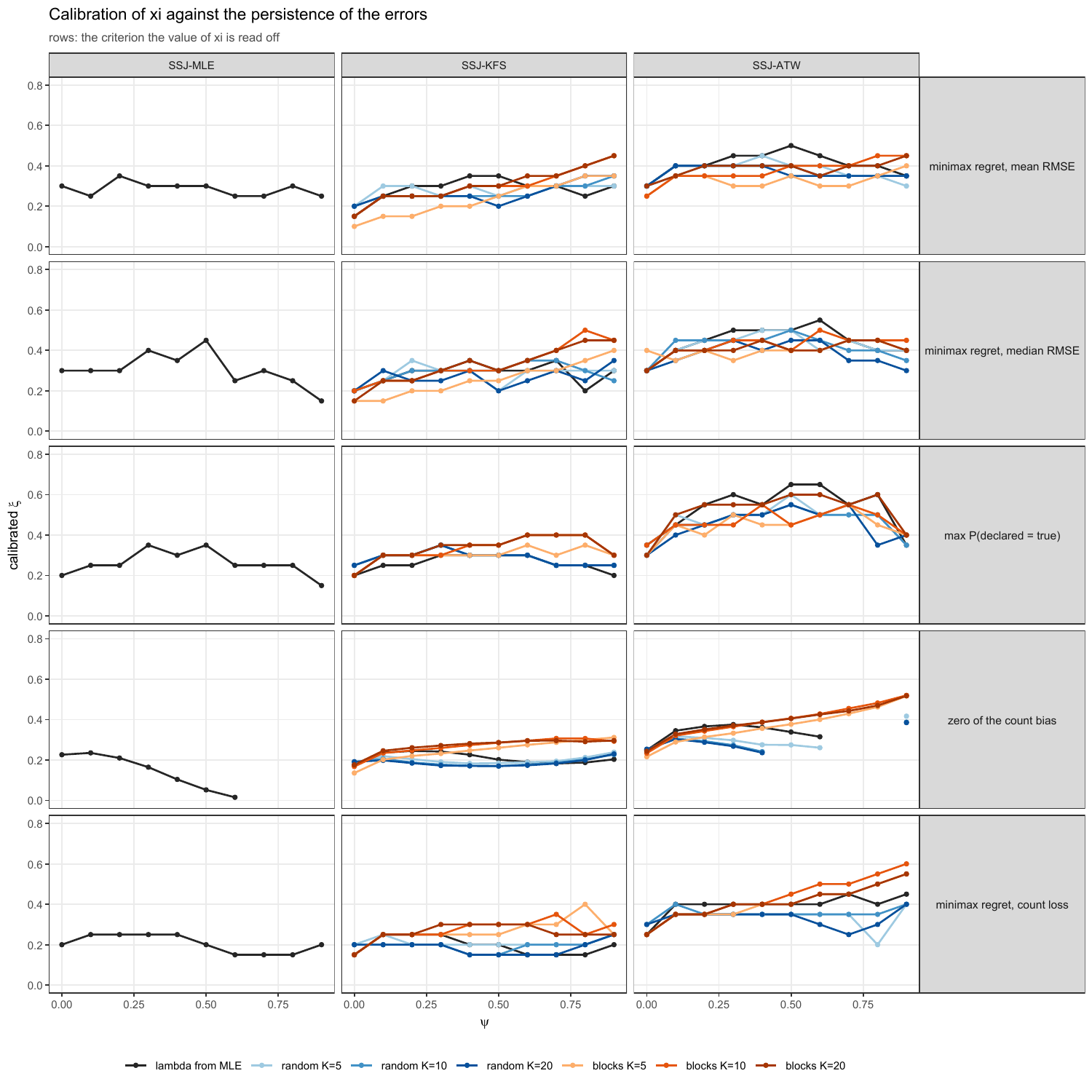}
\caption{Scenario 2: the calibrated $\xi$ against $\psi$ under five criteria. Rows one and two apply minimax regret to the mean and to the median cell summary; row three takes the maximum of exact recovery; row four the zero of the signed count error, kept as the check of Section~\ref{sec:xi_rules}; row five the minimax regret in the count error, which is the second rule of the tables. Rows four and five move together and are visibly smoother than the first two: a criterion built on the count is better conditioned than one that locates the minimum of a flat valley. A gap in row four means the signed error has no zero --- the arm under-declares at every $\xi$ --- where row five is still defined.}
\label{fig:cal_err}
\end{figure}

\begin{figure}[htbp]\centering
\includegraphics[width=0.90\linewidth]{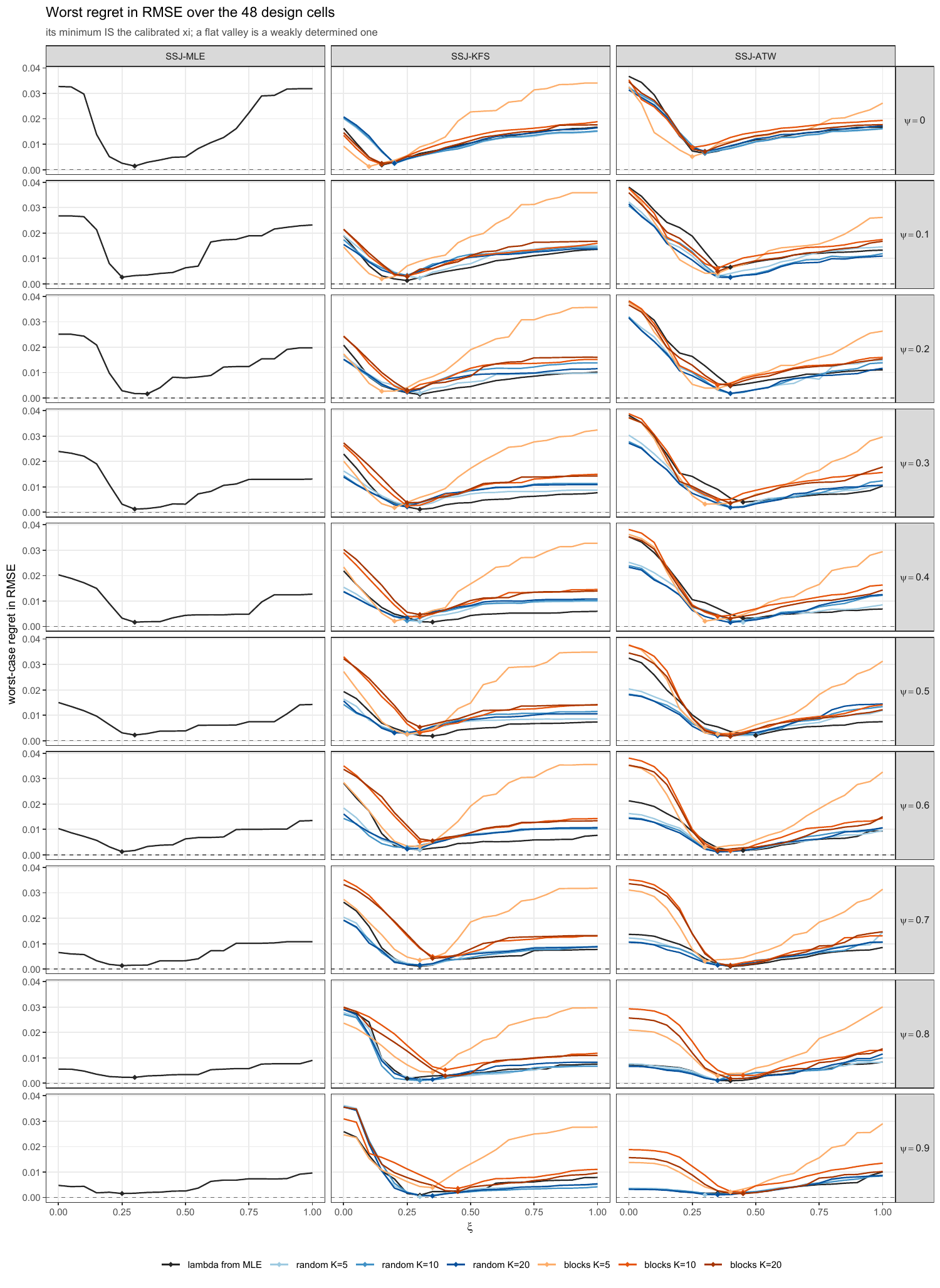}
\caption{Scenario 2, worst-case regret, one level of $\psi$ per row. The valley flattens as $\psi$ grows, which is why the minimax rule becomes weakly determined there and why the count criterion was introduced.}
\label{fig:reg_err}
\end{figure}

\begin{figure}[htbp]\centering
\includegraphics[width=0.88\linewidth]{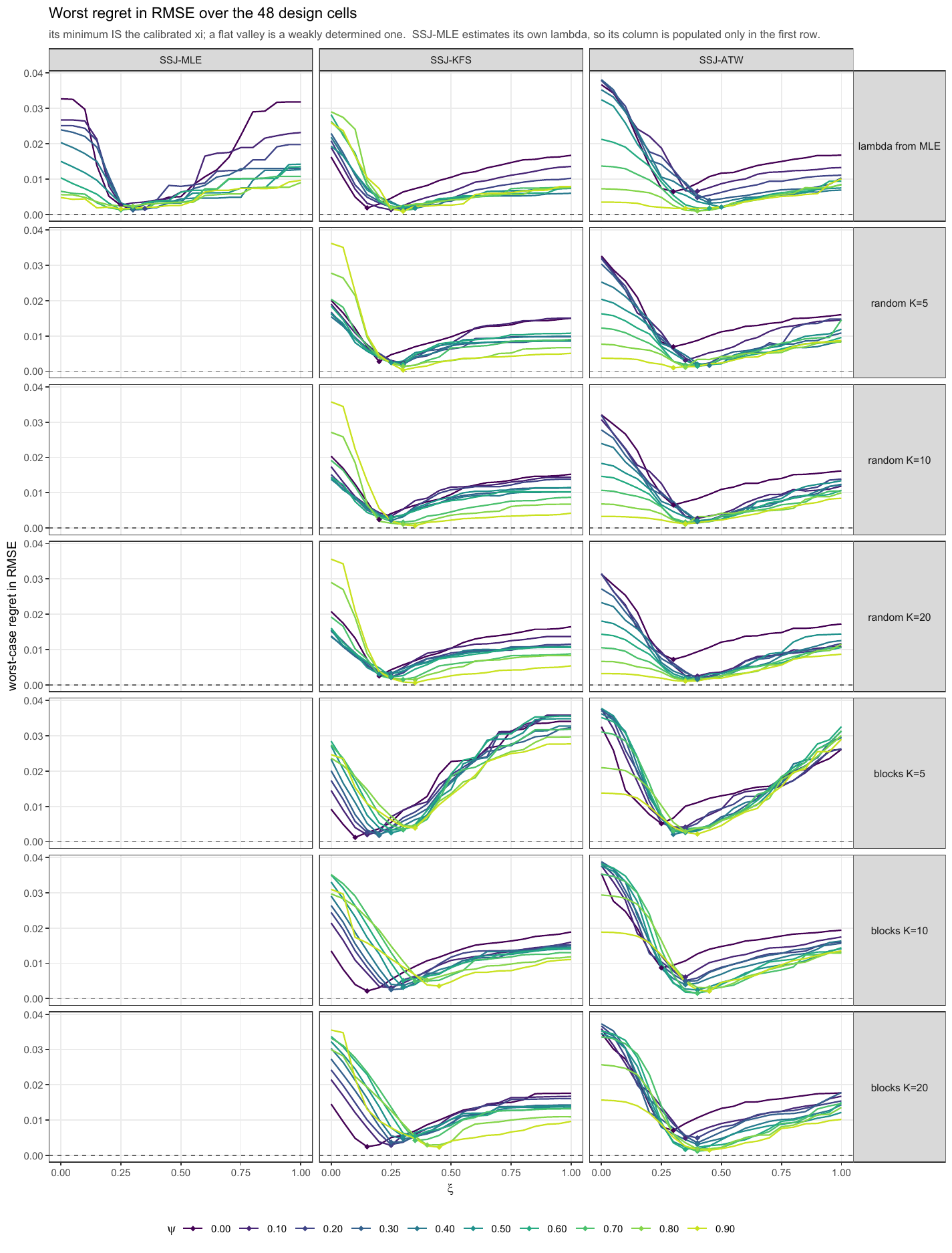}
\caption{Scenario 2, worst-case regret; color is $\psi$.}
\label{fig:reg_err_c}
\end{figure}

\begin{figure}[htbp]\centering
\includegraphics[width=0.88\linewidth]{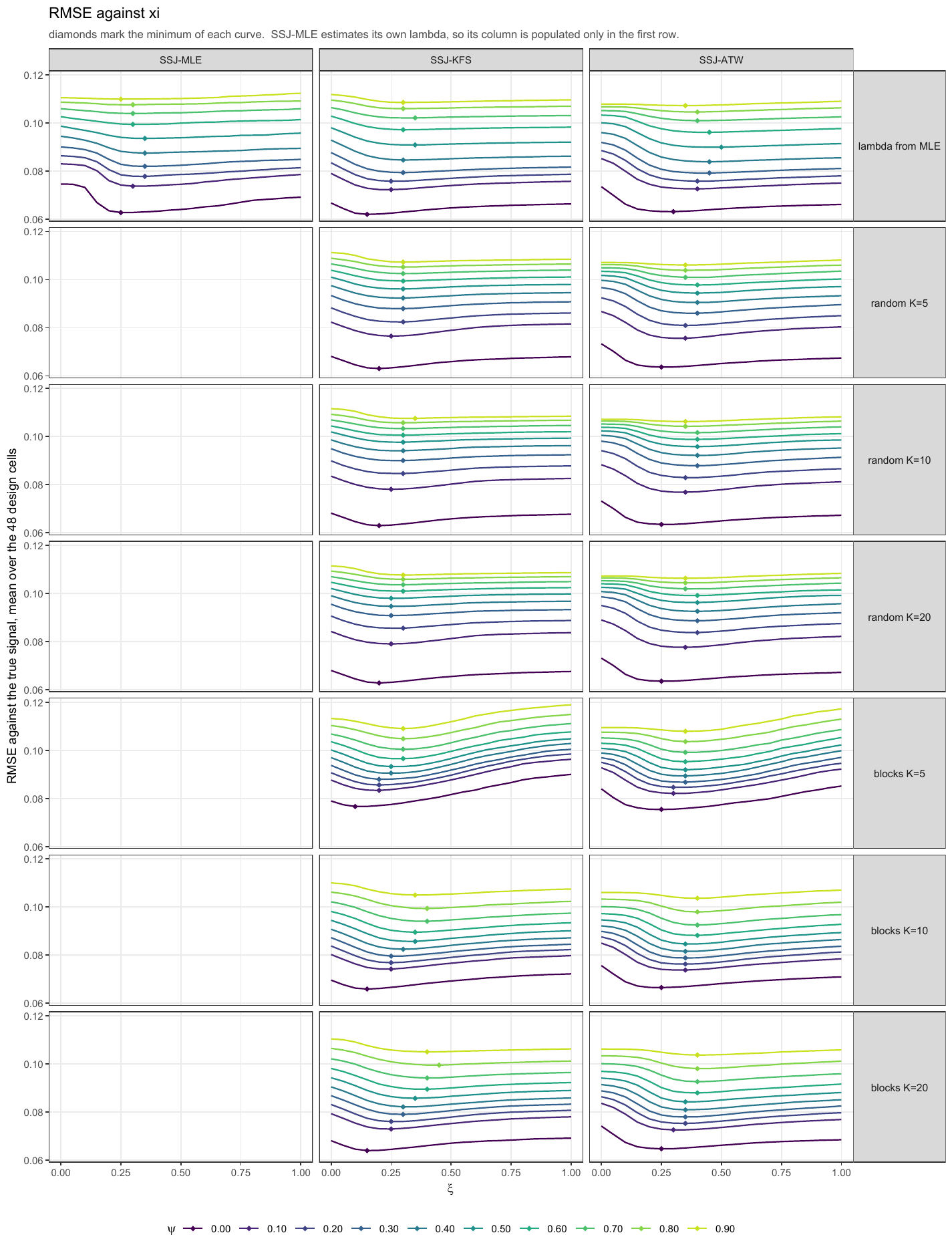}
\caption{Scenario 2, RMSE; color is $\psi$.}
\label{fig:rmse_err_c}
\end{figure}

\begin{figure}[htbp]\centering
\includegraphics[width=0.90\linewidth]{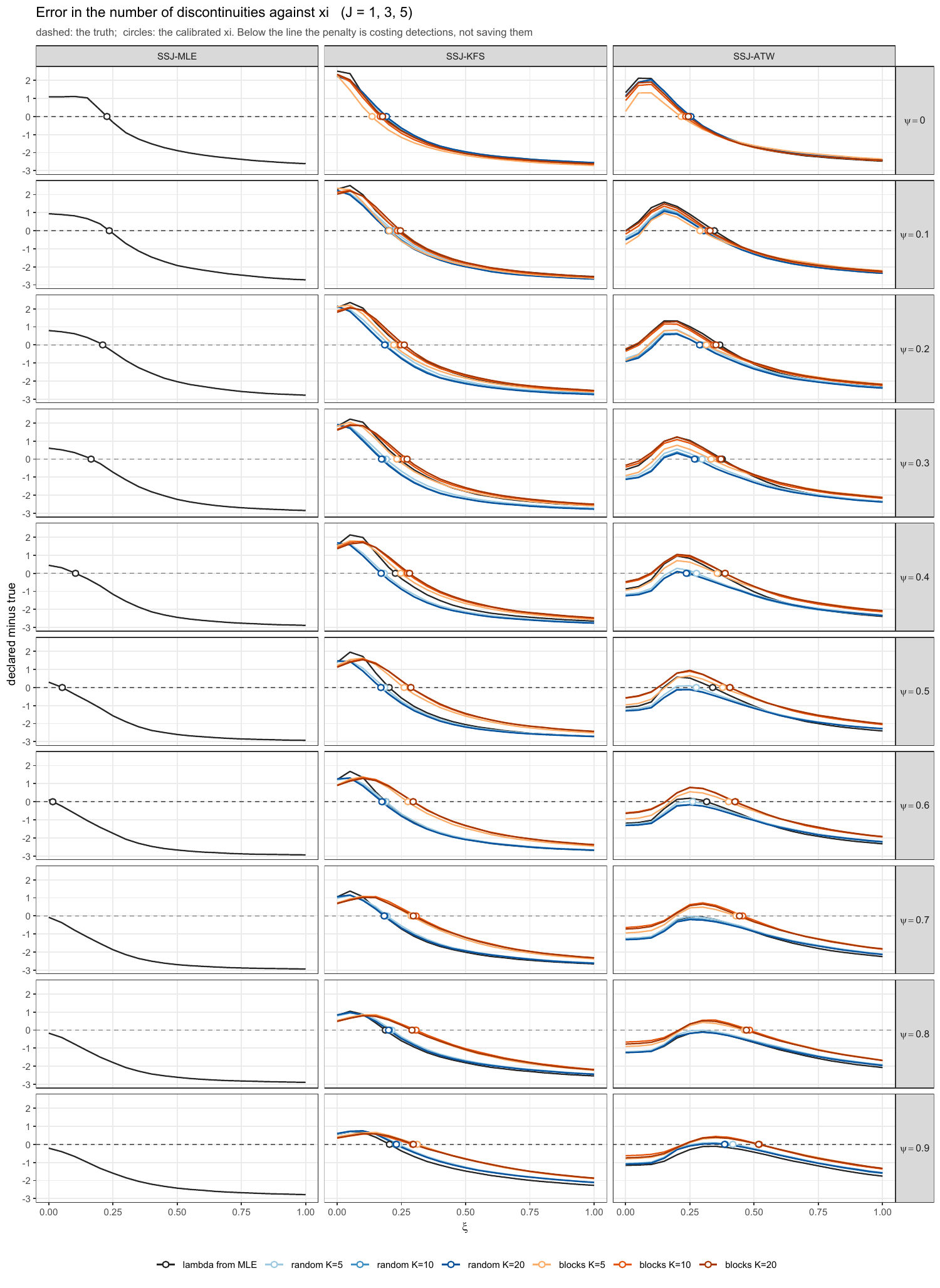}
\caption{Scenario 2, the same, one level of $\psi$ per row. At $\psi=0$ the seven curves lie on top of one another --- every fold layout behaves identically under independence --- and from $\psi=0.5$ onwards the block curves lie above the random ones throughout, for both SSJ--KFS and SSJ--ATW. The reading is that with random folds the fitted curve has already absorbed the discontinuities, so fewer are left to declare at any $\xi$; the block curves stay higher because the smoothing constant they select leaves the discontinuities in the residual, where the budget can still buy them.}
\label{fig:bias_err_m}
\end{figure}

\begin{figure}[htbp]\centering
\includegraphics[width=0.88\linewidth]{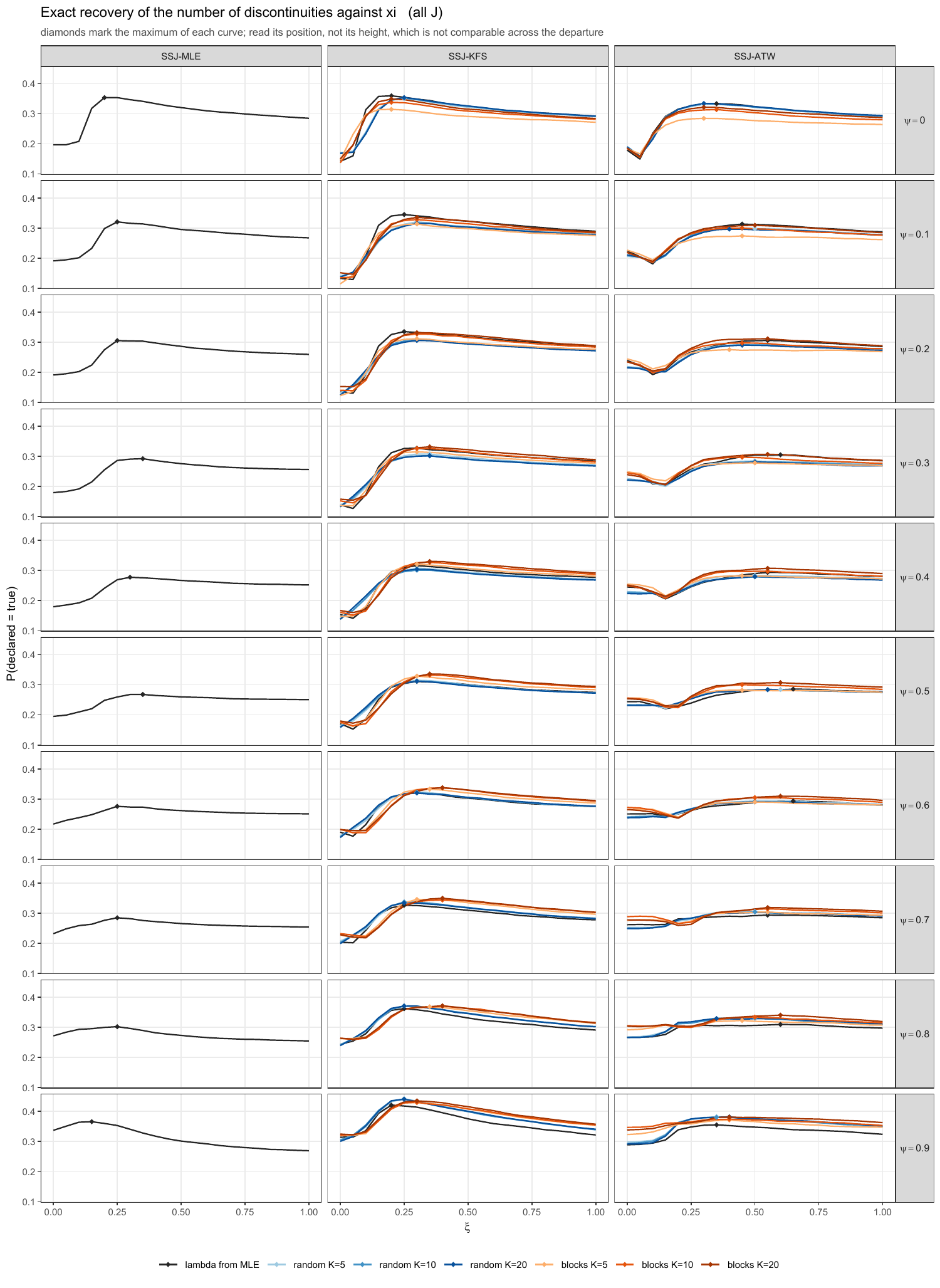}
\caption{Scenario 2, exact recovery over the whole design, one level of $\psi$ per row. The apparent improvement at large $\psi$ is largely the composition effect described above: the estimator declares almost nothing and so is right in the $q=0$ cells.}
\label{fig:hit_err_m}
\end{figure}
\clearpage

\subsection{Scenario 3: a stochastic-trend signal}\label{sec:res_signal}

\begin{table}[htbp]
\centering\footnotesize
\caption{Scenario 3. The calibrated $\xi$ against the weight $\omega$ of the stochastic-trend component under the two rules, minimax regret in RMSE and minimax regret in the count error, both pooled over sample size. Beneath each, in small type, the central half of the same criterion computed cell by cell.}
\label{tab:xi_signal_uni}
\setlength{\tabcolsep}{1.5pt}
\begin{tabular}{ll*{5}{c}@{\enspace}*{5}{c}}
\toprule
Estimator & $\lambda$ from & \multicolumn{5}{c}{minimax regret in RMSE} & \multicolumn{5}{c}{minimax regret in the count error} \\
\cmidrule(lr){3-7}\cmidrule(lr){8-12}
 &  & \multicolumn{5}{c}{$\omega$} & \multicolumn{5}{c}{$\omega$} \\
 &  & 0.00 & 0.25 & 0.50 & 0.75 & 1.00 & 0.00 & 0.25 & 0.50 & 0.75 & 1.00 \\
\midrule
SSJ--MLE & MLE & \begin{tabular}[t]{@{}c@{}}0.30\\[-1.5pt]{\tiny[.25,.35]}\end{tabular} & \begin{tabular}[t]{@{}c@{}}0.30\\[-1.5pt]{\tiny[.25,.31]}\end{tabular} & \begin{tabular}[t]{@{}c@{}}0.30\\[-1.5pt]{\tiny[.25,.35]}\end{tabular} & \begin{tabular}[t]{@{}c@{}}0.35\\[-1.5pt]{\tiny[.25,.36]}\end{tabular} & \begin{tabular}[t]{@{}c@{}}0.45\\[-1.5pt]{\tiny[.30,.45]}\end{tabular} & \begin{tabular}[t]{@{}c@{}}0.20\\[-1.5pt]{\tiny[.00,.42]}\end{tabular} & \begin{tabular}[t]{@{}c@{}}0.20\\[-1.5pt]{\tiny[.00,.41]}\end{tabular} & \begin{tabular}[t]{@{}c@{}}0.25\\[-1.5pt]{\tiny[.00,.46]}\end{tabular} & \begin{tabular}[t]{@{}c@{}}0.30\\[-1.5pt]{\tiny[.04,.84]}\end{tabular} & \begin{tabular}[t]{@{}c@{}}0.40\\[-1.5pt]{\tiny[.00,.95]}\end{tabular} \\
\addlinespace
SSJ--KFS & MLE & \begin{tabular}[t]{@{}c@{}}0.15\\[-1.5pt]{\tiny[.10,.20]}\end{tabular} & \begin{tabular}[t]{@{}c@{}}0.15\\[-1.5pt]{\tiny[.10,.20]}\end{tabular} & \begin{tabular}[t]{@{}c@{}}0.15\\[-1.5pt]{\tiny[.14,.20]}\end{tabular} & \begin{tabular}[t]{@{}c@{}}0.15\\[-1.5pt]{\tiny[.10,.20]}\end{tabular} & \begin{tabular}[t]{@{}c@{}}0.15\\[-1.5pt]{\tiny[.10,.20]}\end{tabular} & \begin{tabular}[t]{@{}c@{}}0.15\\[-1.5pt]{\tiny[.00,.50]}\end{tabular} & \begin{tabular}[t]{@{}c@{}}0.15\\[-1.5pt]{\tiny[.00,.61]}\end{tabular} & \begin{tabular}[t]{@{}c@{}}0.20\\[-1.5pt]{\tiny[.00,.81]}\end{tabular} & \begin{tabular}[t]{@{}c@{}}0.25\\[-1.5pt]{\tiny[.09,.91]}\end{tabular} & \begin{tabular}[t]{@{}c@{}}0.45\\[-1.5pt]{\tiny[.05,1.00]}\end{tabular} \\
 & random $K{=}5$ & \begin{tabular}[t]{@{}c@{}}0.20\\[-1.5pt]{\tiny[.15,.25]}\end{tabular} & \begin{tabular}[t]{@{}c@{}}0.20\\[-1.5pt]{\tiny[.15,.25]}\end{tabular} & \begin{tabular}[t]{@{}c@{}}0.20\\[-1.5pt]{\tiny[.14,.21]}\end{tabular} & \begin{tabular}[t]{@{}c@{}}0.15\\[-1.5pt]{\tiny[.10,.25]}\end{tabular} & \begin{tabular}[t]{@{}c@{}}0.20\\[-1.5pt]{\tiny[.10,.20]}\end{tabular} & \begin{tabular}[t]{@{}c@{}}0.20\\[-1.5pt]{\tiny[.00,.60]}\end{tabular} & \begin{tabular}[t]{@{}c@{}}0.15\\[-1.5pt]{\tiny[.00,.63]}\end{tabular} & \begin{tabular}[t]{@{}c@{}}0.20\\[-1.5pt]{\tiny[.00,.70]}\end{tabular} & \begin{tabular}[t]{@{}c@{}}0.20\\[-1.5pt]{\tiny[.05,.85]}\end{tabular} & \begin{tabular}[t]{@{}c@{}}0.25\\[-1.5pt]{\tiny[.05,.95]}\end{tabular} \\
 & random $K{=}10$ & \begin{tabular}[t]{@{}c@{}}0.20\\[-1.5pt]{\tiny[.14,.25]}\end{tabular} & \begin{tabular}[t]{@{}c@{}}0.20\\[-1.5pt]{\tiny[.14,.21]}\end{tabular} & \begin{tabular}[t]{@{}c@{}}0.20\\[-1.5pt]{\tiny[.14,.25]}\end{tabular} & \begin{tabular}[t]{@{}c@{}}0.20\\[-1.5pt]{\tiny[.10,.25]}\end{tabular} & \begin{tabular}[t]{@{}c@{}}0.15\\[-1.5pt]{\tiny[.10,.25]}\end{tabular} & \begin{tabular}[t]{@{}c@{}}0.20\\[-1.5pt]{\tiny[.00,.60]}\end{tabular} & \begin{tabular}[t]{@{}c@{}}0.20\\[-1.5pt]{\tiny[.00,.56]}\end{tabular} & \begin{tabular}[t]{@{}c@{}}0.20\\[-1.5pt]{\tiny[.00,.71]}\end{tabular} & \begin{tabular}[t]{@{}c@{}}0.20\\[-1.5pt]{\tiny[.05,.81]}\end{tabular} & \begin{tabular}[t]{@{}c@{}}0.25\\[-1.5pt]{\tiny[.10,.95]}\end{tabular} \\
 & random $K{=}20$ & \begin{tabular}[t]{@{}c@{}}0.20\\[-1.5pt]{\tiny[.15,.25]}\end{tabular} & \begin{tabular}[t]{@{}c@{}}0.20\\[-1.5pt]{\tiny[.15,.25]}\end{tabular} & \begin{tabular}[t]{@{}c@{}}0.20\\[-1.5pt]{\tiny[.14,.26]}\end{tabular} & \begin{tabular}[t]{@{}c@{}}0.20\\[-1.5pt]{\tiny[.10,.25]}\end{tabular} & \begin{tabular}[t]{@{}c@{}}0.20\\[-1.5pt]{\tiny[.10,.25]}\end{tabular} & \begin{tabular}[t]{@{}c@{}}0.20\\[-1.5pt]{\tiny[.00,.60]}\end{tabular} & \begin{tabular}[t]{@{}c@{}}0.20\\[-1.5pt]{\tiny[.00,.58]}\end{tabular} & \begin{tabular}[t]{@{}c@{}}0.20\\[-1.5pt]{\tiny[.00,.65]}\end{tabular} & \begin{tabular}[t]{@{}c@{}}0.20\\[-1.5pt]{\tiny[.05,.85]}\end{tabular} & \begin{tabular}[t]{@{}c@{}}0.20\\[-1.5pt]{\tiny[.12,.95]}\end{tabular} \\
 & blocks $K{=}5$ & \begin{tabular}[t]{@{}c@{}}0.10\\[-1.5pt]{\tiny[.10,.20]}\end{tabular} & \begin{tabular}[t]{@{}c@{}}0.10\\[-1.5pt]{\tiny[.10,.16]}\end{tabular} & \begin{tabular}[t]{@{}c@{}}0.10\\[-1.5pt]{\tiny[.10,.20]}\end{tabular} & \begin{tabular}[t]{@{}c@{}}0.10\\[-1.5pt]{\tiny[.10,.15]}\end{tabular} & \begin{tabular}[t]{@{}c@{}}0.15\\[-1.5pt]{\tiny[.10,.20]}\end{tabular} & \begin{tabular}[t]{@{}c@{}}0.15\\[-1.5pt]{\tiny[.00,.76]}\end{tabular} & \begin{tabular}[t]{@{}c@{}}0.15\\[-1.5pt]{\tiny[.00,.70]}\end{tabular} & \begin{tabular}[t]{@{}c@{}}0.20\\[-1.5pt]{\tiny[.00,.86]}\end{tabular} & \begin{tabular}[t]{@{}c@{}}0.25\\[-1.5pt]{\tiny[.00,.95]}\end{tabular} & \begin{tabular}[t]{@{}c@{}}0.30\\[-1.5pt]{\tiny[.04,1.00]}\end{tabular} \\
 & blocks $K{=}10$ & \begin{tabular}[t]{@{}c@{}}0.15\\[-1.5pt]{\tiny[.10,.21]}\end{tabular} & \begin{tabular}[t]{@{}c@{}}0.15\\[-1.5pt]{\tiny[.10,.20]}\end{tabular} & \begin{tabular}[t]{@{}c@{}}0.15\\[-1.5pt]{\tiny[.10,.20]}\end{tabular} & \begin{tabular}[t]{@{}c@{}}0.15\\[-1.5pt]{\tiny[.10,.20]}\end{tabular} & \begin{tabular}[t]{@{}c@{}}0.15\\[-1.5pt]{\tiny[.10,.20]}\end{tabular} & \begin{tabular}[t]{@{}c@{}}0.15\\[-1.5pt]{\tiny[.00,.68]}\end{tabular} & \begin{tabular}[t]{@{}c@{}}0.15\\[-1.5pt]{\tiny[.00,.81]}\end{tabular} & \begin{tabular}[t]{@{}c@{}}0.15\\[-1.5pt]{\tiny[.00,.80]}\end{tabular} & \begin{tabular}[t]{@{}c@{}}0.20\\[-1.5pt]{\tiny[.00,.95]}\end{tabular} & \begin{tabular}[t]{@{}c@{}}0.25\\[-1.5pt]{\tiny[.10,.96]}\end{tabular} \\
 & blocks $K{=}20$ & \begin{tabular}[t]{@{}c@{}}0.15\\[-1.5pt]{\tiny[.10,.25]}\end{tabular} & \begin{tabular}[t]{@{}c@{}}0.15\\[-1.5pt]{\tiny[.10,.25]}\end{tabular} & \begin{tabular}[t]{@{}c@{}}0.15\\[-1.5pt]{\tiny[.10,.25]}\end{tabular} & \begin{tabular}[t]{@{}c@{}}0.15\\[-1.5pt]{\tiny[.10,.25]}\end{tabular} & \begin{tabular}[t]{@{}c@{}}0.20\\[-1.5pt]{\tiny[.10,.21]}\end{tabular} & \begin{tabular}[t]{@{}c@{}}0.15\\[-1.5pt]{\tiny[.00,.51]}\end{tabular} & \begin{tabular}[t]{@{}c@{}}0.15\\[-1.5pt]{\tiny[.00,.63]}\end{tabular} & \begin{tabular}[t]{@{}c@{}}0.15\\[-1.5pt]{\tiny[.00,.80]}\end{tabular} & \begin{tabular}[t]{@{}c@{}}0.20\\[-1.5pt]{\tiny[.05,.81]}\end{tabular} & \begin{tabular}[t]{@{}c@{}}0.25\\[-1.5pt]{\tiny[.19,.96]}\end{tabular} \\
\addlinespace
SSJ--ATW & MLE & \begin{tabular}[t]{@{}c@{}}0.30\\[-1.5pt]{\tiny[.15,.31]}\end{tabular} & \begin{tabular}[t]{@{}c@{}}0.25\\[-1.5pt]{\tiny[.15,.30]}\end{tabular} & \begin{tabular}[t]{@{}c@{}}0.25\\[-1.5pt]{\tiny[.15,.30]}\end{tabular} & \begin{tabular}[t]{@{}c@{}}0.20\\[-1.5pt]{\tiny[.15,.26]}\end{tabular} & \begin{tabular}[t]{@{}c@{}}0.15\\[-1.5pt]{\tiny[.10,.20]}\end{tabular} & \begin{tabular}[t]{@{}c@{}}0.25\\[-1.5pt]{\tiny[.00,.56]}\end{tabular} & \begin{tabular}[t]{@{}c@{}}0.30\\[-1.5pt]{\tiny[.00,.58]}\end{tabular} & \begin{tabular}[t]{@{}c@{}}0.30\\[-1.5pt]{\tiny[.00,.85]}\end{tabular} & \begin{tabular}[t]{@{}c@{}}0.35\\[-1.5pt]{\tiny[.00,1.00]}\end{tabular} & \begin{tabular}[t]{@{}c@{}}0.50\\[-1.5pt]{\tiny[.04,.95]}\end{tabular} \\
 & random $K{=}5$ & \begin{tabular}[t]{@{}c@{}}0.30\\[-1.5pt]{\tiny[.15,.30]}\end{tabular} & \begin{tabular}[t]{@{}c@{}}0.25\\[-1.5pt]{\tiny[.15,.31]}\end{tabular} & \begin{tabular}[t]{@{}c@{}}0.25\\[-1.5pt]{\tiny[.15,.30]}\end{tabular} & \begin{tabular}[t]{@{}c@{}}0.25\\[-1.5pt]{\tiny[.15,.30]}\end{tabular} & \begin{tabular}[t]{@{}c@{}}0.25\\[-1.5pt]{\tiny[.15,.26]}\end{tabular} & \begin{tabular}[t]{@{}c@{}}0.25\\[-1.5pt]{\tiny[.00,.44]}\end{tabular} & \begin{tabular}[t]{@{}c@{}}0.30\\[-1.5pt]{\tiny[.00,.69]}\end{tabular} & \begin{tabular}[t]{@{}c@{}}0.25\\[-1.5pt]{\tiny[.00,.70]}\end{tabular} & \begin{tabular}[t]{@{}c@{}}0.35\\[-1.5pt]{\tiny[.00,.86]}\end{tabular} & \begin{tabular}[t]{@{}c@{}}0.35\\[-1.5pt]{\tiny[.14,.91]}\end{tabular} \\
 & random $K{=}10$ & \begin{tabular}[t]{@{}c@{}}0.30\\[-1.5pt]{\tiny[.15,.30]}\end{tabular} & \begin{tabular}[t]{@{}c@{}}0.25\\[-1.5pt]{\tiny[.15,.30]}\end{tabular} & \begin{tabular}[t]{@{}c@{}}0.25\\[-1.5pt]{\tiny[.15,.30]}\end{tabular} & \begin{tabular}[t]{@{}c@{}}0.25\\[-1.5pt]{\tiny[.15,.30]}\end{tabular} & \begin{tabular}[t]{@{}c@{}}0.25\\[-1.5pt]{\tiny[.15,.25]}\end{tabular} & \begin{tabular}[t]{@{}c@{}}0.30\\[-1.5pt]{\tiny[.00,.42]}\end{tabular} & \begin{tabular}[t]{@{}c@{}}0.30\\[-1.5pt]{\tiny[.00,.73]}\end{tabular} & \begin{tabular}[t]{@{}c@{}}0.30\\[-1.5pt]{\tiny[.00,.81]}\end{tabular} & \begin{tabular}[t]{@{}c@{}}0.35\\[-1.5pt]{\tiny[.00,.86]}\end{tabular} & \begin{tabular}[t]{@{}c@{}}0.40\\[-1.5pt]{\tiny[.10,.90]}\end{tabular} \\
 & random $K{=}20$ & \begin{tabular}[t]{@{}c@{}}0.30\\[-1.5pt]{\tiny[.15,.30]}\end{tabular} & \begin{tabular}[t]{@{}c@{}}0.25\\[-1.5pt]{\tiny[.15,.30]}\end{tabular} & \begin{tabular}[t]{@{}c@{}}0.25\\[-1.5pt]{\tiny[.15,.25]}\end{tabular} & \begin{tabular}[t]{@{}c@{}}0.25\\[-1.5pt]{\tiny[.15,.30]}\end{tabular} & \begin{tabular}[t]{@{}c@{}}0.25\\[-1.5pt]{\tiny[.15,.25]}\end{tabular} & \begin{tabular}[t]{@{}c@{}}0.30\\[-1.5pt]{\tiny[.00,.41]}\end{tabular} & \begin{tabular}[t]{@{}c@{}}0.30\\[-1.5pt]{\tiny[.00,.63]}\end{tabular} & \begin{tabular}[t]{@{}c@{}}0.30\\[-1.5pt]{\tiny[.00,.73]}\end{tabular} & \begin{tabular}[t]{@{}c@{}}0.35\\[-1.5pt]{\tiny[.00,.86]}\end{tabular} & \begin{tabular}[t]{@{}c@{}}0.40\\[-1.5pt]{\tiny[.14,.90]}\end{tabular} \\
 & blocks $K{=}5$ & \begin{tabular}[t]{@{}c@{}}0.25\\[-1.5pt]{\tiny[.14,.35]}\end{tabular} & \begin{tabular}[t]{@{}c@{}}0.25\\[-1.5pt]{\tiny[.14,.30]}\end{tabular} & \begin{tabular}[t]{@{}c@{}}0.25\\[-1.5pt]{\tiny[.10,.25]}\end{tabular} & \begin{tabular}[t]{@{}c@{}}0.20\\[-1.5pt]{\tiny[.15,.25]}\end{tabular} & \begin{tabular}[t]{@{}c@{}}0.20\\[-1.5pt]{\tiny[.10,.20]}\end{tabular} & \begin{tabular}[t]{@{}c@{}}0.25\\[-1.5pt]{\tiny[.00,.78]}\end{tabular} & \begin{tabular}[t]{@{}c@{}}0.25\\[-1.5pt]{\tiny[.00,.66]}\end{tabular} & \begin{tabular}[t]{@{}c@{}}0.25\\[-1.5pt]{\tiny[.00,.90]}\end{tabular} & \begin{tabular}[t]{@{}c@{}}0.30\\[-1.5pt]{\tiny[.00,.90]}\end{tabular} & \begin{tabular}[t]{@{}c@{}}0.45\\[-1.5pt]{\tiny[.05,.96]}\end{tabular} \\
 & blocks $K{=}10$ & \begin{tabular}[t]{@{}c@{}}0.25\\[-1.5pt]{\tiny[.15,.30]}\end{tabular} & \begin{tabular}[t]{@{}c@{}}0.25\\[-1.5pt]{\tiny[.15,.35]}\end{tabular} & \begin{tabular}[t]{@{}c@{}}0.25\\[-1.5pt]{\tiny[.15,.25]}\end{tabular} & \begin{tabular}[t]{@{}c@{}}0.20\\[-1.5pt]{\tiny[.15,.25]}\end{tabular} & \begin{tabular}[t]{@{}c@{}}0.20\\[-1.5pt]{\tiny[.10,.20]}\end{tabular} & \begin{tabular}[t]{@{}c@{}}0.25\\[-1.5pt]{\tiny[.00,.60]}\end{tabular} & \begin{tabular}[t]{@{}c@{}}0.25\\[-1.5pt]{\tiny[.00,.65]}\end{tabular} & \begin{tabular}[t]{@{}c@{}}0.25\\[-1.5pt]{\tiny[.00,.86]}\end{tabular} & \begin{tabular}[t]{@{}c@{}}0.35\\[-1.5pt]{\tiny[.00,.83]}\end{tabular} & \begin{tabular}[t]{@{}c@{}}0.40\\[-1.5pt]{\tiny[.05,.95]}\end{tabular} \\
 & blocks $K{=}20$ & \begin{tabular}[t]{@{}c@{}}0.30\\[-1.5pt]{\tiny[.15,.31]}\end{tabular} & \begin{tabular}[t]{@{}c@{}}0.25\\[-1.5pt]{\tiny[.15,.26]}\end{tabular} & \begin{tabular}[t]{@{}c@{}}0.25\\[-1.5pt]{\tiny[.15,.26]}\end{tabular} & \begin{tabular}[t]{@{}c@{}}0.25\\[-1.5pt]{\tiny[.15,.25]}\end{tabular} & \begin{tabular}[t]{@{}c@{}}0.20\\[-1.5pt]{\tiny[.15,.25]}\end{tabular} & \begin{tabular}[t]{@{}c@{}}0.25\\[-1.5pt]{\tiny[.00,.46]}\end{tabular} & \begin{tabular}[t]{@{}c@{}}0.30\\[-1.5pt]{\tiny[.00,.70]}\end{tabular} & \begin{tabular}[t]{@{}c@{}}0.30\\[-1.5pt]{\tiny[.00,.78]}\end{tabular} & \begin{tabular}[t]{@{}c@{}}0.30\\[-1.5pt]{\tiny[.00,.81]}\end{tabular} & \begin{tabular}[t]{@{}c@{}}0.35\\[-1.5pt]{\tiny[.10,.95]}\end{tabular} \\
\bottomrule
\end{tabular}
\end{table}

Here the two criteria disagree. On reconstruction the baseline calibration transfers unchanged: across all fifteen arms and all five levels of $\omega$ the minimax value stays between $0.15$ and $0.30$ for SSJ--ATW, $0.10$ and $0.20$ for SSJ--KFS and $0.30$ and $0.45$ for SSJ--MLE, so the calibration is not an artifact of the particular curve used, and only SSJ--MLE (upward) drifts. The reconstruction even improves at each arm's calibrated value: a signal drawn from the estimator's own prior is one it reconstructs well.

The declared count meanwhile deteriorates markedly. At $\omega=1$, with nothing to find, the estimators with $\lambda$ from SSJ--MLE declare several discontinuities per replication where at $\omega=0$ they declared almost none, and the same happens where there is something to find, so this is inflation of the count and not a failure confined to the null cells. The count rule therefore moves a long way: from $0.25$ to $0.50$ for SSJ--ATW at the maximum likelihood $\lambda$, from $0.15$ to $0.45$ for SSJ--KFS and from $0.20$ to $0.40$ for SSJ--MLE; with a cross-validated $\lambda$ the movement is smaller, to $0.35$--$0.45$ and to $0.20$--$0.30$, but in the same direction for fourteen of the fifteen arms (SSJ--KFS with random folds at $K=20$ does not move). The rise is not gradual: flat to $\omega=0.5$ and steep thereafter, so a series only partly trend-like needs little correction.

Exact recovery improves, but only at a higher $\xi$. Its maximum over $\xi$ rises for nearly every arm, and the value at which it is attained roughly doubles, to about $0.5$. This is not a composition effect of the $q=0$ cells becoming easier: restricted to the cells where there is something to find, the probability of recovering the count exactly at the calibrated value is substantially higher at $\omega=1$ than at $\omega=0$, in every arm.

The general reading is that in the presence of stochastic trends a single fixed constant might be wrong. A criterion built on reconstruction error is nearly blind to a change that only affects how a fitted warping sequence is turned into a list of discontinuities, and at $\omega=1$ the estimator's view of the signal is exactly right, so nothing can be blamed on misspecification. What is left is the warping budget and the rule that reads it. At $\omega=1$ the criterion buys more warping than at $\omega=0$: in the cells with $q=0$, at each arm's calibrated value, the selected budget rises from $0.05$ to $0.14$ for SSJ--MLE and from $0.14$ to $0.22$ for SSJ--ATW at $\lambda$ from SSJ--MLE, the number of active warps from about one and a half to two and from three and a half to six, and for SSJ--KFS at that $\lambda$ from under one to more than six, because a wandering trend is cheaper to follow by deforming the domain at a few places than by paying curvature everywhere (i.e., the effective degrees of freedom actually \emph{fall}, from $0.19n$ to about $0.09n$). The declaration rule then asks only whether a warp holds more than a fixed share of the total warping, and one or two of those warps do, so they are reported as discontinuities on a signal that has none. The rule cannot distinguish warping bought to track a wandering trend from warping bought to represent a jump, and that is the whole of the effect.

\begin{figure}[htbp]\centering
\includegraphics[width=0.90\linewidth]{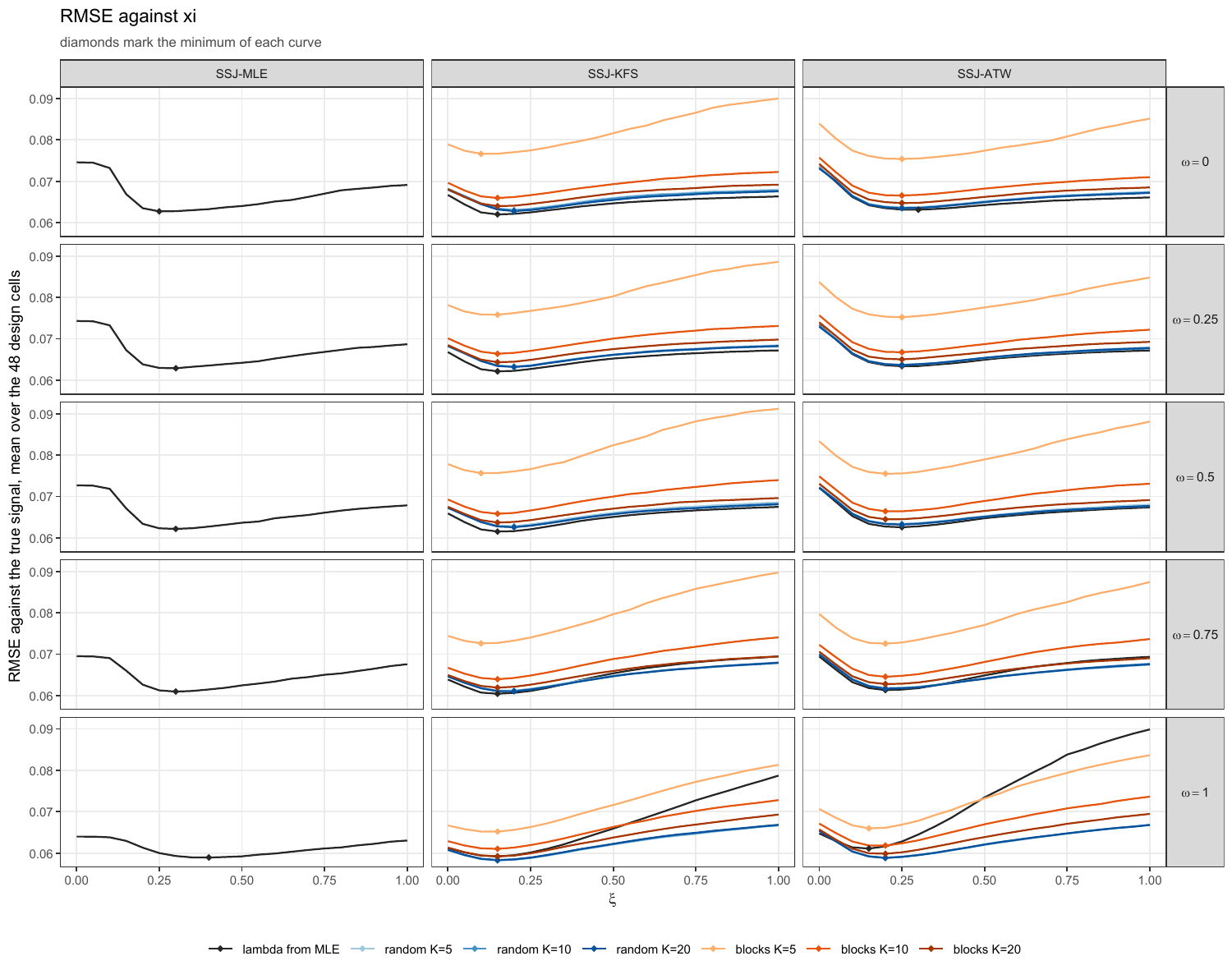}
\caption{Scenario 3, the same. The ordering of the fold layouts is the reverse of Figure~\ref{fig:rmse_err}: with independent errors the random-fold curves lie below the block ones at every $\omega$, and blocks at $K=5$ are the worst of the seven --- a fifth of the sample is too long a stretch to hold out, and the criterion is then minimized by a constant that over-smooths.}
\label{fig:rmse_sig}
\end{figure}

\begin{figure}[htbp]\centering
\includegraphics[width=0.90\linewidth]{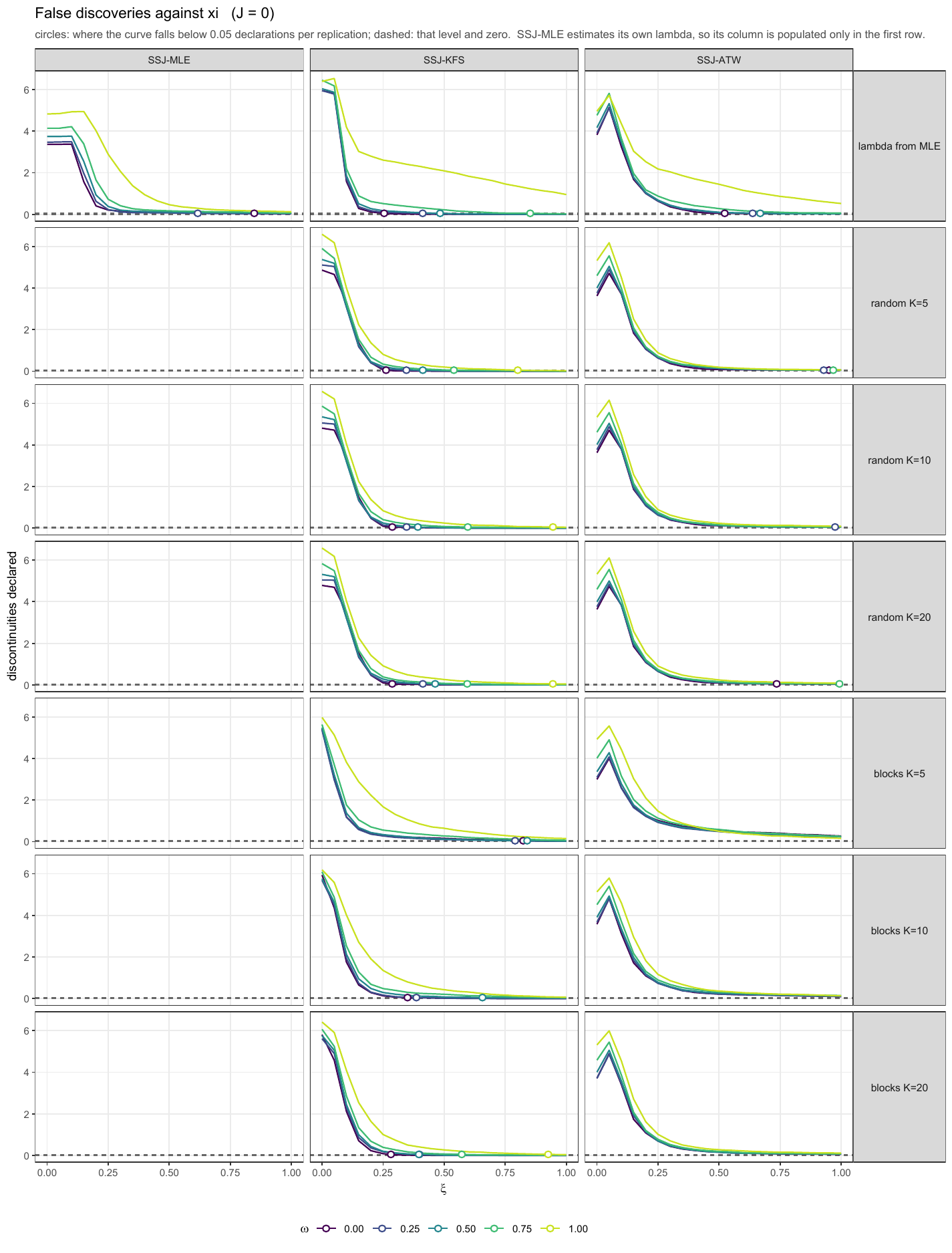}
\caption{Scenario 3, the same against $\omega$. The $\omega=1$ curve sits well above the rest and decays far more slowly: SSJ--KFS at $\lambda$ from maximum likelihood still declares about one spurious discontinuity per replication at $\xi=1$, where at $\omega\le0.5$ it is below one in twenty by $\xi=0.5$.}
\label{fig:false_sig}
\end{figure}

\begin{figure}[htbp]\centering
\includegraphics[width=0.90\linewidth]{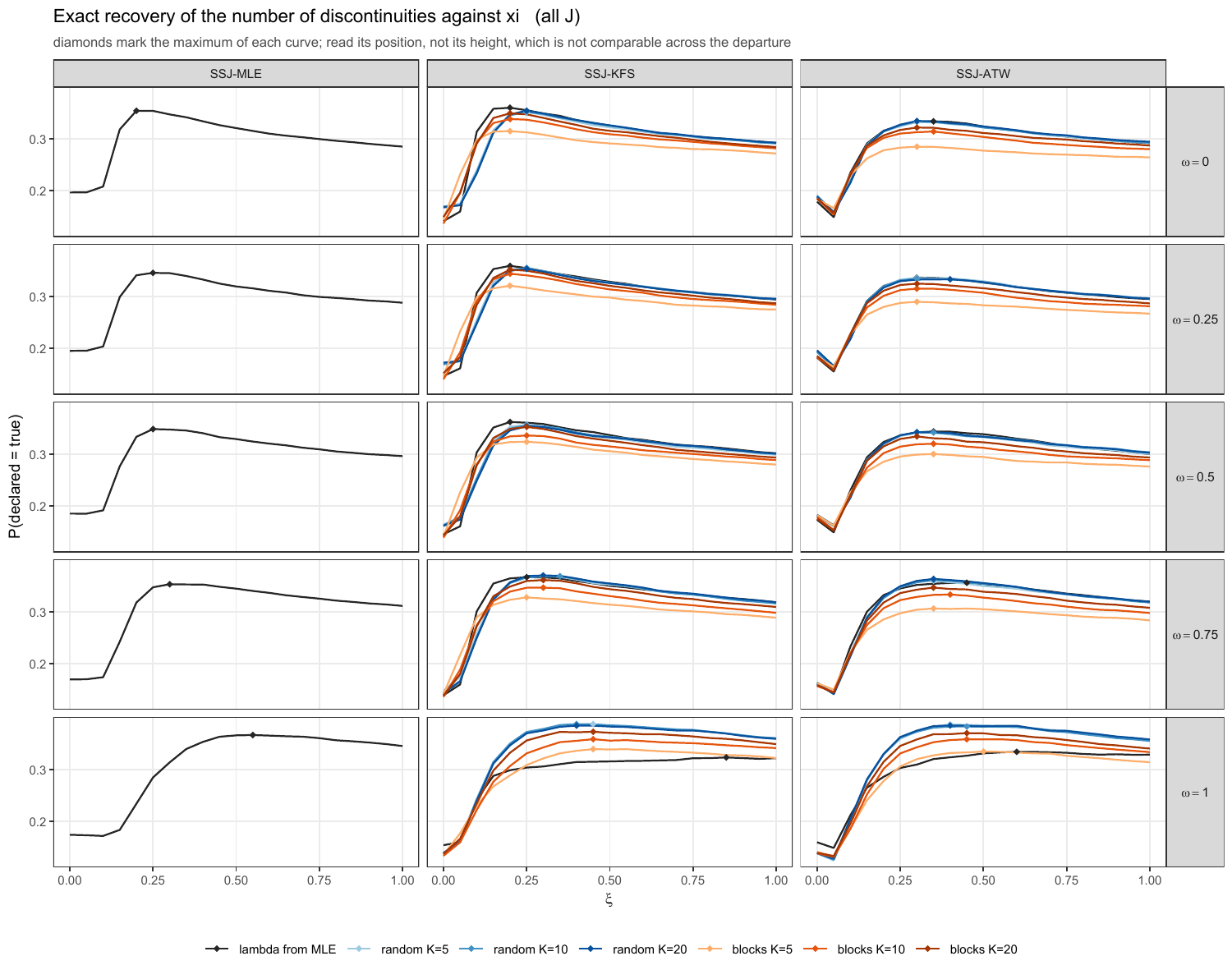}
\caption{Scenario 3, the probability of declaring exactly the right number over the whole design, one level of $\omega$ per row. Read the position of the maximum rather than its height: a quarter of the cells have $q=0$ and become easier as the departure smooths the fit, so the level is not comparable across rows. The position moves steadily to the right as $\omega$ grows.}
\label{fig:hit_sig_m}
\end{figure}

\begin{figure}[htbp]\centering
\includegraphics[width=0.90\linewidth]{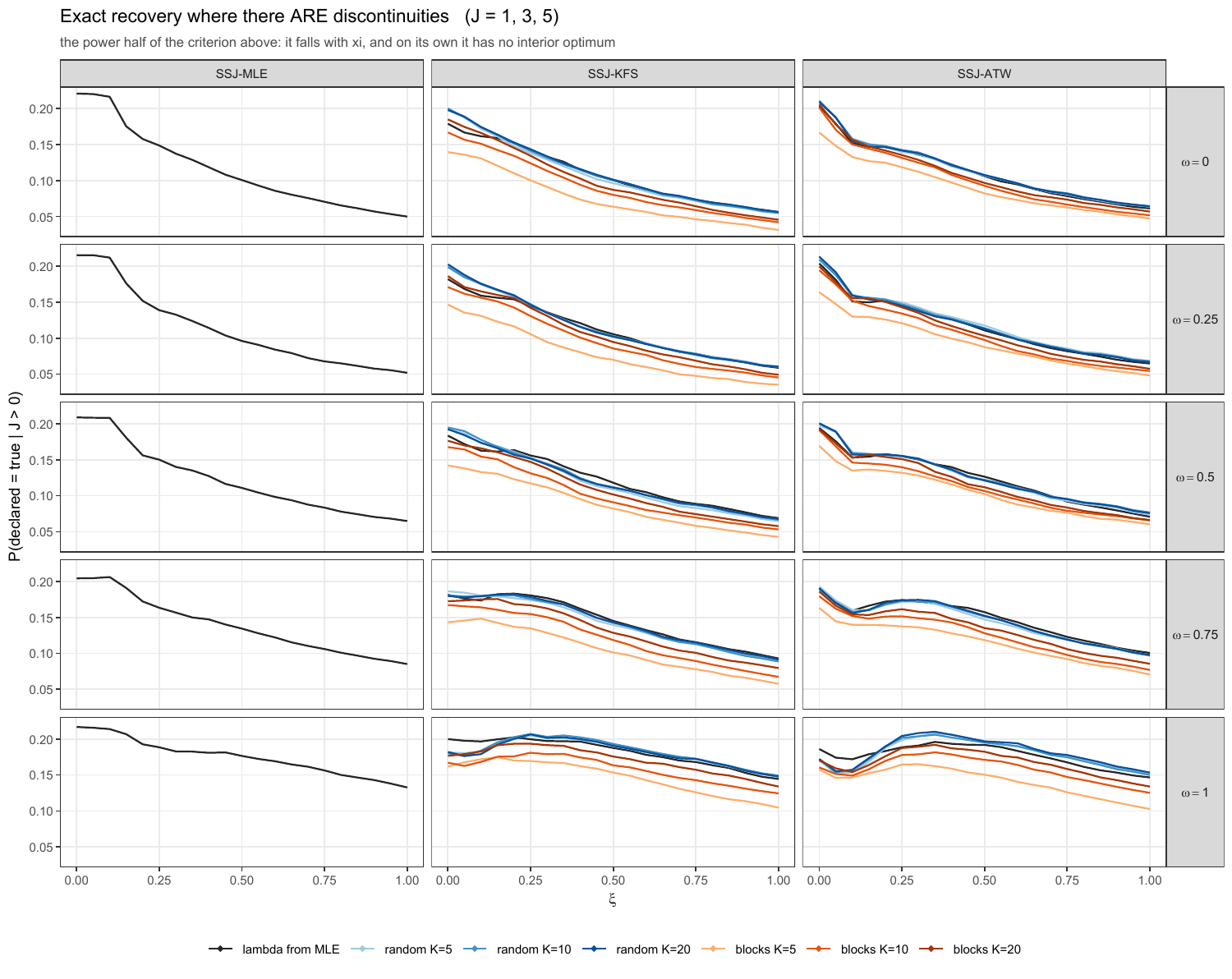}
\caption{Scenario 3, exact recovery restricted to the cells with $q>0$: the power half of the criterion. At $\omega=0$ it falls steeply from a maximum at $\xi=0$ and has no interior optimum; at $\omega=1$ it is nearly flat and develops a mild interior bump. This is the panel that shows the improvement at $\omega=1$ is real rather than a composition effect --- at $\xi=0.45$ the curve is between $1.7$ and $2.4$ times higher at $\omega=1$ than at $\omega=0$, in every arm.}
\label{fig:hitpos_sig_m}
\end{figure}

\begin{figure}[htbp]\centering
\includegraphics[width=0.90\linewidth]{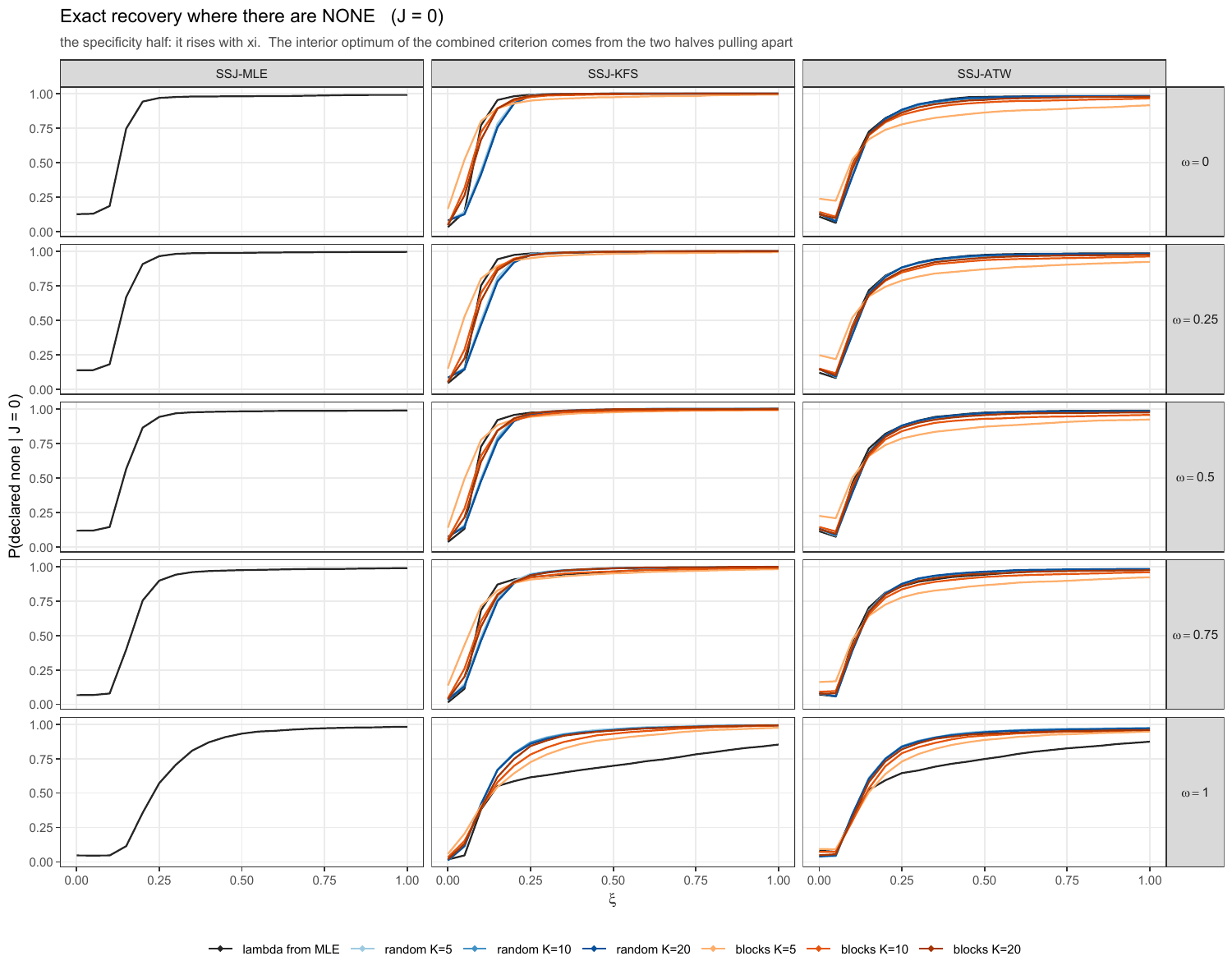}
\caption{Scenario 3, exact recovery restricted to the cells with $q=0$: the specificity half, which rises monotonically in $\xi$. The interior optimum of the combined criterion in Figure~\ref{fig:hit_sig_m} exists only because this half and the previous one pull in opposite directions.}
\label{fig:hitnul_sig_m}
\end{figure}

The numbers behind these readings are the following.
\begin{itemize}
\item \textbf{The reconstruction improves.} At each arm's own calibrated value the RMSE \emph{falls}: $0.0628$ to $0.0591$ for SSJ--MLE, $0.0636$ to $0.0591$ for SSJ--ATW with random folds at $K=20$. A signal drawn from the estimator's own prior is one it reconstructs well (Figure~\ref{fig:rmse_sig}).

\item \textbf{The declared count deteriorates markedly with $\lambda$ from SSJ--MLE.} At $\omega=1$, with nothing to find, SSJ--KFS at a smoothing constant from SSJ--MLE declares $2.19$ discontinuities per replication against $0.28$ at $\omega=0$, SSJ--ATW at the same constant $1.46$ against $0.64$, and SSJ--MLE $0.94$ against $0.41$. The same happens where there is something to find, so this is inflation of the count and not a failure confined to the null cells (Figure~\ref{fig:false_sig}).

\item \textbf{Exact recovery improves, but only at a higher $\xi$.} The maximum over $\xi$ of the probability of declaring exactly the right number rises from $0.29$--$0.36$ at $\omega=0$ to $0.32$--$0.39$ at $\omega=1$ for thirteen of the fifteen arms, and the value at which it is attained moves from $0.20$--$0.35$ to $0.40$--$0.60$ ($0.85$ for SSJ--KFS at $\lambda$ from SSJ--MLE). This is not a composition effect of the $q=0$ cells becoming easier: restricted to the cells where there is something to find, the probability at $\xi=0.45$ is between $1.7$ and $2.4$ times higher at $\omega=1$ than at $\omega=0$, in all fifteen arms without exception. What changes is the shape rather than the height (at $\omega=0$ the curve falls steeply from a maximum near $\xi=0$, at $\omega=1$ it is nearly flat) so the two are comparable at small $\xi$ and far apart at the values the count criterion recommends (Figures~\ref{fig:hit_sig_m} to \ref{fig:hitnul_sig_m}).

\item \textbf{The source of $\lambda$ matters more than the fold layout.} Cross-validating $\lambda$ protects SSJ--ATW entirely: its count at $q=0$ is $0.37$ at both $\omega=0$ and $\omega=1$ with random folds at $K=20$ and falls from $0.69$ to $0.50$ with contiguous blocks, against $0.64$ to $1.46$ when the constant is inherited from SSJ--MLE. Between the two layouts there is little to choose, and in RMSE the ordering is the exact reverse of Scenario 2: here random folds lie below blocks at every $\omega$, which is the premium referred to above.
\end{itemize}

\begin{figure}[htbp]\centering
\includegraphics[width=0.97\linewidth]{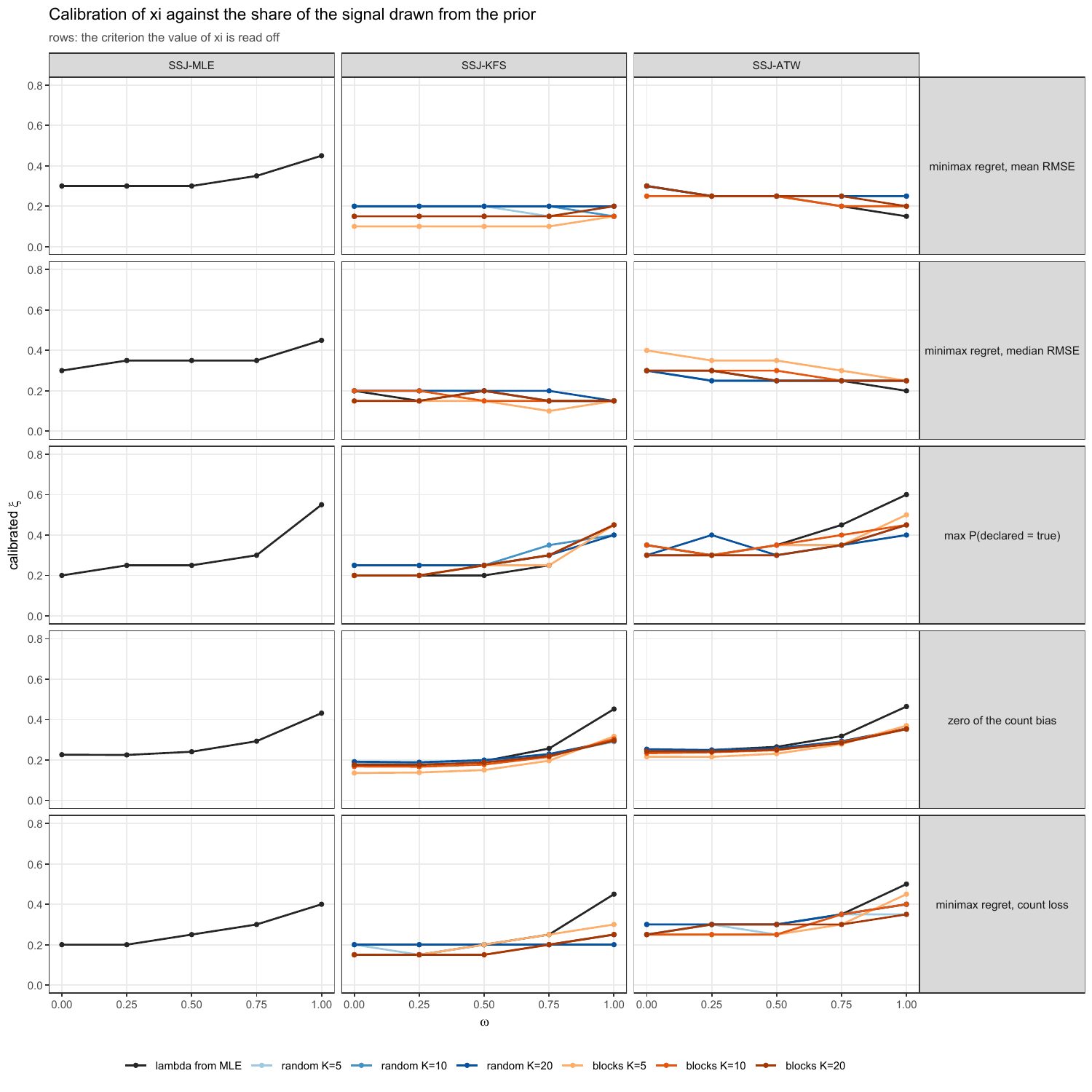}
\caption{Scenario 3: the same five criteria against $\omega$. The first two rows are nearly flat and the last two are not, which is the central observation of Section~\ref{sec:res_signal}: what $\omega$ costs is detection accuracy, not reconstruction, and a criterion built on RMSE cannot see it.}
\label{fig:cal_sig}
\end{figure}

\begin{figure}[htbp]\centering
\includegraphics[width=0.90\linewidth]{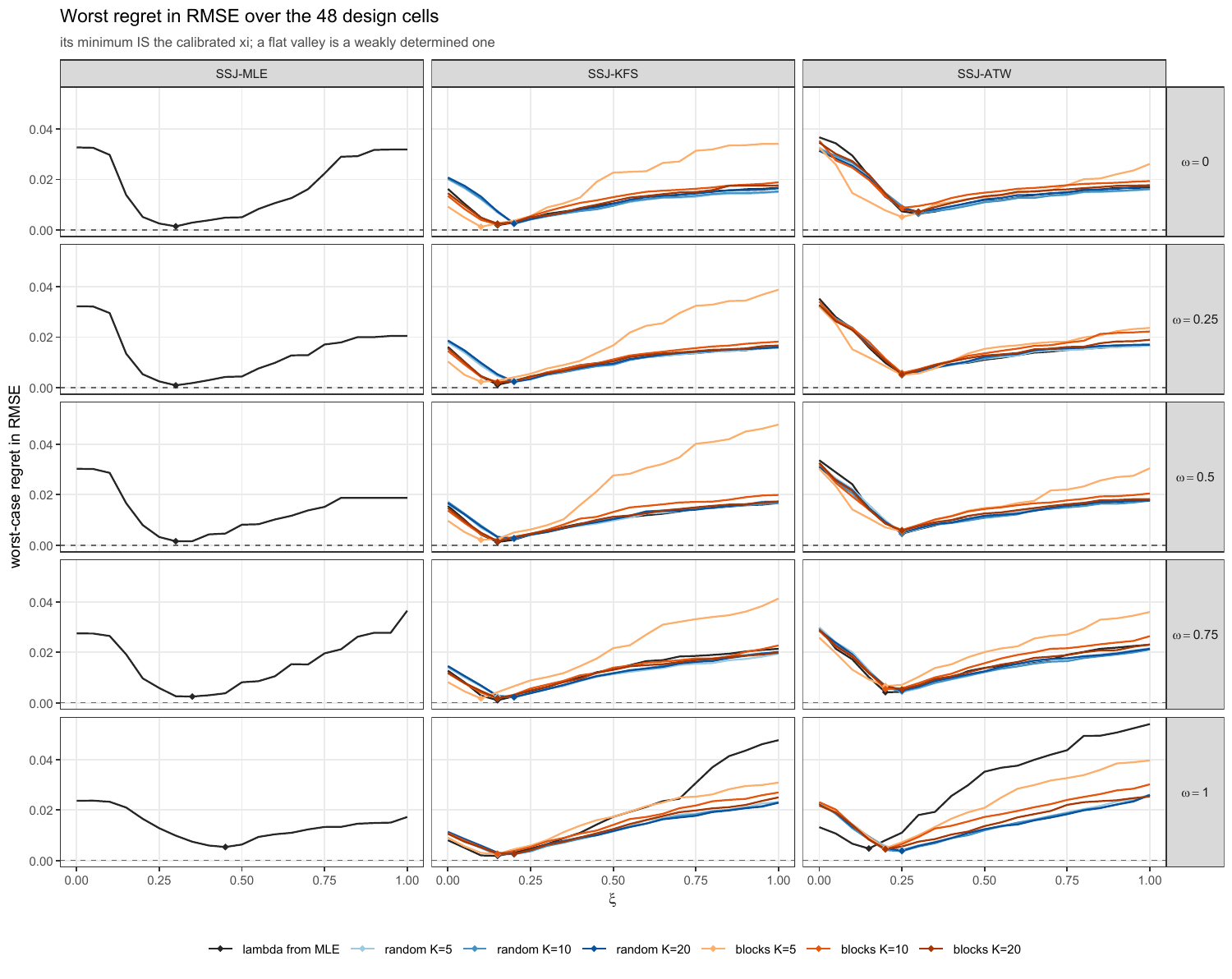}
\caption{Scenario 3, worst-case regret, one level of $\omega$ per row. At $\omega=1$ the SSJ--MLE valley is essentially level between $0.35$ and $0.55$: the argmin there is a point drawn out of a plateau.}
\label{fig:reg_sig}
\end{figure}

\begin{figure}[htbp]\centering
\includegraphics[width=0.90\linewidth]{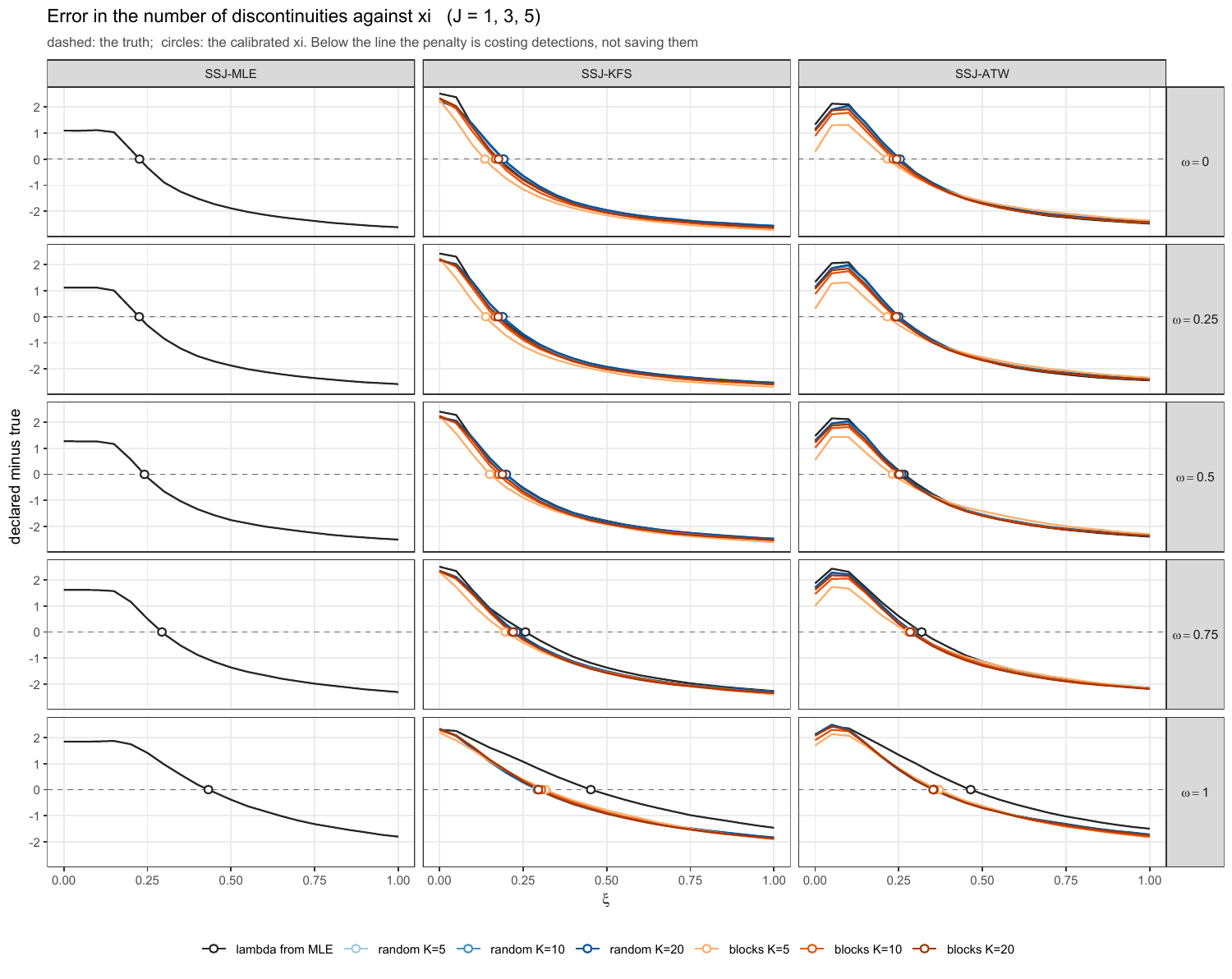}
\caption{Scenario 3, one level of $\omega$ per row. The zero of the curve barely moves up to $\omega=0.5$ and then travels sharply to the right. At $\omega=1$ SSJ--ATW crosses zero at about $0.36$ with a cross-validated $\lambda$ and at $0.46$ with $\lambda$ from SSJ--MLE, SSJ--KFS at about $0.30$ and $0.45$, and SSJ--MLE at $0.43$. Notice that the count rule of Table~\ref{tab:xi_signal_uni} is not the same quantity, since it protects the worst cell rather than the average, and at $\omega=1$ it gives $0.35$ to $0.50$ for SSJ--ATW, $0.40$ for SSJ--MLE and $0.20$ to $0.45$ for SSJ--KFS depending on the source of $\lambda$. At $\omega=0.75$, by contrast, a value of $0.30$ or a little below is still adequate for every arm.}
\label{fig:bias_sig_m}
\end{figure}

\begin{figure}[htbp]\centering
\includegraphics[width=0.88\linewidth]{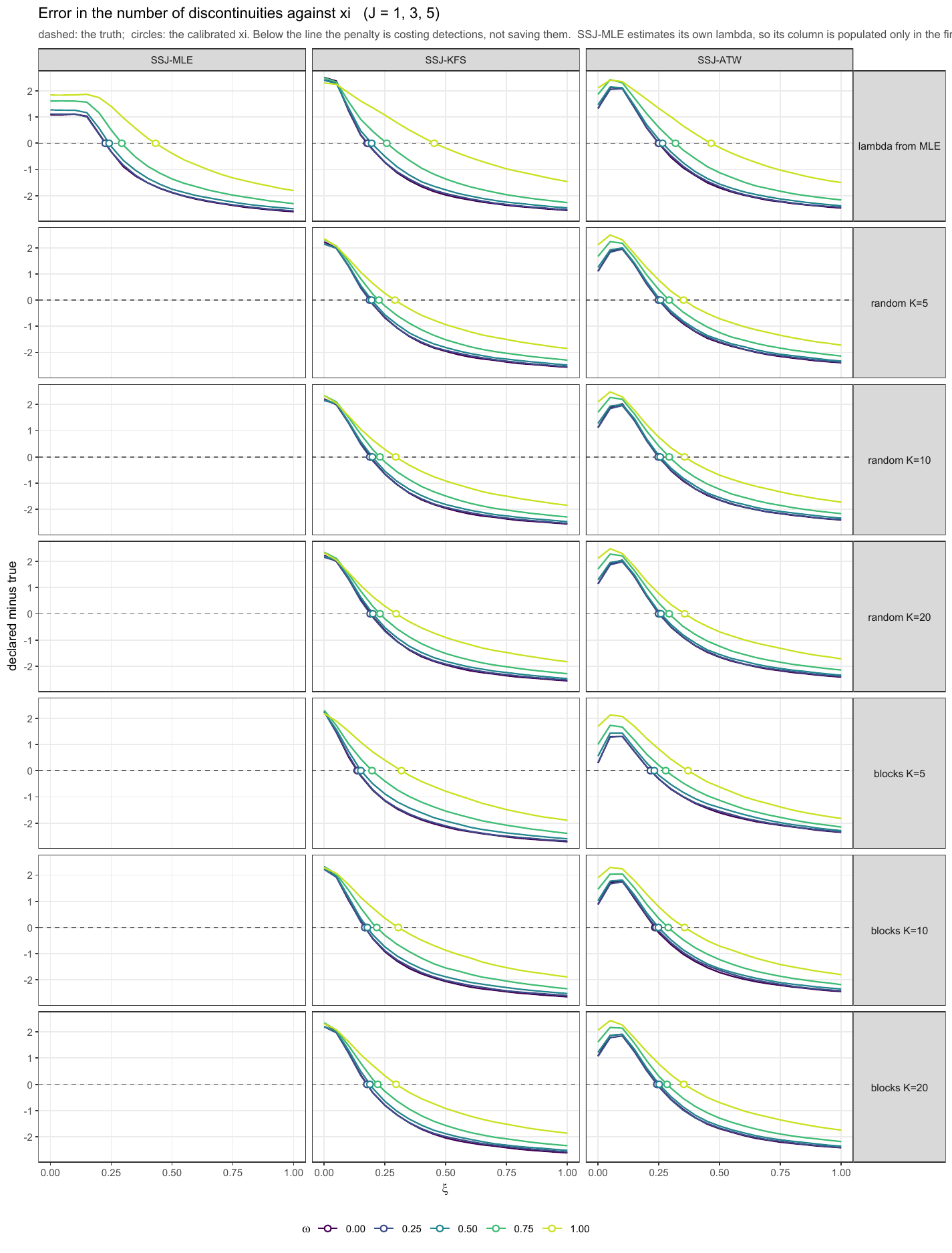}
\caption{Scenario 3, declared minus true; color is $\omega$.}
\label{fig:bias_sig_c}
\end{figure}

\clearpage

\subsection{Dependence on the length of the series}\label{sec:byn}
The calibrated value is not the same at every sample size. Table~\ref{tab:by_n} reports it at each $n$ for the baseline, for every arm and under both rules. Table~\ref{tab:byn_scen} does the same for seven representative arms at every level of the two departures, under the reconstruction rule.

\noindent The count rule is not shown there because at a single sample size it is decided by one cell: with nine cells left instead of thirty-six, the maximum is attained by a $q=5$ cell in which the estimator under-declares at every $\xi$, and the rule then returns the floor of the grid. Those entries at the floor are consistent with the rule of Section~\ref{sec:charge}, of which they are the limit: none of them occurs at $n=50$ and $25$ of the $30$ occur at $n=200$ and $n=400$, where the $\mathrm{df}\log n$ term already levies most of the charge per discontinuity and, on the nine cells of one sample size, nothing is left for the extended term to supply, so the most permissive penalty minimizes the worst-case count error.

\begin{table}[htbp]
\centering\footnotesize
\caption{Scenario 1, by sample size, for every arm: minimax regret in RMSE on the left and in the count error on the right, each computed over the cells of one sample size. Beneath each value, in small type, the central half of the same criterion computed cell by cell. A $0.00$ under the count rule is the floor of the grid: at that $n$ the worst cell is one with five discontinuities in which the estimator under-declares at every $\xi$.}
\label{tab:by_n}
\setlength{\tabcolsep}{2pt}
\begin{tabular}{ll*{4}{c}@{\qquad}*{4}{c}}
\toprule
Estimator & $\lambda$ from & \multicolumn{4}{c}{minimax regret in RMSE} & \multicolumn{4}{c}{minimax regret in the count error} \\
\cmidrule(lr){3-6}\cmidrule(lr){7-10}
 &  & \multicolumn{4}{c}{$n$} & \multicolumn{4}{c}{$n$} \\
 &  & 50 & 100 & 200 & 400 & 50 & 100 & 200 & 400 \\
\midrule
SSJ--MLE & MLE & \begin{tabular}[t]{@{}c@{}}0.30\\[-1.5pt]{\tiny[.29,.38]}\end{tabular} & \begin{tabular}[t]{@{}c@{}}0.30\\[-1.5pt]{\tiny[.25,.30]}\end{tabular} & \begin{tabular}[t]{@{}c@{}}0.25\\[-1.5pt]{\tiny[.25,.30]}\end{tabular} & \begin{tabular}[t]{@{}c@{}}0.25\\[-1.5pt]{\tiny[.20,.36]}\end{tabular} & \begin{tabular}[t]{@{}c@{}}0.20\\[-1.5pt]{\tiny[.00,.70]}\end{tabular} & \begin{tabular}[t]{@{}c@{}}0.20\\[-1.5pt]{\tiny[.00,.65]}\end{tabular} & \begin{tabular}[t]{@{}c@{}}0.00\\[-1.5pt]{\tiny[.00,.25]}\end{tabular} & \begin{tabular}[t]{@{}c@{}}0.00\\[-1.5pt]{\tiny[.00,.25]}\end{tabular} \\
\addlinespace
SSJ--KFS & MLE & \begin{tabular}[t]{@{}c@{}}0.20\\[-1.5pt]{\tiny[.19,.26]}\end{tabular} & \begin{tabular}[t]{@{}c@{}}0.20\\[-1.5pt]{\tiny[.14,.21]}\end{tabular} & \begin{tabular}[t]{@{}c@{}}0.15\\[-1.5pt]{\tiny[.10,.15]}\end{tabular} & \begin{tabular}[t]{@{}c@{}}0.10\\[-1.5pt]{\tiny[.10,.15]}\end{tabular} & \begin{tabular}[t]{@{}c@{}}0.15\\[-1.5pt]{\tiny[.05,.40]}\end{tabular} & \begin{tabular}[t]{@{}c@{}}0.20\\[-1.5pt]{\tiny[.05,.50]}\end{tabular} & \begin{tabular}[t]{@{}c@{}}0.20\\[-1.5pt]{\tiny[.00,.60]}\end{tabular} & \begin{tabular}[t]{@{}c@{}}0.25\\[-1.5pt]{\tiny[.00,.05]}\end{tabular} \\
 & random $K{=}5$ & \begin{tabular}[t]{@{}c@{}}0.20\\[-1.5pt]{\tiny[.15,.26]}\end{tabular} & \begin{tabular}[t]{@{}c@{}}0.20\\[-1.5pt]{\tiny[.15,.25]}\end{tabular} & \begin{tabular}[t]{@{}c@{}}0.20\\[-1.5pt]{\tiny[.14,.20]}\end{tabular} & \begin{tabular}[t]{@{}c@{}}0.20\\[-1.5pt]{\tiny[.10,.21]}\end{tabular} & \begin{tabular}[t]{@{}c@{}}0.20\\[-1.5pt]{\tiny[.05,.60]}\end{tabular} & \begin{tabular}[t]{@{}c@{}}0.20\\[-1.5pt]{\tiny[.05,.60]}\end{tabular} & \begin{tabular}[t]{@{}c@{}}0.20\\[-1.5pt]{\tiny[.00,.65]}\end{tabular} & \begin{tabular}[t]{@{}c@{}}0.25\\[-1.5pt]{\tiny[.00,.15]}\end{tabular} \\
 & random $K{=}10$ & \begin{tabular}[t]{@{}c@{}}0.20\\[-1.5pt]{\tiny[.14,.30]}\end{tabular} & \begin{tabular}[t]{@{}c@{}}0.25\\[-1.5pt]{\tiny[.15,.25]}\end{tabular} & \begin{tabular}[t]{@{}c@{}}0.20\\[-1.5pt]{\tiny[.14,.22]}\end{tabular} & \begin{tabular}[t]{@{}c@{}}0.20\\[-1.5pt]{\tiny[.10,.21]}\end{tabular} & \begin{tabular}[t]{@{}c@{}}0.20\\[-1.5pt]{\tiny[.05,.45]}\end{tabular} & \begin{tabular}[t]{@{}c@{}}0.20\\[-1.5pt]{\tiny[.00,.60]}\end{tabular} & \begin{tabular}[t]{@{}c@{}}0.20\\[-1.5pt]{\tiny[.00,.75]}\end{tabular} & \begin{tabular}[t]{@{}c@{}}0.25\\[-1.5pt]{\tiny[.00,.15]}\end{tabular} \\
 & random $K{=}20$ & \begin{tabular}[t]{@{}c@{}}0.20\\[-1.5pt]{\tiny[.15,.30]}\end{tabular} & \begin{tabular}[t]{@{}c@{}}0.20\\[-1.5pt]{\tiny[.15,.26]}\end{tabular} & \begin{tabular}[t]{@{}c@{}}0.20\\[-1.5pt]{\tiny[.14,.20]}\end{tabular} & \begin{tabular}[t]{@{}c@{}}0.20\\[-1.5pt]{\tiny[.10,.21]}\end{tabular} & \begin{tabular}[t]{@{}c@{}}0.15\\[-1.5pt]{\tiny[.00,.45]}\end{tabular} & \begin{tabular}[t]{@{}c@{}}0.20\\[-1.5pt]{\tiny[.00,.60]}\end{tabular} & \begin{tabular}[t]{@{}c@{}}0.20\\[-1.5pt]{\tiny[.00,.75]}\end{tabular} & \begin{tabular}[t]{@{}c@{}}0.25\\[-1.5pt]{\tiny[.00,.15]}\end{tabular} \\
 & blocks $K{=}5$ & \begin{tabular}[t]{@{}c@{}}0.10\\[-1.5pt]{\tiny[.05,.15]}\end{tabular} & \begin{tabular}[t]{@{}c@{}}0.20\\[-1.5pt]{\tiny[.15,.25]}\end{tabular} & \begin{tabular}[t]{@{}c@{}}0.15\\[-1.5pt]{\tiny[.10,.16]}\end{tabular} & \begin{tabular}[t]{@{}c@{}}0.10\\[-1.5pt]{\tiny[.09,.19]}\end{tabular} & \begin{tabular}[t]{@{}c@{}}0.15\\[-1.5pt]{\tiny[.05,.80]}\end{tabular} & \begin{tabular}[t]{@{}c@{}}0.15\\[-1.5pt]{\tiny[.00,.85]}\end{tabular} & \begin{tabular}[t]{@{}c@{}}0.20\\[-1.5pt]{\tiny[.00,.80]}\end{tabular} & \begin{tabular}[t]{@{}c@{}}0.15\\[-1.5pt]{\tiny[.00,.60]}\end{tabular} \\
 & blocks $K{=}10$ & \begin{tabular}[t]{@{}c@{}}0.15\\[-1.5pt]{\tiny[.10,.25]}\end{tabular} & \begin{tabular}[t]{@{}c@{}}0.20\\[-1.5pt]{\tiny[.15,.21]}\end{tabular} & \begin{tabular}[t]{@{}c@{}}0.15\\[-1.5pt]{\tiny[.10,.20]}\end{tabular} & \begin{tabular}[t]{@{}c@{}}0.15\\[-1.5pt]{\tiny[.10,.18]}\end{tabular} & \begin{tabular}[t]{@{}c@{}}0.15\\[-1.5pt]{\tiny[.05,.65]}\end{tabular} & \begin{tabular}[t]{@{}c@{}}0.20\\[-1.5pt]{\tiny[.00,.75]}\end{tabular} & \begin{tabular}[t]{@{}c@{}}0.20\\[-1.5pt]{\tiny[.00,.65]}\end{tabular} & \begin{tabular}[t]{@{}c@{}}0.20\\[-1.5pt]{\tiny[.00,.60]}\end{tabular} \\
 & blocks $K{=}20$ & \begin{tabular}[t]{@{}c@{}}0.20\\[-1.5pt]{\tiny[.14,.31]}\end{tabular} & \begin{tabular}[t]{@{}c@{}}0.20\\[-1.5pt]{\tiny[.15,.26]}\end{tabular} & \begin{tabular}[t]{@{}c@{}}0.20\\[-1.5pt]{\tiny[.10,.21]}\end{tabular} & \begin{tabular}[t]{@{}c@{}}0.15\\[-1.5pt]{\tiny[.10,.16]}\end{tabular} & \begin{tabular}[t]{@{}c@{}}0.15\\[-1.5pt]{\tiny[.05,.45]}\end{tabular} & \begin{tabular}[t]{@{}c@{}}0.20\\[-1.5pt]{\tiny[.00,.50]}\end{tabular} & \begin{tabular}[t]{@{}c@{}}0.20\\[-1.5pt]{\tiny[.00,.55]}\end{tabular} & \begin{tabular}[t]{@{}c@{}}0.20\\[-1.5pt]{\tiny[.00,.15]}\end{tabular} \\
\addlinespace
SSJ--ATW & MLE & \begin{tabular}[t]{@{}c@{}}0.40\\[-1.5pt]{\tiny[.34,.51]}\end{tabular} & \begin{tabular}[t]{@{}c@{}}0.25\\[-1.5pt]{\tiny[.20,.30]}\end{tabular} & \begin{tabular}[t]{@{}c@{}}0.15\\[-1.5pt]{\tiny[.15,.20]}\end{tabular} & \begin{tabular}[t]{@{}c@{}}0.10\\[-1.5pt]{\tiny[.10,.16]}\end{tabular} & \begin{tabular}[t]{@{}c@{}}0.30\\[-1.5pt]{\tiny[.00,.60]}\end{tabular} & \begin{tabular}[t]{@{}c@{}}0.25\\[-1.5pt]{\tiny[.00,.60]}\end{tabular} & \begin{tabular}[t]{@{}c@{}}0.20\\[-1.5pt]{\tiny[.00,.15]}\end{tabular} & \begin{tabular}[t]{@{}c@{}}0.00\\[-1.5pt]{\tiny[.00,.10]}\end{tabular} \\
 & random $K{=}5$ & \begin{tabular}[t]{@{}c@{}}0.35\\[-1.5pt]{\tiny[.34,.52]}\end{tabular} & \begin{tabular}[t]{@{}c@{}}0.25\\[-1.5pt]{\tiny[.20,.30]}\end{tabular} & \begin{tabular}[t]{@{}c@{}}0.15\\[-1.5pt]{\tiny[.15,.16]}\end{tabular} & \begin{tabular}[t]{@{}c@{}}0.15\\[-1.5pt]{\tiny[.10,.16]}\end{tabular} & \begin{tabular}[t]{@{}c@{}}0.30\\[-1.5pt]{\tiny[.00,.90]}\end{tabular} & \begin{tabular}[t]{@{}c@{}}0.25\\[-1.5pt]{\tiny[.00,.65]}\end{tabular} & \begin{tabular}[t]{@{}c@{}}0.20\\[-1.5pt]{\tiny[.00,.15]}\end{tabular} & \begin{tabular}[t]{@{}c@{}}0.00\\[-1.5pt]{\tiny[.00,.10]}\end{tabular} \\
 & random $K{=}10$ & \begin{tabular}[t]{@{}c@{}}0.35\\[-1.5pt]{\tiny[.34,.52]}\end{tabular} & \begin{tabular}[t]{@{}c@{}}0.25\\[-1.5pt]{\tiny[.24,.30]}\end{tabular} & \begin{tabular}[t]{@{}c@{}}0.15\\[-1.5pt]{\tiny[.15,.16]}\end{tabular} & \begin{tabular}[t]{@{}c@{}}0.15\\[-1.5pt]{\tiny[.10,.16]}\end{tabular} & \begin{tabular}[t]{@{}c@{}}0.30\\[-1.5pt]{\tiny[.00,.80]}\end{tabular} & \begin{tabular}[t]{@{}c@{}}0.25\\[-1.5pt]{\tiny[.00,.65]}\end{tabular} & \begin{tabular}[t]{@{}c@{}}0.25\\[-1.5pt]{\tiny[.00,.15]}\end{tabular} & \begin{tabular}[t]{@{}c@{}}0.00\\[-1.5pt]{\tiny[.00,.10]}\end{tabular} \\
 & random $K{=}20$ & \begin{tabular}[t]{@{}c@{}}0.35\\[-1.5pt]{\tiny[.34,.58]}\end{tabular} & \begin{tabular}[t]{@{}c@{}}0.25\\[-1.5pt]{\tiny[.20,.30]}\end{tabular} & \begin{tabular}[t]{@{}c@{}}0.15\\[-1.5pt]{\tiny[.15,.16]}\end{tabular} & \begin{tabular}[t]{@{}c@{}}0.15\\[-1.5pt]{\tiny[.10,.16]}\end{tabular} & \begin{tabular}[t]{@{}c@{}}0.30\\[-1.5pt]{\tiny[.00,.75]}\end{tabular} & \begin{tabular}[t]{@{}c@{}}0.25\\[-1.5pt]{\tiny[.00,.65]}\end{tabular} & \begin{tabular}[t]{@{}c@{}}0.25\\[-1.5pt]{\tiny[.00,.20]}\end{tabular} & \begin{tabular}[t]{@{}c@{}}0.00\\[-1.5pt]{\tiny[.00,.10]}\end{tabular} \\
 & blocks $K{=}5$ & \begin{tabular}[t]{@{}c@{}}0.30\\[-1.5pt]{\tiny[.04,.35]}\end{tabular} & \begin{tabular}[t]{@{}c@{}}0.35\\[-1.5pt]{\tiny[.30,.41]}\end{tabular} & \begin{tabular}[t]{@{}c@{}}0.15\\[-1.5pt]{\tiny[.15,.31]}\end{tabular} & \begin{tabular}[t]{@{}c@{}}0.10\\[-1.5pt]{\tiny[.10,.15]}\end{tabular} & \begin{tabular}[t]{@{}c@{}}0.35\\[-1.5pt]{\tiny[.00,.95]}\end{tabular} & \begin{tabular}[t]{@{}c@{}}0.30\\[-1.5pt]{\tiny[.00,.85]}\end{tabular} & \begin{tabular}[t]{@{}c@{}}0.20\\[-1.5pt]{\tiny[.05,.75]}\end{tabular} & \begin{tabular}[t]{@{}c@{}}0.20\\[-1.5pt]{\tiny[.00,.10]}\end{tabular} \\
 & blocks $K{=}10$ & \begin{tabular}[t]{@{}c@{}}0.35\\[-1.5pt]{\tiny[.30,.61]}\end{tabular} & \begin{tabular}[t]{@{}c@{}}0.25\\[-1.5pt]{\tiny[.20,.26]}\end{tabular} & \begin{tabular}[t]{@{}c@{}}0.15\\[-1.5pt]{\tiny[.15,.20]}\end{tabular} & \begin{tabular}[t]{@{}c@{}}0.10\\[-1.5pt]{\tiny[.10,.15]}\end{tabular} & \begin{tabular}[t]{@{}c@{}}0.30\\[-1.5pt]{\tiny[.00,.95]}\end{tabular} & \begin{tabular}[t]{@{}c@{}}0.25\\[-1.5pt]{\tiny[.00,.90]}\end{tabular} & \begin{tabular}[t]{@{}c@{}}0.20\\[-1.5pt]{\tiny[.00,.20]}\end{tabular} & \begin{tabular}[t]{@{}c@{}}0.25\\[-1.5pt]{\tiny[.00,.15]}\end{tabular} \\
 & blocks $K{=}20$ & \begin{tabular}[t]{@{}c@{}}0.35\\[-1.5pt]{\tiny[.34,.70]}\end{tabular} & \begin{tabular}[t]{@{}c@{}}0.25\\[-1.5pt]{\tiny[.24,.30]}\end{tabular} & \begin{tabular}[t]{@{}c@{}}0.15\\[-1.5pt]{\tiny[.15,.20]}\end{tabular} & \begin{tabular}[t]{@{}c@{}}0.10\\[-1.5pt]{\tiny[.10,.16]}\end{tabular} & \begin{tabular}[t]{@{}c@{}}0.30\\[-1.5pt]{\tiny[.00,.60]}\end{tabular} & \begin{tabular}[t]{@{}c@{}}0.25\\[-1.5pt]{\tiny[.00,.60]}\end{tabular} & \begin{tabular}[t]{@{}c@{}}0.20\\[-1.5pt]{\tiny[.00,.20]}\end{tabular} & \begin{tabular}[t]{@{}c@{}}0.25\\[-1.5pt]{\tiny[.00,.10]}\end{tabular} \\
\bottomrule
\end{tabular}
\end{table}

At the baseline the two rules disagree about the direction. The reconstruction rule falls with $n$: for SSJ--ATW from $0.40$ at $n=50$ to $0.10$ at $n=400$, for SSJ--KFS from $0.20$ to $0.10$, for SSJ--MLE from $0.30$ to $0.25$. The count rule rises with $n$ for SSJ--KFS, from $0.15$ to $0.25$, and drifts down for SSJ--ATW, from $0.30$ to $0.20$, until at $n=400$ its worst cell becomes one that under-declares and the rule drops to the floor.

\noindent A single constant is therefore a compromise in $n$ under either rule, and in opposite directions for the two.

\begin{table}[htbp]
\centering\footnotesize
\caption{The calibration by sample size, in both scenarios: minimax regret in RMSE, for seven representative arms. The first column is the reference case; then the persistence $\psi$ of the errors and the weight $\omega$ of the stochastic-trend component. Beneath each value, in small type, the central half of the same criterion over the twelve cells of that sample size.}
\label{tab:byn_scen}
\setlength{\tabcolsep}{2pt}
\begin{tabular}{ll c @{\enspace} *{5}{c}@{\enspace}*{4}{c}}
\toprule
Estimator & & Scenario 1 & \multicolumn{5}{c}{Scenario 2: dependent errors} & \multicolumn{4}{c}{Scenario 3: stochastic trend} \\
\cmidrule(lr){3-3}\cmidrule(lr){4-8}\cmidrule(lr){9-12}
 &  & $\psi{=}\omega{=}0$ & \multicolumn{5}{c}{$\psi$} & \multicolumn{4}{c}{$\omega$} \\
 &  & & 0.20 & 0.40 & 0.60 & 0.80 & 0.90 & 0.25 & 0.50 & 0.75 & 1.00 \\
\midrule
SSJ--MLE, MLE & $n{=}50$ & \begin{tabular}[t]{@{}c@{}}0.30\\[-1.5pt]{\tiny[.29,.38]}\end{tabular} & \begin{tabular}[t]{@{}c@{}}0.25\\[-1.5pt]{\tiny[.25,.35]}\end{tabular} & \begin{tabular}[t]{@{}c@{}}0.25\\[-1.5pt]{\tiny[.25,.35]}\end{tabular} & \begin{tabular}[t]{@{}c@{}}0.25\\[-1.5pt]{\tiny[.24,.31]}\end{tabular} & \begin{tabular}[t]{@{}c@{}}0.30\\[-1.5pt]{\tiny[.20,.35]}\end{tabular} & \begin{tabular}[t]{@{}c@{}}0.40\\[-1.5pt]{\tiny[.24,.35]}\end{tabular} & \begin{tabular}[t]{@{}c@{}}0.30\\[-1.5pt]{\tiny[.29,.33]}\end{tabular} & \begin{tabular}[t]{@{}c@{}}0.35\\[-1.5pt]{\tiny[.29,.36]}\end{tabular} & \begin{tabular}[t]{@{}c@{}}0.35\\[-1.5pt]{\tiny[.29,.40]}\end{tabular} & \begin{tabular}[t]{@{}c@{}}0.45\\[-1.5pt]{\tiny[.18,.49]}\end{tabular} \\
 & $n{=}100$ & \begin{tabular}[t]{@{}c@{}}0.30\\[-1.5pt]{\tiny[.25,.30]}\end{tabular} & \begin{tabular}[t]{@{}c@{}}0.35\\[-1.5pt]{\tiny[.30,.45]}\end{tabular} & \begin{tabular}[t]{@{}c@{}}0.30\\[-1.5pt]{\tiny[.30,.41]}\end{tabular} & \begin{tabular}[t]{@{}c@{}}0.25\\[-1.5pt]{\tiny[.30,.40]}\end{tabular} & \begin{tabular}[t]{@{}c@{}}0.20\\[-1.5pt]{\tiny[.19,.30]}\end{tabular} & \begin{tabular}[t]{@{}c@{}}0.15\\[-1.5pt]{\tiny[.09,.35]}\end{tabular} & \begin{tabular}[t]{@{}c@{}}0.30\\[-1.5pt]{\tiny[.29,.35]}\end{tabular} & \begin{tabular}[t]{@{}c@{}}0.30\\[-1.5pt]{\tiny[.29,.35]}\end{tabular} & \begin{tabular}[t]{@{}c@{}}0.35\\[-1.5pt]{\tiny[.30,.35]}\end{tabular} & \begin{tabular}[t]{@{}c@{}}0.45\\[-1.5pt]{\tiny[.30,.52]}\end{tabular} \\
 & $n{=}200$ & \begin{tabular}[t]{@{}c@{}}0.25\\[-1.5pt]{\tiny[.25,.30]}\end{tabular} & \begin{tabular}[t]{@{}c@{}}0.35\\[-1.5pt]{\tiny[.34,.41]}\end{tabular} & \begin{tabular}[t]{@{}c@{}}0.35\\[-1.5pt]{\tiny[.35,.45]}\end{tabular} & \begin{tabular}[t]{@{}c@{}}0.30\\[-1.5pt]{\tiny[.30,.52]}\end{tabular} & \begin{tabular}[t]{@{}c@{}}0.30\\[-1.5pt]{\tiny[.25,.40]}\end{tabular} & \begin{tabular}[t]{@{}c@{}}0.20\\[-1.5pt]{\tiny[.08,.35]}\end{tabular} & \begin{tabular}[t]{@{}c@{}}0.25\\[-1.5pt]{\tiny[.25,.30]}\end{tabular} & \begin{tabular}[t]{@{}c@{}}0.25\\[-1.5pt]{\tiny[.25,.38]}\end{tabular} & \begin{tabular}[t]{@{}c@{}}0.30\\[-1.5pt]{\tiny[.25,.36]}\end{tabular} & \begin{tabular}[t]{@{}c@{}}0.40\\[-1.5pt]{\tiny[.30,.45]}\end{tabular} \\
 & $n{=}400$ & \begin{tabular}[t]{@{}c@{}}0.25\\[-1.5pt]{\tiny[.20,.36]}\end{tabular} & \begin{tabular}[t]{@{}c@{}}0.30\\[-1.5pt]{\tiny[.34,.45]}\end{tabular} & \begin{tabular}[t]{@{}c@{}}0.40\\[-1.5pt]{\tiny[.34,.42]}\end{tabular} & \begin{tabular}[t]{@{}c@{}}0.30\\[-1.5pt]{\tiny[.24,.46]}\end{tabular} & \begin{tabular}[t]{@{}c@{}}0.25\\[-1.5pt]{\tiny[.25,.58]}\end{tabular} & \begin{tabular}[t]{@{}c@{}}0.20\\[-1.5pt]{\tiny[.19,.50]}\end{tabular} & \begin{tabular}[t]{@{}c@{}}0.25\\[-1.5pt]{\tiny[.25,.33]}\end{tabular} & \begin{tabular}[t]{@{}c@{}}0.25\\[-1.5pt]{\tiny[.24,.36]}\end{tabular} & \begin{tabular}[t]{@{}c@{}}0.25\\[-1.5pt]{\tiny[.25,.36]}\end{tabular} & \begin{tabular}[t]{@{}c@{}}0.35\\[-1.5pt]{\tiny[.34,.36]}\end{tabular} \\
\addlinespace
SSJ--KFS, MLE & $n{=}50$ & \begin{tabular}[t]{@{}c@{}}0.20\\[-1.5pt]{\tiny[.19,.26]}\end{tabular} & \begin{tabular}[t]{@{}c@{}}0.25\\[-1.5pt]{\tiny[.20,.31]}\end{tabular} & \begin{tabular}[t]{@{}c@{}}0.30\\[-1.5pt]{\tiny[.24,.30]}\end{tabular} & \begin{tabular}[t]{@{}c@{}}0.25\\[-1.5pt]{\tiny[.20,.30]}\end{tabular} & \begin{tabular}[t]{@{}c@{}}0.25\\[-1.5pt]{\tiny[.20,.35]}\end{tabular} & \begin{tabular}[t]{@{}c@{}}0.30\\[-1.5pt]{\tiny[.25,.35]}\end{tabular} & \begin{tabular}[t]{@{}c@{}}0.20\\[-1.5pt]{\tiny[.15,.20]}\end{tabular} & \begin{tabular}[t]{@{}c@{}}0.20\\[-1.5pt]{\tiny[.15,.21]}\end{tabular} & \begin{tabular}[t]{@{}c@{}}0.15\\[-1.5pt]{\tiny[.15,.21]}\end{tabular} & \begin{tabular}[t]{@{}c@{}}0.15\\[-1.5pt]{\tiny[.14,.20]}\end{tabular} \\
 & $n{=}100$ & \begin{tabular}[t]{@{}c@{}}0.20\\[-1.5pt]{\tiny[.14,.21]}\end{tabular} & \begin{tabular}[t]{@{}c@{}}0.30\\[-1.5pt]{\tiny[.25,.36]}\end{tabular} & \begin{tabular}[t]{@{}c@{}}0.35\\[-1.5pt]{\tiny[.25,.36]}\end{tabular} & \begin{tabular}[t]{@{}c@{}}0.35\\[-1.5pt]{\tiny[.29,.40]}\end{tabular} & \begin{tabular}[t]{@{}c@{}}0.30\\[-1.5pt]{\tiny[.30,.35]}\end{tabular} & \begin{tabular}[t]{@{}c@{}}0.25\\[-1.5pt]{\tiny[.25,.35]}\end{tabular} & \begin{tabular}[t]{@{}c@{}}0.20\\[-1.5pt]{\tiny[.15,.20]}\end{tabular} & \begin{tabular}[t]{@{}c@{}}0.20\\[-1.5pt]{\tiny[.15,.20]}\end{tabular} & \begin{tabular}[t]{@{}c@{}}0.15\\[-1.5pt]{\tiny[.10,.25]}\end{tabular} & \begin{tabular}[t]{@{}c@{}}0.10\\[-1.5pt]{\tiny[.14,.21]}\end{tabular} \\
 & $n{=}200$ & \begin{tabular}[t]{@{}c@{}}0.15\\[-1.5pt]{\tiny[.10,.15]}\end{tabular} & \begin{tabular}[t]{@{}c@{}}0.30\\[-1.5pt]{\tiny[.25,.31]}\end{tabular} & \begin{tabular}[t]{@{}c@{}}0.35\\[-1.5pt]{\tiny[.30,.40]}\end{tabular} & \begin{tabular}[t]{@{}c@{}}0.40\\[-1.5pt]{\tiny[.29,.48]}\end{tabular} & \begin{tabular}[t]{@{}c@{}}0.40\\[-1.5pt]{\tiny[.25,.41]}\end{tabular} & \begin{tabular}[t]{@{}c@{}}0.30\\[-1.5pt]{\tiny[.24,.35]}\end{tabular} & \begin{tabular}[t]{@{}c@{}}0.15\\[-1.5pt]{\tiny[.10,.20]}\end{tabular} & \begin{tabular}[t]{@{}c@{}}0.15\\[-1.5pt]{\tiny[.10,.16]}\end{tabular} & \begin{tabular}[t]{@{}c@{}}0.15\\[-1.5pt]{\tiny[.10,.15]}\end{tabular} & \begin{tabular}[t]{@{}c@{}}0.15\\[-1.5pt]{\tiny[.10,.20]}\end{tabular} \\
 & $n{=}400$ & \begin{tabular}[t]{@{}c@{}}0.10\\[-1.5pt]{\tiny[.10,.15]}\end{tabular} & \begin{tabular}[t]{@{}c@{}}0.25\\[-1.5pt]{\tiny[.20,.26]}\end{tabular} & \begin{tabular}[t]{@{}c@{}}0.40\\[-1.5pt]{\tiny[.25,.39]}\end{tabular} & \begin{tabular}[t]{@{}c@{}}0.30\\[-1.5pt]{\tiny[.29,.36]}\end{tabular} & \begin{tabular}[t]{@{}c@{}}0.35\\[-1.5pt]{\tiny[.29,.35]}\end{tabular} & \begin{tabular}[t]{@{}c@{}}0.40\\[-1.5pt]{\tiny[.31,.46]}\end{tabular} & \begin{tabular}[t]{@{}c@{}}0.10\\[-1.5pt]{\tiny[.10,.15]}\end{tabular} & \begin{tabular}[t]{@{}c@{}}0.10\\[-1.5pt]{\tiny[.10,.15]}\end{tabular} & \begin{tabular}[t]{@{}c@{}}0.10\\[-1.5pt]{\tiny[.10,.15]}\end{tabular} & \begin{tabular}[t]{@{}c@{}}0.15\\[-1.5pt]{\tiny[.10,.20]}\end{tabular} \\
\addlinespace
SSJ--KFS, random $K{=}20$ & $n{=}50$ & \begin{tabular}[t]{@{}c@{}}0.20\\[-1.5pt]{\tiny[.15,.30]}\end{tabular} & \begin{tabular}[t]{@{}c@{}}0.20\\[-1.5pt]{\tiny[.14,.25]}\end{tabular} & \begin{tabular}[t]{@{}c@{}}0.20\\[-1.5pt]{\tiny[.15,.25]}\end{tabular} & \begin{tabular}[t]{@{}c@{}}0.25\\[-1.5pt]{\tiny[.20,.30]}\end{tabular} & \begin{tabular}[t]{@{}c@{}}0.35\\[-1.5pt]{\tiny[.25,.35]}\end{tabular} & \begin{tabular}[t]{@{}c@{}}0.35\\[-1.5pt]{\tiny[.29,.40]}\end{tabular} & \begin{tabular}[t]{@{}c@{}}0.20\\[-1.5pt]{\tiny[.15,.25]}\end{tabular} & \begin{tabular}[t]{@{}c@{}}0.20\\[-1.5pt]{\tiny[.15,.30]}\end{tabular} & \begin{tabular}[t]{@{}c@{}}0.20\\[-1.5pt]{\tiny[.14,.25]}\end{tabular} & \begin{tabular}[t]{@{}c@{}}0.20\\[-1.5pt]{\tiny[.10,.25]}\end{tabular} \\
 & $n{=}100$ & \begin{tabular}[t]{@{}c@{}}0.20\\[-1.5pt]{\tiny[.15,.26]}\end{tabular} & \begin{tabular}[t]{@{}c@{}}0.30\\[-1.5pt]{\tiny[.20,.40]}\end{tabular} & \begin{tabular}[t]{@{}c@{}}0.30\\[-1.5pt]{\tiny[.25,.40]}\end{tabular} & \begin{tabular}[t]{@{}c@{}}0.35\\[-1.5pt]{\tiny[.29,.40]}\end{tabular} & \begin{tabular}[t]{@{}c@{}}0.30\\[-1.5pt]{\tiny[.25,.41]}\end{tabular} & \begin{tabular}[t]{@{}c@{}}0.30\\[-1.5pt]{\tiny[.25,.33]}\end{tabular} & \begin{tabular}[t]{@{}c@{}}0.20\\[-1.5pt]{\tiny[.15,.25]}\end{tabular} & \begin{tabular}[t]{@{}c@{}}0.20\\[-1.5pt]{\tiny[.15,.26]}\end{tabular} & \begin{tabular}[t]{@{}c@{}}0.20\\[-1.5pt]{\tiny[.15,.26]}\end{tabular} & \begin{tabular}[t]{@{}c@{}}0.15\\[-1.5pt]{\tiny[.10,.21]}\end{tabular} \\
 & $n{=}200$ & \begin{tabular}[t]{@{}c@{}}0.20\\[-1.5pt]{\tiny[.14,.20]}\end{tabular} & \begin{tabular}[t]{@{}c@{}}0.35\\[-1.5pt]{\tiny[.24,.36]}\end{tabular} & \begin{tabular}[t]{@{}c@{}}0.35\\[-1.5pt]{\tiny[.29,.41]}\end{tabular} & \begin{tabular}[t]{@{}c@{}}0.35\\[-1.5pt]{\tiny[.30,.45]}\end{tabular} & \begin{tabular}[t]{@{}c@{}}0.35\\[-1.5pt]{\tiny[.30,.40]}\end{tabular} & \begin{tabular}[t]{@{}c@{}}0.30\\[-1.5pt]{\tiny[.30,.41]}\end{tabular} & \begin{tabular}[t]{@{}c@{}}0.20\\[-1.5pt]{\tiny[.14,.22]}\end{tabular} & \begin{tabular}[t]{@{}c@{}}0.20\\[-1.5pt]{\tiny[.10,.22]}\end{tabular} & \begin{tabular}[t]{@{}c@{}}0.15\\[-1.5pt]{\tiny[.10,.18]}\end{tabular} & \begin{tabular}[t]{@{}c@{}}0.15\\[-1.5pt]{\tiny[.10,.19]}\end{tabular} \\
 & $n{=}400$ & \begin{tabular}[t]{@{}c@{}}0.20\\[-1.5pt]{\tiny[.10,.21]}\end{tabular} & \begin{tabular}[t]{@{}c@{}}0.35\\[-1.5pt]{\tiny[.20,.35]}\end{tabular} & \begin{tabular}[t]{@{}c@{}}0.25\\[-1.5pt]{\tiny[.20,.26]}\end{tabular} & \begin{tabular}[t]{@{}c@{}}0.30\\[-1.5pt]{\tiny[.24,.30]}\end{tabular} & \begin{tabular}[t]{@{}c@{}}0.35\\[-1.5pt]{\tiny[.25,.31]}\end{tabular} & \begin{tabular}[t]{@{}c@{}}0.50\\[-1.5pt]{\tiny[.29,.40]}\end{tabular} & \begin{tabular}[t]{@{}c@{}}0.20\\[-1.5pt]{\tiny[.10,.21]}\end{tabular} & \begin{tabular}[t]{@{}c@{}}0.15\\[-1.5pt]{\tiny[.10,.21]}\end{tabular} & \begin{tabular}[t]{@{}c@{}}0.15\\[-1.5pt]{\tiny[.10,.20]}\end{tabular} & \begin{tabular}[t]{@{}c@{}}0.15\\[-1.5pt]{\tiny[.10,.21]}\end{tabular} \\
\addlinespace
SSJ--KFS, blocks $K{=}20$ & $n{=}50$ & \begin{tabular}[t]{@{}c@{}}0.20\\[-1.5pt]{\tiny[.14,.31]}\end{tabular} & \begin{tabular}[t]{@{}c@{}}0.20\\[-1.5pt]{\tiny[.19,.25]}\end{tabular} & \begin{tabular}[t]{@{}c@{}}0.15\\[-1.5pt]{\tiny[.15,.30]}\end{tabular} & \begin{tabular}[t]{@{}c@{}}0.25\\[-1.5pt]{\tiny[.20,.25]}\end{tabular} & \begin{tabular}[t]{@{}c@{}}0.30\\[-1.5pt]{\tiny[.25,.35]}\end{tabular} & \begin{tabular}[t]{@{}c@{}}0.30\\[-1.5pt]{\tiny[.25,.40]}\end{tabular} & \begin{tabular}[t]{@{}c@{}}0.15\\[-1.5pt]{\tiny[.10,.25]}\end{tabular} & \begin{tabular}[t]{@{}c@{}}0.15\\[-1.5pt]{\tiny[.10,.26]}\end{tabular} & \begin{tabular}[t]{@{}c@{}}0.15\\[-1.5pt]{\tiny[.10,.26]}\end{tabular} & \begin{tabular}[t]{@{}c@{}}0.20\\[-1.5pt]{\tiny[.10,.26]}\end{tabular} \\
 & $n{=}100$ & \begin{tabular}[t]{@{}c@{}}0.20\\[-1.5pt]{\tiny[.15,.26]}\end{tabular} & \begin{tabular}[t]{@{}c@{}}0.30\\[-1.5pt]{\tiny[.20,.35]}\end{tabular} & \begin{tabular}[t]{@{}c@{}}0.30\\[-1.5pt]{\tiny[.25,.40]}\end{tabular} & \begin{tabular}[t]{@{}c@{}}0.35\\[-1.5pt]{\tiny[.29,.36]}\end{tabular} & \begin{tabular}[t]{@{}c@{}}0.30\\[-1.5pt]{\tiny[.30,.40]}\end{tabular} & \begin{tabular}[t]{@{}c@{}}0.30\\[-1.5pt]{\tiny[.25,.35]}\end{tabular} & \begin{tabular}[t]{@{}c@{}}0.20\\[-1.5pt]{\tiny[.15,.25]}\end{tabular} & \begin{tabular}[t]{@{}c@{}}0.20\\[-1.5pt]{\tiny[.15,.25]}\end{tabular} & \begin{tabular}[t]{@{}c@{}}0.15\\[-1.5pt]{\tiny[.15,.21]}\end{tabular} & \begin{tabular}[t]{@{}c@{}}0.15\\[-1.5pt]{\tiny[.10,.21]}\end{tabular} \\
 & $n{=}200$ & \begin{tabular}[t]{@{}c@{}}0.20\\[-1.5pt]{\tiny[.10,.21]}\end{tabular} & \begin{tabular}[t]{@{}c@{}}0.30\\[-1.5pt]{\tiny[.24,.38]}\end{tabular} & \begin{tabular}[t]{@{}c@{}}0.35\\[-1.5pt]{\tiny[.25,.40]}\end{tabular} & \begin{tabular}[t]{@{}c@{}}0.40\\[-1.5pt]{\tiny[.30,.45]}\end{tabular} & \begin{tabular}[t]{@{}c@{}}0.40\\[-1.5pt]{\tiny[.39,.46]}\end{tabular} & \begin{tabular}[t]{@{}c@{}}0.40\\[-1.5pt]{\tiny[.35,.42]}\end{tabular} & \begin{tabular}[t]{@{}c@{}}0.15\\[-1.5pt]{\tiny[.14,.22]}\end{tabular} & \begin{tabular}[t]{@{}c@{}}0.15\\[-1.5pt]{\tiny[.10,.18]}\end{tabular} & \begin{tabular}[t]{@{}c@{}}0.15\\[-1.5pt]{\tiny[.10,.18]}\end{tabular} & \begin{tabular}[t]{@{}c@{}}0.15\\[-1.5pt]{\tiny[.10,.21]}\end{tabular} \\
 & $n{=}400$ & \begin{tabular}[t]{@{}c@{}}0.15\\[-1.5pt]{\tiny[.10,.16]}\end{tabular} & \begin{tabular}[t]{@{}c@{}}0.30\\[-1.5pt]{\tiny[.20,.31]}\end{tabular} & \begin{tabular}[t]{@{}c@{}}0.35\\[-1.5pt]{\tiny[.25,.40]}\end{tabular} & \begin{tabular}[t]{@{}c@{}}0.40\\[-1.5pt]{\tiny[.30,.51]}\end{tabular} & \begin{tabular}[t]{@{}c@{}}0.55\\[-1.5pt]{\tiny[.40,.58]}\end{tabular} & \begin{tabular}[t]{@{}c@{}}0.55\\[-1.5pt]{\tiny[.40,.61]}\end{tabular} & \begin{tabular}[t]{@{}c@{}}0.15\\[-1.5pt]{\tiny[.10,.18]}\end{tabular} & \begin{tabular}[t]{@{}c@{}}0.15\\[-1.5pt]{\tiny[.10,.16]}\end{tabular} & \begin{tabular}[t]{@{}c@{}}0.15\\[-1.5pt]{\tiny[.10,.19]}\end{tabular} & \begin{tabular}[t]{@{}c@{}}0.15\\[-1.5pt]{\tiny[.09,.20]}\end{tabular} \\
\bottomrule
\end{tabular}
\end{table}
\begin{table}[htbp]
\addtocounter{table}{-1}
\centering\footnotesize
\caption{(continued)}
\setlength{\tabcolsep}{2pt}
\begin{tabular}{ll c @{\enspace} *{5}{c}@{\enspace}*{4}{c}}
\toprule
Estimator & & Scenario 1 & \multicolumn{5}{c}{Scenario 2: dependent errors} & \multicolumn{4}{c}{Scenario 3: stochastic trend} \\
\cmidrule(lr){3-3}\cmidrule(lr){4-8}\cmidrule(lr){9-12}
 &  & $\psi{=}\omega{=}0$ & \multicolumn{5}{c}{$\psi$} & \multicolumn{4}{c}{$\omega$} \\
 &  & & 0.20 & 0.40 & 0.60 & 0.80 & 0.90 & 0.25 & 0.50 & 0.75 & 1.00 \\
\midrule
SSJ--ATW, MLE & $n{=}50$ & \begin{tabular}[t]{@{}c@{}}0.40\\[-1.5pt]{\tiny[.34,.51]}\end{tabular} & \begin{tabular}[t]{@{}c@{}}0.50\\[-1.5pt]{\tiny[.19,.70]}\end{tabular} & \begin{tabular}[t]{@{}c@{}}0.45\\[-1.5pt]{\tiny[.15,.61]}\end{tabular} & \begin{tabular}[t]{@{}c@{}}0.45\\[-1.5pt]{\tiny[.05,.56]}\end{tabular} & \begin{tabular}[t]{@{}c@{}}0.05\\[-1.5pt]{\tiny[.00,.18]}\end{tabular} & \begin{tabular}[t]{@{}c@{}}0.05\\[-1.5pt]{\tiny[.00,.08]}\end{tabular} & \begin{tabular}[t]{@{}c@{}}0.35\\[-1.5pt]{\tiny[.34,.41]}\end{tabular} & \begin{tabular}[t]{@{}c@{}}0.35\\[-1.5pt]{\tiny[.35,.40]}\end{tabular} & \begin{tabular}[t]{@{}c@{}}0.35\\[-1.5pt]{\tiny[.29,.35]}\end{tabular} & \begin{tabular}[t]{@{}c@{}}0.25\\[-1.5pt]{\tiny[.20,.25]}\end{tabular} \\
 & $n{=}100$ & \begin{tabular}[t]{@{}c@{}}0.25\\[-1.5pt]{\tiny[.20,.30]}\end{tabular} & \begin{tabular}[t]{@{}c@{}}0.40\\[-1.5pt]{\tiny[.39,.40]}\end{tabular} & \begin{tabular}[t]{@{}c@{}}0.45\\[-1.5pt]{\tiny[.45,.50]}\end{tabular} & \begin{tabular}[t]{@{}c@{}}0.45\\[-1.5pt]{\tiny[.45,.50]}\end{tabular} & \begin{tabular}[t]{@{}c@{}}0.45\\[-1.5pt]{\tiny[.09,.46]}\end{tabular} & \begin{tabular}[t]{@{}c@{}}0.35\\[-1.5pt]{\tiny[.19,.46]}\end{tabular} & \begin{tabular}[t]{@{}c@{}}0.25\\[-1.5pt]{\tiny[.24,.30]}\end{tabular} & \begin{tabular}[t]{@{}c@{}}0.25\\[-1.5pt]{\tiny[.24,.26]}\end{tabular} & \begin{tabular}[t]{@{}c@{}}0.20\\[-1.5pt]{\tiny[.20,.25]}\end{tabular} & \begin{tabular}[t]{@{}c@{}}0.20\\[-1.5pt]{\tiny[.15,.20]}\end{tabular} \\
 & $n{=}200$ & \begin{tabular}[t]{@{}c@{}}0.15\\[-1.5pt]{\tiny[.15,.20]}\end{tabular} & \begin{tabular}[t]{@{}c@{}}0.30\\[-1.5pt]{\tiny[.25,.35]}\end{tabular} & \begin{tabular}[t]{@{}c@{}}0.35\\[-1.5pt]{\tiny[.34,.45]}\end{tabular} & \begin{tabular}[t]{@{}c@{}}0.40\\[-1.5pt]{\tiny[.40,.46]}\end{tabular} & \begin{tabular}[t]{@{}c@{}}0.35\\[-1.5pt]{\tiny[.35,.41]}\end{tabular} & \begin{tabular}[t]{@{}c@{}}0.25\\[-1.5pt]{\tiny[.35,.40]}\end{tabular} & \begin{tabular}[t]{@{}c@{}}0.15\\[-1.5pt]{\tiny[.15,.15]}\end{tabular} & \begin{tabular}[t]{@{}c@{}}0.15\\[-1.5pt]{\tiny[.15,.21]}\end{tabular} & \begin{tabular}[t]{@{}c@{}}0.15\\[-1.5pt]{\tiny[.15,.16]}\end{tabular} & \begin{tabular}[t]{@{}c@{}}0.15\\[-1.5pt]{\tiny[.10,.15]}\end{tabular} \\
 & $n{=}400$ & \begin{tabular}[t]{@{}c@{}}0.10\\[-1.5pt]{\tiny[.10,.16]}\end{tabular} & \begin{tabular}[t]{@{}c@{}}0.25\\[-1.5pt]{\tiny[.24,.30]}\end{tabular} & \begin{tabular}[t]{@{}c@{}}0.30\\[-1.5pt]{\tiny[.29,.45]}\end{tabular} & \begin{tabular}[t]{@{}c@{}}0.35\\[-1.5pt]{\tiny[.30,.40]}\end{tabular} & \begin{tabular}[t]{@{}c@{}}0.25\\[-1.5pt]{\tiny[.30,.58]}\end{tabular} & \begin{tabular}[t]{@{}c@{}}0.20\\[-1.5pt]{\tiny[.25,.45]}\end{tabular} & \begin{tabular}[t]{@{}c@{}}0.15\\[-1.5pt]{\tiny[.10,.16]}\end{tabular} & \begin{tabular}[t]{@{}c@{}}0.10\\[-1.5pt]{\tiny[.10,.16]}\end{tabular} & \begin{tabular}[t]{@{}c@{}}0.10\\[-1.5pt]{\tiny[.10,.15]}\end{tabular} & \begin{tabular}[t]{@{}c@{}}0.10\\[-1.5pt]{\tiny[.09,.11]}\end{tabular} \\
\addlinespace
SSJ--ATW, random $K{=}20$ & $n{=}50$ & \begin{tabular}[t]{@{}c@{}}0.35\\[-1.5pt]{\tiny[.34,.58]}\end{tabular} & \begin{tabular}[t]{@{}c@{}}0.40\\[-1.5pt]{\tiny[.19,.46]}\end{tabular} & \begin{tabular}[t]{@{}c@{}}0.40\\[-1.5pt]{\tiny[.05,.46]}\end{tabular} & \begin{tabular}[t]{@{}c@{}}0.35\\[-1.5pt]{\tiny[.11,.41]}\end{tabular} & \begin{tabular}[t]{@{}c@{}}0.20\\[-1.5pt]{\tiny[.00,.22]}\end{tabular} & \begin{tabular}[t]{@{}c@{}}0.05\\[-1.5pt]{\tiny[.00,.06]}\end{tabular} & \begin{tabular}[t]{@{}c@{}}0.35\\[-1.5pt]{\tiny[.32,.42]}\end{tabular} & \begin{tabular}[t]{@{}c@{}}0.30\\[-1.5pt]{\tiny[.35,.42]}\end{tabular} & \begin{tabular}[t]{@{}c@{}}0.30\\[-1.5pt]{\tiny[.30,.36]}\end{tabular} & \begin{tabular}[t]{@{}c@{}}0.30\\[-1.5pt]{\tiny[.20,.35]}\end{tabular} \\
 & $n{=}100$ & \begin{tabular}[t]{@{}c@{}}0.25\\[-1.5pt]{\tiny[.20,.30]}\end{tabular} & \begin{tabular}[t]{@{}c@{}}0.35\\[-1.5pt]{\tiny[.35,.41]}\end{tabular} & \begin{tabular}[t]{@{}c@{}}0.35\\[-1.5pt]{\tiny[.39,.41]}\end{tabular} & \begin{tabular}[t]{@{}c@{}}0.40\\[-1.5pt]{\tiny[.39,.45]}\end{tabular} & \begin{tabular}[t]{@{}c@{}}0.45\\[-1.5pt]{\tiny[.35,.45]}\end{tabular} & \begin{tabular}[t]{@{}c@{}}0.35\\[-1.5pt]{\tiny[.26,.40]}\end{tabular} & \begin{tabular}[t]{@{}c@{}}0.25\\[-1.5pt]{\tiny[.20,.26]}\end{tabular} & \begin{tabular}[t]{@{}c@{}}0.25\\[-1.5pt]{\tiny[.20,.25]}\end{tabular} & \begin{tabular}[t]{@{}c@{}}0.20\\[-1.5pt]{\tiny[.20,.26]}\end{tabular} & \begin{tabular}[t]{@{}c@{}}0.20\\[-1.5pt]{\tiny[.15,.21]}\end{tabular} \\
 & $n{=}200$ & \begin{tabular}[t]{@{}c@{}}0.15\\[-1.5pt]{\tiny[.15,.16]}\end{tabular} & \begin{tabular}[t]{@{}c@{}}0.35\\[-1.5pt]{\tiny[.29,.35]}\end{tabular} & \begin{tabular}[t]{@{}c@{}}0.35\\[-1.5pt]{\tiny[.35,.48]}\end{tabular} & \begin{tabular}[t]{@{}c@{}}0.35\\[-1.5pt]{\tiny[.35,.48]}\end{tabular} & \begin{tabular}[t]{@{}c@{}}0.40\\[-1.5pt]{\tiny[.34,.55]}\end{tabular} & \begin{tabular}[t]{@{}c@{}}0.40\\[-1.5pt]{\tiny[.35,.46]}\end{tabular} & \begin{tabular}[t]{@{}c@{}}0.15\\[-1.5pt]{\tiny[.15,.20]}\end{tabular} & \begin{tabular}[t]{@{}c@{}}0.15\\[-1.5pt]{\tiny[.15,.16]}\end{tabular} & \begin{tabular}[t]{@{}c@{}}0.15\\[-1.5pt]{\tiny[.15,.16]}\end{tabular} & \begin{tabular}[t]{@{}c@{}}0.15\\[-1.5pt]{\tiny[.15,.20]}\end{tabular} \\
 & $n{=}400$ & \begin{tabular}[t]{@{}c@{}}0.15\\[-1.5pt]{\tiny[.10,.16]}\end{tabular} & \begin{tabular}[t]{@{}c@{}}0.30\\[-1.5pt]{\tiny[.24,.39]}\end{tabular} & \begin{tabular}[t]{@{}c@{}}0.45\\[-1.5pt]{\tiny[.30,.41]}\end{tabular} & \begin{tabular}[t]{@{}c@{}}0.35\\[-1.5pt]{\tiny[.30,.48]}\end{tabular} & \begin{tabular}[t]{@{}c@{}}0.30\\[-1.5pt]{\tiny[.35,.45]}\end{tabular} & \begin{tabular}[t]{@{}c@{}}0.35\\[-1.5pt]{\tiny[.35,.41]}\end{tabular} & \begin{tabular}[t]{@{}c@{}}0.15\\[-1.5pt]{\tiny[.10,.15]}\end{tabular} & \begin{tabular}[t]{@{}c@{}}0.15\\[-1.5pt]{\tiny[.10,.16]}\end{tabular} & \begin{tabular}[t]{@{}c@{}}0.15\\[-1.5pt]{\tiny[.10,.15]}\end{tabular} & \begin{tabular}[t]{@{}c@{}}0.10\\[-1.5pt]{\tiny[.10,.15]}\end{tabular} \\
\addlinespace
SSJ--ATW, blocks $K{=}20$ & $n{=}50$ & \begin{tabular}[t]{@{}c@{}}0.35\\[-1.5pt]{\tiny[.34,.70]}\end{tabular} & \begin{tabular}[t]{@{}c@{}}0.45\\[-1.5pt]{\tiny[.26,.51]}\end{tabular} & \begin{tabular}[t]{@{}c@{}}0.40\\[-1.5pt]{\tiny[.00,.46]}\end{tabular} & \begin{tabular}[t]{@{}c@{}}0.35\\[-1.5pt]{\tiny[.08,.40]}\end{tabular} & \begin{tabular}[t]{@{}c@{}}0.20\\[-1.5pt]{\tiny[.04,.15]}\end{tabular} & \begin{tabular}[t]{@{}c@{}}0.00\\[-1.5pt]{\tiny[.00,.05]}\end{tabular} & \begin{tabular}[t]{@{}c@{}}0.35\\[-1.5pt]{\tiny[.25,.50]}\end{tabular} & \begin{tabular}[t]{@{}c@{}}0.30\\[-1.5pt]{\tiny[.29,.50]}\end{tabular} & \begin{tabular}[t]{@{}c@{}}0.35\\[-1.5pt]{\tiny[.29,.38]}\end{tabular} & \begin{tabular}[t]{@{}c@{}}0.25\\[-1.5pt]{\tiny[.22,.31]}\end{tabular} \\
 & $n{=}100$ & \begin{tabular}[t]{@{}c@{}}0.25\\[-1.5pt]{\tiny[.24,.30]}\end{tabular} & \begin{tabular}[t]{@{}c@{}}0.35\\[-1.5pt]{\tiny[.35,.40]}\end{tabular} & \begin{tabular}[t]{@{}c@{}}0.40\\[-1.5pt]{\tiny[.35,.45]}\end{tabular} & \begin{tabular}[t]{@{}c@{}}0.40\\[-1.5pt]{\tiny[.39,.45]}\end{tabular} & \begin{tabular}[t]{@{}c@{}}0.40\\[-1.5pt]{\tiny[.40,.50]}\end{tabular} & \begin{tabular}[t]{@{}c@{}}0.35\\[-1.5pt]{\tiny[.29,.45]}\end{tabular} & \begin{tabular}[t]{@{}c@{}}0.25\\[-1.5pt]{\tiny[.20,.30]}\end{tabular} & \begin{tabular}[t]{@{}c@{}}0.20\\[-1.5pt]{\tiny[.20,.26]}\end{tabular} & \begin{tabular}[t]{@{}c@{}}0.20\\[-1.5pt]{\tiny[.20,.25]}\end{tabular} & \begin{tabular}[t]{@{}c@{}}0.20\\[-1.5pt]{\tiny[.20,.21]}\end{tabular} \\
 & $n{=}200$ & \begin{tabular}[t]{@{}c@{}}0.15\\[-1.5pt]{\tiny[.15,.20]}\end{tabular} & \begin{tabular}[t]{@{}c@{}}0.25\\[-1.5pt]{\tiny[.25,.33]}\end{tabular} & \begin{tabular}[t]{@{}c@{}}0.30\\[-1.5pt]{\tiny[.30,.40]}\end{tabular} & \begin{tabular}[t]{@{}c@{}}0.40\\[-1.5pt]{\tiny[.35,.41]}\end{tabular} & \begin{tabular}[t]{@{}c@{}}0.45\\[-1.5pt]{\tiny[.40,.45]}\end{tabular} & \begin{tabular}[t]{@{}c@{}}0.40\\[-1.5pt]{\tiny[.40,.56]}\end{tabular} & \begin{tabular}[t]{@{}c@{}}0.15\\[-1.5pt]{\tiny[.15,.16]}\end{tabular} & \begin{tabular}[t]{@{}c@{}}0.15\\[-1.5pt]{\tiny[.15,.16]}\end{tabular} & \begin{tabular}[t]{@{}c@{}}0.15\\[-1.5pt]{\tiny[.15,.16]}\end{tabular} & \begin{tabular}[t]{@{}c@{}}0.15\\[-1.5pt]{\tiny[.14,.16]}\end{tabular} \\
 & $n{=}400$ & \begin{tabular}[t]{@{}c@{}}0.10\\[-1.5pt]{\tiny[.10,.16]}\end{tabular} & \begin{tabular}[t]{@{}c@{}}0.25\\[-1.5pt]{\tiny[.20,.28]}\end{tabular} & \begin{tabular}[t]{@{}c@{}}0.30\\[-1.5pt]{\tiny[.25,.31]}\end{tabular} & \begin{tabular}[t]{@{}c@{}}0.40\\[-1.5pt]{\tiny[.34,.45]}\end{tabular} & \begin{tabular}[t]{@{}c@{}}0.45\\[-1.5pt]{\tiny[.35,.51]}\end{tabular} & \begin{tabular}[t]{@{}c@{}}0.50\\[-1.5pt]{\tiny[.44,.58]}\end{tabular} & \begin{tabular}[t]{@{}c@{}}0.15\\[-1.5pt]{\tiny[.10,.15]}\end{tabular} & \begin{tabular}[t]{@{}c@{}}0.10\\[-1.5pt]{\tiny[.10,.15]}\end{tabular} & \begin{tabular}[t]{@{}c@{}}0.10\\[-1.5pt]{\tiny[.10,.16]}\end{tabular} & \begin{tabular}[t]{@{}c@{}}0.10\\[-1.5pt]{\tiny[.10,.15]}\end{tabular} \\
\bottomrule
\end{tabular}
\end{table}

\begin{table}[htbp]
\centering\footnotesize
\caption{The calibration by sample size under minimax regret in the count error, over the nine cells with $q>0$ of each sample size. A $0.00$ is the floor of the grid: at that sample size and level the worst cell is one with five discontinuities in which the estimator under-declares at every $\xi$, so the most permissive penalty minimizes the worst-case count error.}
\label{tab:byn_scen_count}
\setlength{\tabcolsep}{2pt}
\begin{tabular}{ll c @{\enspace} *{5}{c}@{\enspace}*{4}{c}}
\toprule
Estimator & & Scenario 1 & \multicolumn{5}{c}{Scenario 2: dependent errors} & \multicolumn{4}{c}{Scenario 3: stochastic trend} \\
\cmidrule(lr){3-3}\cmidrule(lr){4-8}\cmidrule(lr){9-12}
 &  & $\psi{=}\omega{=}0$ & \multicolumn{5}{c}{$\psi$} & \multicolumn{4}{c}{$\omega$} \\
 &  & & 0.20 & 0.40 & 0.60 & 0.80 & 0.90 & 0.25 & 0.50 & 0.75 & 1.00 \\
\midrule
SSJ--MLE, MLE & $n{=}50$ & \begin{tabular}[t]{@{}c@{}}0.20\\[-1.5pt]{\tiny[.00,.70]}\end{tabular} & \begin{tabular}[t]{@{}c@{}}0.25\\[-1.5pt]{\tiny[.00,.75]}\end{tabular} & \begin{tabular}[t]{@{}c@{}}0.25\\[-1.5pt]{\tiny[.00,.70]}\end{tabular} & \begin{tabular}[t]{@{}c@{}}0.20\\[-1.5pt]{\tiny[.00,.70]}\end{tabular} & \begin{tabular}[t]{@{}c@{}}0.20\\[-1.5pt]{\tiny[.00,.80]}\end{tabular} & \begin{tabular}[t]{@{}c@{}}0.20\\[-1.5pt]{\tiny[.00,.50]}\end{tabular} & \begin{tabular}[t]{@{}c@{}}0.20\\[-1.5pt]{\tiny[.00,.70]}\end{tabular} & \begin{tabular}[t]{@{}c@{}}0.25\\[-1.5pt]{\tiny[.00,.80]}\end{tabular} & \begin{tabular}[t]{@{}c@{}}0.30\\[-1.5pt]{\tiny[.00,.95]}\end{tabular} & \begin{tabular}[t]{@{}c@{}}0.40\\[-1.5pt]{\tiny[.00,.95]}\end{tabular} \\
 & $n{=}100$ & \begin{tabular}[t]{@{}c@{}}0.20\\[-1.5pt]{\tiny[.00,.65]}\end{tabular} & \begin{tabular}[t]{@{}c@{}}0.25\\[-1.5pt]{\tiny[.00,.80]}\end{tabular} & \begin{tabular}[t]{@{}c@{}}0.20\\[-1.5pt]{\tiny[.00,.75]}\end{tabular} & \begin{tabular}[t]{@{}c@{}}0.20\\[-1.5pt]{\tiny[.00,.60]}\end{tabular} & \begin{tabular}[t]{@{}c@{}}0.15\\[-1.5pt]{\tiny[.00,.65]}\end{tabular} & \begin{tabular}[t]{@{}c@{}}0.05\\[-1.5pt]{\tiny[.00,.55]}\end{tabular} & \begin{tabular}[t]{@{}c@{}}0.20\\[-1.5pt]{\tiny[.00,.60]}\end{tabular} & \begin{tabular}[t]{@{}c@{}}0.25\\[-1.5pt]{\tiny[.00,.90]}\end{tabular} & \begin{tabular}[t]{@{}c@{}}0.25\\[-1.5pt]{\tiny[.00,.95]}\end{tabular} & \begin{tabular}[t]{@{}c@{}}0.40\\[-1.5pt]{\tiny[.10,.95]}\end{tabular} \\
 & $n{=}200$ & \begin{tabular}[t]{@{}c@{}}0.00\\[-1.5pt]{\tiny[.00,.25]}\end{tabular} & \begin{tabular}[t]{@{}c@{}}0.00\\[-1.5pt]{\tiny[.00,.25]}\end{tabular} & \begin{tabular}[t]{@{}c@{}}0.00\\[-1.5pt]{\tiny[.00,.05]}\end{tabular} & \begin{tabular}[t]{@{}c@{}}0.05\\[-1.5pt]{\tiny[.00,.00]}\end{tabular} & \begin{tabular}[t]{@{}c@{}}0.00\\[-1.5pt]{\tiny[.00,.00]}\end{tabular} & \begin{tabular}[t]{@{}c@{}}0.00\\[-1.5pt]{\tiny[.00,.00]}\end{tabular} & \begin{tabular}[t]{@{}c@{}}0.00\\[-1.5pt]{\tiny[.00,.30]}\end{tabular} & \begin{tabular}[t]{@{}c@{}}0.25\\[-1.5pt]{\tiny[.00,.25]}\end{tabular} & \begin{tabular}[t]{@{}c@{}}0.25\\[-1.5pt]{\tiny[.10,.25]}\end{tabular} & \begin{tabular}[t]{@{}c@{}}0.40\\[-1.5pt]{\tiny[.00,.95]}\end{tabular} \\
 & $n{=}400$ & \begin{tabular}[t]{@{}c@{}}0.00\\[-1.5pt]{\tiny[.00,.25]}\end{tabular} & \begin{tabular}[t]{@{}c@{}}0.10\\[-1.5pt]{\tiny[.05,.30]}\end{tabular} & \begin{tabular}[t]{@{}c@{}}0.00\\[-1.5pt]{\tiny[.00,.00]}\end{tabular} & \begin{tabular}[t]{@{}c@{}}0.05\\[-1.5pt]{\tiny[.00,.00]}\end{tabular} & \begin{tabular}[t]{@{}c@{}}0.00\\[-1.5pt]{\tiny[.00,.00]}\end{tabular} & \begin{tabular}[t]{@{}c@{}}0.00\\[-1.5pt]{\tiny[.00,.00]}\end{tabular} & \begin{tabular}[t]{@{}c@{}}0.10\\[-1.5pt]{\tiny[.00,.20]}\end{tabular} & \begin{tabular}[t]{@{}c@{}}0.00\\[-1.5pt]{\tiny[.00,.20]}\end{tabular} & \begin{tabular}[t]{@{}c@{}}0.00\\[-1.5pt]{\tiny[.10,.40]}\end{tabular} & \begin{tabular}[t]{@{}c@{}}0.55\\[-1.5pt]{\tiny[.00,.65]}\end{tabular} \\
\addlinespace
SSJ--KFS, MLE & $n{=}50$ & \begin{tabular}[t]{@{}c@{}}0.15\\[-1.5pt]{\tiny[.05,.40]}\end{tabular} & \begin{tabular}[t]{@{}c@{}}0.20\\[-1.5pt]{\tiny[.05,.70]}\end{tabular} & \begin{tabular}[t]{@{}c@{}}0.25\\[-1.5pt]{\tiny[.05,.65]}\end{tabular} & \begin{tabular}[t]{@{}c@{}}0.25\\[-1.5pt]{\tiny[.05,.55]}\end{tabular} & \begin{tabular}[t]{@{}c@{}}0.25\\[-1.5pt]{\tiny[.00,.85]}\end{tabular} & \begin{tabular}[t]{@{}c@{}}0.25\\[-1.5pt]{\tiny[.10,.65]}\end{tabular} & \begin{tabular}[t]{@{}c@{}}0.15\\[-1.5pt]{\tiny[.05,.60]}\end{tabular} & \begin{tabular}[t]{@{}c@{}}0.15\\[-1.5pt]{\tiny[.00,.85]}\end{tabular} & \begin{tabular}[t]{@{}c@{}}0.25\\[-1.5pt]{\tiny[.05,.95]}\end{tabular} & \begin{tabular}[t]{@{}c@{}}0.45\\[-1.5pt]{\tiny[.05,1.00]}\end{tabular} \\
 & $n{=}100$ & \begin{tabular}[t]{@{}c@{}}0.20\\[-1.5pt]{\tiny[.05,.50]}\end{tabular} & \begin{tabular}[t]{@{}c@{}}0.25\\[-1.5pt]{\tiny[.00,.65]}\end{tabular} & \begin{tabular}[t]{@{}c@{}}0.25\\[-1.5pt]{\tiny[.00,.75]}\end{tabular} & \begin{tabular}[t]{@{}c@{}}0.25\\[-1.5pt]{\tiny[.00,.75]}\end{tabular} & \begin{tabular}[t]{@{}c@{}}0.30\\[-1.5pt]{\tiny[.00,.95]}\end{tabular} & \begin{tabular}[t]{@{}c@{}}0.35\\[-1.5pt]{\tiny[.00,.90]}\end{tabular} & \begin{tabular}[t]{@{}c@{}}0.20\\[-1.5pt]{\tiny[.10,.65]}\end{tabular} & \begin{tabular}[t]{@{}c@{}}0.25\\[-1.5pt]{\tiny[.05,.80]}\end{tabular} & \begin{tabular}[t]{@{}c@{}}0.25\\[-1.5pt]{\tiny[.10,.90]}\end{tabular} & \begin{tabular}[t]{@{}c@{}}0.50\\[-1.5pt]{\tiny[.05,1.00]}\end{tabular} \\
 & $n{=}200$ & \begin{tabular}[t]{@{}c@{}}0.20\\[-1.5pt]{\tiny[.00,.60]}\end{tabular} & \begin{tabular}[t]{@{}c@{}}0.25\\[-1.5pt]{\tiny[.00,.65]}\end{tabular} & \begin{tabular}[t]{@{}c@{}}0.20\\[-1.5pt]{\tiny[.00,.70]}\end{tabular} & \begin{tabular}[t]{@{}c@{}}0.00\\[-1.5pt]{\tiny[.00,.10]}\end{tabular} & \begin{tabular}[t]{@{}c@{}}0.00\\[-1.5pt]{\tiny[.00,.15]}\end{tabular} & \begin{tabular}[t]{@{}c@{}}0.00\\[-1.5pt]{\tiny[.00,.10]}\end{tabular} & \begin{tabular}[t]{@{}c@{}}0.20\\[-1.5pt]{\tiny[.00,.85]}\end{tabular} & \begin{tabular}[t]{@{}c@{}}0.20\\[-1.5pt]{\tiny[.00,.90]}\end{tabular} & \begin{tabular}[t]{@{}c@{}}0.25\\[-1.5pt]{\tiny[.15,.90]}\end{tabular} & \begin{tabular}[t]{@{}c@{}}0.40\\[-1.5pt]{\tiny[.00,.95]}\end{tabular} \\
 & $n{=}400$ & \begin{tabular}[t]{@{}c@{}}0.25\\[-1.5pt]{\tiny[.00,.05]}\end{tabular} & \begin{tabular}[t]{@{}c@{}}0.30\\[-1.5pt]{\tiny[.00,.20]}\end{tabular} & \begin{tabular}[t]{@{}c@{}}0.00\\[-1.5pt]{\tiny[.00,.15]}\end{tabular} & \begin{tabular}[t]{@{}c@{}}0.00\\[-1.5pt]{\tiny[.00,.10]}\end{tabular} & \begin{tabular}[t]{@{}c@{}}0.00\\[-1.5pt]{\tiny[.05,.10]}\end{tabular} & \begin{tabular}[t]{@{}c@{}}0.00\\[-1.5pt]{\tiny[.05,.10]}\end{tabular} & \begin{tabular}[t]{@{}c@{}}0.20\\[-1.5pt]{\tiny[.00,.10]}\end{tabular} & \begin{tabular}[t]{@{}c@{}}0.25\\[-1.5pt]{\tiny[.00,.05]}\end{tabular} & \begin{tabular}[t]{@{}c@{}}0.35\\[-1.5pt]{\tiny[.00,.30]}\end{tabular} & \begin{tabular}[t]{@{}c@{}}0.50\\[-1.5pt]{\tiny[.20,.70]}\end{tabular} \\
\addlinespace
SSJ--KFS, random $K{=}20$ & $n{=}50$ & \begin{tabular}[t]{@{}c@{}}0.15\\[-1.5pt]{\tiny[.00,.45]}\end{tabular} & \begin{tabular}[t]{@{}c@{}}0.25\\[-1.5pt]{\tiny[.05,.75]}\end{tabular} & \begin{tabular}[t]{@{}c@{}}0.25\\[-1.5pt]{\tiny[.10,.80]}\end{tabular} & \begin{tabular}[t]{@{}c@{}}0.25\\[-1.5pt]{\tiny[.05,.60]}\end{tabular} & \begin{tabular}[t]{@{}c@{}}0.30\\[-1.5pt]{\tiny[.00,.55]}\end{tabular} & \begin{tabular}[t]{@{}c@{}}0.25\\[-1.5pt]{\tiny[.10,.65]}\end{tabular} & \begin{tabular}[t]{@{}c@{}}0.15\\[-1.5pt]{\tiny[.00,.45]}\end{tabular} & \begin{tabular}[t]{@{}c@{}}0.20\\[-1.5pt]{\tiny[.00,.60]}\end{tabular} & \begin{tabular}[t]{@{}c@{}}0.20\\[-1.5pt]{\tiny[.05,.80]}\end{tabular} & \begin{tabular}[t]{@{}c@{}}0.20\\[-1.5pt]{\tiny[.05,.85]}\end{tabular} \\
 & $n{=}100$ & \begin{tabular}[t]{@{}c@{}}0.20\\[-1.5pt]{\tiny[.00,.60]}\end{tabular} & \begin{tabular}[t]{@{}c@{}}0.25\\[-1.5pt]{\tiny[.00,.70]}\end{tabular} & \begin{tabular}[t]{@{}c@{}}0.20\\[-1.5pt]{\tiny[.00,.75]}\end{tabular} & \begin{tabular}[t]{@{}c@{}}0.20\\[-1.5pt]{\tiny[.00,.70]}\end{tabular} & \begin{tabular}[t]{@{}c@{}}0.25\\[-1.5pt]{\tiny[.00,.90]}\end{tabular} & \begin{tabular}[t]{@{}c@{}}0.30\\[-1.5pt]{\tiny[.10,.95]}\end{tabular} & \begin{tabular}[t]{@{}c@{}}0.20\\[-1.5pt]{\tiny[.00,.65]}\end{tabular} & \begin{tabular}[t]{@{}c@{}}0.20\\[-1.5pt]{\tiny[.00,.95]}\end{tabular} & \begin{tabular}[t]{@{}c@{}}0.25\\[-1.5pt]{\tiny[.05,.85]}\end{tabular} & \begin{tabular}[t]{@{}c@{}}0.25\\[-1.5pt]{\tiny[.05,.95]}\end{tabular} \\
 & $n{=}200$ & \begin{tabular}[t]{@{}c@{}}0.20\\[-1.5pt]{\tiny[.00,.75]}\end{tabular} & \begin{tabular}[t]{@{}c@{}}0.15\\[-1.5pt]{\tiny[.00,.75]}\end{tabular} & \begin{tabular}[t]{@{}c@{}}0.10\\[-1.5pt]{\tiny[.00,.35]}\end{tabular} & \begin{tabular}[t]{@{}c@{}}0.10\\[-1.5pt]{\tiny[.00,.10]}\end{tabular} & \begin{tabular}[t]{@{}c@{}}0.15\\[-1.5pt]{\tiny[.05,.20]}\end{tabular} & \begin{tabular}[t]{@{}c@{}}0.00\\[-1.5pt]{\tiny[.00,.15]}\end{tabular} & \begin{tabular}[t]{@{}c@{}}0.20\\[-1.5pt]{\tiny[.00,.85]}\end{tabular} & \begin{tabular}[t]{@{}c@{}}0.20\\[-1.5pt]{\tiny[.00,.85]}\end{tabular} & \begin{tabular}[t]{@{}c@{}}0.25\\[-1.5pt]{\tiny[.10,.85]}\end{tabular} & \begin{tabular}[t]{@{}c@{}}0.30\\[-1.5pt]{\tiny[.15,1.00]}\end{tabular} \\
 & $n{=}400$ & \begin{tabular}[t]{@{}c@{}}0.25\\[-1.5pt]{\tiny[.00,.15]}\end{tabular} & \begin{tabular}[t]{@{}c@{}}0.05\\[-1.5pt]{\tiny[.00,.70]}\end{tabular} & \begin{tabular}[t]{@{}c@{}}0.05\\[-1.5pt]{\tiny[.00,.20]}\end{tabular} & \begin{tabular}[t]{@{}c@{}}0.05\\[-1.5pt]{\tiny[.00,.20]}\end{tabular} & \begin{tabular}[t]{@{}c@{}}0.05\\[-1.5pt]{\tiny[.05,.30]}\end{tabular} & \begin{tabular}[t]{@{}c@{}}0.10\\[-1.5pt]{\tiny[.10,.30]}\end{tabular} & \begin{tabular}[t]{@{}c@{}}0.25\\[-1.5pt]{\tiny[.00,.05]}\end{tabular} & \begin{tabular}[t]{@{}c@{}}0.25\\[-1.5pt]{\tiny[.00,.10]}\end{tabular} & \begin{tabular}[t]{@{}c@{}}0.30\\[-1.5pt]{\tiny[.00,.35]}\end{tabular} & \begin{tabular}[t]{@{}c@{}}0.45\\[-1.5pt]{\tiny[.20,1.00]}\end{tabular} \\
\addlinespace
SSJ--KFS, blocks $K{=}20$ & $n{=}50$ & \begin{tabular}[t]{@{}c@{}}0.15\\[-1.5pt]{\tiny[.05,.45]}\end{tabular} & \begin{tabular}[t]{@{}c@{}}0.25\\[-1.5pt]{\tiny[.05,.60]}\end{tabular} & \begin{tabular}[t]{@{}c@{}}0.25\\[-1.5pt]{\tiny[.05,.60]}\end{tabular} & \begin{tabular}[t]{@{}c@{}}0.25\\[-1.5pt]{\tiny[.10,.85]}\end{tabular} & \begin{tabular}[t]{@{}c@{}}0.30\\[-1.5pt]{\tiny[.05,.80]}\end{tabular} & \begin{tabular}[t]{@{}c@{}}0.30\\[-1.5pt]{\tiny[.05,.60]}\end{tabular} & \begin{tabular}[t]{@{}c@{}}0.15\\[-1.5pt]{\tiny[.05,.60]}\end{tabular} & \begin{tabular}[t]{@{}c@{}}0.15\\[-1.5pt]{\tiny[.00,.80]}\end{tabular} & \begin{tabular}[t]{@{}c@{}}0.20\\[-1.5pt]{\tiny[.00,.65]}\end{tabular} & \begin{tabular}[t]{@{}c@{}}0.25\\[-1.5pt]{\tiny[.00,1.00]}\end{tabular} \\
 & $n{=}100$ & \begin{tabular}[t]{@{}c@{}}0.20\\[-1.5pt]{\tiny[.00,.50]}\end{tabular} & \begin{tabular}[t]{@{}c@{}}0.25\\[-1.5pt]{\tiny[.00,.75]}\end{tabular} & \begin{tabular}[t]{@{}c@{}}0.25\\[-1.5pt]{\tiny[.00,.70]}\end{tabular} & \begin{tabular}[t]{@{}c@{}}0.25\\[-1.5pt]{\tiny[.00,.75]}\end{tabular} & \begin{tabular}[t]{@{}c@{}}0.30\\[-1.5pt]{\tiny[.00,.90]}\end{tabular} & \begin{tabular}[t]{@{}c@{}}0.35\\[-1.5pt]{\tiny[.05,1.00]}\end{tabular} & \begin{tabular}[t]{@{}c@{}}0.20\\[-1.5pt]{\tiny[.00,.70]}\end{tabular} & \begin{tabular}[t]{@{}c@{}}0.20\\[-1.5pt]{\tiny[.00,.70]}\end{tabular} & \begin{tabular}[t]{@{}c@{}}0.20\\[-1.5pt]{\tiny[.05,.85]}\end{tabular} & \begin{tabular}[t]{@{}c@{}}0.30\\[-1.5pt]{\tiny[.05,.95]}\end{tabular} \\
 & $n{=}200$ & \begin{tabular}[t]{@{}c@{}}0.20\\[-1.5pt]{\tiny[.00,.55]}\end{tabular} & \begin{tabular}[t]{@{}c@{}}0.25\\[-1.5pt]{\tiny[.00,.65]}\end{tabular} & \begin{tabular}[t]{@{}c@{}}0.30\\[-1.5pt]{\tiny[.00,.10]}\end{tabular} & \begin{tabular}[t]{@{}c@{}}0.00\\[-1.5pt]{\tiny[.00,.25]}\end{tabular} & \begin{tabular}[t]{@{}c@{}}0.00\\[-1.5pt]{\tiny[.00,.10]}\end{tabular} & \begin{tabular}[t]{@{}c@{}}0.05\\[-1.5pt]{\tiny[.00,.20]}\end{tabular} & \begin{tabular}[t]{@{}c@{}}0.20\\[-1.5pt]{\tiny[.00,.85]}\end{tabular} & \begin{tabular}[t]{@{}c@{}}0.20\\[-1.5pt]{\tiny[.00,.80]}\end{tabular} & \begin{tabular}[t]{@{}c@{}}0.25\\[-1.5pt]{\tiny[.05,.80]}\end{tabular} & \begin{tabular}[t]{@{}c@{}}0.35\\[-1.5pt]{\tiny[.15,.95]}\end{tabular} \\
 & $n{=}400$ & \begin{tabular}[t]{@{}c@{}}0.20\\[-1.5pt]{\tiny[.00,.15]}\end{tabular} & \begin{tabular}[t]{@{}c@{}}0.30\\[-1.5pt]{\tiny[.00,.20]}\end{tabular} & \begin{tabular}[t]{@{}c@{}}0.00\\[-1.5pt]{\tiny[.00,.25]}\end{tabular} & \begin{tabular}[t]{@{}c@{}}0.05\\[-1.5pt]{\tiny[.05,.25]}\end{tabular} & \begin{tabular}[t]{@{}c@{}}0.10\\[-1.5pt]{\tiny[.05,.25]}\end{tabular} & \begin{tabular}[t]{@{}c@{}}0.15\\[-1.5pt]{\tiny[.05,.35]}\end{tabular} & \begin{tabular}[t]{@{}c@{}}0.20\\[-1.5pt]{\tiny[.00,.15]}\end{tabular} & \begin{tabular}[t]{@{}c@{}}0.20\\[-1.5pt]{\tiny[.00,.15]}\end{tabular} & \begin{tabular}[t]{@{}c@{}}0.30\\[-1.5pt]{\tiny[.00,.25]}\end{tabular} & \begin{tabular}[t]{@{}c@{}}0.40\\[-1.5pt]{\tiny[.30,.95]}\end{tabular} \\
\bottomrule
\end{tabular}
\end{table}
\begin{table}[htbp]
\addtocounter{table}{-1}
\centering\footnotesize
\caption{(continued)}
\setlength{\tabcolsep}{2pt}
\begin{tabular}{ll c @{\enspace} *{5}{c}@{\enspace}*{4}{c}}
\toprule
Estimator & & Scenario 1 & \multicolumn{5}{c}{Scenario 2: dependent errors} & \multicolumn{4}{c}{Scenario 3: stochastic trend} \\
\cmidrule(lr){3-3}\cmidrule(lr){4-8}\cmidrule(lr){9-12}
 &  & $\psi{=}\omega{=}0$ & \multicolumn{5}{c}{$\psi$} & \multicolumn{4}{c}{$\omega$} \\
 &  & & 0.20 & 0.40 & 0.60 & 0.80 & 0.90 & 0.25 & 0.50 & 0.75 & 1.00 \\
\midrule
SSJ--ATW, MLE & $n{=}50$ & \begin{tabular}[t]{@{}c@{}}0.30\\[-1.5pt]{\tiny[.00,.60]}\end{tabular} & \begin{tabular}[t]{@{}c@{}}0.45\\[-1.5pt]{\tiny[.00,.95]}\end{tabular} & \begin{tabular}[t]{@{}c@{}}0.50\\[-1.5pt]{\tiny[.00,.85]}\end{tabular} & \begin{tabular}[t]{@{}c@{}}0.50\\[-1.5pt]{\tiny[.00,1.00]}\end{tabular} & \begin{tabular}[t]{@{}c@{}}0.60\\[-1.5pt]{\tiny[.00,1.00]}\end{tabular} & \begin{tabular}[t]{@{}c@{}}0.70\\[-1.5pt]{\tiny[.05,.95]}\end{tabular} & \begin{tabular}[t]{@{}c@{}}0.30\\[-1.5pt]{\tiny[.00,.80]}\end{tabular} & \begin{tabular}[t]{@{}c@{}}0.35\\[-1.5pt]{\tiny[.05,.95]}\end{tabular} & \begin{tabular}[t]{@{}c@{}}0.40\\[-1.5pt]{\tiny[.00,1.00]}\end{tabular} & \begin{tabular}[t]{@{}c@{}}0.55\\[-1.5pt]{\tiny[.15,.95]}\end{tabular} \\
 & $n{=}100$ & \begin{tabular}[t]{@{}c@{}}0.25\\[-1.5pt]{\tiny[.00,.60]}\end{tabular} & \begin{tabular}[t]{@{}c@{}}0.00\\[-1.5pt]{\tiny[.00,.40]}\end{tabular} & \begin{tabular}[t]{@{}c@{}}0.00\\[-1.5pt]{\tiny[.05,.40]}\end{tabular} & \begin{tabular}[t]{@{}c@{}}0.10\\[-1.5pt]{\tiny[.00,.35]}\end{tabular} & \begin{tabular}[t]{@{}c@{}}0.20\\[-1.5pt]{\tiny[.00,.40]}\end{tabular} & \begin{tabular}[t]{@{}c@{}}0.20\\[-1.5pt]{\tiny[.05,.45]}\end{tabular} & \begin{tabular}[t]{@{}c@{}}0.25\\[-1.5pt]{\tiny[.00,.65]}\end{tabular} & \begin{tabular}[t]{@{}c@{}}0.25\\[-1.5pt]{\tiny[.00,.85]}\end{tabular} & \begin{tabular}[t]{@{}c@{}}0.30\\[-1.5pt]{\tiny[.00,.90]}\end{tabular} & \begin{tabular}[t]{@{}c@{}}0.45\\[-1.5pt]{\tiny[.25,.95]}\end{tabular} \\
 & $n{=}200$ & \begin{tabular}[t]{@{}c@{}}0.20\\[-1.5pt]{\tiny[.00,.15]}\end{tabular} & \begin{tabular}[t]{@{}c@{}}0.05\\[-1.5pt]{\tiny[.05,.20]}\end{tabular} & \begin{tabular}[t]{@{}c@{}}0.40\\[-1.5pt]{\tiny[.00,.25]}\end{tabular} & \begin{tabular}[t]{@{}c@{}}0.40\\[-1.5pt]{\tiny[.10,.30]}\end{tabular} & \begin{tabular}[t]{@{}c@{}}0.45\\[-1.5pt]{\tiny[.15,.35]}\end{tabular} & \begin{tabular}[t]{@{}c@{}}0.60\\[-1.5pt]{\tiny[.20,.40]}\end{tabular} & \begin{tabular}[t]{@{}c@{}}0.20\\[-1.5pt]{\tiny[.00,.15]}\end{tabular} & \begin{tabular}[t]{@{}c@{}}0.20\\[-1.5pt]{\tiny[.00,.95]}\end{tabular} & \begin{tabular}[t]{@{}c@{}}0.25\\[-1.5pt]{\tiny[.00,1.00]}\end{tabular} & \begin{tabular}[t]{@{}c@{}}0.35\\[-1.5pt]{\tiny[.05,.95]}\end{tabular} \\
 & $n{=}400$ & \begin{tabular}[t]{@{}c@{}}0.00\\[-1.5pt]{\tiny[.00,.10]}\end{tabular} & \begin{tabular}[t]{@{}c@{}}0.40\\[-1.5pt]{\tiny[.00,.15]}\end{tabular} & \begin{tabular}[t]{@{}c@{}}0.30\\[-1.5pt]{\tiny[.05,.20]}\end{tabular} & \begin{tabular}[t]{@{}c@{}}0.25\\[-1.5pt]{\tiny[.10,.20]}\end{tabular} & \begin{tabular}[t]{@{}c@{}}0.25\\[-1.5pt]{\tiny[.15,.25]}\end{tabular} & \begin{tabular}[t]{@{}c@{}}0.30\\[-1.5pt]{\tiny[.20,.30]}\end{tabular} & \begin{tabular}[t]{@{}c@{}}0.00\\[-1.5pt]{\tiny[.00,.15]}\end{tabular} & \begin{tabular}[t]{@{}c@{}}0.25\\[-1.5pt]{\tiny[.00,.10]}\end{tabular} & \begin{tabular}[t]{@{}c@{}}0.35\\[-1.5pt]{\tiny[.00,.15]}\end{tabular} & \begin{tabular}[t]{@{}c@{}}0.45\\[-1.5pt]{\tiny[.00,.10]}\end{tabular} \\
\addlinespace
SSJ--ATW, random $K{=}20$ & $n{=}50$ & \begin{tabular}[t]{@{}c@{}}0.30\\[-1.5pt]{\tiny[.00,.75]}\end{tabular} & \begin{tabular}[t]{@{}c@{}}0.45\\[-1.5pt]{\tiny[.00,1.00]}\end{tabular} & \begin{tabular}[t]{@{}c@{}}0.50\\[-1.5pt]{\tiny[.00,1.00]}\end{tabular} & \begin{tabular}[t]{@{}c@{}}0.55\\[-1.5pt]{\tiny[.00,1.00]}\end{tabular} & \begin{tabular}[t]{@{}c@{}}0.65\\[-1.5pt]{\tiny[.15,1.00]}\end{tabular} & \begin{tabular}[t]{@{}c@{}}0.65\\[-1.5pt]{\tiny[.10,1.00]}\end{tabular} & \begin{tabular}[t]{@{}c@{}}0.30\\[-1.5pt]{\tiny[.00,.60]}\end{tabular} & \begin{tabular}[t]{@{}c@{}}0.30\\[-1.5pt]{\tiny[.10,.70]}\end{tabular} & \begin{tabular}[t]{@{}c@{}}0.35\\[-1.5pt]{\tiny[.15,.90]}\end{tabular} & \begin{tabular}[t]{@{}c@{}}0.40\\[-1.5pt]{\tiny[.20,.90]}\end{tabular} \\
 & $n{=}100$ & \begin{tabular}[t]{@{}c@{}}0.25\\[-1.5pt]{\tiny[.00,.65]}\end{tabular} & \begin{tabular}[t]{@{}c@{}}0.50\\[-1.5pt]{\tiny[.00,.35]}\end{tabular} & \begin{tabular}[t]{@{}c@{}}0.00\\[-1.5pt]{\tiny[.00,.45]}\end{tabular} & \begin{tabular}[t]{@{}c@{}}0.10\\[-1.5pt]{\tiny[.00,.35]}\end{tabular} & \begin{tabular}[t]{@{}c@{}}0.15\\[-1.5pt]{\tiny[.00,.40]}\end{tabular} & \begin{tabular}[t]{@{}c@{}}0.20\\[-1.5pt]{\tiny[.00,.40]}\end{tabular} & \begin{tabular}[t]{@{}c@{}}0.25\\[-1.5pt]{\tiny[.00,.70]}\end{tabular} & \begin{tabular}[t]{@{}c@{}}0.25\\[-1.5pt]{\tiny[.00,.90]}\end{tabular} & \begin{tabular}[t]{@{}c@{}}0.30\\[-1.5pt]{\tiny[.00,.85]}\end{tabular} & \begin{tabular}[t]{@{}c@{}}0.35\\[-1.5pt]{\tiny[.15,.90]}\end{tabular} \\
 & $n{=}200$ & \begin{tabular}[t]{@{}c@{}}0.25\\[-1.5pt]{\tiny[.00,.20]}\end{tabular} & \begin{tabular}[t]{@{}c@{}}0.40\\[-1.5pt]{\tiny[.05,.20]}\end{tabular} & \begin{tabular}[t]{@{}c@{}}0.45\\[-1.5pt]{\tiny[.05,.30]}\end{tabular} & \begin{tabular}[t]{@{}c@{}}0.50\\[-1.5pt]{\tiny[.10,.35]}\end{tabular} & \begin{tabular}[t]{@{}c@{}}0.50\\[-1.5pt]{\tiny[.15,.35]}\end{tabular} & \begin{tabular}[t]{@{}c@{}}0.55\\[-1.5pt]{\tiny[.15,.40]}\end{tabular} & \begin{tabular}[t]{@{}c@{}}0.20\\[-1.5pt]{\tiny[.00,.15]}\end{tabular} & \begin{tabular}[t]{@{}c@{}}0.20\\[-1.5pt]{\tiny[.00,.15]}\end{tabular} & \begin{tabular}[t]{@{}c@{}}0.25\\[-1.5pt]{\tiny[.00,.80]}\end{tabular} & \begin{tabular}[t]{@{}c@{}}0.35\\[-1.5pt]{\tiny[.15,.90]}\end{tabular} \\
 & $n{=}400$ & \begin{tabular}[t]{@{}c@{}}0.00\\[-1.5pt]{\tiny[.00,.10]}\end{tabular} & \begin{tabular}[t]{@{}c@{}}0.25\\[-1.5pt]{\tiny[.10,.15]}\end{tabular} & \begin{tabular}[t]{@{}c@{}}0.25\\[-1.5pt]{\tiny[.15,.20]}\end{tabular} & \begin{tabular}[t]{@{}c@{}}0.20\\[-1.5pt]{\tiny[.10,.20]}\end{tabular} & \begin{tabular}[t]{@{}c@{}}0.20\\[-1.5pt]{\tiny[.15,.25]}\end{tabular} & \begin{tabular}[t]{@{}c@{}}0.25\\[-1.5pt]{\tiny[.20,.25]}\end{tabular} & \begin{tabular}[t]{@{}c@{}}0.00\\[-1.5pt]{\tiny[.00,.10]}\end{tabular} & \begin{tabular}[t]{@{}c@{}}0.00\\[-1.5pt]{\tiny[.00,.10]}\end{tabular} & \begin{tabular}[t]{@{}c@{}}0.35\\[-1.5pt]{\tiny[.00,.10]}\end{tabular} & \begin{tabular}[t]{@{}c@{}}0.40\\[-1.5pt]{\tiny[.00,.45]}\end{tabular} \\
\addlinespace
SSJ--ATW, blocks $K{=}20$ & $n{=}50$ & \begin{tabular}[t]{@{}c@{}}0.30\\[-1.5pt]{\tiny[.00,.60]}\end{tabular} & \begin{tabular}[t]{@{}c@{}}0.45\\[-1.5pt]{\tiny[.00,.90]}\end{tabular} & \begin{tabular}[t]{@{}c@{}}0.50\\[-1.5pt]{\tiny[.00,1.00]}\end{tabular} & \begin{tabular}[t]{@{}c@{}}0.55\\[-1.5pt]{\tiny[.00,.95]}\end{tabular} & \begin{tabular}[t]{@{}c@{}}0.60\\[-1.5pt]{\tiny[.10,1.00]}\end{tabular} & \begin{tabular}[t]{@{}c@{}}0.60\\[-1.5pt]{\tiny[.20,1.00]}\end{tabular} & \begin{tabular}[t]{@{}c@{}}0.30\\[-1.5pt]{\tiny[.05,.70]}\end{tabular} & \begin{tabular}[t]{@{}c@{}}0.35\\[-1.5pt]{\tiny[.15,1.00]}\end{tabular} & \begin{tabular}[t]{@{}c@{}}0.35\\[-1.5pt]{\tiny[.15,.85]}\end{tabular} & \begin{tabular}[t]{@{}c@{}}0.40\\[-1.5pt]{\tiny[.15,1.00]}\end{tabular} \\
 & $n{=}100$ & \begin{tabular}[t]{@{}c@{}}0.25\\[-1.5pt]{\tiny[.00,.60]}\end{tabular} & \begin{tabular}[t]{@{}c@{}}0.35\\[-1.5pt]{\tiny[.00,.30]}\end{tabular} & \begin{tabular}[t]{@{}c@{}}0.00\\[-1.5pt]{\tiny[.05,.40]}\end{tabular} & \begin{tabular}[t]{@{}c@{}}0.00\\[-1.5pt]{\tiny[.00,.50]}\end{tabular} & \begin{tabular}[t]{@{}c@{}}0.15\\[-1.5pt]{\tiny[.00,.35]}\end{tabular} & \begin{tabular}[t]{@{}c@{}}0.20\\[-1.5pt]{\tiny[.00,.35]}\end{tabular} & \begin{tabular}[t]{@{}c@{}}0.25\\[-1.5pt]{\tiny[.00,.70]}\end{tabular} & \begin{tabular}[t]{@{}c@{}}0.25\\[-1.5pt]{\tiny[.00,.75]}\end{tabular} & \begin{tabular}[t]{@{}c@{}}0.30\\[-1.5pt]{\tiny[.00,.85]}\end{tabular} & \begin{tabular}[t]{@{}c@{}}0.35\\[-1.5pt]{\tiny[.15,1.00]}\end{tabular} \\
 & $n{=}200$ & \begin{tabular}[t]{@{}c@{}}0.20\\[-1.5pt]{\tiny[.00,.20]}\end{tabular} & \begin{tabular}[t]{@{}c@{}}0.05\\[-1.5pt]{\tiny[.00,.20]}\end{tabular} & \begin{tabular}[t]{@{}c@{}}0.10\\[-1.5pt]{\tiny[.05,.30]}\end{tabular} & \begin{tabular}[t]{@{}c@{}}0.45\\[-1.5pt]{\tiny[.15,.35]}\end{tabular} & \begin{tabular}[t]{@{}c@{}}0.55\\[-1.5pt]{\tiny[.15,.35]}\end{tabular} & \begin{tabular}[t]{@{}c@{}}0.70\\[-1.5pt]{\tiny[.15,.40]}\end{tabular} & \begin{tabular}[t]{@{}c@{}}0.20\\[-1.5pt]{\tiny[.00,.15]}\end{tabular} & \begin{tabular}[t]{@{}c@{}}0.20\\[-1.5pt]{\tiny[.05,.85]}\end{tabular} & \begin{tabular}[t]{@{}c@{}}0.20\\[-1.5pt]{\tiny[.00,.80]}\end{tabular} & \begin{tabular}[t]{@{}c@{}}0.30\\[-1.5pt]{\tiny[.15,.90]}\end{tabular} \\
 & $n{=}400$ & \begin{tabular}[t]{@{}c@{}}0.25\\[-1.5pt]{\tiny[.00,.10]}\end{tabular} & \begin{tabular}[t]{@{}c@{}}0.40\\[-1.5pt]{\tiny[.00,.15]}\end{tabular} & \begin{tabular}[t]{@{}c@{}}0.40\\[-1.5pt]{\tiny[.05,.20]}\end{tabular} & \begin{tabular}[t]{@{}c@{}}0.50\\[-1.5pt]{\tiny[.10,.25]}\end{tabular} & \begin{tabular}[t]{@{}c@{}}0.20\\[-1.5pt]{\tiny[.15,.30]}\end{tabular} & \begin{tabular}[t]{@{}c@{}}0.25\\[-1.5pt]{\tiny[.15,.35]}\end{tabular} & \begin{tabular}[t]{@{}c@{}}0.00\\[-1.5pt]{\tiny[.00,.15]}\end{tabular} & \begin{tabular}[t]{@{}c@{}}0.25\\[-1.5pt]{\tiny[.00,.15]}\end{tabular} & \begin{tabular}[t]{@{}c@{}}0.30\\[-1.5pt]{\tiny[.00,.15]}\end{tabular} & \begin{tabular}[t]{@{}c@{}}0.40\\[-1.5pt]{\tiny[.00,.25]}\end{tabular} \\
\bottomrule
\end{tabular}
\end{table}

The two departures act on that pattern in opposite ways.

\noindent The stochastic trend leaves the baseline pattern untouched: at every level of $\omega$ the value falls with $n$ as in the reference column, and the rows of the right-hand block of Table~\ref{tab:byn_scen} are nearly indistinguishable from one another (i.e., SSJ--ATW with blocks $0.35$, $0.25$, $0.15$, $0.10$ in the reference case and $0.25$, $0.20$, $0.15$, $0.10$ at $\omega=1$; SSJ--KFS at the maximum likelihood $\lambda$ $0.20$, $0.20$, $0.15$, $0.10$ and $0.15$, $0.10$, $0.15$, $0.15$), and the rows at intermediate $\omega$ differ from these by at most one grid step, save at $n=50$ for SSJ--ATW at the maximum likelihood $\lambda$, where the value drifts down from $0.40$ to $0.25$ across the range. The departure that moves the count criterion a long way moves this one nowhere at any sample size. The one exception is again SSJ--MLE, which shifts upward at $\omega=1$, to $0.45$, $0.45$, $0.40$, $0.35$ against $0.30$, $0.30$, $0.25$, $0.25$, while still falling with the sample size.

\noindent The dependence reverses that pattern. Under strong persistence the fall of the reconstruction optimum turns into a rise: a longer series supplies more observations but not proportionally more independent information, so the penalty has to increase with $n$ to compensate. Under independence the calibrated value falls with the length of the series, for the reason Section~\ref{sec:charge} makes precise (i.e., the plain BIC term charges more per discontinuity as $n$ grows and less is left for the extended term to supply): SSJ--ATW with contiguous blocks runs $0.35$, $0.25$, $0.15$, $0.10$ over $n=50$, $100$, $200$, $400$ at $\psi=0$. At $\psi=0.4$ the fall is already shallower, $0.40$, $0.40$, $0.30$, $0.30$, and at $\psi=0.9$ it has turned into a rise from $n=100$ onward, $0.35$, $0.40$, $0.50$, the entry at $n=50$ being the floor of the grid; SSJ--KFS with contiguous blocks goes the same way, from $0.20$, $0.20$, $0.20$, $0.15$ at $\psi=0$ to $0.30$, $0.30$, $0.40$, $0.55$ at $\psi=0.9$. With $\lambda$ from maximum likelihood the rise is weaker, and for SSJ--ATW it is absent at $\psi=0.9$, where the estimator has lost most of its power. Thus, a longer series supplies more observations but not proportionally more independent information, so the penalty has to increase to compensate rather than fall as the charge argument alone would predict. The two effects work against each other, and which of them dominates is a property of the data rather than of the estimator.

\subsection{What the EBIC charges per discontinuity}\label{sec:charge}
Two of the results above look like anomalies and are not. Every value calibrated in this document sits below the nominal consistency floor of about $0.50$ recalled in Section~\ref{sec:xi_what}, and at the baseline the value that minimizes reconstruction error is not a constant but falls with the length of the series. Both dissolve once one asks what the criterion charges per discontinuity in total, rather than what $\xi$ alone charges.

Because $\mathrm{df}$ grows with the number of active warps, the plain BIC term of Equation~\eqref{eq:ebic_app} already charges each declared discontinuity before the extended term is added. Writing $s$ for the increase in $\mathrm{df}$ that one activated warp produces (small for SSJ--MLE, larger for SSJ--KFS, largest for SSJ--ATW) the total charge per discontinuity is
\begin{equation}\label{eq:charge}
  \Phi(n,\xi) \;=\; s\,\log(n) \;+\; 2\,\xi\,\log(n-1),
\end{equation}
and if what the data ask for is a charge $C$ that does not depend on $n$, then the constant that delivers it is
\begin{equation}\label{eq:xirule}
  \xi(n) \;=\; \frac{C - s\log(n)}{2\log(n-1)} .
\end{equation}

\noindent Whether the optimal charge is in fact flat in $n$ is a testable statement: for each arm the reconstruction-optimal $\xi$ is found cell by cell over the $36$ cells with $q\ge1$, turned into a charge by Equation~\eqref{eq:charge}, and regressed on $\log n$; the charge is taken as flat when the slope is not distinguishable from zero at the $10\%$ level. It holds for SSJ--KFS with $\lambda$ from maximum likelihood and from contiguous blocks at $K=20$, the two arms of Table~\ref{tab:fixed} for which it is tested, and for SSJ--KFS with random folds at $K=5$; with random folds at $K=10$ and $K=20$ and with blocks at $K=5$ and $K=10$ the slope is positive, as it is for SSJ--MLE, and for the SSJ--ATW arms it is negative, that is, the charge these arms ask for falls with $n$ faster than the plain BIC term supplies it, except with contiguous blocks at $K=5$ and $K=20$, where it is not distinguishable from zero. Where the charge is flat, Equation~\eqref{eq:xirule} tracks the observed optima closely at every sample size. Table~\ref{tab:rule_all} sets the rule against the observed optima for every arm whose charge is flat, and Table~\ref{tab:rule} gives it for the two SSJ--KFS arms of the table of adopted values, at the sample sizes of the design and beyond them.

\begin{table}[htbp]
\centering\footnotesize
\caption{The rule of Equation~\eqref{eq:xirule} for SSJ--KFS, with $\lambda$ from maximum likelihood and from contiguous blocks at $K=20$: $\hat C$ is the mean optimal charge over the $36$ cells with $q\ge1$, and the columns give $(\hat C - s\log n)/\{2\log(n-1)\}$ at the four sample sizes of the design and at two beyond it.}
\label{tab:rule}
\setlength{\tabcolsep}{4pt}
\begin{tabular}{l c *{6}{c}}
\toprule
Arm & $\hat C$ & $n{=}50$ & $n{=}100$ & $n{=}200$ & $n{=}400$ & $n{=}800$ & $n{=}1600$ \\
\midrule
SSJ--KFS, MLE & 2.05 & 0.20 & 0.16 & 0.13 & 0.11 & 0.09 & 0.07 \\
SSJ--KFS, blocks $K{=}20$ & 2.09 & 0.20 & 0.16 & 0.13 & 0.11 & 0.09 & 0.08 \\
\bottomrule
\end{tabular}
\end{table}

\begin{table}[htbp]
\centering\footnotesize
\caption{The rule of Equation~\eqref{eq:xirule} for the arms whose optimal charge is flat in $n$: $\hat C$ is the mean charge over the $36$ cells, ``observed'' the reconstruction-optimal $\xi$ averaged over the nine cells of that sample size, and ``rule'' the value $(\hat C - s\log n)/\{2\log(n-1)\}$, also at two sample sizes beyond the design.}
\label{tab:rule_all}
\setlength{\tabcolsep}{4pt}
\begin{tabular}{l c l *{6}{c}}
\toprule
Arm & $\hat C$ & & $n{=}50$ & $n{=}100$ & $n{=}200$ & $n{=}400$ & $n{=}800$ & $n{=}1600$ \\
\midrule
\multirow{2}{*}{SSJ--KFS, MLE} & \multirow{2}{*}{2.05} & observed & 0.211 & 0.167 & 0.122 & 0.100 & -- & -- \\
 &  & rule & 0.199 & 0.159 & 0.130 & 0.107 & 0.089 & 0.075 \\
\addlinespace
\multirow{2}{*}{SSJ--KFS, random $K{=}5$} & \multirow{2}{*}{2.16} & observed & 0.183 & 0.172 & 0.144 & 0.128 & -- & -- \\
 &  & rule & 0.212 & 0.170 & 0.139 & 0.115 & 0.097 & 0.082 \\
\addlinespace
\multirow{2}{*}{SSJ--KFS, blocks $K{=}20$} & \multirow{2}{*}{2.09} & observed & 0.183 & 0.172 & 0.133 & 0.117 & -- & -- \\
 &  & rule & 0.204 & 0.163 & 0.133 & 0.110 & 0.092 & 0.077 \\
\addlinespace
\multirow{2}{*}{SSJ--ATW, blocks $K{=}5$} & \multirow{2}{*}{3.76} & observed & 0.206 & 0.339 & 0.189 & 0.106 & -- & -- \\
 &  & rule & 0.304 & 0.230 & 0.177 & 0.136 & 0.103 & 0.077 \\
\addlinespace
\multirow{2}{*}{SSJ--ATW, blocks $K{=}20$} & \multirow{2}{*}{3.70} & observed & 0.339 & 0.239 & 0.150 & 0.111 & -- & -- \\
 &  & rule & 0.297 & 0.224 & 0.171 & 0.131 & 0.099 & 0.073 \\
\bottomrule
\end{tabular}
\end{table}

Four things follow. First, the consistency floor is not violated: what the theory of \citet{CheChe2008} constrains is the total charge per discontinuity, and the extended term supplies only what the $\mathrm{df}\log n$ term has not already levied, so a calibrated $\xi$ well below $0.50$ is consistent with a floor stated for $\xi$ alone because part of the charge is paid elsewhere. On the $\xi$ scale that part is largest for SSJ--ATW, whose effective constant comes close to the floor, and small for SSJ--KFS and SSJ--MLE; what the decomposition leaves is finite-sample, the floor being an asymptotic requirement for selection consistency while the calibration targets risk at the sample sizes of the design. Second, the fall of $\xi$ with $n$ at the baseline is not drift but the arithmetic of Equation~\eqref{eq:xirule}: holding the charge fixed while $\log n$ grows requires exactly that. Third, the common charge can be read as follows: on the $-2\log L$ scale, SSJ--ATW ends up charging each discontinuity about what the Akaike criterion charges for one free parameter, and SSJ--KFS about half of that; the difference tracks $s$, since SSJ--ATW's degrees of freedom respond most strongly to an activated warp and more of its total charge is therefore already levied by the BIC term. Fourth, the reversal under persistence is the argument's own premise failing rather than a contradiction of it: a fixed charge buys a fixed protection only if a longer series carries proportionally more information, and persistent errors are precisely the case in which it does not, so the charge, and with it $\xi$, has to rise with $n$ instead.

In numbers. Expressed on the $\xi$ scale, the implicit charge $s\log n/\{2\log(n-1)\}$ is $0.04$ for SSJ--MLE ($s=0.07$), $0.07$ for SSJ--KFS ($s=0.13$) and $0.18$ for SSJ--ATW ($s=0.36$) at $n=100$, so the effective constant is about $0.43$ for SSJ--ATW at its reference value of $0.25$, against the floor of $0.50$, but only $0.22$ for SSJ--KFS at $0.15$ and $0.24$ to $0.34$ for SSJ--MLE at $0.20$ to $0.30$. The common charge is a reading of its own: it is about $2.1$ for SSJ--KFS and, at the reference sample sizes, about $3.7$ to $3.8$ for SSJ--ATW; on the $-2\log L$ scale a charge of $4$ is two log-likelihood units per discontinuity, which is what the Akaike criterion charges for one free parameter, so SSJ--ATW's calibrated penalty is close to charging each discontinuity as a single parameter and SSJ--KFS's about half of that. And at $\psi=0.9$ the realized lag-one autocorrelation is about $0.87$, so consecutive observations share most of their information.

For practice this means that where $n$ is far from the middle of the range studied here, Equation~\eqref{eq:xirule} should be preferred to any single constant, and that a comparison between the criterion and cross-validation which holds $\xi$ fixed across sample sizes understates the criterion.

\subsection{What the count rule costs in reconstruction}\label{sec:cost}
Where the two rules disagree, the reconstruction rule is the less demanding one, because the RMSE is nearly flat in $\xi$ near its minimum; following the count rule therefore costs little in reconstruction almost everywhere, and Table~\ref{tab:cost} reports what it costs, as the increase in mean RMSE at the count-calibrated $\xi$ relative to the reconstruction-calibrated one. The cost is at most about $1\%$ in the reference case and $1.5\%$ under serial dependence, for every estimator. It is material in one place only, a signal close to a stochastic trend with $\lambda$ inherited from SSJ--MLE, where it reaches $9\%$ for SSJ--KFS and $20\%$ for SSJ--ATW; with $\lambda$ cross-validated on contiguous blocks the same estimators pay $1\%$ and $3\%$. That is the one setting in which the two objectives of the paper genuinely conflict, and it does so because at $\omega=1$ the reconstruction rule wants a smaller $\xi$ than in the reference case while the count rule wants a larger one; the table shows the price of resolving the conflict in favor of the count.

\begin{table}[htbp]
\centering\footnotesize
\caption{The reconstruction cost of the count rule: the increase in mean RMSE at the count-calibrated $\xi$ relative to the reconstruction-calibrated one, pooled over the $48$ cells, in \%, as a range over the arms of each estimator. A negative value means the count value does marginally better on the mean, which the minimax rule does not optimize.}
\label{tab:cost}
\setlength{\tabcolsep}{4pt}
\begin{tabular}{l ccc}
\toprule
 & \multicolumn{3}{c}{increase in mean RMSE at the count value, \%} \\
\cmidrule(lr){2-4}
Character of the series & SSJ--KFS & SSJ--ATW & SSJ--MLE \\
\midrule
Smooth signal, independent errors & $0.0$ & $-0.1$--$0.1$ & $1.1$ \\
Dependent errors, blocked folds & $-0.1$--$1.4$ & $0.0$--$1.6$ & $0.1$--$1.1$ \\
Dependent errors, random folds & $0.1$--$0.8$ & $-0.0$--$0.9$ & $0.1$--$1.1$ \\
Stochastic-trend signal & $0.0$--$9.2$ & $1.6$--$20.2$ & $-0.2$ \\
\bottomrule
\end{tabular}
\end{table}

\subsection{Behavior by the number of true discontinuities}\label{sec:byq}
Table~\ref{tab:byq} shows, for seven representative arms, the mean number of discontinuities declared, the share of replications that get the number exactly right, and beneath them the median and the central half of the replication distribution, separately at $q=0$, $1$, $3$ and $5$. Each arm is read at its own calibrated $\xi$ from the count criterion, printed in the third column of the table.

\begin{table}[htbp]
\centering\footnotesize
\caption{Mean number of discontinuities declared and, in round brackets, the share of replications declaring exactly the right number; beneath, in small type, the median of the same replication distribution and its central half. Each arm is read at its own calibrated $\xi$ under the count rule, printed in the third column. Five settings: the baseline, two levels of the persistence and two of the stochastic-trend weight. $2{,}400$ replications behind each entry.}
\label{tab:byq}
\setlength{\tabcolsep}{4pt}
\begin{tabular}{llc*{4}{c}}
\toprule
Estimator & $\lambda$ from & $\xi$ & \multicolumn{4}{c}{true number of discontinuities $q$} \\
\cmidrule(lr){4-7}
 &  &  & $0$ & $1$ & $3$ & $5$ \\
\midrule
\multicolumn{7}{l}{\emph{Scenario 1: $\psi=\omega=0$}}\\
SSJ--MLE & MLE & $0.20$ & \begin{tabular}[t]{@{}c@{}}0.41 {\scriptsize(94\%)}\\[-1.5pt]{\tiny 0 [0,0]}\end{tabular} & \begin{tabular}[t]{@{}c@{}}2.12 {\scriptsize(3\%)}\\[-1.5pt]{\tiny 0 [0,4]}\end{tabular} & \begin{tabular}[t]{@{}c@{}}3.54 {\scriptsize(18\%)}\\[-1.5pt]{\tiny 4 [0,5]}\end{tabular} & \begin{tabular}[t]{@{}c@{}}4.43 {\scriptsize(27\%)}\\[-1.5pt]{\tiny 5 [4,6]}\end{tabular} \\
SSJ--KFS & MLE & $0.15$ & \begin{tabular}[t]{@{}c@{}}0.28 {\scriptsize(95\%)}\\[-1.5pt]{\tiny 0 [0,0]}\end{tabular} & \begin{tabular}[t]{@{}c@{}}2.00 {\scriptsize(1\%)}\\[-1.5pt]{\tiny 0 [0,4]}\end{tabular} & \begin{tabular}[t]{@{}c@{}}3.49 {\scriptsize(16\%)}\\[-1.5pt]{\tiny 4 [0,5]}\end{tabular} & \begin{tabular}[t]{@{}c@{}}4.42 {\scriptsize(30\%)}\\[-1.5pt]{\tiny 5 [4,6]}\end{tabular} \\
SSJ--KFS & random $K{=}20$ & $0.20$ & \begin{tabular}[t]{@{}c@{}}0.47 {\scriptsize(93\%)}\\[-1.5pt]{\tiny 0 [0,0]}\end{tabular} & \begin{tabular}[t]{@{}c@{}}1.82 {\scriptsize(2\%)}\\[-1.5pt]{\tiny 0 [0,4]}\end{tabular} & \begin{tabular}[t]{@{}c@{}}2.95 {\scriptsize(17\%)}\\[-1.5pt]{\tiny 3 [0,5]}\end{tabular} & \begin{tabular}[t]{@{}c@{}}3.90 {\scriptsize(27\%)}\\[-1.5pt]{\tiny 5 [3,6]}\end{tabular} \\
SSJ--KFS & blocks $K{=}20$ & $0.15$ & \begin{tabular}[t]{@{}c@{}}0.71 {\scriptsize(89\%)}\\[-1.5pt]{\tiny 0 [0,0]}\end{tabular} & \begin{tabular}[t]{@{}c@{}}2.27 {\scriptsize(2\%)}\\[-1.5pt]{\tiny 0 [0,5]}\end{tabular} & \begin{tabular}[t]{@{}c@{}}3.52 {\scriptsize(16\%)}\\[-1.5pt]{\tiny 4 [0,5]}\end{tabular} & \begin{tabular}[t]{@{}c@{}}4.32 {\scriptsize(29\%)}\\[-1.5pt]{\tiny 5 [4,6]}\end{tabular} \\
SSJ--ATW & MLE & $0.25$ & \begin{tabular}[t]{@{}c@{}}0.64 {\scriptsize(88\%)}\\[-1.5pt]{\tiny 0 [0,0]}\end{tabular} & \begin{tabular}[t]{@{}c@{}}1.62 {\scriptsize(1\%)}\\[-1.5pt]{\tiny 0 [0,3]}\end{tabular} & \begin{tabular}[t]{@{}c@{}}3.09 {\scriptsize(17\%)}\\[-1.5pt]{\tiny 3 [0,5]}\end{tabular} & \begin{tabular}[t]{@{}c@{}}4.25 {\scriptsize(24\%)}\\[-1.5pt]{\tiny 5 [3,6]}\end{tabular} \\
SSJ--ATW & random $K{=}20$ & $0.30$ & \begin{tabular}[t]{@{}c@{}}0.37 {\scriptsize(92\%)}\\[-1.5pt]{\tiny 0 [0,0]}\end{tabular} & \begin{tabular}[t]{@{}c@{}}1.23 {\scriptsize(2\%)}\\[-1.5pt]{\tiny 0 [0,2]}\end{tabular} & \begin{tabular}[t]{@{}c@{}}2.54 {\scriptsize(16\%)}\\[-1.5pt]{\tiny 3 [0,4]}\end{tabular} & \begin{tabular}[t]{@{}c@{}}3.70 {\scriptsize(23\%)}\\[-1.5pt]{\tiny 4 [0,5]}\end{tabular} \\
SSJ--ATW & blocks $K{=}20$ & $0.25$ & \begin{tabular}[t]{@{}c@{}}0.69 {\scriptsize(86\%)}\\[-1.5pt]{\tiny 0 [0,0]}\end{tabular} & \begin{tabular}[t]{@{}c@{}}1.71 {\scriptsize(2\%)}\\[-1.5pt]{\tiny 0 [0,4]}\end{tabular} & \begin{tabular}[t]{@{}c@{}}2.99 {\scriptsize(16\%)}\\[-1.5pt]{\tiny 3 [0,5]}\end{tabular} & \begin{tabular}[t]{@{}c@{}}4.07 {\scriptsize(23\%)}\\[-1.5pt]{\tiny 5 [3,6]}\end{tabular} \\
\midrule
\multicolumn{7}{l}{\emph{Scenario 2: $\psi=0.6$}}\\
SSJ--MLE & MLE & $0.15$ & \begin{tabular}[t]{@{}c@{}}1.37 {\scriptsize(64\%)}\\[-1.5pt]{\tiny 0 [0,2]}\end{tabular} & \begin{tabular}[t]{@{}c@{}}1.69 {\scriptsize(11\%)}\\[-1.5pt]{\tiny 0 [0,3]}\end{tabular} & \begin{tabular}[t]{@{}c@{}}2.00 {\scriptsize(12\%)}\\[-1.5pt]{\tiny 1 [0,4]}\end{tabular} & \begin{tabular}[t]{@{}c@{}}2.17 {\scriptsize(12\%)}\\[-1.5pt]{\tiny 0 [0,4]}\end{tabular} \\
SSJ--KFS & MLE & $0.15$ & \begin{tabular}[t]{@{}c@{}}1.84 {\scriptsize(68\%)}\\[-1.5pt]{\tiny 0 [0,4]}\end{tabular} & \begin{tabular}[t]{@{}c@{}}2.84 {\scriptsize(3\%)}\\[-1.5pt]{\tiny 2 [0,5]}\end{tabular} & \begin{tabular}[t]{@{}c@{}}3.51 {\scriptsize(14\%)}\\[-1.5pt]{\tiny 4 [0,6]}\end{tabular} & \begin{tabular}[t]{@{}c@{}}4.28 {\scriptsize(25\%)}\\[-1.5pt]{\tiny 5 [3,6]}\end{tabular} \\
SSJ--KFS & random $K{=}20$ & $0.15$ & \begin{tabular}[t]{@{}c@{}}1.74 {\scriptsize(70\%)}\\[-1.5pt]{\tiny 0 [0,4]}\end{tabular} & \begin{tabular}[t]{@{}c@{}}2.50 {\scriptsize(3\%)}\\[-1.5pt]{\tiny 1 [0,5]}\end{tabular} & \begin{tabular}[t]{@{}c@{}}3.34 {\scriptsize(14\%)}\\[-1.5pt]{\tiny 4 [0,5]}\end{tabular} & \begin{tabular}[t]{@{}c@{}}4.00 {\scriptsize(25\%)}\\[-1.5pt]{\tiny 5 [2,6]}\end{tabular} \\
SSJ--KFS & blocks $K{=}20$ & $0.30$ & \begin{tabular}[t]{@{}c@{}}0.78 {\scriptsize(85\%)}\\[-1.5pt]{\tiny 0 [0,0]}\end{tabular} & \begin{tabular}[t]{@{}c@{}}2.00 {\scriptsize(2\%)}\\[-1.5pt]{\tiny 0 [0,4]}\end{tabular} & \begin{tabular}[t]{@{}c@{}}2.99 {\scriptsize(17\%)}\\[-1.5pt]{\tiny 3 [0,5]}\end{tabular} & \begin{tabular}[t]{@{}c@{}}3.89 {\scriptsize(25\%)}\\[-1.5pt]{\tiny 4 [3,5]}\end{tabular} \\
SSJ--ATW & MLE & $0.40$ & \begin{tabular}[t]{@{}c@{}}0.98 {\scriptsize(76\%)}\\[-1.5pt]{\tiny 0 [0,0]}\end{tabular} & \begin{tabular}[t]{@{}c@{}}1.75 {\scriptsize(3\%)}\\[-1.5pt]{\tiny 0 [0,4]}\end{tabular} & \begin{tabular}[t]{@{}c@{}}2.60 {\scriptsize(13\%)}\\[-1.5pt]{\tiny 3 [0,5]}\end{tabular} & \begin{tabular}[t]{@{}c@{}}3.34 {\scriptsize(20\%)}\\[-1.5pt]{\tiny 4 [0,5]}\end{tabular} \\
SSJ--ATW & random $K{=}20$ & $0.30$ & \begin{tabular}[t]{@{}c@{}}1.26 {\scriptsize(68\%)}\\[-1.5pt]{\tiny 0 [0,2]}\end{tabular} & \begin{tabular}[t]{@{}c@{}}2.04 {\scriptsize(7\%)}\\[-1.5pt]{\tiny 1 [0,4]}\end{tabular} & \begin{tabular}[t]{@{}c@{}}2.81 {\scriptsize(15\%)}\\[-1.5pt]{\tiny 3 [0,5]}\end{tabular} & \begin{tabular}[t]{@{}c@{}}3.42 {\scriptsize(21\%)}\\[-1.5pt]{\tiny 4 [1,5]}\end{tabular} \\
SSJ--ATW & blocks $K{=}20$ & $0.45$ & \begin{tabular}[t]{@{}c@{}}0.85 {\scriptsize(78\%)}\\[-1.5pt]{\tiny 0 [0,0]}\end{tabular} & \begin{tabular}[t]{@{}c@{}}1.71 {\scriptsize(4\%)}\\[-1.5pt]{\tiny 0 [0,4]}\end{tabular} & \begin{tabular}[t]{@{}c@{}}2.93 {\scriptsize(16\%)}\\[-1.5pt]{\tiny 3 [0,5]}\end{tabular} & \begin{tabular}[t]{@{}c@{}}3.90 {\scriptsize(23\%)}\\[-1.5pt]{\tiny 4 [3,5]}\end{tabular} \\
\midrule
\multicolumn{7}{l}{\emph{Scenario 2: $\psi=0.9$}}\\
SSJ--MLE & MLE & $0.20$ & \begin{tabular}[t]{@{}c@{}}0.37 {\scriptsize(88\%)}\\[-1.5pt]{\tiny 0 [0,0]}\end{tabular} & \begin{tabular}[t]{@{}c@{}}0.94 {\scriptsize(17\%)}\\[-1.5pt]{\tiny 0 [0,1]}\end{tabular} & \begin{tabular}[t]{@{}c@{}}1.80 {\scriptsize(23\%)}\\[-1.5pt]{\tiny 2 [0,3]}\end{tabular} & \begin{tabular}[t]{@{}c@{}}2.28 {\scriptsize(17\%)}\\[-1.5pt]{\tiny 2 [0,4]}\end{tabular} \\
SSJ--KFS & MLE & $0.20$ & \begin{tabular}[t]{@{}c@{}}0.47 {\scriptsize(90\%)}\\[-1.5pt]{\tiny 0 [0,0]}\end{tabular} & \begin{tabular}[t]{@{}c@{}}1.86 {\scriptsize(8\%)}\\[-1.5pt]{\tiny 1 [0,3]}\end{tabular} & \begin{tabular}[t]{@{}c@{}}3.07 {\scriptsize(33\%)}\\[-1.5pt]{\tiny 3 [2,4]}\end{tabular} & \begin{tabular}[t]{@{}c@{}}4.16 {\scriptsize(37\%)}\\[-1.5pt]{\tiny 5 [4,5]}\end{tabular} \\
SSJ--KFS & random $K{=}20$ & $0.25$ & \begin{tabular}[t]{@{}c@{}}0.20 {\scriptsize(96\%)}\\[-1.5pt]{\tiny 0 [0,0]}\end{tabular} & \begin{tabular}[t]{@{}c@{}}1.49 {\scriptsize(9\%)}\\[-1.5pt]{\tiny 0 [0,3]}\end{tabular} & \begin{tabular}[t]{@{}c@{}}2.93 {\scriptsize(33\%)}\\[-1.5pt]{\tiny 3 [2,4]}\end{tabular} & \begin{tabular}[t]{@{}c@{}}4.14 {\scriptsize(37\%)}\\[-1.5pt]{\tiny 5 [4,5]}\end{tabular} \\
SSJ--KFS & blocks $K{=}20$ & $0.25$ & \begin{tabular}[t]{@{}c@{}}0.53 {\scriptsize(87\%)}\\[-1.5pt]{\tiny 0 [0,0]}\end{tabular} & \begin{tabular}[t]{@{}c@{}}2.10 {\scriptsize(14\%)}\\[-1.5pt]{\tiny 2 [0,4]}\end{tabular} & \begin{tabular}[t]{@{}c@{}}3.26 {\scriptsize(36\%)}\\[-1.5pt]{\tiny 3 [3,4]}\end{tabular} & \begin{tabular}[t]{@{}c@{}}4.38 {\scriptsize(36\%)}\\[-1.5pt]{\tiny 5 [4,5]}\end{tabular} \\
SSJ--ATW & MLE & $0.45$ & \begin{tabular}[t]{@{}c@{}}0.86 {\scriptsize(76\%)}\\[-1.5pt]{\tiny 0 [0,0]}\end{tabular} & \begin{tabular}[t]{@{}c@{}}1.67 {\scriptsize(9\%)}\\[-1.5pt]{\tiny 0 [0,3]}\end{tabular} & \begin{tabular}[t]{@{}c@{}}2.73 {\scriptsize(26\%)}\\[-1.5pt]{\tiny 3 [0,4]}\end{tabular} & \begin{tabular}[t]{@{}c@{}}3.74 {\scriptsize(28\%)}\\[-1.5pt]{\tiny 4 [3,5]}\end{tabular} \\
SSJ--ATW & random $K{=}20$ & $0.40$ & \begin{tabular}[t]{@{}c@{}}0.91 {\scriptsize(74\%)}\\[-1.5pt]{\tiny 0 [0,1]}\end{tabular} & \begin{tabular}[t]{@{}c@{}}1.80 {\scriptsize(13\%)}\\[-1.5pt]{\tiny 1 [0,3]}\end{tabular} & \begin{tabular}[t]{@{}c@{}}3.04 {\scriptsize(30\%)}\\[-1.5pt]{\tiny 3 [2,4]}\end{tabular} & \begin{tabular}[t]{@{}c@{}}4.11 {\scriptsize(35\%)}\\[-1.5pt]{\tiny 5 [3,5]}\end{tabular} \\
SSJ--ATW & blocks $K{=}20$ & $0.55$ & \begin{tabular}[t]{@{}c@{}}0.67 {\scriptsize(81\%)}\\[-1.5pt]{\tiny 0 [0,0]}\end{tabular} & \begin{tabular}[t]{@{}c@{}}1.61 {\scriptsize(9\%)}\\[-1.5pt]{\tiny 0 [0,3]}\end{tabular} & \begin{tabular}[t]{@{}c@{}}2.96 {\scriptsize(30\%)}\\[-1.5pt]{\tiny 3 [2,4]}\end{tabular} & \begin{tabular}[t]{@{}c@{}}4.11 {\scriptsize(32\%)}\\[-1.5pt]{\tiny 4 [3,5]}\end{tabular} \\
\bottomrule
\end{tabular}
\end{table}
\begin{table}[htbp]
\addtocounter{table}{-1}
\centering\footnotesize
\caption{(continued)}
\setlength{\tabcolsep}{4pt}
\begin{tabular}{llc*{4}{c}}
\toprule
Estimator & $\lambda$ from & $\xi$ & \multicolumn{4}{c}{true number of discontinuities $q$} \\
\cmidrule(lr){4-7}
 &  &  & $0$ & $1$ & $3$ & $5$ \\
\midrule
\multicolumn{7}{l}{\emph{Scenario 3: $\omega=0.5$}}\\
SSJ--MLE & MLE & $0.25$ & \begin{tabular}[t]{@{}c@{}}0.38 {\scriptsize(94\%)}\\[-1.5pt]{\tiny 0 [0,0]}\end{tabular} & \begin{tabular}[t]{@{}c@{}}1.73 {\scriptsize(2\%)}\\[-1.5pt]{\tiny 0 [0,4]}\end{tabular} & \begin{tabular}[t]{@{}c@{}}2.98 {\scriptsize(17\%)}\\[-1.5pt]{\tiny 3 [0,5]}\end{tabular} & \begin{tabular}[t]{@{}c@{}}3.92 {\scriptsize(26\%)}\\[-1.5pt]{\tiny 5 [0,6]}\end{tabular} \\
SSJ--KFS & MLE & $0.20$ & \begin{tabular}[t]{@{}c@{}}0.27 {\scriptsize(96\%)}\\[-1.5pt]{\tiny 0 [0,0]}\end{tabular} & \begin{tabular}[t]{@{}c@{}}1.74 {\scriptsize(1\%)}\\[-1.5pt]{\tiny 0 [0,4]}\end{tabular} & \begin{tabular}[t]{@{}c@{}}3.04 {\scriptsize(19\%)}\\[-1.5pt]{\tiny 3 [0,5]}\end{tabular} & \begin{tabular}[t]{@{}c@{}}4.06 {\scriptsize(28\%)}\\[-1.5pt]{\tiny 5 [3,6]}\end{tabular} \\
SSJ--KFS & random $K{=}20$ & $0.20$ & \begin{tabular}[t]{@{}c@{}}0.56 {\scriptsize(91\%)}\\[-1.5pt]{\tiny 0 [0,0]}\end{tabular} & \begin{tabular}[t]{@{}c@{}}1.98 {\scriptsize(2\%)}\\[-1.5pt]{\tiny 0 [0,4]}\end{tabular} & \begin{tabular}[t]{@{}c@{}}3.04 {\scriptsize(18\%)}\\[-1.5pt]{\tiny 3 [0,5]}\end{tabular} & \begin{tabular}[t]{@{}c@{}}3.98 {\scriptsize(27\%)}\\[-1.5pt]{\tiny 5 [3,6]}\end{tabular} \\
SSJ--KFS & blocks $K{=}20$ & $0.15$ & \begin{tabular}[t]{@{}c@{}}1.00 {\scriptsize(84\%)}\\[-1.5pt]{\tiny 0 [0,0]}\end{tabular} & \begin{tabular}[t]{@{}c@{}}2.50 {\scriptsize(2\%)}\\[-1.5pt]{\tiny 2 [0,5]}\end{tabular} & \begin{tabular}[t]{@{}c@{}}3.60 {\scriptsize(18\%)}\\[-1.5pt]{\tiny 4 [2,5]}\end{tabular} & \begin{tabular}[t]{@{}c@{}}4.40 {\scriptsize(28\%)}\\[-1.5pt]{\tiny 5 [4,6]}\end{tabular} \\
SSJ--ATW & MLE & $0.30$ & \begin{tabular}[t]{@{}c@{}}0.44 {\scriptsize(91\%)}\\[-1.5pt]{\tiny 0 [0,0]}\end{tabular} & \begin{tabular}[t]{@{}c@{}}1.42 {\scriptsize(2\%)}\\[-1.5pt]{\tiny 0 [0,3]}\end{tabular} & \begin{tabular}[t]{@{}c@{}}2.72 {\scriptsize(18\%)}\\[-1.5pt]{\tiny 3 [0,5]}\end{tabular} & \begin{tabular}[t]{@{}c@{}}3.87 {\scriptsize(26\%)}\\[-1.5pt]{\tiny 5 [0,6]}\end{tabular} \\
SSJ--ATW & random $K{=}20$ & $0.30$ & \begin{tabular}[t]{@{}c@{}}0.43 {\scriptsize(91\%)}\\[-1.5pt]{\tiny 0 [0,0]}\end{tabular} & \begin{tabular}[t]{@{}c@{}}1.39 {\scriptsize(2\%)}\\[-1.5pt]{\tiny 0 [0,3]}\end{tabular} & \begin{tabular}[t]{@{}c@{}}2.68 {\scriptsize(18\%)}\\[-1.5pt]{\tiny 3 [0,4]}\end{tabular} & \begin{tabular}[t]{@{}c@{}}3.70 {\scriptsize(25\%)}\\[-1.5pt]{\tiny 4 [0,5]}\end{tabular} \\
SSJ--ATW & blocks $K{=}20$ & $0.30$ & \begin{tabular}[t]{@{}c@{}}0.46 {\scriptsize(90\%)}\\[-1.5pt]{\tiny 0 [0,0]}\end{tabular} & \begin{tabular}[t]{@{}c@{}}1.37 {\scriptsize(2\%)}\\[-1.5pt]{\tiny 0 [0,3]}\end{tabular} & \begin{tabular}[t]{@{}c@{}}2.60 {\scriptsize(18\%)}\\[-1.5pt]{\tiny 3 [0,4]}\end{tabular} & \begin{tabular}[t]{@{}c@{}}3.62 {\scriptsize(23\%)}\\[-1.5pt]{\tiny 4 [0,5]}\end{tabular} \\
\midrule
\multicolumn{7}{l}{\emph{Scenario 3: $\omega=1$}}\\
SSJ--MLE & MLE & $0.40$ & \begin{tabular}[t]{@{}c@{}}0.94 {\scriptsize(87\%)}\\[-1.5pt]{\tiny 0 [0,0]}\end{tabular} & \begin{tabular}[t]{@{}c@{}}2.28 {\scriptsize(4\%)}\\[-1.5pt]{\tiny 1 [0,4]}\end{tabular} & \begin{tabular}[t]{@{}c@{}}3.26 {\scriptsize(20\%)}\\[-1.5pt]{\tiny 3 [0,5]}\end{tabular} & \begin{tabular}[t]{@{}c@{}}4.03 {\scriptsize(30\%)}\\[-1.5pt]{\tiny 5 [3,6]}\end{tabular} \\
SSJ--KFS & MLE & $0.45$ & \begin{tabular}[t]{@{}c@{}}2.19 {\scriptsize(68\%)}\\[-1.5pt]{\tiny 0 [0,5]}\end{tabular} & \begin{tabular}[t]{@{}c@{}}2.39 {\scriptsize(7\%)}\\[-1.5pt]{\tiny 1 [0,4]}\end{tabular} & \begin{tabular}[t]{@{}c@{}}2.91 {\scriptsize(24\%)}\\[-1.5pt]{\tiny 3 [0,4]}\end{tabular} & \begin{tabular}[t]{@{}c@{}}3.73 {\scriptsize(27\%)}\\[-1.5pt]{\tiny 4 [2,5]}\end{tabular} \\
SSJ--KFS & random $K{=}20$ & $0.20$ & \begin{tabular}[t]{@{}c@{}}1.42 {\scriptsize(78\%)}\\[-1.5pt]{\tiny 0 [0,0]}\end{tabular} & \begin{tabular}[t]{@{}c@{}}2.74 {\scriptsize(5\%)}\\[-1.5pt]{\tiny 3 [0,5]}\end{tabular} & \begin{tabular}[t]{@{}c@{}}3.68 {\scriptsize(23\%)}\\[-1.5pt]{\tiny 4 [3,5]}\end{tabular} & \begin{tabular}[t]{@{}c@{}}4.53 {\scriptsize(33\%)}\\[-1.5pt]{\tiny 5 [4,6]}\end{tabular} \\
SSJ--KFS & blocks $K{=}20$ & $0.25$ & \begin{tabular}[t]{@{}c@{}}1.00 {\scriptsize(84\%)}\\[-1.5pt]{\tiny 0 [0,0]}\end{tabular} & \begin{tabular}[t]{@{}c@{}}2.39 {\scriptsize(5\%)}\\[-1.5pt]{\tiny 2 [0,4]}\end{tabular} & \begin{tabular}[t]{@{}c@{}}3.34 {\scriptsize(23\%)}\\[-1.5pt]{\tiny 4 [2,5]}\end{tabular} & \begin{tabular}[t]{@{}c@{}}4.24 {\scriptsize(30\%)}\\[-1.5pt]{\tiny 5 [3,6]}\end{tabular} \\
SSJ--ATW & MLE & $0.50$ & \begin{tabular}[t]{@{}c@{}}1.46 {\scriptsize(74\%)}\\[-1.5pt]{\tiny 0 [0,1]}\end{tabular} & \begin{tabular}[t]{@{}c@{}}1.89 {\scriptsize(8\%)}\\[-1.5pt]{\tiny 0 [0,3]}\end{tabular} & \begin{tabular}[t]{@{}c@{}}2.83 {\scriptsize(24\%)}\\[-1.5pt]{\tiny 3 [0,4]}\end{tabular} & \begin{tabular}[t]{@{}c@{}}3.79 {\scriptsize(26\%)}\\[-1.5pt]{\tiny 4 [3,5]}\end{tabular} \\
SSJ--ATW & random $K{=}20$ & $0.40$ & \begin{tabular}[t]{@{}c@{}}0.37 {\scriptsize(92\%)}\\[-1.5pt]{\tiny 0 [0,0]}\end{tabular} & \begin{tabular}[t]{@{}c@{}}1.47 {\scriptsize(8\%)}\\[-1.5pt]{\tiny 0 [0,3]}\end{tabular} & \begin{tabular}[t]{@{}c@{}}2.87 {\scriptsize(26\%)}\\[-1.5pt]{\tiny 3 [0,4]}\end{tabular} & \begin{tabular}[t]{@{}c@{}}3.91 {\scriptsize(27\%)}\\[-1.5pt]{\tiny 4 [3,5]}\end{tabular} \\
SSJ--ATW & blocks $K{=}20$ & $0.35$ & \begin{tabular}[t]{@{}c@{}}0.50 {\scriptsize(89\%)}\\[-1.5pt]{\tiny 0 [0,0]}\end{tabular} & \begin{tabular}[t]{@{}c@{}}1.78 {\scriptsize(7\%)}\\[-1.5pt]{\tiny 1 [0,3]}\end{tabular} & \begin{tabular}[t]{@{}c@{}}3.13 {\scriptsize(24\%)}\\[-1.5pt]{\tiny 3 [1,5]}\end{tabular} & \begin{tabular}[t]{@{}c@{}}4.14 {\scriptsize(27\%)}\\[-1.5pt]{\tiny 5 [3,6]}\end{tabular} \\
\bottomrule
\end{tabular}
\end{table}

Some considerations follow:
\begin{itemize}
\item \textbf{The declared count is close to all-or-nothing.} With a single true discontinuity, more than half of the replications (i.e., between 52\% and 71\% of replications) declare nothing at all, and those that declare anything declare several ($4.3$ to $4.8$ on average). The mean in the table averages those two branches and describes neither, which is why the median at $q=1$ is zero while the central half reaches four. The shape is the same at every $q$: what the truth changes is the probability of the empty branch, which falls steadily from $q=0$ to $q=5$, while the count declared on the non-empty branch stays near five throughout. The information about $q$ is therefore carried almost entirely by whether anything is declared, and hardly at all by how many. The declaration rule of Section~\ref{P-sec:tuning} of the paper explains the shape: a warp is kept only if it clears both a share of the total warping and an absolute fraction of the mean sampling interval, and a fitted warping sequence that clears the absolute threshold at all usually clears it in several places at once.

\item \textbf{Exact recovery rises with $q$.} It is negligible at $q=1$ and appreciable at $q=5$ (i.e., it runs 1\% to 3\% at $q=1$ against 23\% to 30\% at $q=5$), not because the estimators count better when there is more to find but because the non-empty branch lands near five whatever the truth: it is right when the truth is five and wrong when it is one. A declared count of one or two is correspondingly rare. Reading the declared count as an estimate of $q$ is therefore a mistake, and reading a declared count of zero as evidence that there is nothing to find is the same mistake in the other direction.

\item \textbf{Persistence costs SSJ--MLE its power.}
By $\psi=0.6$ SSJ--MLE declares $1.69$, $2.00$ and $2.17$ against $1$, $3$ and $5$, and the empty branch has taken over, at 49\% of replications at $q=3$ and 51\% at $q=5$ against 26\% and 15\% at the baseline. At $\psi=0.9$ it declares $0.94$, $1.80$ and $2.28$. The loss is not a threshold but a slide that begins at the first level of the persistence: at $q=5$ the arm declares $4.43$, $3.50$, $3.02$, $2.53$, $2.00$ and $2.03$ as $\psi$ runs from $0$ to $0.5$, and never recovers (Table~\ref{tab:byq_psi}). SSJ--KFS with contiguous blocks is barely touched at the strongest persistence: it declares $2.10$, $3.26$ and $4.38$ at $\psi=0.9$, with the empty branch at 36\%, 13\% and 5\%, and its exact recovery at $q=3$ actually rises, from 16\% to 36\%, because the persistence damps the over-declaration that costs it accuracy under independence.

\item \textbf{The false discoveries have a hump in the persistence.} On null series SSJ--KFS at the maximum likelihood $\lambda$ declares $0.28$ at $\psi=0$, rises to $1.84$ at $\psi=0.6$, where it is right only 68\% of the time against 95\% at the baseline, and falls back to $0.47$ at $\psi=0.9$; SSJ--MLE peaks at the same level, at $1.37$, and SSJ--KFS with blocks later and lower, at $1.19$ at $\psi=0.7$ (Table~\ref{tab:byq_psi}). The cause is in the calibration itself: the count criterion balances over- against under-declaration, and at the levels where an arm begins to lose power on the series that contain discontinuities it answers by lowering $\xi$, from $0.25$ at $\psi=0.3$ to $0.15$ at $\psi=0.6$ for SSJ--KFS at the maximum likelihood $\lambda$, which is paid for on the null series. A criterion calibrated on the count alone is therefore not a criterion for specificity.

\item \textbf{The stochastic trend costs specificity, and the fold layout buys it back.} At $\omega=1$ SSJ--KFS at the maximum likelihood $\lambda$ declares discontinuities on a sizeable share of the series that have none ($2.19$ discontinuities and is right only 68\% of the time), while the same estimator with contiguous blocks stays close to its baseline ($1.00$ and is right 84\%); SSJ--ATW shows the same contrast ($1.46$ against $0.50$). Where the signal wanders, how $\lambda$ is chosen matters more for the false-discovery rate than the penalty does.

\item \textbf{Half the departure is not half the damage.} At $\omega=0.5$ every arm is still close to its baseline at $q=0$ (i.e., 84\% to 96\% correct against 86\% to 95\%), and the complete grid shows the same at $\omega=0.75$ (i.e., 88\% to 94\% at $\omega=0.75$, Table~\ref{tab:byq_omega}): the specificity loss belongs entirely to the last step of the range, so a series that is partly trend-like needs no correction. Persistence is not like that: it costs SSJ--MLE its power from the first level and has taken half of it by $\psi=0.4$, so a practitioner who judges the departure to be moderate is safe in Scenario 3 and is not safe in Scenario 2.
\end{itemize}

\begin{figure}[p]\centering
\includegraphics[width=0.94\linewidth]{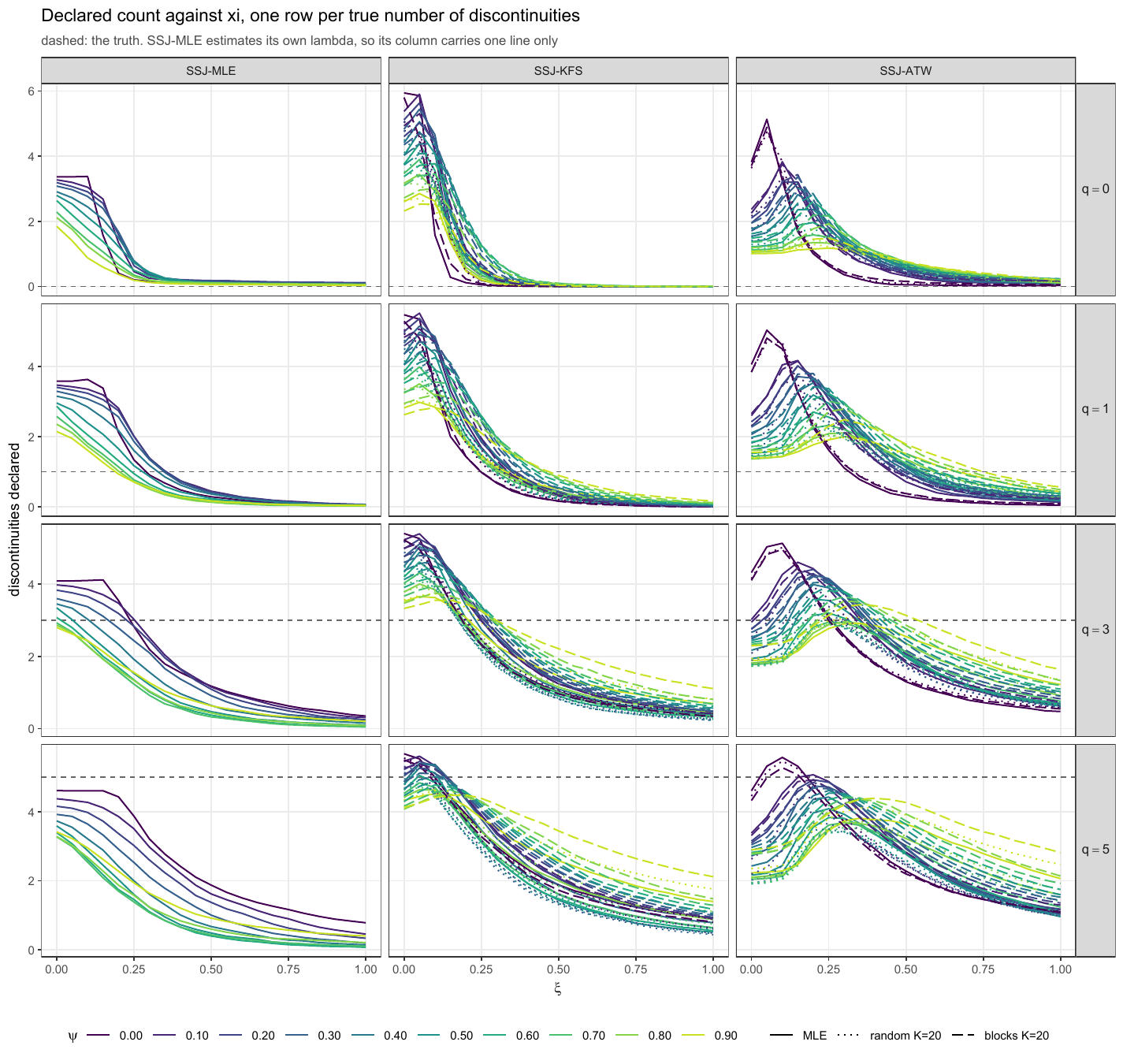}
\caption{Scenario 2. Mean number of discontinuities declared against $\xi$, by the true number $q$. Color is the persistence $\psi$ and line type the source of $\lambda$. Dashed rules mark the truth. Reading down a column shows the progression: at $q=0$ every curve falls to zero and the only question is how soon; at $q=1$ every curve crosses the truth from above, so the signed error has a zero for every arm; at $q=3$ the zero survives only for some; at $q=5$ the SSJ--MLE curve never reaches the line at any $\xi$ once $\psi$ is large, which is the power loss that leaves the pooled criterion without a target.}
\label{fig:byq_err}
\end{figure}

\begin{figure}[p]\centering
\includegraphics[width=0.94\linewidth]{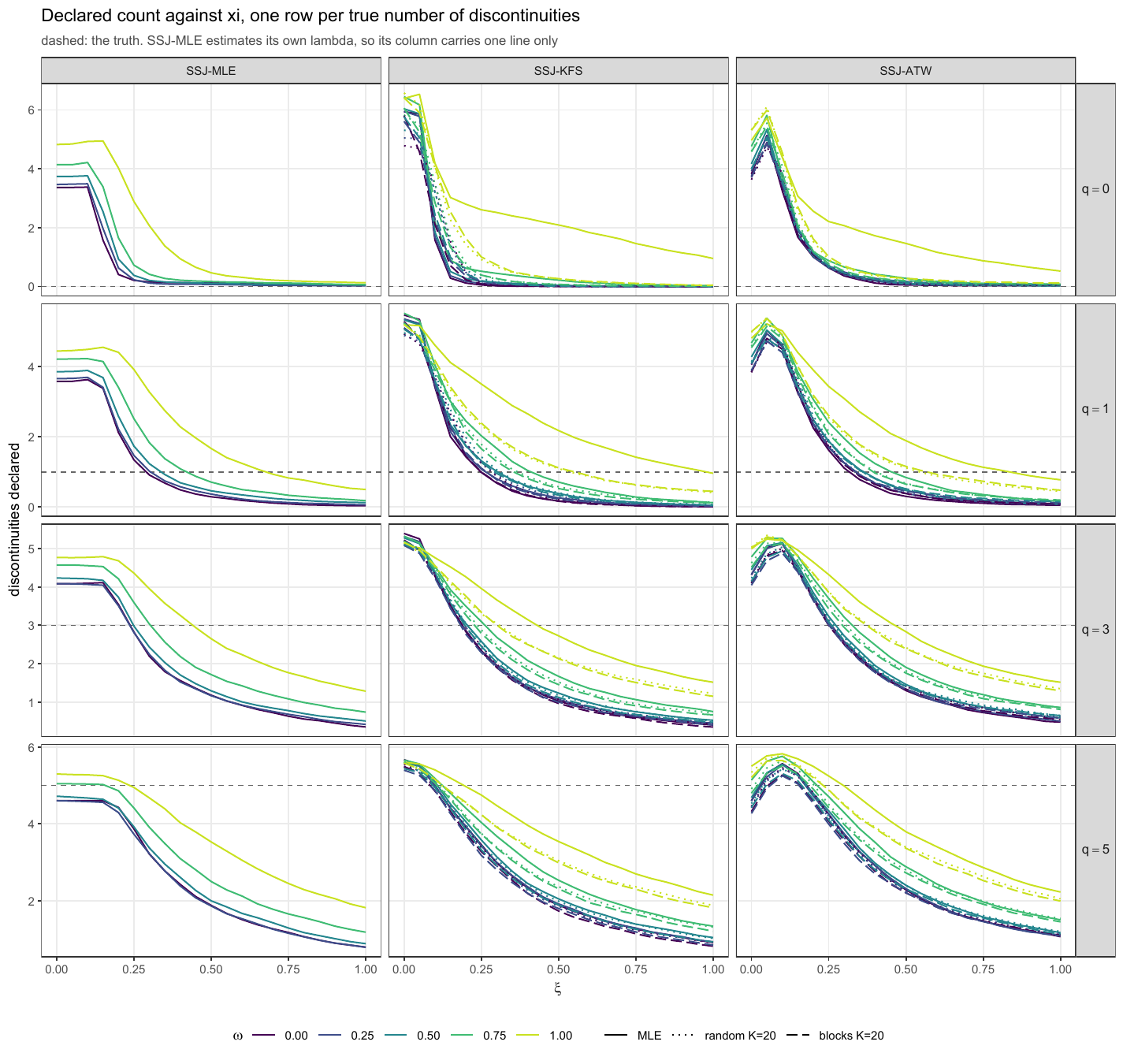}
\caption{Scenario 3, the same against $\omega$. The contrast with Figure~\ref{fig:byq_err} is in the top row: there the curves separate by $\psi$ only at small $\xi$, whereas here the $\omega=1$ curve stays above the rest well past $\xi=0.5$, which is the inflation of the count on a signal that has no discontinuities at all.}
\label{fig:byq_sig}
\end{figure}

Tables~\ref{tab:byq_psi} and \ref{tab:byq_omega} give the whole grid rather than seven settings, so that the level at which each arm turns can be located rather than bracketed, and Figures~\ref{fig:byq_err} and \ref{fig:byq_sig} draw the declared count against $\xi$ rather than at one value of it, one row per true number of discontinuities, so that the opposite directions of the error at $q=1$ and $q=5$, which the criteria of Section~\ref{sec:xi_rules} average away, are visible side by side.

\begin{sidewaystable}[p]
\centering\footnotesize
\caption{Scenario 2, every level of the persistence. Mean number of discontinuities declared and, in brackets, the share of replications declaring exactly the right number, for seven representative arms at each true count $q$ and each $\psi$. Each entry is read at that arm's own calibrated $\xi$ for that level of $\psi$, so a row is not a single penalty held fixed.}
\label{tab:byq_psi}
\setlength{\tabcolsep}{2pt}
\begin{tabular}{ll*{10}{c}}
\toprule
Estimator & & \multicolumn{10}{c}{$\psi$} \\
\cmidrule(lr){3-12}
 &  & 0.00 & 0.10 & 0.20 & 0.30 & 0.40 & 0.50 & 0.60 & 0.70 & 0.80 & 0.90 \\
\midrule
SSJ--MLE, MLE & $q{=}0$ & 0.41 {\scriptsize(94\%)} & 0.46 {\scriptsize(92\%)} & 0.64 {\scriptsize(89\%)} & 0.76 {\scriptsize(86\%)} & 0.79 {\scriptsize(84\%)} & 1.23 {\scriptsize(69\%)} & 1.37 {\scriptsize(64\%)} & 1.06 {\scriptsize(71\%)} & 0.89 {\scriptsize(75\%)} & 0.37 {\scriptsize(88\%)} \\
 & $q{=}1$ & 2.12 {\scriptsize(3\%)} & 2.02 {\scriptsize(2\%)} & 2.01 {\scriptsize(4\%)} & 1.86 {\scriptsize(5\%)} & 1.61 {\scriptsize(6\%)} & 1.66 {\scriptsize(9\%)} & 1.69 {\scriptsize(11\%)} & 1.50 {\scriptsize(11\%)} & 1.39 {\scriptsize(13\%)} & 0.94 {\scriptsize(17\%)} \\
 & $q{=}3$ & 3.54 {\scriptsize(18\%)} & 3.02 {\scriptsize(14\%)} & 2.83 {\scriptsize(14\%)} & 2.43 {\scriptsize(12\%)} & 1.91 {\scriptsize(10\%)} & 1.91 {\scriptsize(10\%)} & 2.00 {\scriptsize(12\%)} & 1.92 {\scriptsize(13\%)} & 1.92 {\scriptsize(16\%)} & 1.80 {\scriptsize(23\%)} \\
 & $q{=}5$ & 4.43 {\scriptsize(27\%)} & 3.50 {\scriptsize(20\%)} & 3.02 {\scriptsize(15\%)} & 2.53 {\scriptsize(11\%)} & 2.00 {\scriptsize(8\%)} & 2.03 {\scriptsize(10\%)} & 2.17 {\scriptsize(12\%)} & 2.09 {\scriptsize(11\%)} & 2.28 {\scriptsize(14\%)} & 2.28 {\scriptsize(17\%)} \\
\addlinespace
SSJ--KFS, MLE & $q{=}0$ & 0.28 {\scriptsize(95\%)} & 0.18 {\scriptsize(97\%)} & 0.33 {\scriptsize(94\%)} & 0.45 {\scriptsize(93\%)} & 1.14 {\scriptsize(81\%)} & 1.03 {\scriptsize(82\%)} & 1.84 {\scriptsize(68\%)} & 1.60 {\scriptsize(71\%)} & 1.55 {\scriptsize(71\%)} & 0.47 {\scriptsize(90\%)} \\
 & $q{=}1$ & 2.00 {\scriptsize(1\%)} & 1.69 {\scriptsize(1\%)} & 1.85 {\scriptsize(1\%)} & 1.95 {\scriptsize(1\%)} & 2.55 {\scriptsize(2\%)} & 2.22 {\scriptsize(2\%)} & 2.84 {\scriptsize(3\%)} & 2.62 {\scriptsize(3\%)} & 2.49 {\scriptsize(6\%)} & 1.86 {\scriptsize(8\%)} \\
 & $q{=}3$ & 3.49 {\scriptsize(16\%)} & 2.97 {\scriptsize(15\%)} & 3.07 {\scriptsize(13\%)} & 2.99 {\scriptsize(13\%)} & 3.41 {\scriptsize(12\%)} & 3.05 {\scriptsize(11\%)} & 3.51 {\scriptsize(14\%)} & 3.36 {\scriptsize(17\%)} & 3.39 {\scriptsize(23\%)} & 3.07 {\scriptsize(33\%)} \\
 & $q{=}5$ & 4.42 {\scriptsize(30\%)} & 3.86 {\scriptsize(25\%)} & 3.90 {\scriptsize(25\%)} & 3.78 {\scriptsize(23\%)} & 4.05 {\scriptsize(24\%)} & 3.81 {\scriptsize(21\%)} & 4.28 {\scriptsize(25\%)} & 4.21 {\scriptsize(26\%)} & 4.31 {\scriptsize(31\%)} & 4.16 {\scriptsize(37\%)} \\
\addlinespace
SSJ--KFS, random $K{=}20$ & $q{=}0$ & 0.47 {\scriptsize(93\%)} & 1.13 {\scriptsize(80\%)} & 0.99 {\scriptsize(83\%)} & 0.87 {\scriptsize(85\%)} & 1.78 {\scriptsize(70\%)} & 1.76 {\scriptsize(70\%)} & 1.74 {\scriptsize(70\%)} & 1.69 {\scriptsize(69\%)} & 0.76 {\scriptsize(86\%)} & 0.20 {\scriptsize(96\%)} \\
 & $q{=}1$ & 1.82 {\scriptsize(2\%)} & 2.23 {\scriptsize(4\%)} & 2.07 {\scriptsize(2\%)} & 1.94 {\scriptsize(1\%)} & 2.63 {\scriptsize(2\%)} & 2.57 {\scriptsize(2\%)} & 2.50 {\scriptsize(3\%)} & 2.54 {\scriptsize(4\%)} & 1.93 {\scriptsize(6\%)} & 1.49 {\scriptsize(9\%)} \\
 & $q{=}3$ & 2.95 {\scriptsize(17\%)} & 3.00 {\scriptsize(12\%)} & 2.82 {\scriptsize(10\%)} & 2.66 {\scriptsize(9\%)} & 3.29 {\scriptsize(10\%)} & 3.29 {\scriptsize(11\%)} & 3.34 {\scriptsize(14\%)} & 3.40 {\scriptsize(18\%)} & 3.03 {\scriptsize(22\%)} & 2.93 {\scriptsize(33\%)} \\
 & $q{=}5$ & 3.90 {\scriptsize(27\%)} & 3.70 {\scriptsize(22\%)} & 3.50 {\scriptsize(21\%)} & 3.33 {\scriptsize(20\%)} & 3.87 {\scriptsize(22\%)} & 3.89 {\scriptsize(23\%)} & 4.00 {\scriptsize(25\%)} & 4.19 {\scriptsize(28\%)} & 4.04 {\scriptsize(31\%)} & 4.14 {\scriptsize(37\%)} \\
\addlinespace
SSJ--KFS, blocks $K{=}20$ & $q{=}0$ & 0.71 {\scriptsize(89\%)} & 0.51 {\scriptsize(92\%)} & 0.69 {\scriptsize(89\%)} & 0.43 {\scriptsize(93\%)} & 0.57 {\scriptsize(90\%)} & 0.65 {\scriptsize(88\%)} & 0.78 {\scriptsize(85\%)} & 1.19 {\scriptsize(75\%)} & 0.87 {\scriptsize(80\%)} & 0.53 {\scriptsize(87\%)} \\
 & $q{=}1$ & 2.27 {\scriptsize(2\%)} & 1.96 {\scriptsize(2\%)} & 2.20 {\scriptsize(2\%)} & 1.72 {\scriptsize(1\%)} & 1.84 {\scriptsize(2\%)} & 1.97 {\scriptsize(2\%)} & 2.00 {\scriptsize(2\%)} & 2.46 {\scriptsize(5\%)} & 2.28 {\scriptsize(8\%)} & 2.10 {\scriptsize(14\%)} \\
 & $q{=}3$ & 3.52 {\scriptsize(16\%)} & 2.99 {\scriptsize(14\%)} & 3.16 {\scriptsize(14\%)} & 2.74 {\scriptsize(13\%)} & 2.85 {\scriptsize(14\%)} & 2.87 {\scriptsize(16\%)} & 2.99 {\scriptsize(17\%)} & 3.29 {\scriptsize(22\%)} & 3.25 {\scriptsize(26\%)} & 3.26 {\scriptsize(36\%)} \\
 & $q{=}5$ & 4.32 {\scriptsize(29\%)} & 3.89 {\scriptsize(24\%)} & 4.00 {\scriptsize(25\%)} & 3.61 {\scriptsize(24\%)} & 3.72 {\scriptsize(23\%)} & 3.76 {\scriptsize(24\%)} & 3.89 {\scriptsize(25\%)} & 4.29 {\scriptsize(27\%)} & 4.28 {\scriptsize(30\%)} & 4.38 {\scriptsize(36\%)} \\
\addlinespace
SSJ--ATW, MLE & $q{=}0$ & 0.64 {\scriptsize(88\%)} & 0.60 {\scriptsize(86\%)} & 0.69 {\scriptsize(83\%)} & 0.78 {\scriptsize(81\%)} & 0.86 {\scriptsize(79\%)} & 0.97 {\scriptsize(77\%)} & 0.98 {\scriptsize(76\%)} & 0.79 {\scriptsize(79\%)} & 0.96 {\scriptsize(75\%)} & 0.86 {\scriptsize(76\%)} \\
 & $q{=}1$ & 1.62 {\scriptsize(1\%)} & 1.33 {\scriptsize(1\%)} & 1.54 {\scriptsize(2\%)} & 1.71 {\scriptsize(2\%)} & 1.80 {\scriptsize(2\%)} & 1.78 {\scriptsize(2\%)} & 1.75 {\scriptsize(3\%)} & 1.49 {\scriptsize(3\%)} & 1.77 {\scriptsize(7\%)} & 1.67 {\scriptsize(9\%)} \\
 & $q{=}3$ & 3.09 {\scriptsize(17\%)} & 2.56 {\scriptsize(14\%)} & 2.77 {\scriptsize(13\%)} & 2.90 {\scriptsize(12\%)} & 2.79 {\scriptsize(12\%)} & 2.64 {\scriptsize(11\%)} & 2.60 {\scriptsize(13\%)} & 2.36 {\scriptsize(14\%)} & 2.77 {\scriptsize(18\%)} & 2.73 {\scriptsize(26\%)} \\
 & $q{=}5$ & 4.25 {\scriptsize(24\%)} & 3.64 {\scriptsize(22\%)} & 3.77 {\scriptsize(22\%)} & 3.76 {\scriptsize(21\%)} & 3.50 {\scriptsize(19\%)} & 3.35 {\scriptsize(18\%)} & 3.34 {\scriptsize(20\%)} & 3.22 {\scriptsize(19\%)} & 3.57 {\scriptsize(24\%)} & 3.74 {\scriptsize(28\%)} \\
\addlinespace
SSJ--ATW, random $K{=}20$ & $q{=}0$ & 0.37 {\scriptsize(92\%)} & 0.77 {\scriptsize(83\%)} & 0.85 {\scriptsize(81\%)} & 0.95 {\scriptsize(77\%)} & 1.01 {\scriptsize(75\%)} & 1.00 {\scriptsize(75\%)} & 1.26 {\scriptsize(68\%)} & 1.46 {\scriptsize(61\%)} & 1.19 {\scriptsize(67\%)} & 0.91 {\scriptsize(74\%)} \\
 & $q{=}1$ & 1.23 {\scriptsize(2\%)} & 1.66 {\scriptsize(2\%)} & 1.67 {\scriptsize(3\%)} & 1.68 {\scriptsize(3\%)} & 1.67 {\scriptsize(4\%)} & 1.75 {\scriptsize(4\%)} & 2.04 {\scriptsize(7\%)} & 2.07 {\scriptsize(13\%)} & 2.04 {\scriptsize(13\%)} & 1.80 {\scriptsize(13\%)} \\
 & $q{=}3$ & 2.54 {\scriptsize(16\%)} & 2.70 {\scriptsize(13\%)} & 2.60 {\scriptsize(12\%)} & 2.53 {\scriptsize(11\%)} & 2.55 {\scriptsize(10\%)} & 2.57 {\scriptsize(12\%)} & 2.81 {\scriptsize(15\%)} & 2.88 {\scriptsize(18\%)} & 2.92 {\scriptsize(23\%)} & 3.04 {\scriptsize(30\%)} \\
 & $q{=}5$ & 3.70 {\scriptsize(23\%)} & 3.44 {\scriptsize(19\%)} & 3.29 {\scriptsize(18\%)} & 3.23 {\scriptsize(18\%)} & 3.24 {\scriptsize(19\%)} & 3.25 {\scriptsize(19\%)} & 3.42 {\scriptsize(21\%)} & 3.44 {\scriptsize(21\%)} & 3.68 {\scriptsize(27\%)} & 4.11 {\scriptsize(35\%)} \\
\addlinespace
SSJ--ATW, blocks $K{=}20$ & $q{=}0$ & 0.69 {\scriptsize(86\%)} & 0.76 {\scriptsize(82\%)} & 0.93 {\scriptsize(79\%)} & 0.83 {\scriptsize(79\%)} & 0.92 {\scriptsize(78\%)} & 0.96 {\scriptsize(76\%)} & 0.85 {\scriptsize(78\%)} & 0.92 {\scriptsize(76\%)} & 0.75 {\scriptsize(79\%)} & 0.67 {\scriptsize(81\%)} \\
 & $q{=}1$ & 1.71 {\scriptsize(2\%)} & 1.67 {\scriptsize(2\%)} & 1.92 {\scriptsize(3\%)} & 1.71 {\scriptsize(3\%)} & 1.79 {\scriptsize(3\%)} & 1.92 {\scriptsize(4\%)} & 1.71 {\scriptsize(4\%)} & 1.87 {\scriptsize(5\%)} & 1.75 {\scriptsize(6\%)} & 1.61 {\scriptsize(9\%)} \\
 & $q{=}3$ & 2.99 {\scriptsize(16\%)} & 2.88 {\scriptsize(14\%)} & 3.11 {\scriptsize(16\%)} & 2.79 {\scriptsize(15\%)} & 3.00 {\scriptsize(16\%)} & 3.15 {\scriptsize(16\%)} & 2.93 {\scriptsize(16\%)} & 3.01 {\scriptsize(18\%)} & 2.86 {\scriptsize(22\%)} & 2.96 {\scriptsize(30\%)} \\
 & $q{=}5$ & 4.07 {\scriptsize(23\%)} & 3.83 {\scriptsize(23\%)} & 4.00 {\scriptsize(25\%)} & 3.78 {\scriptsize(23\%)} & 3.91 {\scriptsize(23\%)} & 4.04 {\scriptsize(24\%)} & 3.90 {\scriptsize(23\%)} & 4.00 {\scriptsize(25\%)} & 3.93 {\scriptsize(28\%)} & 4.11 {\scriptsize(32\%)} \\
\bottomrule
\end{tabular}
\end{sidewaystable}

\begin{table}[htbp]
\centering\footnotesize
\caption{Scenario 3, every level of the stochastic-trend weight. As Table~\ref{tab:byq_psi}, against $\omega$.}
\label{tab:byq_omega}
\setlength{\tabcolsep}{2pt}
\begin{tabular}{ll*{5}{c}}
\toprule
Estimator & & \multicolumn{5}{c}{$\omega$} \\
\cmidrule(lr){3-7}
 &  & 0.00 & 0.25 & 0.50 & 0.75 & 1.00 \\
\midrule
SSJ--MLE, MLE & $q{=}0$ & 0.41 {\scriptsize(94\%)} & 0.63 {\scriptsize(91\%)} & 0.38 {\scriptsize(94\%)} & 0.41 {\scriptsize(94\%)} & 0.94 {\scriptsize(87\%)} \\
 & $q{=}1$ & 2.12 {\scriptsize(3\%)} & 2.22 {\scriptsize(3\%)} & 1.73 {\scriptsize(2\%)} & 1.83 {\scriptsize(2\%)} & 2.28 {\scriptsize(4\%)} \\
 & $q{=}3$ & 3.54 {\scriptsize(18\%)} & 3.51 {\scriptsize(18\%)} & 2.98 {\scriptsize(17\%)} & 3.03 {\scriptsize(19\%)} & 3.26 {\scriptsize(20\%)} \\
 & $q{=}5$ & 4.43 {\scriptsize(27\%)} & 4.29 {\scriptsize(25\%)} & 3.92 {\scriptsize(26\%)} & 3.91 {\scriptsize(26\%)} & 4.03 {\scriptsize(30\%)} \\
\addlinespace
SSJ--KFS, MLE & $q{=}0$ & 0.28 {\scriptsize(95\%)} & 0.37 {\scriptsize(94\%)} & 0.27 {\scriptsize(96\%)} & 0.52 {\scriptsize(92\%)} & 2.19 {\scriptsize(68\%)} \\
 & $q{=}1$ & 2.00 {\scriptsize(1\%)} & 2.15 {\scriptsize(2\%)} & 1.74 {\scriptsize(1\%)} & 2.03 {\scriptsize(2\%)} & 2.39 {\scriptsize(7\%)} \\
 & $q{=}3$ & 3.49 {\scriptsize(16\%)} & 3.44 {\scriptsize(16\%)} & 3.04 {\scriptsize(19\%)} & 3.10 {\scriptsize(23\%)} & 2.91 {\scriptsize(24\%)} \\
 & $q{=}5$ & 4.42 {\scriptsize(30\%)} & 4.40 {\scriptsize(29\%)} & 4.06 {\scriptsize(28\%)} & 4.04 {\scriptsize(30\%)} & 3.73 {\scriptsize(27\%)} \\
\addlinespace
SSJ--KFS, random $K{=}20$ & $q{=}0$ & 0.47 {\scriptsize(93\%)} & 0.48 {\scriptsize(92\%)} & 0.56 {\scriptsize(91\%)} & 0.79 {\scriptsize(88\%)} & 1.42 {\scriptsize(78\%)} \\
 & $q{=}1$ & 1.82 {\scriptsize(2\%)} & 1.80 {\scriptsize(3\%)} & 1.98 {\scriptsize(2\%)} & 2.27 {\scriptsize(2\%)} & 2.74 {\scriptsize(5\%)} \\
 & $q{=}3$ & 2.95 {\scriptsize(17\%)} & 2.95 {\scriptsize(17\%)} & 3.04 {\scriptsize(18\%)} & 3.36 {\scriptsize(22\%)} & 3.68 {\scriptsize(23\%)} \\
 & $q{=}5$ & 3.90 {\scriptsize(27\%)} & 3.83 {\scriptsize(28\%)} & 3.98 {\scriptsize(27\%)} & 4.21 {\scriptsize(30\%)} & 4.53 {\scriptsize(33\%)} \\
\addlinespace
SSJ--KFS, blocks $K{=}20$ & $q{=}0$ & 0.71 {\scriptsize(89\%)} & 0.86 {\scriptsize(86\%)} & 1.00 {\scriptsize(84\%)} & 0.68 {\scriptsize(89\%)} & 1.00 {\scriptsize(84\%)} \\
 & $q{=}1$ & 2.27 {\scriptsize(2\%)} & 2.28 {\scriptsize(3\%)} & 2.50 {\scriptsize(2\%)} & 2.14 {\scriptsize(2\%)} & 2.39 {\scriptsize(5\%)} \\
 & $q{=}3$ & 3.52 {\scriptsize(16\%)} & 3.46 {\scriptsize(17\%)} & 3.60 {\scriptsize(18\%)} & 3.27 {\scriptsize(20\%)} & 3.34 {\scriptsize(23\%)} \\
 & $q{=}5$ & 4.32 {\scriptsize(29\%)} & 4.27 {\scriptsize(28\%)} & 4.40 {\scriptsize(28\%)} & 4.15 {\scriptsize(28\%)} & 4.24 {\scriptsize(30\%)} \\
\addlinespace
SSJ--ATW, MLE & $q{=}0$ & 0.64 {\scriptsize(88\%)} & 0.40 {\scriptsize(92\%)} & 0.44 {\scriptsize(91\%)} & 0.54 {\scriptsize(91\%)} & 1.46 {\scriptsize(74\%)} \\
 & $q{=}1$ & 1.62 {\scriptsize(1\%)} & 1.26 {\scriptsize(2\%)} & 1.42 {\scriptsize(2\%)} & 1.54 {\scriptsize(2\%)} & 1.89 {\scriptsize(8\%)} \\
 & $q{=}3$ & 3.09 {\scriptsize(17\%)} & 2.57 {\scriptsize(17\%)} & 2.72 {\scriptsize(18\%)} & 2.81 {\scriptsize(22\%)} & 2.83 {\scriptsize(24\%)} \\
 & $q{=}5$ & 4.25 {\scriptsize(24\%)} & 3.74 {\scriptsize(24\%)} & 3.87 {\scriptsize(26\%)} & 3.87 {\scriptsize(27\%)} & 3.79 {\scriptsize(26\%)} \\
\addlinespace
SSJ--ATW, random $K{=}20$ & $q{=}0$ & 0.37 {\scriptsize(92\%)} & 0.43 {\scriptsize(92\%)} & 0.43 {\scriptsize(91\%)} & 0.33 {\scriptsize(93\%)} & 0.37 {\scriptsize(92\%)} \\
 & $q{=}1$ & 1.23 {\scriptsize(2\%)} & 1.28 {\scriptsize(2\%)} & 1.39 {\scriptsize(2\%)} & 1.29 {\scriptsize(2\%)} & 1.47 {\scriptsize(8\%)} \\
 & $q{=}3$ & 2.54 {\scriptsize(16\%)} & 2.58 {\scriptsize(16\%)} & 2.68 {\scriptsize(18\%)} & 2.62 {\scriptsize(23\%)} & 2.87 {\scriptsize(26\%)} \\
 & $q{=}5$ & 3.70 {\scriptsize(23\%)} & 3.64 {\scriptsize(23\%)} & 3.70 {\scriptsize(25\%)} & 3.64 {\scriptsize(26\%)} & 3.91 {\scriptsize(27\%)} \\
\addlinespace
SSJ--ATW, blocks $K{=}20$ & $q{=}0$ & 0.69 {\scriptsize(86\%)} & 0.49 {\scriptsize(89\%)} & 0.46 {\scriptsize(90\%)} & 0.52 {\scriptsize(90\%)} & 0.50 {\scriptsize(89\%)} \\
 & $q{=}1$ & 1.71 {\scriptsize(2\%)} & 1.25 {\scriptsize(2\%)} & 1.37 {\scriptsize(2\%)} & 1.63 {\scriptsize(2\%)} & 1.78 {\scriptsize(7\%)} \\
 & $q{=}3$ & 2.99 {\scriptsize(16\%)} & 2.49 {\scriptsize(16\%)} & 2.60 {\scriptsize(18\%)} & 2.95 {\scriptsize(20\%)} & 3.13 {\scriptsize(24\%)} \\
 & $q{=}5$ & 4.07 {\scriptsize(23\%)} & 3.49 {\scriptsize(23\%)} & 3.62 {\scriptsize(23\%)} & 3.98 {\scriptsize(25\%)} & 4.14 {\scriptsize(27\%)} \\
\bottomrule
\end{tabular}
\end{table}

\clearpage

\section{Suggested calibrated values and concluding remarks}\label{sec:concl}
Under a departure the calibration runs over several levels of $\psi$ or $\omega$, and the series a single constant has to serve may sit at any of them. The two rules are therefore \emph{pooled} over the levels of a setting before a value is read off: the pooled rule is the minimax regret over the cells of every level at once, that is, the value of $\xi$ whose worst regret over the $4\times48$ cells (RMSE) or the $4\times36$ cells (count) with $\psi\ge0.6$ is smallest, so that it returns one value per arm and not a range over levels. In the reference case and at $\omega=1$ there is a single level and the pooled rule is the value of Tables~\ref{tab:base_all} and \ref{tab:xi_signal_uni}. Table~\ref{tab:reco} reports the range over the arms of the calibrated values by the character of the series and by estimator, under the two pooled rules.

\begin{table}[htbp]
\centering\small
\caption{Calibrated values of $\xi$ by the character of the series and by estimator, under minimax regret in RMSE and in the count error, each pooled over the levels of the setting. Each entry is the range over the ways of supplying $\lambda$; the second row uses contiguous cross-validation folds, SSJ--MLE entering it with its own $\lambda$. A series with both properties was not simulated: it is bracketed by the last two rows rather than resolved by either.}
\label{tab:reco}
\setlength{\tabcolsep}{1.75pt}
\begin{tabular}{l cc @{\hspace{2pt}} cc @{\hspace{2pt}} cc}
\toprule
 & \multicolumn{2}{c}{SSJ--KFS} & \multicolumn{2}{c}{SSJ--ATW} & \multicolumn{2}{c}{SSJ--MLE} \\
\cmidrule(lr){2-3}\cmidrule(lr){4-5}\cmidrule(lr){6-7}
Character of the series & RMSE & count & RMSE & count & RMSE & count \\
\midrule
Smooth signal, independent errors & $0.10$--$0.20$ & $0.15$--$0.20$ & $0.25$--$0.30$ & $0.25$--$0.30$ & $0.30$ & $0.20$ \\
Dependent errors, blocked folds & $0.30$--$0.40$ & $0.30$--$0.35$ & $0.35$--$0.45$ & $0.45$--$0.50$ & $0.30$ & $0.15$ \\
Dependent errors, random folds & $0.25$--$0.30$ & $0.15$--$0.20$ & $0.35$ & $0.30$--$0.35$ & $0.30$ & $0.15$ \\
Stochastic-trend signal & $0.15$--$0.20$ & $0.20$--$0.45$ & $0.15$--$0.25$ & $0.35$--$0.50$ & $0.45$ & $0.40$ \\
\bottomrule
\end{tabular}
\end{table}

\noindent Three regularities are read off Table~\ref{tab:reco}.

\noindent First, in the reference case the two rules coincide for SSJ--KFS and SSJ--ATW and the ranges over the ways of supplying $\lambda$ are at most two grid steps wide, so the reference calibration is settled: $0.15$--$0.20$ for SSJ--KFS, $0.25$--$0.30$ for SSJ--ATW, and $0.20$ or $0.30$ for SSJ--MLE according to the rule.

\noindent Second, under dependent errors both rules move upward with contiguous blocks ($0.30$--$0.40$ for SSJ--KFS and to $0.35$--$0.50$ for SSJ--ATW, that is, three to five grid steps above the reference case). With random folds the reconstruction rule moves as well (to $0.25$--$0.30$ and $0.35$), while the count rule stays at its reference level, because a constant chosen on random folds has already absorbed the persistence into the fit and no penalty on the count can recover it. SSJ--MLE moves under the reconstruction rule only, to $0.30$, its count rule falling to $0.15$ because the estimator loses power rather than specificity as the persistence grows.

\noindent Third, under a stochastic trend the two rules separate: the reconstruction rule stays at the reference values or below, while the count rule rises to $0.20$--$0.45$ for SSJ--KFS and $0.35$--$0.50$ for SSJ--ATW, the upper ends of both ranges being the arms with $\lambda$ inherited from SSJ--MLE. In every row SSJ--KFS lies below SSJ--ATW, so the constant is a property of the estimator as much as of the data.

\noindent To synthesize the interval values from Table~\ref{tab:reco} into a single value for each setting we consider a \emph{midpoint} between the value calibrated according to the RMSE minimax and the count minimax (i.e., the average of the two pooled values rounded up to the grid).

Table~\ref{tab:fixed} fixes, for each scenario and for each source of $\lambda$ (i.e., from SSJ--MLE or by cross-validation at $K=5$, $10$ and $20$, with folds drawn at random or as contiguous blocks), the midpoint of the two pooled rules; the lower panel reports the two rules themselves. The number of folds is not what the value depends on: between $K=5$ and $K=20$ it moves by at most one grid step in every cell, and $K=10$ lies within one step of either except with contiguous blocks under dependent errors. The paper reports $K=5$ and $K=20$, with the latter being the better choice With contiguous blocks since holding out a fifth of the series (i.e., $K=5$) in one stretch over-smooths (Section~\ref{sec:res_errors}).

Recall that the values are pooled over the sample sizes of the design and depend on $n$: Table~\ref{tab:byn_scen} gives the calibration at each sample size and at every level of the departures.

\begin{table}[htbp]
\centering\small
\caption{Calibrated values of $\xi$ by the character of the series and by estimator, under the two syntheses of the pooled rules: the midpoint, rounded up to the grid, and the larger of the two. Each entry is the range over the ways of supplying $\lambda$; the second row uses contiguous cross-validation folds, SSJ--MLE entering it with its own $\lambda$.}
\label{tab:reco_single}
\setlength{\tabcolsep}{1.75pt}
\begin{tabular}{l cc @{\hspace{2pt}} cc @{\hspace{2pt}} cc}
\toprule
 & \multicolumn{2}{c}{SSJ--KFS} & \multicolumn{2}{c}{SSJ--ATW} & \multicolumn{2}{c}{SSJ--MLE} \\
\cmidrule(lr){2-3}\cmidrule(lr){4-5}\cmidrule(lr){6-7}
Character of the series & midpoint & larger & midpoint & larger & midpoint & larger \\
\midrule
Smooth signal, independent errors & $0.15$--$0.20$ & $0.15$--$0.20$ & $0.25$--$0.30$ & $0.25$--$0.30$ & $0.25$ & $0.30$ \\
Dependent errors, blocked folds & $0.30$--$0.40$ & $0.30$--$0.40$ & $0.40$--$0.50$ & $0.45$--$0.50$ & $0.25$ & $0.30$ \\
Dependent errors, random folds & $0.20$--$0.25$ & $0.25$--$0.30$ & $0.35$ & $0.35$ & $0.25$ & $0.30$ \\
Stochastic-trend signal & $0.20$--$0.30$ & $0.20$--$0.45$ & $0.30$--$0.35$ & $0.35$--$0.50$ & $0.45$ & $0.45$ \\
\bottomrule
\end{tabular}
\end{table}

\begin{table}[htbp]
\centering\footnotesize
\caption{Calibrated values of $\xi$ for each scenario and arm. The upper panel reports the value adopted for each source of $\lambda$ (i.e., from SSJ--MLE or by cross-validation at $K=5$, $10$ and $20$, with folds drawn at random or as contiguous blocks): the midpoint of minimax regret in RMSE and in the count error, each pooled over the levels of the setting, rounded up to the grid. The values are pooled over the four sample sizes of the design and depend on the length of the series. The lower panel reports the two pooled rules in the reference case, over $\psi\ge0.6$ under dependent errors, and at $\omega=1$ under a stochastic trend.}
\label{tab:fixed}
\begin{tabular}{@{}c@{}}
\emph{(a) Values adopted, by the character of the series} \\[3pt]
\setlength{\tabcolsep}{6pt}%
\begin{tabular}{ll ccc}
\toprule
Estimator & $\lambda$ from & reference case & dependent errors & stochastic trend \\
\midrule
SSJ--KFS & blocks $K{=}5$ & $0.15$ & $0.30$ & $0.25$ \\
SSJ--KFS & blocks $K{=}10$ & $0.15$ & $0.40$ & $0.20$ \\
SSJ--KFS & blocks $K{=}20$ & $0.15$ & $0.35$ & $0.25$ \\
SSJ--KFS & random $K{=}5$ & $0.20$ & $0.25$ & $0.25$ \\
SSJ--KFS & random $K{=}10$ & $0.20$ & $0.25$ & $0.20$ \\
SSJ--KFS & random $K{=}20$ & $0.20$ & $0.20$ & $0.20$ \\
SSJ--KFS & MLE & $0.15$ & $0.20$ & $0.30$ \\
\addlinespace
SSJ--ATW & blocks $K{=}5$ & $0.25$ & $0.40$ & $0.35$ \\
SSJ--ATW & blocks $K{=}10$ & $0.25$ & $0.50$ & $0.30$ \\
SSJ--ATW & blocks $K{=}20$ & $0.30$ & $0.45$ & $0.30$ \\
SSJ--ATW & random $K{=}5$ & $0.30$ & $0.35$ & $0.30$ \\
SSJ--ATW & random $K{=}10$ & $0.30$ & $0.35$ & $0.35$ \\
SSJ--ATW & random $K{=}20$ & $0.30$ & $0.35$ & $0.35$ \\
SSJ--ATW & MLE & $0.30$ & $0.45$ & $0.35$ \\
\addlinespace
SSJ--MLE & MLE & $0.25$ & $0.25$ & $0.45$ \\
\bottomrule
\end{tabular}
\\[9pt]
\emph{(b) The two rules behind the values adopted, pooled over the levels of the setting} \\[3pt]
\setlength{\tabcolsep}{4pt}
\begin{tabular}{ll cc cc cc}
\toprule
 & & \multicolumn{2}{c}{reference case} & \multicolumn{2}{c}{dependent errors, $\psi\ge0.6$} & \multicolumn{2}{c}{stochastic trend, $\omega=1$} \\
\cmidrule(lr){3-4}\cmidrule(lr){5-6}\cmidrule(lr){7-8}
Estimator & $\lambda$ from & RMSE & count & RMSE & count & RMSE & count \\
\midrule
SSJ--KFS & blocks $K{=}5$ & $0.10$ & $0.15$ & $0.30$ & $0.30$ & $0.15$ & $0.30$ \\
SSJ--KFS & blocks $K{=}10$ & $0.15$ & $0.15$ & $0.40$ & $0.35$ & $0.15$ & $0.25$ \\
SSJ--KFS & blocks $K{=}20$ & $0.15$ & $0.15$ & $0.35$ & $0.30$ & $0.20$ & $0.25$ \\
SSJ--KFS & random $K{=}5$ & $0.20$ & $0.20$ & $0.30$ & $0.20$ & $0.20$ & $0.25$ \\
SSJ--KFS & random $K{=}10$ & $0.20$ & $0.20$ & $0.25$ & $0.20$ & $0.15$ & $0.25$ \\
SSJ--KFS & random $K{=}20$ & $0.20$ & $0.20$ & $0.25$ & $0.15$ & $0.20$ & $0.20$ \\
SSJ--KFS & MLE & $0.15$ & $0.15$ & $0.25$ & $0.15$ & $0.15$ & $0.45$ \\
\addlinespace
SSJ--ATW & blocks $K{=}5$ & $0.25$ & $0.25$ & $0.35$ & $0.45$ & $0.20$ & $0.45$ \\
SSJ--ATW & blocks $K{=}10$ & $0.25$ & $0.25$ & $0.45$ & $0.50$ & $0.20$ & $0.40$ \\
SSJ--ATW & blocks $K{=}20$ & $0.30$ & $0.25$ & $0.40$ & $0.45$ & $0.20$ & $0.35$ \\
SSJ--ATW & random $K{=}5$ & $0.30$ & $0.25$ & $0.35$ & $0.35$ & $0.25$ & $0.35$ \\
SSJ--ATW & random $K{=}10$ & $0.30$ & $0.30$ & $0.35$ & $0.35$ & $0.25$ & $0.40$ \\
SSJ--ATW & random $K{=}20$ & $0.30$ & $0.30$ & $0.35$ & $0.30$ & $0.25$ & $0.40$ \\
SSJ--ATW & MLE & $0.30$ & $0.25$ & $0.45$ & $0.40$ & $0.15$ & $0.50$ \\
\addlinespace
SSJ--MLE & MLE & $0.30$ & $0.20$ & $0.30$ & $0.15$ & $0.45$ & $0.40$ \\
\bottomrule
\end{tabular}
\end{tabular}
\end{table}

To testify the robustness of our choice, we also consider alternative synthesis criteria, such as the \emph{separate} rules, and the \emph{larger} of the two, which provides a more conservative penalty (Table~\ref{tab:fixed_rob}). The two agree to within one grid step in the reference case and under dependent errors, where the rules themselves are close; they part company under a stochastic trend for the arms with $\lambda$ inherited from SSJ--MLE, where the count rule alone asks for $0.45$ and $0.50$ and the reconstruction rule for $0.15$, so the midpoint settles at $0.30$--$0.35$ and the larger at the count value, and, less widely, for SSJ--ATW with contiguous blocks at $K=5$ and $10$, where the count rule asks for $0.45$ and $0.40$ against $0.20$ for reconstruction (midpoint $0.35$ and $0.30$, larger $0.45$ and $0.40$). The midpoint is the synthesis adopted below: the two rules protect different things, the accuracy of the curve and the number of discontinuities declared, and the midpoint is the single value that gives up the least of both, since the RMSE is nearly flat around its minimum and the count regret rises slowly on the side of the larger penalty. Table~\ref{tab:fixed_rob} reports the same values under the four syntheses (i.e., each rule alone, the midpoint and the larger of the two) as a robustness check on the decision rule: the midpoint and the larger differ by at most one grid step everywhere except under a stochastic trend for the arms with $\lambda$ from SSJ--MLE and for SSJ--ATW with contiguous blocks at $K=5$ and $10$, where the choice between them is the choice between the reconstruction of the curve and the number of discontinuities declared, at the reconstruction cost of Section~\ref{sec:cost}.

\begin{table}[htbp]
\centering\footnotesize
\caption{Robustness to the decision rule: the values of $\xi$ adopted for each scenario and arm (cross-validation at $K=5$, $10$ and $20$) under the four syntheses of the two pooled rules --- the count rule alone, the RMSE rule alone, their midpoint rounded up to the grid (the synthesis of Table~\ref{tab:fixed}) and the larger of the two. Columns: the reference case, dependent errors ($\psi\ge0.6$) and a stochastic trend ($\omega=1$).}
\label{tab:fixed_rob}
\setlength{\tabcolsep}{2pt}
\begin{tabular}{ll *{4}{@{\hspace{5pt}}ccc}}
\toprule
 & & \multicolumn{3}{c}{count rule} & \multicolumn{3}{c}{RMSE rule} & \multicolumn{3}{c}{midpoint} & \multicolumn{3}{c}{larger of the two} \\
\cmidrule(lr){3-5}\cmidrule(lr){6-8}\cmidrule(lr){9-11}\cmidrule(lr){12-14}
Estimator & $\lambda$ from & ref. & dep. & trend & ref. & dep. & trend & ref. & dep. & trend & ref. & dep. & trend \\
\midrule
SSJ--KFS & blocks $K{=}5$ & $0.15$ & $0.30$ & $0.30$ & $0.10$ & $0.30$ & $0.15$ & $0.15$ & $0.30$ & $0.25$ & $0.15$ & $0.30$ & $0.30$ \\
SSJ--KFS & blocks $K{=}10$ & $0.15$ & $0.35$ & $0.25$ & $0.15$ & $0.40$ & $0.15$ & $0.15$ & $0.40$ & $0.20$ & $0.15$ & $0.40$ & $0.25$ \\
SSJ--KFS & blocks $K{=}20$ & $0.15$ & $0.30$ & $0.25$ & $0.15$ & $0.35$ & $0.20$ & $0.15$ & $0.35$ & $0.25$ & $0.15$ & $0.35$ & $0.25$ \\
SSJ--KFS & random $K{=}5$ & $0.20$ & $0.20$ & $0.25$ & $0.20$ & $0.30$ & $0.20$ & $0.20$ & $0.25$ & $0.25$ & $0.20$ & $0.30$ & $0.25$ \\
SSJ--KFS & random $K{=}10$ & $0.20$ & $0.20$ & $0.25$ & $0.20$ & $0.25$ & $0.15$ & $0.20$ & $0.25$ & $0.20$ & $0.20$ & $0.25$ & $0.25$ \\
SSJ--KFS & random $K{=}20$ & $0.20$ & $0.15$ & $0.20$ & $0.20$ & $0.25$ & $0.20$ & $0.20$ & $0.20$ & $0.20$ & $0.20$ & $0.25$ & $0.20$ \\
SSJ--KFS & MLE & $0.15$ & $0.15$ & $0.45$ & $0.15$ & $0.25$ & $0.15$ & $0.15$ & $0.20$ & $0.30$ & $0.15$ & $0.25$ & $0.45$ \\
\addlinespace
SSJ--ATW & blocks $K{=}5$ & $0.25$ & $0.45$ & $0.45$ & $0.25$ & $0.35$ & $0.20$ & $0.25$ & $0.40$ & $0.35$ & $0.25$ & $0.45$ & $0.45$ \\
SSJ--ATW & blocks $K{=}10$ & $0.25$ & $0.50$ & $0.40$ & $0.25$ & $0.45$ & $0.20$ & $0.25$ & $0.50$ & $0.30$ & $0.25$ & $0.50$ & $0.40$ \\
SSJ--ATW & blocks $K{=}20$ & $0.25$ & $0.45$ & $0.35$ & $0.30$ & $0.40$ & $0.20$ & $0.30$ & $0.45$ & $0.30$ & $0.30$ & $0.45$ & $0.35$ \\
SSJ--ATW & random $K{=}5$ & $0.25$ & $0.35$ & $0.35$ & $0.30$ & $0.35$ & $0.25$ & $0.30$ & $0.35$ & $0.30$ & $0.30$ & $0.35$ & $0.35$ \\
SSJ--ATW & random $K{=}10$ & $0.30$ & $0.35$ & $0.40$ & $0.30$ & $0.35$ & $0.25$ & $0.30$ & $0.35$ & $0.35$ & $0.30$ & $0.35$ & $0.40$ \\
SSJ--ATW & random $K{=}20$ & $0.30$ & $0.30$ & $0.40$ & $0.30$ & $0.35$ & $0.25$ & $0.30$ & $0.35$ & $0.35$ & $0.30$ & $0.35$ & $0.40$ \\
SSJ--ATW & MLE & $0.25$ & $0.40$ & $0.50$ & $0.30$ & $0.45$ & $0.15$ & $0.30$ & $0.45$ & $0.35$ & $0.30$ & $0.45$ & $0.50$ \\
\addlinespace
SSJ--MLE & MLE & $0.20$ & $0.15$ & $0.40$ & $0.30$ & $0.30$ & $0.45$ & $0.25$ & $0.25$ & $0.45$ & $0.30$ & $0.30$ & $0.45$ \\
\bottomrule
\end{tabular}
\end{table}

The findings that carry over to practice are the following:
\begin{itemize}
\item In the reference case the two minimax regret rules largely agree (Table~\ref{tab:base_all}), and supplying $\lambda$ from a preliminary SSJ--MLE fit, which fixes one parameter before the budget is chosen, calls for a penalty no heavier than a cross-validated $\lambda$ does (Section~\ref{sec:res_base});
\item Serial dependence and stochastic trends call for a heavier penalty than the reference case through different channels: persistence acts through $\lambda$, and the penalty responds to it when $\lambda$ is cross-validated on contiguous blocks, and for SSJ--ATW also when $\lambda$ is inherited from SSJ--MLE, but not with random folds (Table~\ref{tab:xi_errors_uni}); a stochastic trend leaves reconstruction unaffected and inflates the declared count, so only the count rule moves (Table~\ref{tab:xi_signal_uni});
\item The value depends on the estimator as much as on the data: under persistence SSJ--ATW needs from a third more to twice what SSJ--KFS needs;
\item Following the count rule costs little in reconstruction except for a stochastic trend (Table~\ref{tab:cost});
\item A single constant is a compromise across sample sizes (Table~\ref{tab:byn_scen}); for SSJ--KFS with $\lambda$ from SSJ--MLE or from contiguous blocks the total charge per discontinuity is flat in $n$ and the compromise has the closed form of Equation~\eqref{eq:xirule} (see Table~\ref{tab:rule} in Section \ref{sec:charge});
\item Exact recovery of the true number rises with $q$, because the declared count is close to \emph{all-or-nothing}; persistence costs SSJ--MLE its power; a stochastic trend costs specificity, which contiguous folds buy back;
\item The property of the extended BIC for which a smoothing constant fixed before the budget is chosen calls for a milder $\xi$ (Section~\ref{sec:tuning}) is confirmed: SSJ--KFS with $\lambda$ from SSJ--MLE asks for less than SSJ--MLE itself in every scenario ($0.15$ against $0.25$ in the reference case, $0.20$ against $0.25$ under dependent errors and $0.30$ against $0.45$ under a stochastic trend, Table~\ref{tab:fixed}), and the same ordering holds at every sample size under the reconstruction rule (Table~\ref{tab:by_n}).
\end{itemize}

\bibliographystyle{agsm}
\bibliography{ssj}


\maketitle

\noindent
This document collects the material that supports the two applications of Section~\ref{P-sec:applications} of the paper.
Section~\ref{sec:supp_app1} reports the temperature--demand application over the whole grid of the extended-BIC constant $\xi$ and compares the estimated curve with the degree-day representation of the temperature response.
Section~\ref{sec:supp_app} gathers everything on the fuel-price application: the check of the calibration of $\xi$ on the policy dates, the 2026 episode, the two excise episodes and the boundary experiment at a readable size, the application re-estimated at the three values of $\xi$ that bracket the one the paper adopts, and the evidence behind the number of cross-validation folds.
The calibration of $\xi$ itself, with every table and figure behind it, is the subject of a separate document, the Calibration Supplement.

\tableofcontents

\newpage

\section{The temperature--demand application}\label{sec:supp_app1}
This section supports Section~\ref{P-sec:elettro} of the paper: Section~\ref{sec:supp_app1_xi} reports the discontinuities declared over the grid of $\xi$, and Section~\ref{sec:supp_app1_dd} compares the estimated curve with the degree-day construct.

\subsection{The declared discontinuities over the grid of $\xi$}\label{sec:supp_app1_xi}
Table~\ref{tab:supp_app1_counts} gives, for the two algorithms of Figure~\ref{P-fig:elettro} of the paper, the number of discontinuities for every value of $\xi$ from $0.10$ to $0.50$. According to the calibration, the recommended values are around $0.30-0.35$ for SSJ--MLE and around $0.25-0.30$ for SSJ--ATW. The number of detected discontinuities is rather stable in the respective ranges and above.

\begin{table}[htbp]
\centering\footnotesize
\caption{Number of temperatures declared as discontinuities by the two estimators of Figure~\ref{P-fig:elettro} of the paper, at each value of $\xi$: warps declared according to the rule described in the main paper.}
\label{tab:supp_app1_counts}
\setlength{\tabcolsep}{3pt}
\begin{tabular}{l *{9}{c}}
\toprule
 & \multicolumn{9}{c}{$\xi$} \\
\cmidrule(lr){2-10}
Estimator & 0.10 & 0.15 & 0.20 & 0.25 & 0.30 & 0.35 & 0.40 & 0.45 & 0.50 \\
\midrule
SSJ--MLE                  & 8 & 8 & 7 & 7 & 3 & 3 & 2 & 2 & 2 \\
SSJ--ATW, $\lambda$ by 5-fold CV & 5 & 5 & 3 & 2 & 2 & 2 & 2 & 2 & 2 \\
\bottomrule
\end{tabular}
\end{table}

\begin{figure}[htb]
    \centering
    \includegraphics[width=\linewidth]{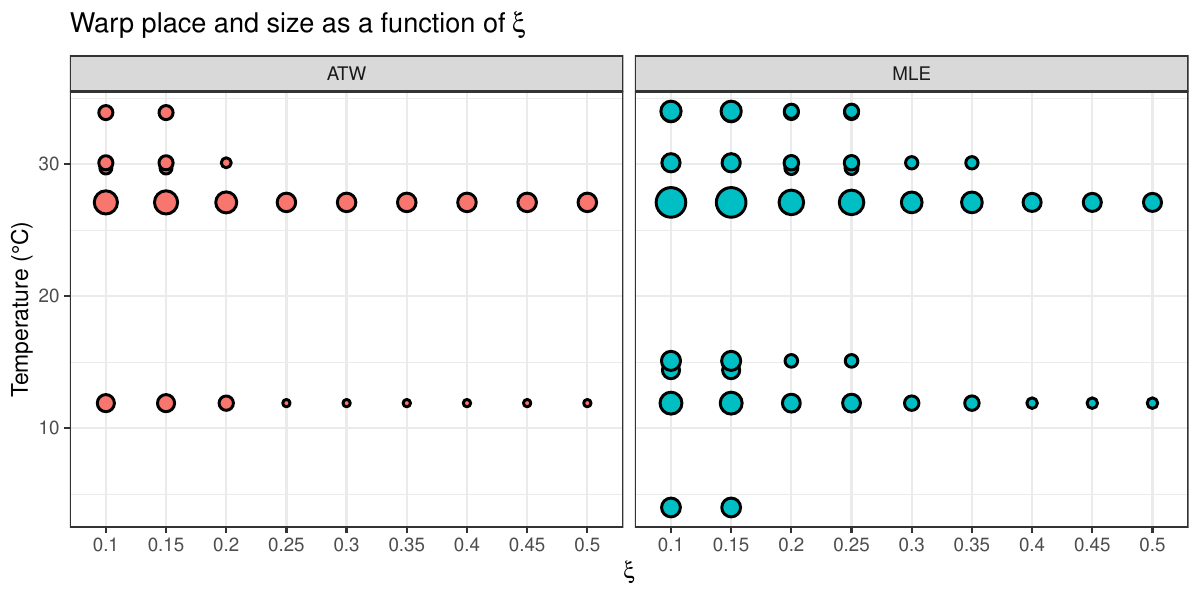}
    \caption{The value of the EBIC parameter $\xi$ is on the $x$-axis, the value of the temperature
    is on the $y$-axis, the size of the point is proportional to the warp identified by each algorithm.}
    \label{fig:warps}
\end{figure}

Figure \ref{fig:warps} shows the position with respect to temperature and the strength of the warps as a function of the EBIC constant $\xi$. As already noticed, the warps' placement stabilizes when the values of $\xi$ are equal to or above the values suggested by the calibration exercises. The discontinuities detected at any value of $\xi$ occur at $11.9$ and $27.1$ degrees Celsius, with the second being much stronger.

\subsection{Comparison with degree days}\label{sec:supp_app1_dd}
It is worth setting the estimated curve against the device that most energy systems use in its place. Heating demand is conventionally summarized by \emph{degree days}: one fixes a base temperature $T_0$ and accumulates, throughout the season, the positive part of $T_0-\bar T_d$, where $\bar T_d$ is the mean outdoor temperature on day $d$. The implied model of the temperature response is exactly two straight lines joined at $T_0$, flat above the base and linear below it, and its single free parameter is fixed by convention rather than estimated. Italy sets the base at $20^{\circ}$C, and does so for a reason that makes it higher than the bases used elsewhere: the figure in DPR 412/1993 is the conventional indoor temperature, differenced directly against the outdoor daily mean, without taking into account the internal and solar gains that the lower foreign bases implicitly subtract. The same regulation partitions the country into climatic zones A--F by the resulting annual totals, and those totals govern how long the heating season may run in each municipality.

Our estimate allows us to ask what the data would have chosen. The curve fitted by SSJ--MLE (Figure~\ref{P-fig:elettro} of the paper) attains its minimum at $18.9^{\circ}$C, and at exactly $20^{\circ}$C, the convention, demand sits within $0.1\%$ of that minimum. The reason $20^{\circ}$C does so well is not that the curve turns there: demand stays within $1\%$ of its minimum throughout the interval $[16.0, 23.4]^\circ$C, so the vertex that the degree-day construct posits is not a point the data can locate, since there is nothing sharp there to locate, and any base between roughly $16$ and $23^\circ$C would describe this sample about equally well. The two arms are also asymmetric: below $14^{\circ}$C each additional degree of cold adds about $270$ MWh, while above $24^{\circ}$C each additional degree of heat adds about $320$, a fifth more, and about $340$ per degree between $27$ and $30^{\circ}$C. A single base temperature cannot serve both sides, which is why the Italian regulation defines degree days for heating only and leaves cooling without a counterpart.

Where the two descriptions truly part company is at the discontinuities. The estimator places jumps at about $11.9^{\circ}$C and, in the cooling regime, at about $27^{\circ}$C and $29.7^{\circ}$C, that is, at temperatures at which demand steps up rather than merely bending upward, as air conditioning is switched on across a large number of premises at once. A degree-day representation cannot express this at any base temperature, because it is continuous by construction; the best it can do is absorb the steps into a steeper slope and spread them over the whole of the cooling range. The estimators of Section~\ref{P-sec:methodology} of the paper do not have to choose between a kink and a step, and here they find both.

\clearpage

\section{The fuel-price application}\label{sec:supp_app}

This section collects everything that supports Section~\ref{P-sec:carburanti} of the paper: the check of the calibration of $\xi$ on the policy dates (Section~\ref{sec:supp_app_sweep}), the 2026 episode (Section~\ref{sec:supp_app_2026}), the two excise episodes at a readable resolution and the boundary experiment (Section~\ref{sec:supp_app_figs}), the same application re-estimated at the other values of the extended-BIC constant that the paper discusses (Section~\ref{sec:supp_app_xi}), and the evidence behind the number of cross-validation folds (Section~\ref{sec:supp_app_folds}).

\subsection{Checking the calibration on the policy dates}\label{sec:supp_app_sweep}
\begin{figure}[p]
    \centering
    \includegraphics[width=\linewidth,height=0.8\textheight,keepaspectratio]{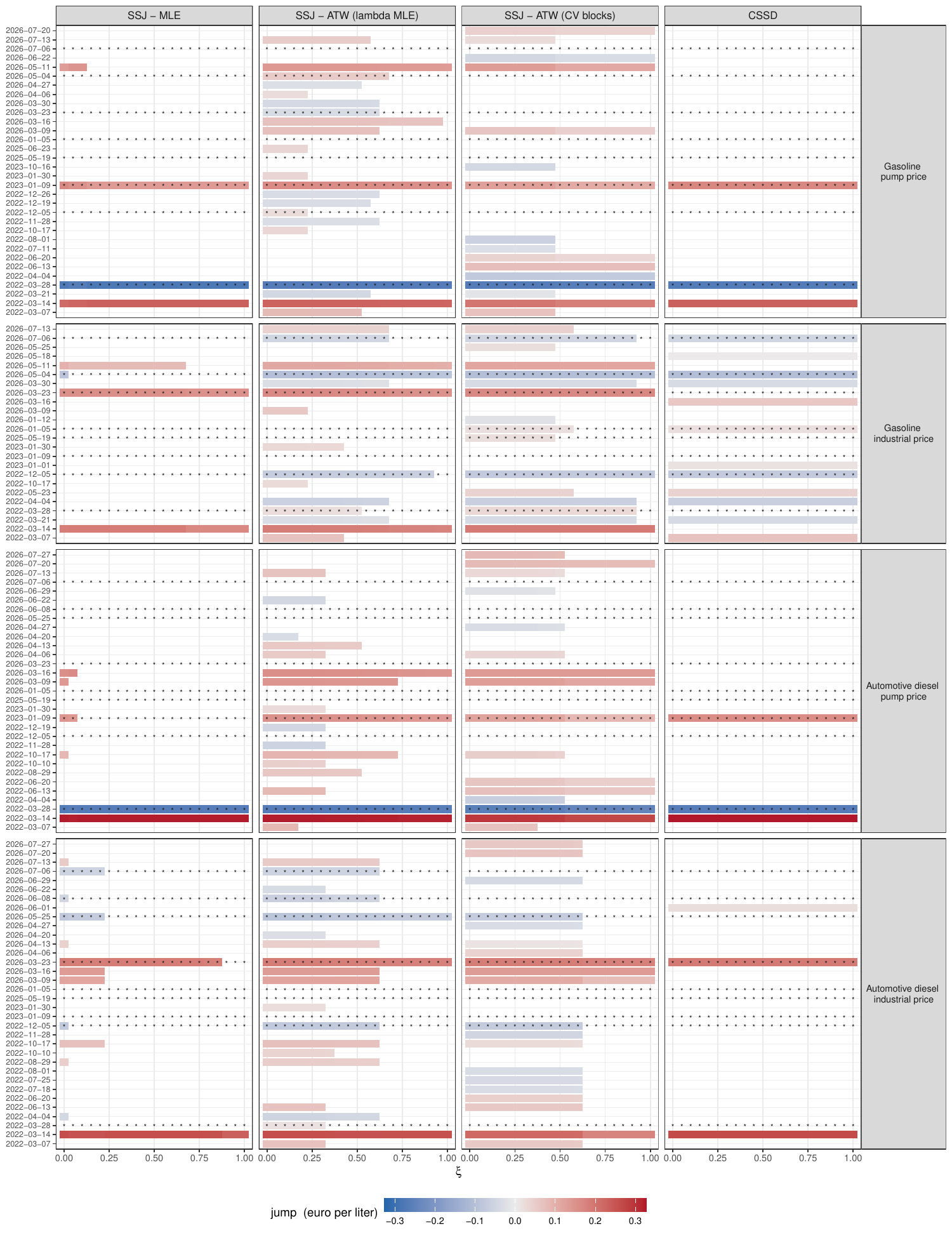}
    \caption{\small Weeks declared as discontinuities by each estimator (columns) on each series (rows), as $\xi$ runs from $0$ to $1$ in steps of $0.05$. Each band is one date and each tile a value of $\xi$ at which it is declared, colored by the size and sign of the level shift. Black asterisks mark the weeks in which the excise duty changed (Table~\ref{P-tab:excise_dates} of the paper): on a colored tile a hit, on white a miss.}
    \label{fig:carburanti_xi}
\end{figure}

Because the dates of the interventions are known, the fuel-price series allow the calibration of $\xi$ summarized in Section~\ref{P-sec:tuning} of the paper to be checked on real data. Figure~\ref{fig:carburanti_xi} sweeps $\xi$ over $[0,1]$ and records every week that each estimator declares as a break, with the enforcement weeks of Table~\ref{P-tab:excise_dates} of the paper marked by black asterisks. Because $\xi$ enters only the selection of the budget and never a fit, the smoothing constants are computed once per series and held fixed, so the figure isolates the penalty and nothing else.

The pattern is the one the simulations predict. At the reference-case value the estimators declare a long tail of dates that are not policy events: SSJ--ATW at the maximum likelihood $\lambda$ reports $16$ such weeks on the two pump-price series alongside $11$ of the $17$ true events, and with contiguous folds the tail is longer still, $20$ weeks alongside $8$ events. Raising $\xi$ to $0.45$ removes much of it while costing little of the signal: at the maximum likelihood $\lambda$ the estimator then reports $10$ spurious weeks and still recovers $9$ events, so precision rises from $0.41$ to $0.47$, and the two events given up are small ones, the partial restoration of $+0.10$ euro of December 2022 on one fuel and the final $+0.05$ step of July 2026 on the other. Beyond $0.45$ the maximum likelihood arm does not change at all and the blocked one gives up an event, so that is where the gain has been realized (Section~\ref{sec:supp_app_xi} gives the counts).

The declared sets are nested in $\xi$: every date enters at $\xi=0$ and, once it leaves, does not return, for all $107$ dates declared anywhere in the four series, so each occupies one uninterrupted band whose length measures how much penalty it can absorb. The great majority are transient and disappear within one or two grid steps of the reference-case value; the survivors at every admissible penalty are 14 March 2022, 28 March 2022 and 9 January 2023 (the market shock, the excise cut and its restoration) on either fuel. Four of the seventeen events are recovered by no estimator at any $\xi$, and they are the same four throughout, namely the two realignment steps on both fuels, which move the duty by $1.5$ and $4$ cents against a weekly standard deviation of the pump price several times larger, and which are visible in the figure as rows of asterisks on white across every column. On the industrial price, from which the duty is excluded, both SSJ--ATW variants declare a discontinuity at the 2022 policy date at the smallest penalties and both still do at $0.45$; the one with contiguous folds keeps it up to $0.90$, so a penalty above that would be needed before neither does. Here the two objectives of the calibration differ, and the gap is the price of insisting on no false alarms rather than on an accurate curve.

Two asymmetries in the detection map are also worth recording. SSJ--ATW with a block cross-validated $\lambda$ retains far more of its pump-price dates across the whole grid than the other three estimators of the map: that constant is the smallest of those reported, the budget it supports is correspondingly large, and the extended term has comparatively little purchase on it, so $\xi$ buys less there than elsewhere---which is visible in the figure as a column of long unbroken bands, and is why the block-cross-validated variant is the one whose precision barely improves when the penalty is raised. And CSSD, which is the conservative benchmark on the pump prices with six dates, declares $15$ on the industrial prices and keeps all of them, more than any SSJ estimator retains; its parsimony is a property of the series it is applied to rather than of the estimator.

\clearpage

\subsection{The 2026 episode}\label{sec:supp_app_2026}
The 2026 episode inverts the pattern of 2022 and illustrates what the decomposition into gross and net prices buys. The pump price barely reacts to the cut of 23 March 2026: it falls by $0.040$ euro per liter for gasoline and by $0.010$ for diesel against a mechanical $0.244$, and no discontinuity is declared on it, while a large one appears on the industrial price, estimated at $+0.157$ to $+0.164$ for gasoline and $+0.179$ to $+0.191$ for diesel. This reading deserves caution. The two weeks preceding the cut carry a violent market surge, with the industrial price rising by about $0.06$ euro per liter per week on gasoline and $0.13$ on diesel, that is, of the same order as the fiscal measure. What we measure at the policy date is the one-step change of the fitted signal, which superimposes the discontinuity on one week of the prevailing drift; netting out a drift of that size would raise the implied transmission to roughly one half on gasoline and two thirds on diesel. The qualitative conclusion, that the second intervention reached the pump far less completely than the first, is robust to the correction, but the magnitude is not, and we do not press it further. Panel~(b) of Figure~\ref{fig:carburanti_zoom} shows the episode week by week.

\subsection{The two excise episodes and the boundary experiment}\label{sec:supp_app_figs}
\begin{figure}[p]
    \centering
    \includegraphics[width=\linewidth]{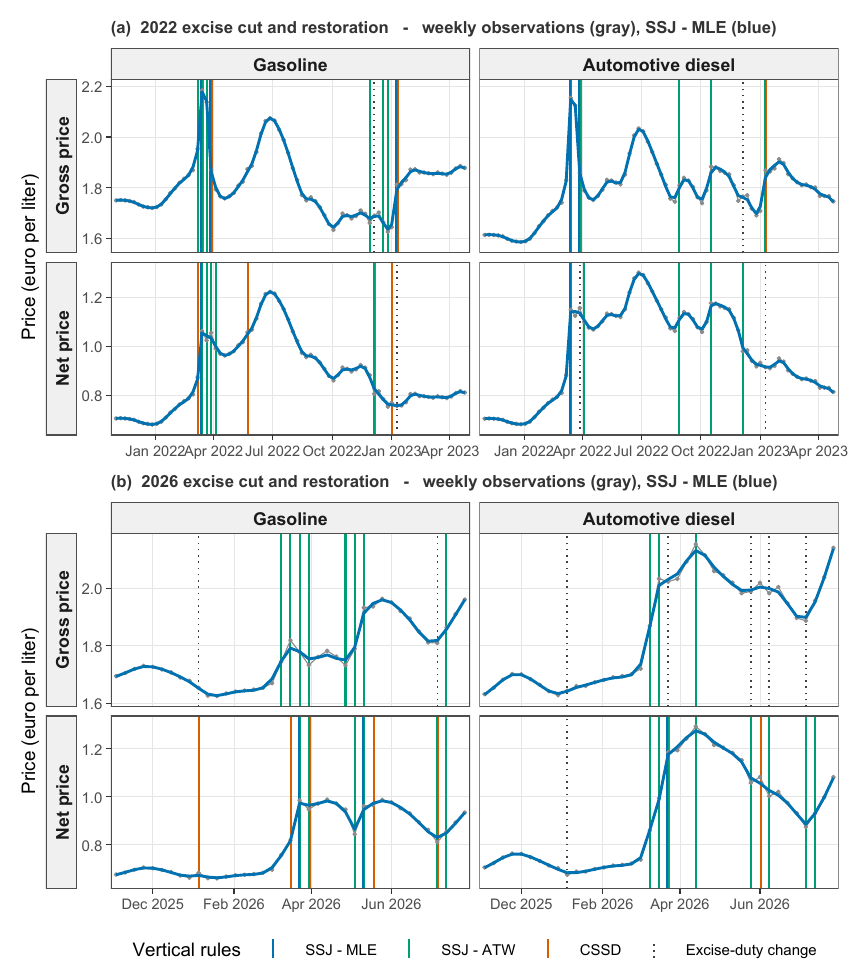}
    \caption{\small Zoom-in of the two excise-duty episodes. Rows: price paid at the pump and industrial price; columns: the two fuels. Points are the weekly observations and the blue line the signal extracted by SSJ--MLE; vertical rules as in Figure~\ref{P-fig:carburanti} of the paper, with the sideways nudge reduced in proportion to the horizontal scale. Panel (a): the cut of 28 March 2022, its partial withdrawal in December 2022, and its full restoration in January 2023. Panel (b): the cut of 23 March 2026 and its staggered restoration.}
    \label{fig:carburanti_zoom}
\end{figure}

Figure~\ref{fig:carburanti_zoom} expands the two excise episodes, at a resolution at which the individual weeks can be separated and the agreement between the three estimators can be checked one date at a time.

\begin{figure}[p]
    \centering
    \includegraphics[width=\linewidth]{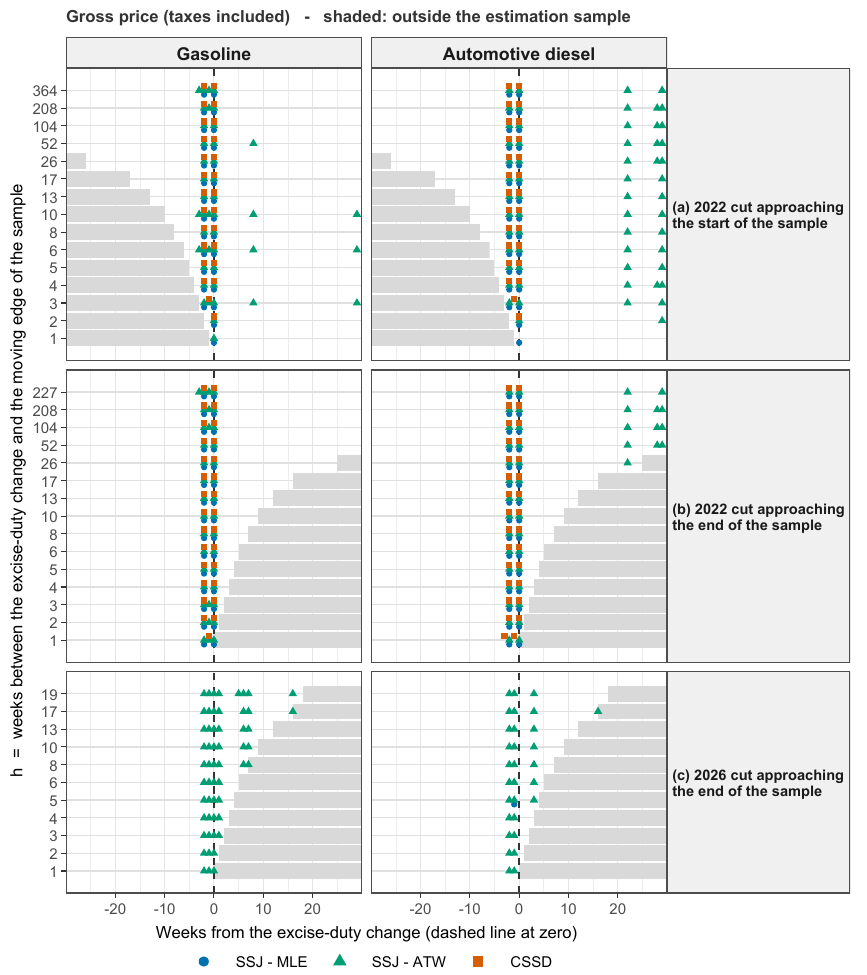}
    \caption{\small Boundary robustness on the pump (gross) price. Panels (a) and (b) shrink the sample toward the cut of 28 March 2022 from the left and from the right respectively; panel (c) repeats the exercise on the cut of 23 March 2026.}
    \label{fig:carburanti_det}
\end{figure}

\begin{figure}[p]
    \centering
    \includegraphics[width=\linewidth]{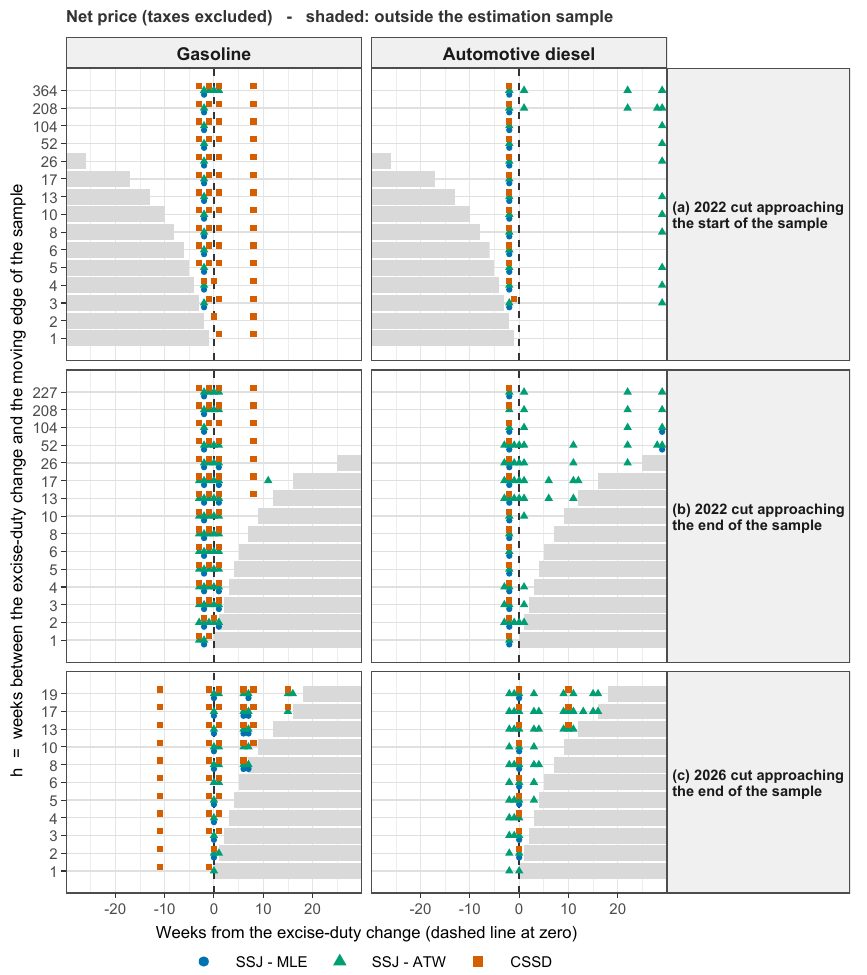}
    \caption{\small Boundary robustness on the industrial (net) price. Panels (a) and (b) shrink the sample toward the cut of 28 March 2022 from the left and from the right respectively; panel (c) repeats the exercise on the cut of 23 March 2026.}
    \label{fig:carburanti_det_net}
\end{figure}

Figures~\ref{fig:carburanti_det} and \ref{fig:carburanti_det_net} report the boundary experiment of Section~\ref{P-sec:carburanti} of the paper, on the pump price and on the industrial price respectively. A change point that falls in the middle of a long sample is an easy target. The formulation of Section~\ref{P-sec:methodology} of the paper places no restriction on where a jump may occur, since $\gamma_i$ is defined on every one of the $n-1$ intervals; near the boundary, however, the spline has almost no data on one side, and the roughness penalty alone must arbitrate between a discontinuity and a steep continuous transition. Because the date of the March 2022 cut is known, this can be put to the test directly. We build three families of nested subsamples: one in which the first observation is moved forward until the cut becomes the second observation of the sample, one in which the last observation is moved backward until the cut becomes the last, and one repeating the second exercise on the March 2026 cut; all three procedures are then re-estimated on each of the $41$ windows. Every horizontal line is one estimation, run on a different subsample; the shaded blocks are the weeks that estimation could not see, so that moving down a panel the sample is progressively cut back toward the excise-duty change, which sits at zero on the horizontal axis. The vertical axis reports $h$, the number of weekly observations still separating the change from the moving edge of the sample, from the full sample at the top to $h=1$ at the bottom. The symbols are the discontinuities found in that estimation, positioned in weeks from the excise-duty change and split into one sub-row per estimator, so that an estimator insensitive to the position of the break in the sample produces a vertical stack of symbols at zero all the way down.

Four patterns recur. A column at zero is the excise-duty change itself. The column two weeks earlier, in the 2022 panels, is the market shock of 14 March 2022: on the pump price it is a second and smaller discontinuity, on the industrial price it is the only one, since the 2022 cut was passed on almost in full and left the industrial price alone, which is why Figure~\ref{fig:carburanti_det_net} has its main column at $-2$ and nothing at $0$. Isolated symbols away from zero are further genuine discontinuities that a particular window happens to contain, such as the movements in the industrial price of diesel in late August and mid-October 2022; they disappear as soon as the window no longer reaches them, which is what makes them look isolated. Finally, the short runs of adjacent symbols visible in the shortest windows are the warping spread over a few consecutive weeks rather than concentrated on one: SSJ--ATW does this near the boundary, and it is worth stressing that it still places the jump in the right neighborhood rather than displacing it.

\begin{figure}[p]
    \centering
    \includegraphics[width=\linewidth]{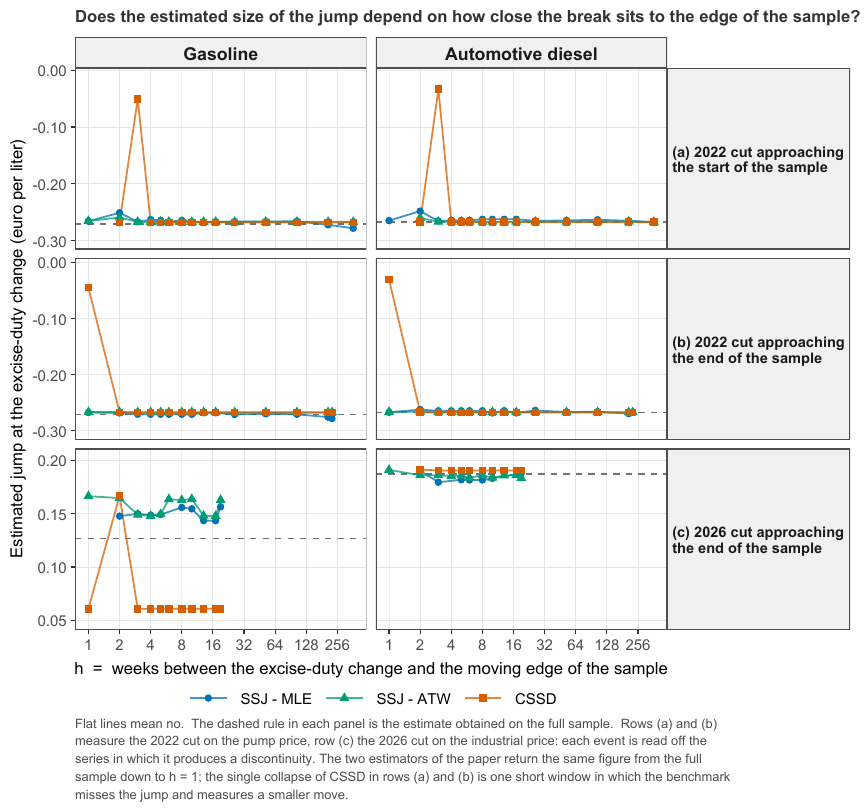}
    \caption{\small Estimated size of the discontinuity at the excise-duty change, in each window of the boundary experiment. Axes as in Figures~\ref{fig:carburanti_det} and \ref{fig:carburanti_det_net}: the vertical axis is the number of weekly observations still separating the change from the moving edge of the sample. A horizontal series of symbols is an estimator whose measurement of the jump does not move as the sample is cut back toward it.}
    \label{fig:carburanti_size}
\end{figure}

Figure~\ref{fig:carburanti_size} reports the estimated size of the recovered discontinuity in each of those windows, which is the quantity the paper's claim of stability rests on: on the pump price it stays at $-0.267$ euro per liter from the full sample down to the shortest window, so the boundary neither displaces the jump nor attenuates it.

\clearpage

\subsection{The application at other values of $\xi$}\label{sec:supp_app_xi}
Section~\ref{P-sec:carburanti} of the paper sets $\xi = 0.45$, for reasons given there: the two simulation experiments bracket $0.35$ to $0.50$ for a series that is both serially dependent and closer to a stochastic trend than to a fixed smooth curve, and $0.45$ lies inside that band without being the value that maximizes detection accuracy on these data. This section re-runs the whole application at the other three values, so that the reader can see what each of them would have produced. Everything else is held fixed: the same data, the same budget grid, the same $\lambda$ for each variant, the same seed.

Table~\ref{tab:supp_xi_app} scores the outcome against the seventeen excise events on the two pump-price series, counting a run of adjacent declarations around one event as one recovered event and allowing a tolerance of one week between the legal date and the price response.

\begin{table}[htbp]
\centering
\small
\caption{The fuel-price application at four values of $\xi$. For each estimator and each $\xi$: the total number of discontinuities declared on the two pump-price series (\emph{all}), the number of the seventeen excise events recovered (\emph{hit}), and the number of declared weeks not within one week of any excise event (\emph{false}).}
\label{tab:supp_xi_app}
\setlength{\tabcolsep}{4pt}
\begin{tabular}{l*{4}{ccc}}
\toprule
 & \multicolumn{3}{c}{$\xi=0.25$} & \multicolumn{3}{c}{$\xi=0.35$} & \multicolumn{3}{c}{$\xi=0.45$} & \multicolumn{3}{c}{$\xi=0.50$} \\
\cmidrule(lr){2-4}\cmidrule(lr){5-7}\cmidrule(lr){8-10}\cmidrule(lr){11-13}
Estimator & all & hit & false & all & hit & false & all & hit & false & all & hit & false \\
\midrule
SSJ--MLE & 5 & 3 & 2 & 5 & 3 & 2 & 5 & 3 & 2 & 5 & 3 & 2 \\
SSJ--ATW (lambda MLE) & 32 & 11 & 16 & 24 & 9 & 10 & 24 & 9 & 10 & 24 & 9 & 10 \\
SSJ--ATW (CV blocks) & 32 & 8 & 20 & 32 & 8 & 20 & 31 & 8 & 19 & 24 & 7 & 15 \\
CSSD & 6 & 4 & 2 & 6 & 4 & 2 & 6 & 4 & 2 & 6 & 4 & 2 \\
\bottomrule
\end{tabular}
\end{table}

Three things are visible.

\begin{itemize}

\item \textbf{The gain is concentrated between $0.25$ and $0.35$, and it is large.} For SSJ--ATW at the maximum likelihood $\lambda$ the declared set falls from $32$ weeks to $24$ and the false detections from $16$ to $10$, while the events recovered fall only from $11$ to $9$. Precision rises from $0.41$ to $0.47$. The two events given up are small ones, the partial restoration of $+0.10$ euro of December 2022 on one fuel and the final $+0.05$ step of July 2026 on the other; the realignment steps of $1.5$ and $4$ cents, which move the pump price by less than its weekly standard deviation, are recovered at no value of $\xi$ (Section~\ref{sec:supp_app_sweep}).

\item \textbf{Beyond $0.35$ the penalty has nothing left to remove.} That arm declares exactly the same $24$ weeks at $0.35$, at the $0.45$ the paper adopts and at $0.50$, so the reading is insensitive to the choice within the band the calibration brackets; the blocked arm, which is still declaring $31$ weeks at $0.45$, is the only one that moves at $0.50$, and it moves by giving up an event as well as five false detections.

\item \textbf{Two estimators do not respond at all, for opposite reasons.}
SSJ--MLE and CSSD declare five and six dates respectively and keep them at every $\xi$: they are already parsimonious and the penalty has nothing to remove. SSJ--ATW with a block cross-validated $\lambda$ also barely moves, from $32$ weeks to $31$ at $\xi = 0.45$, but here the cause is the opposite --- the blocked criterion selects the smallest smoothing constant of the variants reported and the warping budget it supports is correspondingly large, so the extended term has comparatively little purchase and $\xi$ buys less there than elsewhere. That variant is the one whose precision stays near $0.30$ throughout, and it is the reason the paper leads with SSJ--ATW at the maximum likelihood $\lambda$.
\end{itemize}

\clearpage

\subsection{Sensitivity of the cross-validated $\lambda$ to the number of
folds}\label{sec:supp_app_folds}

Section~\ref{P-sec:carburanti} of the paper uses $K = 20$ contiguous blocks for the cross-validated variant of SSJ--ATW. On simulated data with a known amount of dependence (Scenario~2 of the Calibration Supplement) the number of folds barely matters and the layout matters a great deal: at $\psi = 0.5$ the RMSE against the true signal is $0.086$ with blocks at either $K = 10$ or $K = 20$, against $0.098$ with random folds. On the fuel prices that conclusion does not transfer. The simulated design has $n$ between $50$ and $400$ and stationary errors; these are $591$ weekly prices whose level is close to a random walk, and the length of a block interacts with that. Selecting $\lambda$ on all four series at each $K$ gives, in effective degrees of freedom out of $591$:

\begin{center}
\small
\begin{tabular}{lccc@{\qquad}ccc}
\toprule
& \multicolumn{3}{c}{contiguous blocks} & \multicolumn{3}{c}{random folds} \\
\cmidrule(lr){2-4}\cmidrule(lr){5-7}
Series & $K=5$ & $K=10$ & $K=20$ & $K=5$ & $K=10$ & $K=20$ \\
\midrule
Gasoline, pump        &  47.8 &  12.9 & 105.0 & 576 & 576 & 576 \\
Gasoline, industrial  &  33.5 &  84.8 & 577.0 & 316 & 316 & 316 \\
Diesel, pump          &  39.5 &  59.2 & 104.0 & 576 & 576 & 576 \\
Diesel, industrial    & 124.7 &  87.0 &  87.0 & 493 & 379 & 379 \\
\bottomrule
\end{tabular}
\end{center}

\noindent
For the random scheme the choice is immaterial (the same $\lambda$ is selected at $K = 5$, $10$ and $20$ on every series) and the criterion runs to the floor of the search grid on both pump-price series whatever $K$ is, which is the failure that Scenario~2 of the Calibration Supplement predicts and which is why the blocked variant was introduced. For the blocked scheme the choice is not immaterial at all, and each value of $K$ fails somewhere: at $K = 10$ the criterion over-smooths the gasoline pump price so severely that the variant declares no discontinuities at all on the series carrying the largest fiscal shock in the sample; at $K = 5$ it declares nine on the gasoline industrial price, which is the placebo series; at $K = 20$ the criterion runs to the floor of the grid on the gasoline industrial price, although the fit that results still declares only three. $K = 20$ is adopted as the best behaved by outcome, at $4$, $3$, $5$ and $5$ declared discontinuities across the four series.

That a criterion should move by three orders of magnitude between $K = 10$ and $K = 20$ on one series is a real weakness of cross-validating the smoothing constant on data of this kind. It is also the reason the paper's headline estimator is SSJ--ATW at the maximum likelihood $\lambda$, which involves no folds at all, with the cross-validated variants reported beside it rather than in place of it. The value of $\xi$ plays no part in any of this: the selected $\lambda$ is identical at $\xi = 0.25$ and $\xi = 0.45$ in all twenty-four combinations above.